\documentclass[aps,rmp,twocolumn,10pt,longbibliography,unsortedaddress,giveinits,floatfix]{revtex4-2}

\usepackage{graphicx}
\usepackage{amsmath}
\usepackage{amssymb}
\usepackage{dsfont}
\usepackage{xspace}
\usepackage{color}
\usepackage[usenames,dvipsnames]{xcolor}
\usepackage[colorlinks=true,linkcolor=blue,urlcolor=blue,citecolor=blue]{hyperref}
\usepackage[normalem]{ulem}
\usepackage{xfrac}
\usepackage{simplewick}
\usepackage{mathtools}
\usepackage{makecell} 
\usepackage{gensymb}
\usepackage{verbatim}
\usepackage{bm}

\newcommand{\B}[1]{{\bf #1}}

\newcommand{\be}{\begin{eqnarray}}
\newcommand{\ee}{\end{eqnarray}}

\def\ket#1{\mathinner{|{#1}\rangle}}

\newcommand{\veck}{\mathbf k}

\newcommand{\bk}{\mathbf k}

\newcommand{\bp}{\mathbf p}

\newcommand{\vecr}{\mathbf r}
\newcommand{\vecR}{\mathbf R}
\newcommand{\br}{\mathbf r}
\newcommand{\bs}{\mathbf s}

\newcommand{\vecq}{\mathbf q}
\newcommand{\bq}{\mathbf q}

\renewcommand{\be}{\hat b}
\newcommand{\bed}{\hat b^\dagger}

\newcommand{\beq}{\begin{equation}}
\newcommand{\eeq}{\end{equation}}

\definecolor{dgreen}{rgb}{0.0, 0.5, 0.0}
\definecolor{darkblue}{rgb}{0.,0.24,0.51}
\definecolor{britishracinggreen}{rgb}{0.0, 0.26, 0.15}
\definecolor{darkred}{rgb}{0.6,0.0,.}

\begin{document}

\title{Polarons in atomic gases and two-dimensional semiconductors}
\author{Pietro Massignan}
\affiliation{Departament de F\'isica, Universitat Polit\`ecnica de Catalunya, Campus Nord B4-B5, E-08034 Barcelona, Spain}
\author{Richard Schmidt}
\affiliation{Institute for Theoretical Physics, Heidelberg University, Philosophenweg 16, 69120 Heidelberg, Germany}
\author{Grigori E.\ Astrakharchik}
\affiliation{Departament de F\'isica, Universitat Polit\`ecnica de Catalunya, Campus Nord B4-B5, E-08034 Barcelona, Spain}
\author{Ata\c c \.Imamoglu}
\affiliation{Institute for Quantum Electronics, ETH Z\"urich, Z\"urich, Switzerland}
\author{Martin Zwierlein}
\affiliation{MIT-Harvard Center for Ultracold Atoms, Research Laboratory of Electronics,
and Department of Physics, Massachusetts Institute of Technology, Cambridge, Massachusetts 02139, USA}
\author{Jan J. Arlt}
\affiliation{Center for Complex Quantum Systems, Department of Physics and Astronomy, Aarhus University, Ny Munkegade, DK-8000 Aarhus C, Denmark}
\author{Georg M.\ Bruun}
\affiliation{Center for Complex Quantum Systems, Department of Physics and Astronomy, Aarhus University, Ny Munkegade, DK-8000 Aarhus C, Denmark}

\date{\today}

\begin{abstract}
In this work we provide a comprehensive review of theoretical and experimental studies of the properties of polarons formed by mobile impurities strongly interacting with quantum many-body systems. We present a unified perspective on the universal concepts and theoretical techniques used to characterize polarons in two distinct platforms, ultracold atomic gases and atomically-thin transition metal dichalcogenides, which are linked by many deep parallels. We review polarons in both fermionic and bosonic environments, highlighting their similarities and differences including the intricate interplay between few- and many-body physics. Various kinds of polarons with long-range interactions or in magnetic backgrounds are discussed, and the theoretical and experimental progress towards understanding interactions between polarons is described. We outline how polaron physics, regarded as the low density limit of quantum mixtures, provides fundamental insights regarding the phase diagram of complex condensed matter systems. Furthermore, we describe how polarons may serve as  sensors of complex  many-body physics. Our work highlights the open problems, identifies new research directions and provides a comprehensive framework for this rapidly evolving research field.
\end{abstract}

\maketitle

\tableofcontents

\section{Introduction}   \label{sec:Introduction}

Polarons, quasiparticles formed by  mobile impurities interacting with their environment, represent one of the most fascinating, powerful, and versatile concepts in quantum many-body physics. The concept of polarons originated 
in solid-state physics, 
in the seminal works of Landau and Pekar~\cite{Landau1957b,Landau1957,Pekar}, and it was first applied to describe electrons interacting with lattice vibrations in crystals. Since then, polarons became a key ingredient for analyzing quantum systems consisting of many interacting particles. A main reason is that polarons are canonical realizations  of quasiparticles, which (barring any phase transitions) smoothly emerge from bare impurities when their interaction with the environment is adiabatically switched on. Put forward by Landau, this argument was considered a highlight in {\em gedanken} experiments for more than half a century and leads to the powerful theory of quasiparticles, which dramatically simplifies the description of quantum many-body systems, as strong interaction effects can be included via the ``dressing'' of the bare impurities by excitations of the environment~\cite{baym2008landau}. The quasiparticle framework is  used across a vast range of energy scales in physics from ultracold atomic gases, liquid Helium, over condensed matter systems, to atomic nuclei and high energy quark gluon plasmas.

\begin{table*}[bht!]
\centering
\begin{tabular}{|c|c|c|}
\hline
 & {\bf Atomic gases} & {\bf TMD monolayer} \\ \hline
 Dimension & Quasi 1D, quasi 2D, or 3D & 2D \\ \hline
 Impurity & Atom (boson/fermion) & Exciton or polariton ($\sim$boson) \\ \hline
Fermionic environment & Atoms (neutral) & Electrons or holes (charged) \\ \hline
Bosonic environment & Atoms (neutral) & Excitons or polaritons (neutral)\\ \hline
Bound states & Dimer, trimer, etc. & Trion, bi-exciton,  etc.\\ \hline
Density & $10^{12}\sim 10^{14}$cm$^{-3}$ (3D), tunable & $\sim 10^{12}$cm$^{-2}$, tunable  \\ \hline
Bose-Bose interaction & Short range $1/r^6$, tunable & Short range $1/r^6$, tunable \\ \hline
Bose-Fermi interaction & Short range $1/r^6$, tunable & $1/r^4$, tunable \\ \hline
Fermi-Fermi interaction & Short range $1/r^6$, tunable & Screened Coulomb \\ \hline
Temperature range & Quantum degenerate to classical & Quantum degenerate to classical \\ \hline
\end{tabular}
\caption{\label{tab:AtomVsTMDC} Properties of polarons in atomic gases and atomically thin transition metal dichalcogenides (TMDs).}
\end{table*}

In recent years we have witnessed a surge of interest in polarons, driven largely by experimental breakthroughs in ultracold atomic gases and two-dimensional (2D) semiconductors. While polarons in these experimental platforms at first glance seem quite unrelated, with e.g.\ densities and masses differing by many orders of magnitude, they in fact share many properties and can be described using  similar theoretical tools as will become apparent in this review. These striking similarities speak to the universality and general importance of polarons, and while we now have a fairly good understanding of some of their universal properties in these two systems, this convergence of fields opens up new research directions with far-reaching applications in quantum many-body physics, quantum simulation, and quantum sensing.

Given these  exciting developments, it is timely to provide a comprehensive review that brings together the diverse threads of polaron physics. This review  provides a broad perspective on the common concepts used to understand polarons, bridging the gap between different subfields, and presents a unified treatment applicable to ultracold atomic gases, semiconductors, and beyond. It furthermore discusses impurities in both fermionic and bosonic environments, highlighting their similarities and differences. The review is aimed at both the expert researchers as well as PhD students entering this vast and rapidly evolving topic, and it aims to foster new collaborations and research directions.

\subsection{Experimental platforms}   \label{sec:Exp_platforms}
We focus in this review on the realization of polarons in two experimental platforms, which in recent years have experienced substantial breakthroughs: ultracold atomic gases and atomically thin transition metal dichalcogenides. As we shall see, there are many deep parallels between polarons in these two experimental systems, which are compared in Table~\ref{tab:AtomVsTMDC}. Note that here we will not discuss polarons in one spatial dimension, as this problem has quite unique aspects (for example, in 1D quasiparticles have a vanishing residue) and has been covered in a recent  review \cite{Mistakidis2023}.

\subsubsection{Atomic gases}   \label{sec:AtomicGases}
Atomic gases are pristine quantum systems offering precise control over parameters such as the interaction strength, particle statistics, and system geometry~\cite{Dalfovo1999,bloch_many-body_2008,Giorgini2008,Baroni2024review}. They are one of the most powerful quantum simulators available today, solving problems of practical importance 
beyond the reach of classical computers~\cite{Daley2022}. Relevant to this review, they provide an ideal platform for exploring polarons since one can create mixtures of a few impurity atoms immersed in a gas of either fermionic or bosonic majority atoms. By smoothly increasing the interaction between impurities and majority atoms with a Feshbach resonance~\cite{chin_feshbach_2010}, one can then experimentally realize Landau's gedanken experiment and test the emergence of quasiparticles. This flexibility, combined with high precision measurements opened up opportunities to study polarons systematically and in new regimes. Recent experimental progress furthermore enabled the immersion of  Rydberg and ionic impurities in a bath of ultracold atoms. This introduces yet another paradigm in polaron physics, due to the long-range nature of the bath-impurity interactions.

\subsubsection{Two-dimensional semiconductors}   \label{sec:2Dsemiconductors}
In parallel, atomically thin transition metal dichalcogenides (TMDs) such as MoSe$_2$, MoS$_2$, WSe$_2$ and WS$_2$ emerged as a powerful new platform for exploring truly two-dimensional (2D) physics~\cite{Schaibley:2016aa, TMD-exciton-review}. TMDs are direct band-gap semiconductors with a rich set of degrees of freedom, experimental tuning knobs, and measurement techniques, opening up a vast playground for designing novel materials with exciting perspectives both for fundamental science and technology, and with capabilities complementing and often rivalling those of atomic gases. This includes a striking realization of polarons resulting from interactions between excitons playing the role of the impurity particles and itinerant electrons. Polarons have also been created using excitons in different spin states, and pioneering work with TMDs in optical microcavities pushed the boundaries of polaron research by creating novel hybrid light-matter quasiparticles.

\begin{table*}[tbh!]
\centering
\begin{tabular}{|l|l|l|}
\hline
 & {\bf Fermi polarons} & {\bf Bose polarons} \\ \hline
 Compressibility of bath & Small & Large \\ \hline
Number of particles in dressing cloud $\Delta N$& $\mathcal{O}(1)$ & Can be macroscopic \\ \hline
Residue $Z$ & $\mathcal{O}(1)$ (unless large mass)& Can be very small \\ \hline
Minimal interaction parameters & Scattering length $a$ & Scatt.~lengths $a$, $a_b$ + short-range params. \\ \hline 
($n>2$)-body correlations important& Only for light impurities & Always for strong interactions \\ \hline
Orthogonality catastrophe & Only for static impurities & Possible for static and mobile impurities \\ \hline
Temperature effects & Limited & Strong due to BEC transition in the bath\\ \hline
Transition between polaron and dressed dimer & Present & Absent (smooth cross-over) \\ \hline
\end{tabular}
\caption{\label{BoseFermipolaronsTable} General characteristics of Fermi and Bose polarons, which will be discussed in detail throughout this review.}
\end{table*}

\subsection{Bose and Fermi polarons}
The polarons explored in this review generally fall into two broad classes: Bose polarons formed when the majority particles are bosons, and Fermi polarons formed when the majority particles are fermions. As we shall see, whereas many properties of Fermi polarons are by now quite well understood even for strong interactions, many fundamental questions remain open for Bose polarons. The basic reason for this  is that a Bose gas is much more compressible than a Fermi gas so that its density can be strongly modified in the vicinity of an impurity, leading to a dressing cloud that involves many particles. Microscopically, this means that correlations between the impurity and an arbitrary number of bosons may be important, complicating the description significantly, whereas correlations with many fermions are suppressed by the Pauli exclusion principle for short range interactions. Despite these challenges, recent experiments made significant progress in probing Bose polarons, and theoretical approaches provided new insights into their behavior. Table \ref{BoseFermipolaronsTable} compares the most important properties of Bose and Fermi polarons. For a detailed recent review of the Bose polaron, see Ref.~\cite{grusdt2024impuritiespolaronsbosonicquantum}.

\subsection{General properties of polarons}   \label{GeneralProb}

Before going into details, this section describes the generic properties of polarons that are robust and independent of the details of the specific system at hand. The concept of quasiparticles is based on expanding the energy $E$ of a given system in increasing powers of their populations as~\cite{baym2008landau} 
\begin{equation}
E=E_0+\sum_{\mathbf p}\varepsilon_{\mathbf p}n_{\mathbf p}+\frac12\sum_{\mathbf p,\mathbf p'}f_{\mathbf p,\mathbf p'}n_{\mathbf p}n_{\mathbf p'}.
\label{LandauEnergy}
\end{equation}
Here, $E_0$ is the energy of the system when no quasiparticles are present, $\varepsilon_{\mathbf p}$ is the energy of a single quasiparticle with momentum $\mathbf p$, $n_{\mathbf p}$ its occupation number, and $f_{\mathbf p,\mathbf p'}$ is the interaction between quasiparticles. The system volume is taken to be unity throughout this review. Equation~\eqref{LandauEnergy} can be regarded as a Taylor expansion in the number of quasiparticles, and higher order terms neglected here correspond to three- and more-body interactions. One can straightforwardly extend Eq.~\eqref{LandauEnergy} to the case when several kinds of quasiparticle are present by introducing a spin index. 

For most systems, the zero momentum polaron has the smallest energy, and assuming rotational symmetry a Taylor expansion in momentum gives 
\begin{equation}
\varepsilon_{\bf p}=\varepsilon+\frac{p^2}{2m^*},
\end{equation}
where $\varepsilon$ is the polaron energy for zero momentum and $m^*$ defines its effective mass, which generally differs from the bare impurity mass $m$ due to the interactions with the environment. Another important property is the residue, which gives the overlap between the polaron wave function $|\Psi_{\bf p}\rangle$ and the eigenstate $c_{\bf p}^\dagger|\Psi_0\rangle$ for no interactions between the impurity and the majority particles, i.e.\ 
\begin{equation}
Z_{\bf p}=|\langle\Psi_{\bf p}|\hat c_{\bf p}^\dagger|\Psi_0\rangle|^2.
\label{Residue}
\end{equation}
Here, $|\Psi_0\rangle$ is the many-body ground-state of the majority particles, and $\hat c_{\bf p}^\dagger$ creates an impurity particle with momentum $\bf p$. Physically, the residue measures how much the polaron wave function resembles that of a non-interacting (bare) impurity particle. The polaron is a well-defined quasi-particle when $Z_{\bf p}>0$, and the residue moreover affects many observables. 

A key quantity is the impurity spectral function 
\begin{equation}
A({\bf p},\omega)=\int_{-\infty}^\infty\! dt\,e^{i\omega t}\langle \Psi_0|\hat c_{\bf p}(t)\hat c_{\bf p}^\dagger(0)|\Psi_0\rangle,
\label{SpectralFn}
\end{equation} 
where $\hat O(t)=\exp(i \hat Ht)\hat O\exp(-i\hat Ht)$ is the operator $\hat O$ in the Heisenberg picture, evolving under the action of the system Hamiltonian $\hat H$. Throughout this review, we use units in which the (reduced) Planck and Boltzmann constants $\hbar$ and $k_B$ are unity. In Eq.~\eqref{SpectralFn}, we have assumed that there is only a single impurity particle present to simplify the usual expression for the spectral function of a particle in a many-body system~\cite{mahan_many_2000}. The spectral function $A({\bf p},\omega)$ gives the overlap between the eigenstate $\hat c_{\bf p}^\dagger|\Psi_0\rangle$ of a non-interacting impurity and the eigenstates of the interacting system with energy $\omega$ relative to $E_0$. It follows that an undamped polaron with energy $\varepsilon_{\bf p}$ and residue $Z_{\bp}$ yields a contribution $2\pi Z_{\bf p}\delta(\omega-\varepsilon_{\bf p})$ to the spectral function. If the polaron is damped, the corresponding peak in the spectral function has a non-zero width proportional to its decay rate. In addition, the spectral function in general exhibits a continuum corresponding to excited many-body states. 

The spectral function can be measured by radio-frequency (RF) spectroscopy in cold atomic gases and by optical spectroscopy in TMDs and therefore serves as a workhorse providing a wealth of information regarding polarons. One uses RF pulses that are spatially homogeneous over the sample size, to either inject impurity particles from a state that essentially does not interact with the majority particles to one that does, or vice versa to eject the impurities from the interacting state to a non-interacting one. Ejection RF spectroscopy~\cite{schirotzek_observation_2009} probes the interacting ground-state, while injection spectroscopy also gives direct information regarding excited states. The transfer rates $I(\omega)$ obtained by ejection and injection are not independent, but actually linked via the relation $I_{\rm ej} (\omega) = \exp[\beta (\Delta F + \omega)]I_{\rm inj} (-\omega)$, where $\omega$ is the RF frequency measured with respect to the transition frequency of an isolated impurity, $\beta = 1/k_B T$ (with $k_B$ the Boltzmann constant and $T$ the temperature) and $\Delta F = F-F_0$ is the difference in free energy between the states with an interacting and a non-interacting impurity~\cite{Liu2020}.

The real-time dynamics and polaron formation can be probed using Ramsey interferometry, which measures the Fourier transform of the spectral function 
\begin{equation}
S({\bf p},t)=\langle \Psi_0|c_{\bf p}(t)c_{\bf p}^\dagger(0)|\Psi_0\rangle=iG^{>}({\bf p},t),
\label{GreaterGreens}
\end{equation}
where $G^{>}$ is the so-called greater Green's function. Experimentally, $S(t)$ which is sometimes called the Loschmidt amplitude, is measured for $t\ge 0$ and one can then use $S(-t)=S(t)^*$ to obtain this function for negative times as well. For a well-defined polaron with energy $\varepsilon_{\bp}$ and residue $Z_{\bp}$, one has $S({\bf p},t)\rightarrow Z_{\bp}\exp{(-i\varepsilon_{\bp}t)}$ for $t\rightarrow \infty$ when the many-body continuum decoheres. This provides a useful interferometric way to measure the polaron energy complementing RF spectroscopy. Spin-echo interferometry is a more complex and  powerful technique to explore many-body dynamics with strongly reduced noise. It measures $\langle \Psi_0|\hat c_{\bf p}e^{i\hat H_0t}e^{i\hat Ht}e^{-i\hat H_0t}e^{-i\hat Ht}\hat c_{\bf p}^\dagger|\Psi_0\rangle$ where $\hat H_0$ is the non-interacting part of the Hamiltonian. Equations~\eqref{Residue}-\eqref{GreaterGreens} are straightforwardly generalized to non-zero temperature using a thermal average instead of $\langle\Psi_0|\ldots|\Psi_0\rangle$. Further details regarding RF, Ramsey and spin-echo techniques can be found in earlier reviews~\cite{Massignan2014,Schmidt_2018}. 

One can of course calculate these observables directly from the (time-dependent) many-body wave function. They can also (except the spin echo signal) be obtained from the retarded impurity Green's function $G(\bp,t)=-i\theta(t)\langle[\hat c_{\bf p}(t),\hat c_{\bf p}^\dagger(0)]_{\pm}\rangle$ where $[\hat A,\hat B]_{\pm}=\hat A\hat B\pm \hat B\hat A$ is for fermionic/bosonic impurities~\cite{fetter_1971}. Indeed, the energy $\varepsilon_\bp$ of a polaron with momentum $\bp$ can be found by solving 
\begin{equation}\label{PolaronEnergyGreens}
\varepsilon_\bp=\epsilon_\bp+\text{Re}\,\Sigma(\bp,\varepsilon_\bp),
\end{equation}
where $\Sigma(\bp,\omega)$ is the  self-energy and $G(\bp,\omega)=1/[\omega-\epsilon_\bp-\Sigma(\bp,\omega)]$ is the Green's function in momentum/frequency space with $\epsilon_\bp$ is the dispersion of free impurities. In this review, we adopt the notation $\varepsilon$ to denote quasiparticle energies and $\epsilon$ for bare energies. We suppress here and in the following an infinitesimal positive imaginary part of the frequency $\omega$ in the retarded Green's function. The polaron decay rate is $\Gamma_\bp\propto-Z_\bp\text{Im}\Sigma(\bp,\varepsilon_\bp)$, its residue is
\begin{align}
Z_\bp=\frac1{1-\partial_\omega{\rm Re}[\Sigma(\bp,\omega)]|_{\varepsilon_\bp}},
\label{ResidueGreens}
\end{align}
and its effective mass (at zero momentum) is 
\begin{align}
m^*=\frac{m}{Z_{\textbf p=0}[1+\partial_{\epsilon_\bp}{\rm Re}[\Sigma(\bp,\varepsilon_\bp)]|_{p=0}]}.
\label{MassGreens}
\end{align}
The impurity spectral function can be found from the Green's function as $A(\bp,\omega)=-2\text{Im}\,G(\bp,\omega)$. For a single impurity, we moreover have $G(\bp,t)=\theta(t)G^>({\bf p},t)$, which means that its real-time dynamics as probed via Ramsey interferometry can be calculated from the retarded Green's functions with no need to resort to more elaborate non-equilibrium Keldysh Green's functions.

Interactions with the impurity change the density of majority particles in its neighborhood. The total number $\Delta N$ of extra majority particles attracted to the impurity ($\Delta N<0$ if they are repelled), often referred to as the number of particles in its ``dressing cloud''. It can be calculated from thermodynamic arguments by requiring that the density of the majority particles far away from the impurity remains constant. This corresponds to keeping constant the chemical potential $\mu_b$ of the majority particles and gives~\cite{Massignan2005}
\begin{equation}
 \label{DeltaN}
\Delta N\equiv\left(\frac{\partial n}{\partial n_i}\right)_{\mu_b}=-\left(\frac{\partial \varepsilon}{\partial \mu_b}\right)_{n_i}
\end{equation}
where $n$ $(n_i)$ is the density of the majority (impurity) particles. In this review, the subscript $b$ refers to majority (``bath'') particles. When the impurity-majority interaction is short ranged and can be characterised by the associated scattering length $a$, it is useful to consider the contact, which in 3D is given by~\cite{Tan_energetics_2008,tan_large_2008}
\beq \label{contact}
  C=8\pi m_r \, \frac{\partial \varepsilon}{\partial(-1/a)}
\eeq
with $m_r=m_b m/(m_b+m)$ the reduced mass. The contact is proportional to the impurity-majority pair correlation function and describes the likelihood that a particle from the bath is close to the impurity. It also determines the coefficient of the $1/k^4$ tail of the impurity momentum distribution~\cite{Werner2012}. 

Simple scaling arguments show that many of these quantities are tightly linked~\cite{Scazza2022}. For Fermi polarons at a broad Feshbach resonance and at $T=0$, one indeed finds
\beq \label{dimensionalAnalysisFermi}
  \varepsilon+\Delta N\, \epsilon_F+\frac C{16\pi m_ra}=0,
\eeq
with $\epsilon_F$ the Fermi energy of the bath, which leads to $\Delta N = -\varepsilon/\epsilon_F$  at resonance where $1/a=0$. The properties of Bose polarons also depend on the scattering length $a_b$ between the majority bosons, and one obtains
\beq \label{dimensionalAnalysisBose}
  \varepsilon+\frac{3}{2} \Delta N \mu_b+\frac C{16\pi m_ra}+ \frac{a_b}{2} \frac{\partial \varepsilon}{\partial a_b}=0,
\eeq
with $\mu_b=4\pi a_b n/m_b$ for a dilute BEC.

Finally, the interaction between quasiparticles $f_{\mathbf p, \mathbf p'}$ in Eq.~\eqref{LandauEnergy} is key for understanding the thermodynamic and dynamical properties of a collection of polarons. In addition to any direct interaction between the impurities, an inherent source for $f_{\mathbf p, \mathbf p'}$ is the exchange of modulations in the medium between two quasiparticles. As shown in Eq.~\eqref{DeltaN}, one impurity changes the density of majority particles in its surroundings, which is felt by another impurity. While this mediated interaction in general is attractive for two static and therefore distinguishable impurities, the interaction $f_{\mathbf p, \mathbf p'}$ between quasiparticles can be either attractive or repulsive as discussed in Sec.~\ref{QPinteractions}.

The microscopic Hamiltonian describing impurity particles of mass $m$ immersed in a bath of majority particles with mass $m_b$ reads 
\begin{align}
{H } =&\sum_j\frac{{\bf P}_j^2}{2m}+\sum_{j}\frac{{\bf p}_j^2}{2m_b} +\frac12\sum_{j\neq k} V_i({\bf R}_j-{\bf R}_k)+\nonumber\\
& + \frac12\sum_{j\neq k} V_b({\bf r}_j-{\bf r}_k) + \sum_{j,k} V({\bf r}_j-{\bf R}_k),
\label{GeneralHamiltonian}
\end{align}
where ${\bf R}_j$ and ${\bf P}_j$ (${\bf r}_j$ and $\bp_j$) are the positions and momenta of the impurities (of the majority particles). The potentials $V_i$, $V_b$ and $V$ describe, respectively, the interactions between impurity particles, between majority particles and between impurity and majority particles. 

The interaction between neutral atoms has a $\alpha_{\rm vdW}/r^6$ van der Waals form for large separations. Comparing the latter with the kinetic term one obtains a length scale $(\alpha_{\rm vdW} m)^{1/4}\sim 10^2\,a_0$ (with $a_0$ the Bohr radius) which is typically much shorter than the interparticle distance $\gtrsim 5\cdot 10^3\,a_0$ in  atomic gas experiments~\cite{pethick2002}, suggesting that it should possible to describe the polaron in terms of a few parameters characterizing the low energy impurity-majority particle scattering. The interaction between excitons and electrons in TMDs has a classical charge-dipole $1/r^4$ tail, which for many purposes also can be regarded as short range.

The low energy scattering matrix for a pair of particles with center-of-mass momentum $\bf K$, energy $\omega$ and total mass $M=m+m_b$ with vanishing interaction range is 
\begin{equation}
{\mathcal T}_v({\bf K},\omega)=
\begin{cases}
\left[\frac{m_r}{2\pi}\ln\left(\frac{\epsilon_B}{\omega-K^2/2M}\right)\right]^{-1}& \text{2D,}\\
\left(\frac{m_r}{2\pi a}+i\frac{m_r^{3/2}}{\sqrt 2\pi}\sqrt{\omega-\frac{K^2}{2M}}\right)^{-1}&\text{3D}
\end{cases}
\label{Tmatrices}
\end{equation}
with $\ln(-1)=i\pi$.
A two-particle bound state with energy $\epsilon_B<0$ is always present in 2D, while in 3D the $\mathbf K=0$ scattering matrix has a pole at energy $\epsilon_B=-1/2m_r a^2$ only when the 3D scattering length $a$ is positive.

A remarkable feature of atomic gases is that this dimer energy can be controlled by an external magnetic field, yielding so-called Feshbach resonances, which permit to tune the scattering length $a$ to essentially any value~\cite{chin_feshbach_2010}. In a many-body setting, the corresponding interaction strength can be characterised by $k_na$ with $1/k_n$ a typical interparticle spacing~\cite{Giorgini2008,bloch_many-body_2008}. In 3D, strongly-interacting physics takes place near the unitary point where a vacuum dimer appears, and correspondingly $a$ diverges. In 2D, instead, there is always a bound state and therefore no analog of the unitary point: one can define a scattering length from $\epsilon_B=-1/2m_ra^2$, and the strongly interacting regime is found for $\ln(k_na)\sim 0$. While the atomically-thin TMDs are truly 2D, ``quasi-2D'' configurations in atomic gases are created by squeezing one spatial ($z$) direction by means of a tight harmonic trap with frequency $\omega_z$, and this leads to significant corrections to the 2D scattering matrix. A detailed treatment of scattering in quasi-2D may be found in~\cite{Petrov2001,Levinsen2013,Levinsen2DReview2014,Liu2024}. 

A widely used approach to develop a low energy polaron theory is to systematically replace the microscopic interactions in Eq.~\eqref{GeneralHamiltonian} by the scattering matrices in Eq.~\eqref{Tmatrices}, possibly generalized to take into account finite range and trapping effects. This method includes two-body correlations and gives accurate predictions for Fermi polarons in atomic gases and TMDs, as we will see in Secs.~\ref{sec:FermiPolarons} and \ref{TMDsection}, while it less faithfully describes polarons in Bose or multicomponent Fermi environments where correlations between three- and more particles can be important, as we will discuss in Secs.~\ref{sec:BosePolarons} and~\ref{InteractingBaths}.

Finally, we note that the scattering between ultracold atoms is in principle a multichannel problem, since the interaction mixes different hyperfine states. In particular, Feshbach molecules typically have a component in a closed channel. This gives rise to effective range corrections to the scattering matrices given Eq.~\eqref{Tmatrices}, which can be important for narrow Feshbach resonances~\cite{chin_feshbach_2010,Massignan2014}. While the effects of the multichannel nature of the scattering in general are well understood for the Fermi polaron, there are several questions related to this for the Bose polaron as we shall see. In the rest of this review we will mostly use a single channel model where the impurity-bath interaction can be described through the potential $V(\br)$, and explicitly state when a multichannel approach is used. 

\section{Fermi polarons in atomic gases}   \label{sec:FermiPolarons}

\begin{figure}
\centering
\includegraphics[width=0.48\columnwidth]{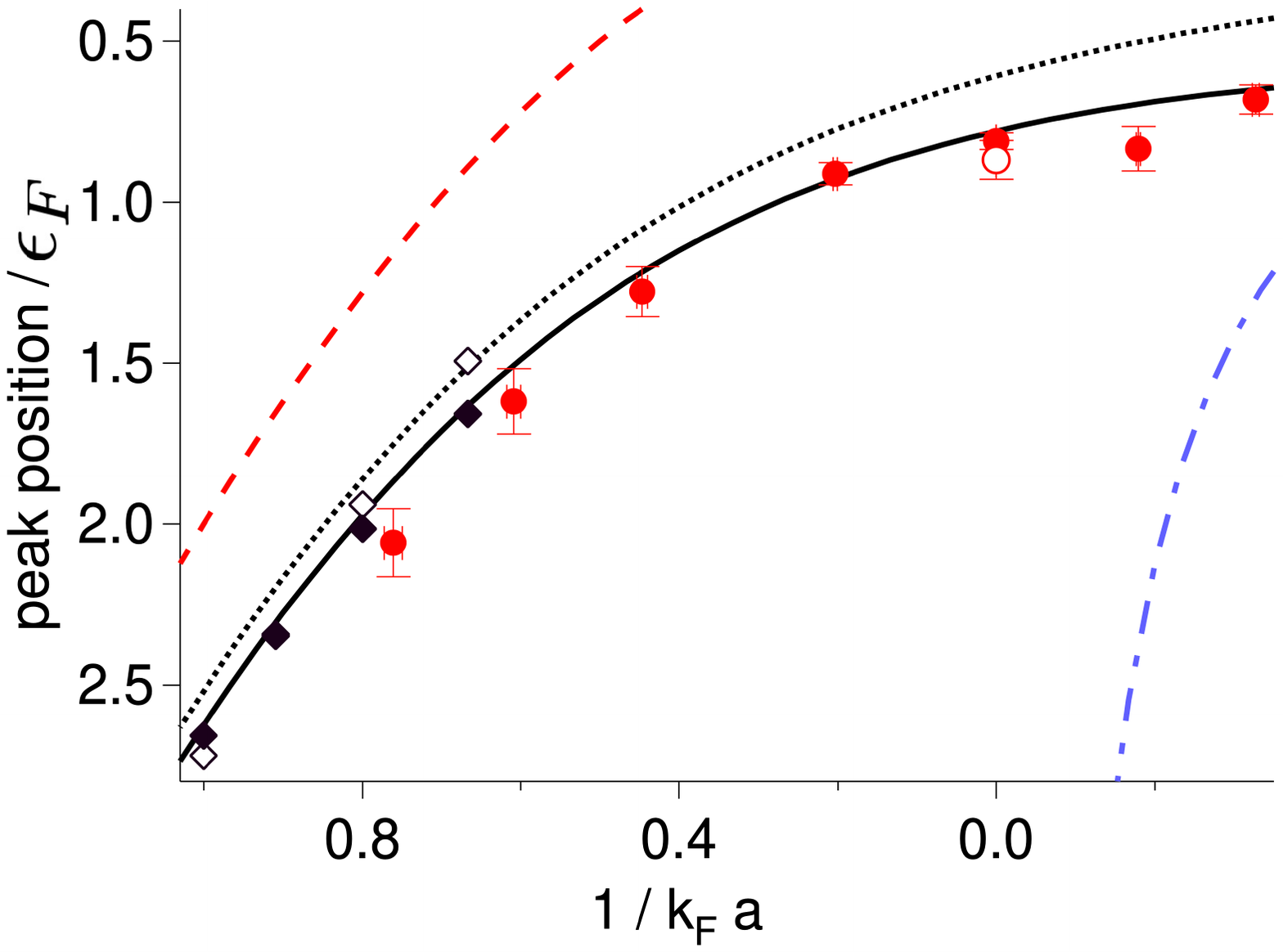}
\includegraphics[width=0.48\columnwidth]{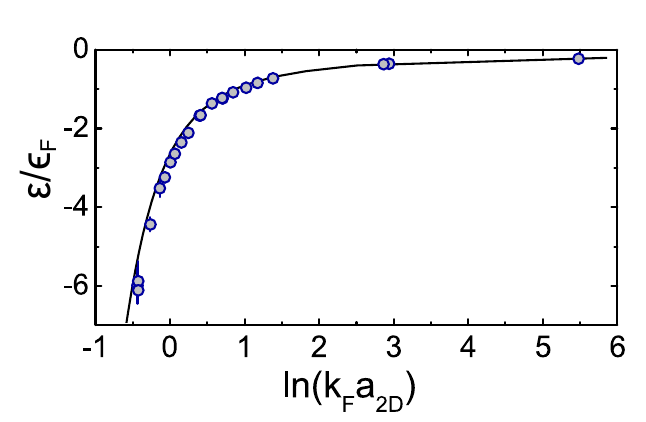}
\caption{\label{fig:attractiveFermiPolaronSchirotzek}
{\bf Attractive Fermi polarons.} (left) Energy of the attractive Fermi polaron in 3D as a function of the impurity-fermion scattering length 
$a$. The open circle shows a measurement with reversed roles of impurity and environment. The dotted/solid line is the  energy of the variational Ansatz \eqref{eq:ChevyAnsatz} excluding/including final state interactions. The dashed line is the dimer energy in vacuum, and the blue dash-dotted line is the mean field energy. Solid/open diamonds are diagrammatic MC polaron/dressed dimer energies~\cite{prok2008}. From \citet{schirotzek_observation_2009}. (right) Attractive polaron energy in 2D as a function of the 2D impurity-fermion scattering length. Circles are 
experimental values and the solid line is the  energy of the variational Ansatz \eqref{eq:ChevyAnsatz}. From \citet{koschorreck_attractive_2012}.}
\end{figure}

In a pioneering experiment,  
 Fermi polarons were created 
by admixing a small number of $^6$Li atoms in one hyperfine state into a large quantum degenerate gas of $^6$Li atoms in another hyperfine state~\cite{schirotzek_observation_2009}. Using a Feshbach resonance to tune the interaction between the two hyperfine components, the Fermi polaron was systematically explored both in the weak and strong coupling regimes, see Fig.~\ref{fig:attractiveFermiPolaronSchirotzek}(left). This inspired several other experimental groups to explore the Fermi polaron in atomic gases~\cite{nascimbene_collective_2009,kohstall_metastability_2012,koschorreck_attractive_2012,Zhang_transitions_2012,wenz2013,cetina_decoherence_2015,Ong2015,cetina_2016,Scazza2016,Oppong2019,Yan2019Fermi,Adlong2020,Ness2020,fritsche2021stability,vivanco2023strongly}, and sparked a large amount of theoretical research. As a result, we have now a good understanding of many aspects of Fermi polarons in their simplest version where the bath is an ideal Fermi gas, even for strong impurity-fermion interactions. The atomic Fermi polaron has been thoroughly discussed in earlier reviews~\cite{chevy_ultra-cold_2010,Massignan2014,Schmidt_2018,Scazza2022,Baroni2024review,Parish2023review}, and in this Section we therefore focus  on its basic properties and theoretical methods it shares with its solid-state counterpart discussed in Sec.~\ref{TMDsection}. We will also discuss recent results not covered in earlier reviews.

As discussed in Sec.~\ref{GeneralProb}, the range of the van der Waals $1/r^6$ interaction is typically small compared to the interparticle distance in atomic gases. It follows that a bath of single component fermions is essentially non-interacting at low temperatures due to the Pauli principle. In this case, one can ignore $V_i({\bf R})$ and $V_b({\bf r})$ in Eq.~\eqref{GeneralHamiltonian}, and the Hamiltonian can be written as  
\begin{equation}
\hat H_{\rm F} = \sum_\bk(\epsilon_{b\bk}\hat{f}_\bk^\dagger\hat{f}_\bk+\epsilon_\bk\hat{c}_\bk^\dagger\hat{c}_\bk)+\sum_{\bk,\bk',\bq}\! V(\bq)\hat{c}_{\bk+\bq}^\dagger\hat{f}_{\bk'-\bq}^\dagger \hat{f}_{\bk'}\hat{c}_\bk,
\label{eq:HF}
\end{equation}
where the operators $\hat{f}^\dagger_\bk$ and $\hat{c}^\dagger_\bk$ create a majority fermion and an impurity with momentum $\bk$ respectively. Here, $\epsilon_{b\bk}=k^2/2m_b$ is the dispersion of the majority particles, and $V(\bq)$ is the Fourier transform of the interaction between the fermions and the impurity. 
Of the many different theoretical techniques that have been successfully applied to describe Fermi polarons (in both ultracold gases and TMDs), this section focuses on two: variational wave functions and Green's functions. 

The incompressibility of the Fermi gas at low temperatures suggests a variational ansatz based on an expansion in the number of particle-hole excitations created by the impurity in the Fermi sea $|{\rm FS}\rangle$. For a single polaron with momentum $\bp$ in a zero-temperature Fermi sea, this expansion  reads~\cite{Chevy2006}
\beq\label{eq:ChevyAnsatz}
|\psi_\bp\rangle=(
\sqrt{Z_{\bp}} \hat{c}^\dagger_\bp
+ \sum_{\bq,\bk}\alpha_{\bp,\bq,\bk}\,\hat{c}^\dagger_{\bp+\bq-\bk}\hat{f}^\dagger_{\bk}\hat{f}_{\bq}+\ldots)|{\rm FS}\rangle
\eeq
where $k_F$ is the Fermi momentum and $|\bq|<k_F<|\bk|$.
 Minimizing the energy of this ansatz with respect to the parameters $Z_{\bp}, \alpha_{\bp,\bq,\bk}\ldots$ leads to closed equations determining the various quasiparticles properties. For most purposes it is sufficient to truncate Eq.~\eqref{eq:ChevyAnsatz} at one particle-hole excitation, commonly called the Chevy ansatz, to obtain accurate results even for strong interactions. This is due to the Pauli principle suppressing $n>2$ body correlations~\cite{Combescot2007}. Notable exceptions appear at large mass imbalance~\cite{Naidon2018,li2023impurity,Petrov2003,efimov1973,Kartavtsev2007, Levinsen2009, Castin2010, Pricoupenko2010, Blume2012, Levinsen2DReview2014, Petrov2017, Petrov2018,Liu2022,Sun2019,Mathy2011}.

For a contact interaction in a continuum system, the Chevy ansatz truncated at one particle-hole excitation is equivalent to the so-called ladder approximation for the self-energy, which at non-zero temperature reads
\begin{equation}\label{eq:SelfenergyLadder}
\Sigma(\bp,\omega)=\sum_{\bq}
\mathcal T(\bp+\bq,\omega+\epsilon_{b\bq})n_F(\epsilon_{b\bq}-\mu).
\end{equation}
Here $n_F(\epsilon)=1/[\exp(\epsilon/T)+1]$ is the Fermi function and 
\begin{equation}
{\mathcal T}(\bk,\omega)=\frac{\mathcal T_0}{1-{\mathcal T_0}[\Pi(\bk,\omega)-\Pi_v(0,0)]}
\label{Tmatrix}
\end{equation}
is the scattering matrix in the ladder approximation with $\mathcal T_0=2\pi a/m_r$. The pair propagator of an impurity and a fermion with total center-of-mass momentum and energy $(\bk,\omega)$ in the presence of a Fermi sea reads
\begin{equation}
\Pi({\mathbf k},\omega)=\sum_{\bp}\frac{1-n_F(\epsilon_{b\bp}-\mu)}{\omega-\epsilon_{{\mathbf k}-\bp }-\epsilon_{b{\mathbf p }}},
\label{Pairpropagator}
\end{equation} 
with its vacuum form $\Pi_v(\bk,\omega)$  obtained by setting $n_F=0$ in Eq.~\eqref{Pairpropagator}. The difference between Eq.~\eqref{Tmatrix} and the 3D vacuum scattering matrix ${\mathcal T}_v$ in Eq.~\eqref{Tmatrices} is that Eq.~\eqref{Pairpropagator} takes into account the Pauli blocking of available scattering states. Indeed, Eq.~\eqref{Tmatrices} is recovered when replacing $\Pi(\bk,\omega)\rightarrow \Pi_v(\bk,\omega)$ in Eq.~\eqref{Tmatrix}. We have assumed a momentum-independent impurity-fermion interaction $V(\bq)=g$, which is replaced by the 3D scattering matrix in Eq.~\eqref{Tmatrices} via the Lippmann-Schwinger equation 
\beq
\mathcal T_0={\mathcal T}_v(0,0)=g+g\Pi_v(0,0){\mathcal T}_v(0,0).
\eeq
Physically, the self-energy $\Sigma(\bk,\omega)$ in Eq.~\eqref{eq:SelfenergyLadder} describes  the energy shift coming from the impurity scattering  fermions from inside to outside the Fermi sea. The Chevy ansatz Eq.~\eqref{eq:ChevyAnsatz} in fact also includes terms describing the scattering of the impurity on holes in the Fermi sea not included in the ladder approximation. While these terms are not important for continuum systems, they can be important when the hole states have significant spectral weight as in lattice systems~\cite{amelio2024polaronformationinsulatorskey}.

\begin{figure}[t]
\centering
\includegraphics[width=\columnwidth]{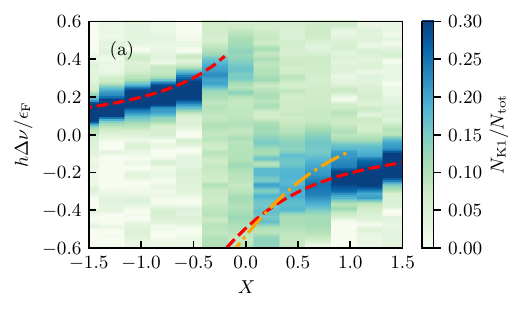}
\caption{\label{fig:FermiPolaronSpectrumExp}
\textbf{Complete Fermi polaron spectrum.} 
Spectral response of few bosonic $^{41}$K impurities in a $^6$Li Fermi sea, as a function of the interaction parameter $X=-1/(k_Fa)$ and the RF detuning $\Delta\nu$. The upper/lower red lines (dashed) show the variational predictions for the repulsive/attractive polarons, and the orange line (dash-dotted) is the prediction for the dressed molecule. From \citet{fritsche2021stability}.}
\end{figure}

A typical experimental RF injection spectrum is shown in Fig.~\ref{fig:FermiPolaronSpectrumExp}, where the impurity spectral function for zero momentum is plotted as a function of the impurity-fermion interaction strength $X=-1/k_Fa$. Two quasiparticle branches are clearly visible: one at negative energies corresponds to the attractive polaron (which is the ground state for $X\gtrsim 0.2$), and one at positive energies corresponds to the repulsive polaron. The repulsive polaron can roughly be thought of as the lowest scattering state and it is continuously connected to the non-interacting impurity particle for $a\rightarrow 0_+$~\cite{Cui2010,massignan_repulsive_2011}. The red dashed lines are the energies of these two quasiparticles as obtained from the Chevy ansatz Eq.~\eqref{eq:ChevyAnsatz} truncated to one particle-hole excitation (ladder approx.), which agree very well with the experimental results. A multichannel model was used to obtain this agreement, since the Feshbach resonance used in this experiment is relatively narrow, giving rise to significant effective range effects [for details, see the earlier review \cite{Massignan2014}]. The orange dash-dotted line shows the energy of another fundamental quasiparticle, i.e., the dimer (formed by the impurity and a bath particle) dressed by particle-hole excitations in the bath, which becomes the ground state for $X\lesssim 0.2$ for the specific case in Fig.~\ref{fig:FermiPolaronSpectrumExp}. This dressed molecule has however a low spectral weight in injection spectra, since its Franck-Condon overlap with a non-interacting impurity is small. The  transition can alternatively be interpreted as the polaron abruptly changing its momentum from zero to the Fermi momentum~\cite{Cui2020}.  Figure \ref{fig:FermiPolaronSpectrumExp} also shows a continuum of many-body states for strong interactions, which consists of states such as a molecule and a hole with total momentum zero. Similar good agreement between experimental data and the ladder approximation was also reported for the attractive polaron in 2D~\cite{koschorreck_attractive_2012}, see Fig.~\ref{fig:attractiveFermiPolaronSchirotzek}(right). 

Further details on Fermi polaron experiments together with the many theoretical techniques used to analyze them have been described in detail in earlier extensive reviews~\cite{Massignan2014,Schmidt_2018,Scazza2022,Parish2023review}.
In general, the ladder approximation provides a description of the energy, residue, and effective mass of individual attractive and repulsive Fermi polarons which agrees remarkably well with most experiments. 
Nonetheless, some interesting open questions remain. First, the damping rate of the Fermi polaron requires a careful analysis of the different decay channels, which for strong interactions and in the presence of a many-body environment (the Fermi sea) represents a formidable challenge. For instance, even for weak coupling a non-self consistent ladder approximation predicts that the energy of a zero momentum ($p=0$) repulsive polaron will be pushed by the mean-field term up into a continuum of $p>0$ bare impurity states, leading to damping, which is unphysical since these states experience the same mean-field shift and therefore have a higher energy. Thus, there is no damping from such processes. Controlled calculations have been performed in two limits: Collisional broadening for weak interactions where the Boltzmann equation can be applied~\cite{BruunTransport2008}, and three-body decay into a Feshbach molecule for weak interactions in a vacuum~\cite{Petrov2003}. 
The damping of polarons and dressed dimers for strong interactions was investigated theoretically and experimentally in Refs.~\cite{kohstall_metastability_2012,cetina_decoherence_2015,Bruun2010,schmidt_excitation_2011}.  \citet{Adlong2020} used a time-dependent variational approach to show that the lifetime of repulsive polarons is dominated by many-body dephasing in both 2D and 3D. 

Other open questions concerning the equilibrium properties of a single Fermi polaron include a detailed understanding and measurement of impurity-bath correlations 
in real space,  the breakdown of the quasiparticle picture as the impurity mass approaches infinity, and  the polaron/molecule transition in lattice models. There are also interesting questions 
concerning the Fermi polaron with a non-zero momentum, and for long range interactions as will discussed in Sec.~\ref{sec:long-ranged_Bose_polarons}. 

In the next two subsections we will discuss recent developments regarding Fermi polarons in atomic gases not covered in earlier reviews: temperature/impurity concentration effects, and non-equilibrium dynamics. In the later Section \ref{QPinteractions} we will discuss the interaction $f_{\bp,\bp'}$ between Fermi polarons, which in some sense is ``the second half'' of Landau's quasiparticle theory crucial for many dynamical and thermodynamic properties, and for which experimental results have been obtained only recently. 

\subsection{Temperature and impurity concentration effects}   \label{TdependenceFermiPolaron}

\begin{figure}[t]
\centering
\includegraphics[width=\columnwidth]{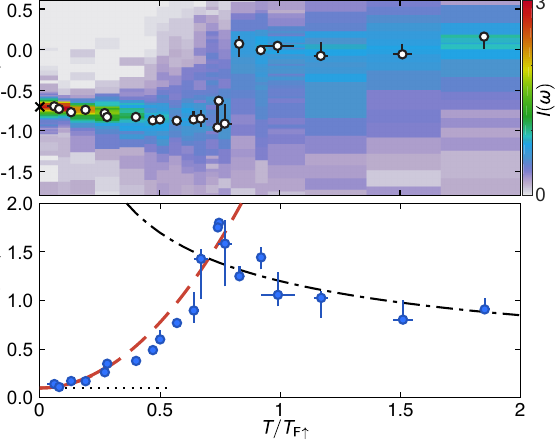}
\caption{\label{fig:FermiPolaronTempDep}
\textbf{Temperature dependence of Fermi polarons.} 
Energy $\varepsilon/\epsilon_F$ (top) and decay rate $\Gamma/\epsilon_F$ (bottom) of unitary Fermi polarons as a function of the bath temperature. The red dashed line shows the Fermi liquid prediction $\Gamma\propto T^2$~\cite{BruunTransport2008}, and the black dash-dotted line indicates the high-temperature behavior $\Gamma\propto 1/\sqrt{T}$~\cite{ENSS2011770,Sun2015}. From \citet{Yan2019Fermi}.}
\end{figure}

Most experiments have explored the Fermi polaron at a low temperature $T\ll T_F$ with $T_F=\epsilon_F$ the Fermi temperature. As the temperature of a Fermi gas containing dilute impurities is raised, one expects that the Fermi-liquid picture of weakly-interacting quasiparticles will eventually stop working. At temperatures well above $T_F$, an accurate description in terms of a classical Boltzmann approach should ultimately be obtainable. The transition between these two regimes was studied in detail in Ref.~\cite{Yan2019Fermi} for unitarity-limited impurity-bath interactions with $1/a=0$. For low temperatures $T\lesssim 0.75T_F$ the spectra showed a sharp peak corresponding to the attractive polaron, whose energy starts from the $T=0$ prediction $\varepsilon=-0.6\epsilon_F$ and lowers gently with increasing temperature. One reason for this is that the Pauli repulsion between bath fermions gradually decreased in agreement with theoretical calculations~\cite{Tajima2018,MULKERIN201929}. Likewise, the width of the polaron peak increases as $T^2$ due to the collisional broadening as expected in Fermi liquids~\cite{BruunTransport2008}. It reaches a width $\gtrsim \epsilon_F$ corresponding to a lifetime smaller than the Fermi time $1/\epsilon_F$ at $T\sim 0.75 T_F$, see Fig.~\ref{fig:FermiPolaronTempDep}. Above this temperature the maximum of the RF spectra suddenly shifts and remains locked at the energy $\omega\sim 0$, and the peak width decreases slowly as $\Gamma \sim 1/\sqrt{T}$, precisely as expected in a classical Boltzmann gas with unitarity limited interactions. In this regime, indeed, one has a cross section $\sigma_{\rm th}\sim 1/k_{\rm th}^2\sim 1/T$ and a typical velocity $v_{\rm th}\sim \sqrt{T}$, giving a scattering rate $\Gamma_{\rm th}=n v_{\rm th}\sigma_{\rm th}\sim 1/\sqrt{T}$, and an energy given to lowest order by the real part of the scattering amplitude, i.e., $\varepsilon\sim 0$. As we shall see in Sec.~\ref{sec:Tempdependence}, similar effects are observed for the Bose polaron in the classical regime. 

This experiment further permitted precise measurements of the temperature dependence of Tan's contact. In the non-degenerate regime, the latter decreased rapidly as predicted by the third-order virial calculation of Ref.~\cite{Liu2010}, and an excellent agreement at all temperatures was obtained using a finite-temperature variational method~\cite{Liu2020radio}. The experiment also confirmed the $T=0$ polaron effective mass $m^*\sim 1.2m$ at unitarity predicted from the ladder approximation, as well as the excess majority fermions around an impurity atom $\Delta N\sim0.6$ in perfect agreement with the value $-\varepsilon/\epsilon_F$ given by Eq.~\eqref{dimensionalAnalysisFermi}. Interestingly, $\Delta N$ displayed no detectable dependency on the impurity density up to concentrations as large as 30\%, indicating an inherent robustness of a Fermi liquid description in terms of polarons at low temperatures.

\begin{figure}[t]
\centering
\includegraphics[width=\columnwidth]{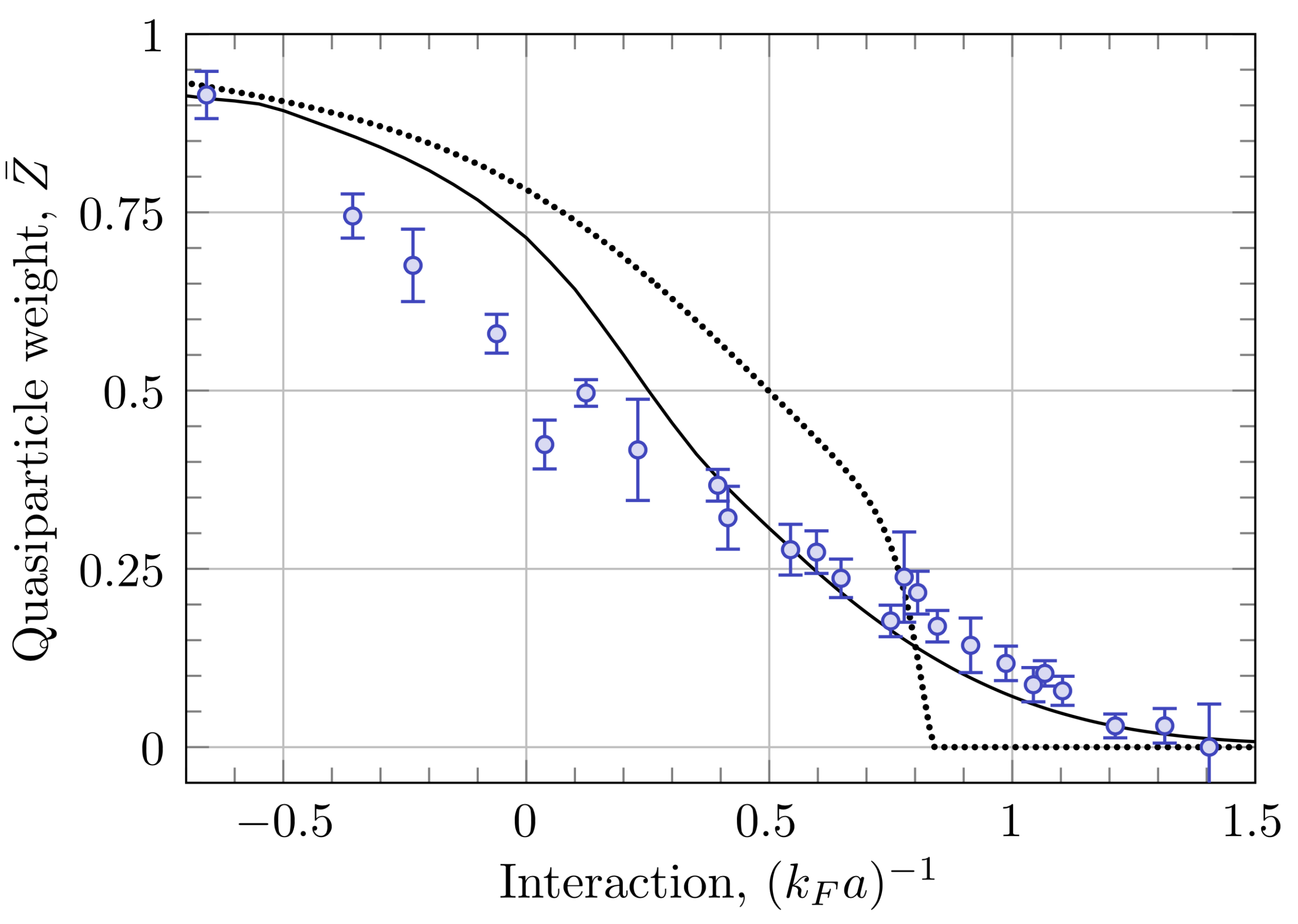}
\caption{\label{SagiFig}
{\bf Smooth polaron/molecule crossover at $T>0$.} Measurement of the polaron's spectral weight averaged over thermally populated states at $T=0.2T_F$
for a non-zero impurity density. 
The solid line is a ladder approximation calculation assuming a non-zero density of thermally occupied polarons, whereas the dotted line assumes a single polaron at $T=0$. 
From \citet{Ness2020}.}
\end{figure}

A single attractive polaron at zero momentum (at a broad resonance and for equal impurity and bath fermion mass) is predicted to be the ground state for $1/k_Fa\lesssim 0.90$, whereas a molecule is the ground state for 
$1/k_Fa\gtrsim 0.90$~\cite{prok2008,VanHoucke2020}. The transition between these two ground states is sharp for a single impurity since, loosely speaking, the two dressed states, have different quantum statistics. It is however smoothed out by a finite impurity concentration and temperature~\cite{Parish2021,Peng2021} since this can populate polaron and molecule states even when they are not the ground state. This was demonstrated experimentally using Raman spectroscopy~\cite{Ness2020}, see Fig.~\ref{SagiFig}.

\subsection{Non-equilibrium dynamics}   \label{NonequilFermipolaron}
A characteristic time-scale for  many-body dynamics is $\tau_F=1/\epsilon_F$. For solid state systems this is typically very short ($\tau_F\simeq 10^{-16}$s), whereas $\tau_F\simeq 10^{-6}$s in atomic gases due to their diluteness and large atomic mass. This makes them well suited for studying quantum many-body dynamics \cite{Schmidt_2018}. Following theoretical proposals~\cite{goold_orthogonality_2011,knap2012}, the dynamical formation of the Fermi polaron was explored in Ref.~\cite{cetina_2016} using a Ramsey scheme. 
$^{40}$K impurity atoms were driven by a $\pi/2$ pulse into an equal superposition of two hyperfine states with one of them interacting with a surrounding bath of fermionic $^6$Li atoms. The system was then allowed to evolve for a time $t$ after which a second $\pi/2$ probe pulse was applied.

\begin{figure}[t]
\centering
\includegraphics[width=\columnwidth]{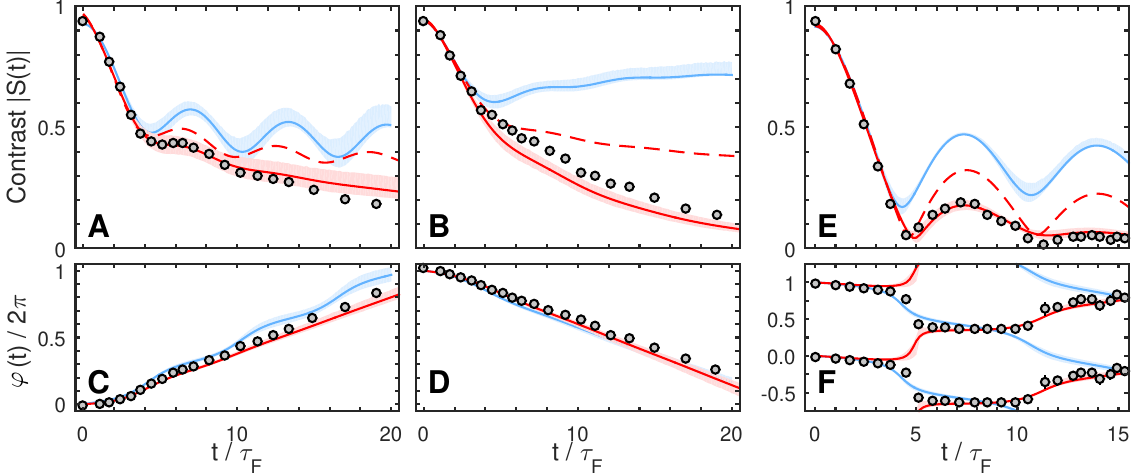}
\caption{\label{GrimmRamseyFig}
{\bf Dynamical formation of Fermi polarons.}
The top and bottom rows show the contrast $|S(t)|$ and the phase $\varphi(t)$ measured by Ramsey spectroscopy as a function of the interaction time $t$. From left to right, the three columns illustrate the results obtained in the repulsive polaron regime ($X=-0.2$, with $X\equiv -1/k_Fa$), in the attractive polaron regime ($X=0.9$), and for resonant interactions ($X=0.1$). The corresponding Chevy ansatz calculations are shown by solid blue lines, and the red lines indicate the results of the FDA calculations (solid: at the measured temperature; dashed: at $T=0$). From \citet{cetina_2016}.}
\end{figure}

The response function probed by this Ramsey scheme is given by Eq.~\eqref{GreaterGreens}, and Fig.~\ref{GrimmRamseyFig} shows its experimentally-measured amplitude and phase as a function of time for different impurity-majority interaction strengths. For short times, the dynamics is governed by high energy $(\epsilon\gg \epsilon_F)$ two-body impurity-fermion scattering, which is independent of the quantum statistics of the bath~\cite{parish2016}. The dynamics can be solved analytically in this regime as discussed in detail for the Bose polaron in Sec.~\ref{sec:NonequilBosepolaron}. For weak interactions, one observes a linear evolution of the phase consistent with $S(t)= Z\exp(-i\varepsilon t)$ coming from the existence of well-defined attractive or repulsive polarons with energy $\varepsilon$ and residue $Z$ as discussed below Eq.~\eqref{GreaterGreens} (the experiment had no momentum resolution). Figure~\ref{GrimmRamseyFig} however shows that the amplitude decays exponentially for long times instead of approaching $Z$, which can be explained quantitatively by decoherence due to polaron-polaron scattering as described by the Boltzmann equation~\cite{cetina_decoherence_2015}. 

For strong interactions close to resonance, the amplitude $|S(t)|$ in Fig.~\ref{GrimmRamseyFig} oscillates strongly while the phase exhibits plateaus as a function of time. 
For early to intermediate times, this behavior is well described by the zero-temperature Chevy Ansatz in Eq.~\eqref{eq:ChevyAnsatz} generalized to time-dependent phenomena~\cite{parish2016}.
For longer evolution times, thermal effects play an important role, and the accuracy of the theoretical prediction was later improved by extending the underlying variational method to finite-temperature dynamics~\cite{Liu2019Dynamics}.
The red lines in Fig.~\ref{GrimmRamseyFig} show an exact solution for a static impurity in a Fermi gas taking into account a non-zero temperature obtained using a functional determinant approach (FDA)~\cite{knap2012,Schmidt_2018}. This approach agrees excellently with the experimental data indicating that recoil effects are small or masked by thermal effects. The oscillations at strong interactions can be attributed to the simultaneous presence of attractive and repulsive polaron peaks in the spectral function, giving rise to a quantum beat between the two polaron states and a revival of the contrast after a time corresponding to their energy difference. Similar effects for the Bose polaron are discussed in Sec.~\ref{sec:NonequilBosepolaron}. 

\begin{figure}[t]
\centering
\includegraphics[width=\columnwidth]{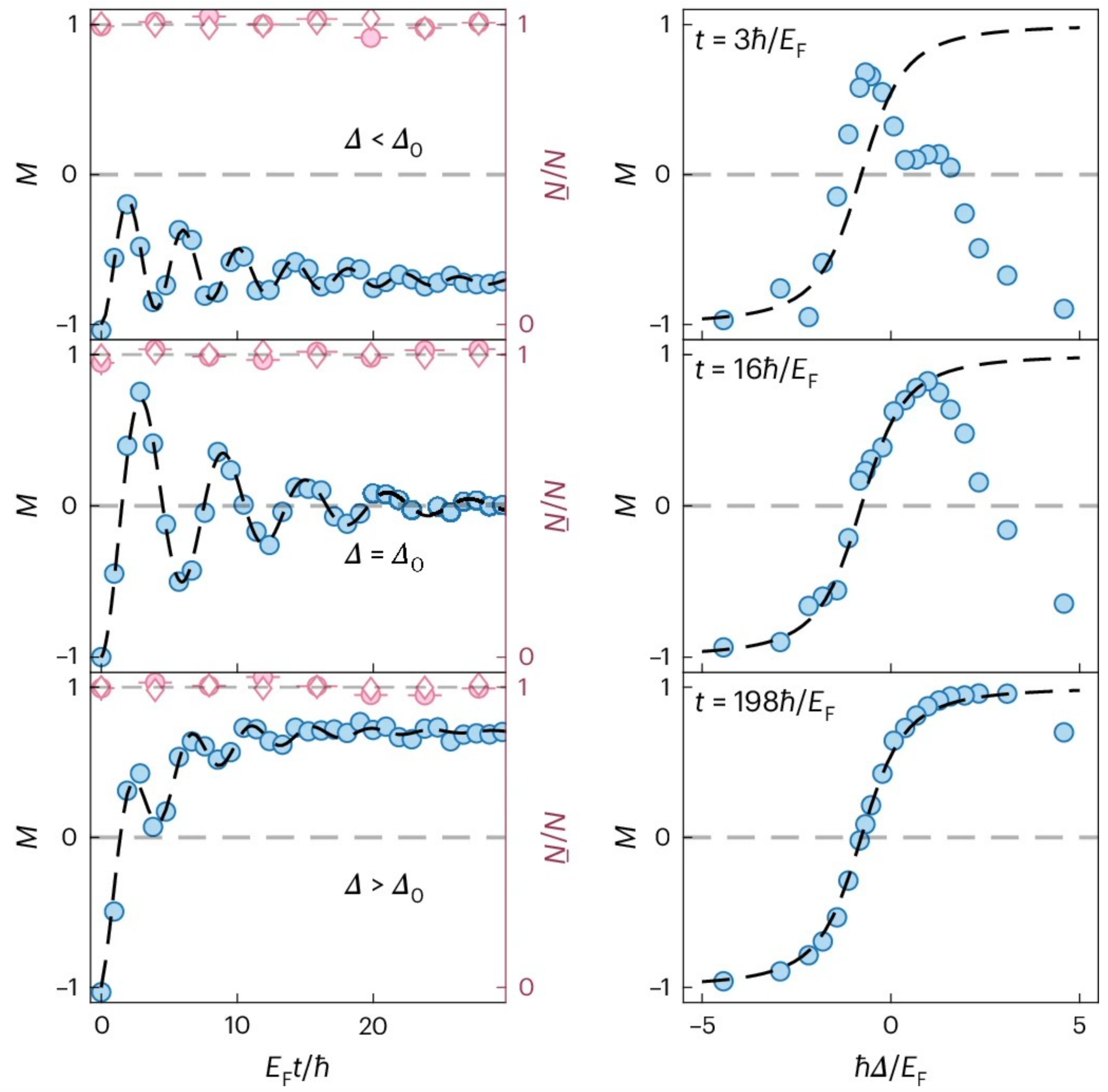}
\caption{\label{NavonFig}
{\bf Rabi oscillations of the Fermi polaron.}
(left) The filled blue circles indicate the time-evolution of the magnetisation $M=(N_\uparrow-N_\downarrow)/((N_\uparrow+N_\downarrow)$ between the populations of two impurity states, respectively interacting resonantly 
($1/k_Fa_\uparrow=0$) and weakly ($k_Fa_\downarrow=0.16$) with a Fermi gas, as a function of the Rabi pulse
duration $t$  at fixed detuning $\Delta$.
From top to bottom, $\Delta/\Delta_0=-0.3, 1, 2.4$. 
(right) The  magnetisation as a function of $\Delta$ for various pulse durations with the steady state value $M=0$  for $\Delta=\Delta_0$. 
In all panels, the Rabi frequency was set to $\Omega=1.3\epsilon_F$.
From \citet{vivanco2023strongly}. 
}
\end{figure}

As discussed above, the equilibrium spectral function can be measured by a weak RF pulse coupling two internal states of the impurity so that linear response applies. For stronger pulses, on the other hand, the impurity performs Rabi oscillations between the two internal states well beyond what can be described by linear response. When the Rabi frequency $\Omega$ is small compared to the Fermi energy $\epsilon_F$, the polaron has time to form and the frequency is reduced by a factor $\Omega\rightarrow \sqrt Z\Omega$ due to the smaller overlap between the polaron wave function and the non-interacting plane wave, see Eq.~\eqref{eq:ChevyAnsatz}. Rabi oscillations have been used to measure the residue and lifetime of Fermi 
polarons~\cite{kohstall_metastability_2012,Scazza2016,Oppong2019,Adlong2020}.  Figure \ref{NavonFig} shows such measured Rabi  oscillations for the drive  
$\Omega/\epsilon_F\simeq 1.3$~\cite{vivanco2023strongly}. For strong driving $\Omega/\epsilon_F\gg 1$, this experiment reported 
 large
 deviations from the weak limit result $\sqrt Z\Omega$ with Rabi frequencies even exceeding  $\Omega$ on resonance. 
 Analysing this  requires solving a challenging non-equilibrium many-body problem of an impurity strongly driven by a Rabi pulse~\cite{Knap2013Dis}, which has been approached using variational, diagrammatic, and quantum kinetic theories~\cite{Adlong2020,Hu2023,Wasak2024}. Thermodynamic relations generalizing Eqs.~\eqref{dimensionalAnalysisFermi}-\eqref{dimensionalAnalysisBose} to the presence of Rabi driving and non-zero temperatures were derived by~\citet{Mulkerin2024}.

\section{Bose polarons in atomic gases}   \label{sec:BosePolarons}

We now turn to the properties of Bose polarons, which in their simplest incarnation emerge when a mobile quantum impurity is immersed into a weakly interacting Bose gas that may have undergone condensation. At first sight, the Bose polaron problem seems very similar to that of the Fermi polaron discussed in the previous section, but as we shall see there are surprisingly many open questions and conflicting theoretical predictions concerning its properties in the strongly interacting regime. 

\begin{figure}[t]
\centering
\includegraphics[width=\columnwidth]{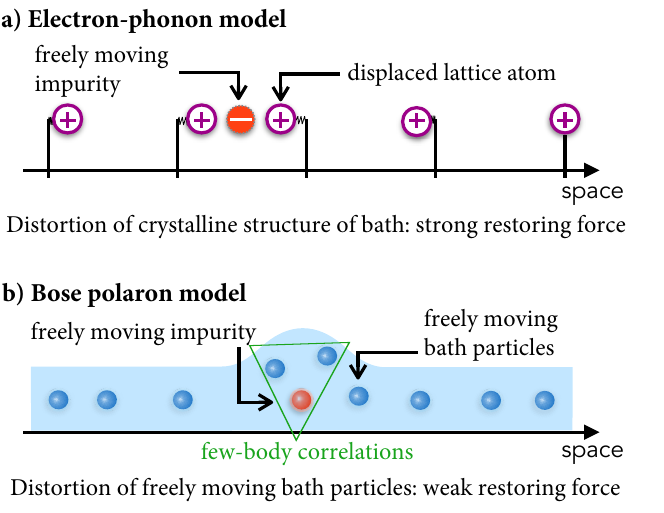}
\caption{\label{Fig:BosepolaronIllustration}
\textbf{The Bose polaron problem.} 
(a) An electron moving through a crystal generates small displacements of the atoms from their equilibrium positions, which are well described by a linear approximation. (b) In a Bose gas, on the other hand, the impurity can strongly distort the surrounding Bose gas, and correlations between the impurity and $n=1, 2, \ldots$ bosons (here exemplified by $3$-body Efimov correlations) may be important.}
\end{figure}

The setting of a mobile impurity interacting with the low energy phonon modes of the bath is reminiscent of electrons interacting with a bath of crystal phonons~\cite{Pekar}. In the latter case, the theoretical model is derived from considering how the negative electron charge displaces the positively charged ions in the crystal of the solid away from their equilibrium positions, see Fig.~\ref{Fig:BosepolaronIllustration}(a). Assuming this displacement to be small, quantization gives rise to bosonic phonons, which couple \textit{linearly} to the electrons. The Bose polaron problem at hand is however distinct from such a linear model. The main reason is that the bosonic particles, which are atoms or molecules in the context of quantum degenerate gases and excitons in the context of TMDs, are \textit{not fixed} in space and thus can move freely, see Fig.~\ref{Fig:BosepolaronIllustration}(b). As a result, the "stiffness" of the environment is much reduced and a linear approximation is bound to fail. This makes the description of Bose polarons a particularly challenging problem. For a detailed review of the Bose polaron including the linear coupling (Fr\"ohlich) case, see~\citet{grusdt2024impuritiespolaronsbosonicquantum}.

The basic Hamiltonian for the Bose polaron problem is given by Eq.~\eqref{GeneralHamiltonian}, which in second quantized form reads
\begin{eqnarray}
\hat H_{\rm B} &= &\sum_\bk(\epsilon_\bk\hat{c}_\bk^\dagger\hat{c}_\bk+\epsilon_{b\bk}\hat{b}_\bk^\dagger\hat{b}_\bk)+\sum_{\bk,\bk',\bq}\frac{V_b(\bq)}2\hat{b}_{\bk+\bq}^\dagger\hat{b}_{\bk'-\bq}^\dagger \hat{b}_{\bk'}\hat{b}_\bk\nonumber\\
&&+\sum_{\bk,\bk',\bq}V(\bq)\hat{c}_{\bk+\bq}^\dagger\hat{b}_{\bk'-\bq}^\dagger \hat{b}_{\bk'}\hat{c}_\bk,
\label{eq:HB}
\end{eqnarray}
where  $\hat{b}_{\bk}^\dagger$  creates a bosonic majority particle.
The second term in Eq.~\eqref{eq:HB} describes the interaction between the majority bosons, which is assumed to be repulsive in order to stabilise the system. In contrast to fermions, the interaction  $V_b$ between the majority bosons, which is assumed to be repulsive, is important even when short range as it stabilises the system. The last term in Eq.~\eqref{eq:HB} may be  written as $\sum_{\veck,\vecq}V(\vecq)\exp\langle{-i \hat{\vecR}\vecq }\rangle \bed_{\veck+\vecq}\be_{\veck}$.

The repulsion between the bosons is typically rather weak, resulting in an environment that, compared to an ideal Fermi gas, has a high compressibility. As a consequence, the density around the impurity can be greatly increased with a large number of bosons in the polaron dressing cloud, see Fig.~\ref{Fig:BosepolaronIllustration}. Equivalently, while the Pauli principle typically suppresses correlations of the impurity with more than one fermion, a priori no such prohibition of $n>2$ body correlations is present in the case of Bose polarons. Indeed, bosonic gases generally support Efimov bound states at strong interactions~\cite{Efimov1970,Braaten2006,Greene2017,Naidon2017}. These three-body bound states have been observed for the first time in cold atoms~\cite{Kraemer2006,Zaccanti2009}, with experimental signs of bound states involving four bosons (tetramers) also reported~\cite{Ferlaino2009} in agreement with theory~\cite{Hammer2007,Stecher2009,schmidt2010fourbody}. Bound states involving even more particles are also predicted~\cite{Stecher2011,Blume2014}. While the physics of weakly bound dimers is universally captured by the scattering lengths, the description of Efimov states requires the specification of another three-body parameter, which intrinsically carries information about the short-distance physics. As a result, the properties of Efimov states, such as their energy and size, depend on short-range physics (the van-der Waals range for cold  atoms~\cite{schmidt2012efimov,wang2012vdwUni, Mestrom2017,chapurin2019precision}). As it turns out, these Efimov states can hybridize with the Bose polaron state further complicating the theoretical description. Contrary to the Fermi polaron, one therefore in general needs length scales in addition to the scattering length $a$ in order to describe Bose polarons like for example the boson-boson scattering length $a_b$, a three body-parameter, or the ranges of the interaction potentials. Related to this, while effective range effects are important only in the vicinity of narrow Feshbach resonances for the Fermi polaron,  they are more important for the Bose polaron. They naturally emerge in two-channel models for Feshbach resonances~\cite{Massignan2014,Levinsen2015,Yoshida2018}, but also in  single channel models where one is often forced to consider potentials with non-vanishing ranges, see for example Refs.~\cite{Massignan2005, Guenther2021,Drescher2020,Christianen2023}. 

\begin{figure*}
\centering
\includegraphics[width=\textwidth]{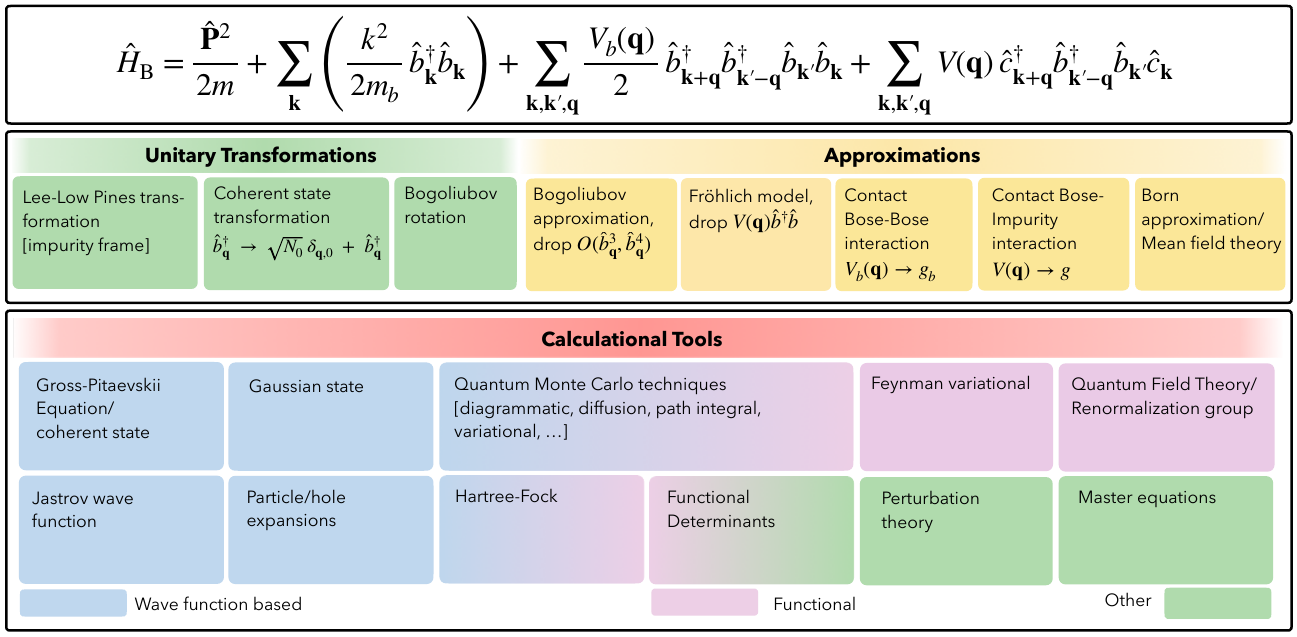}
\caption{\label{Fig:BosePolaronModelOverview}
\textbf{Theoretical approaches for the Bose polaron.}
Different approximations and transformations used to analyse the Bose polaron, and an overview of common theoretical techniques. Their strengths and weaknesses are discussed across Sec.~\ref{sec:BosePolarons}.}
\end{figure*}

Owing to its complexity, the Bose polaron problem has been theoretically studied using many different techniques, as summarized in Fig.~\ref{Fig:BosePolaronModelOverview}. One challenge when comparing these predictions arises from the fact that the corresponding works apply different approximations to the Hamiltonian \eqref{eq:HB}. In particular, it is difficult to compare variational energies arising from different Hamiltonians, and hence there is no notion of ``better'' wave function by comparison of results. As an example, we have already assumed a single channel model for the boson-impurity interaction in Eq.~\eqref{eq:HB}, even though the compressibility of the BEC may make its underlying two-channel Feshbach nature important as we shall see. We now turn to a more detailed discussion of the theoretical and experimental results on Bose polarons.

\subsection{Ladder approximation}   \label{sec:ladder}

Inspired by its accuracy for the Fermi polaron,  the ladder approximation was in a pioneering paper adapted to explore the Bose polaron~\cite{RathSchmidt2013}. While it is not as accurate for the Bose polaron, it describes correctly the essential features of the problem and sets the stage for the ensuing discussion.
\citet{RathSchmidt2013} used the Bogoliubov approximation to describe the weakly interacting bosonic bath. Keeping all terms up to second order in the bosonic creation and annihilation operators, the Hamiltonian in Eq.~\eqref{eq:HB} can be written as 
\begin{align}
H&_{\rm Bog}=\sum_{\bk}E_{\bk}\hat{\gamma}_\bk^\dagger\hat{\gamma}_\bk+\sum_{\bp}(\epsilon_\bp+g n_0)\hat{c}_\bp^\dagger\hat{c}_\bp\label{BeyondFrohlich}\\ \nonumber
&+\underbracket{g\sum_{\mathbf p,\mathbf q}M(\bq)\,\hat c^\dagger_{\mathbf p+\mathbf q}\hat c_\mathbf p(\hat\gamma^\dagger_{-\mathbf q}+\hat\gamma_{\mathbf q})}_{\text{Fr\"ohlich interaction}} +g\sum_{\mathbf p,\mathbf q,\mathbf k}\hat c^\dagger_{\mathbf p+\mathbf q}\hat c_\mathbf p\hat b^\dagger_{\mathbf k-\mathbf q}\hat b_{\mathbf k}.
\end{align}
Here $\hat \gamma_\mathbf p^\dagger=u_{\mathbf p} {\hat b}_{\mathbf b}^\dagger+v_{\mathbf p} {\hat b}_{-\mathbf p}$ creates a Bogoliubov excitation in the BEC with energy $E_{\mathbf p}=\sqrt{\epsilon_{\mathbf pb}(\epsilon_{\mathbf pb}+2\mu)}$, $\mu=g_bn_0$ is the chemical potential, $u^2_{\mathbf p}=1+v^2_{\mathbf p}=[(\epsilon_{\mathbf pb}+\mu)/E_{\mathbf p}+1]/2$, and $n_0$ is the condensate density~\cite{pethick2002,pitaevskii2016bose}. We have assumed weak short-range boson-boson and boson-impurity interactions with strengths $g_b=4\pi a_b/m_b$ and $g= 2\pi a/m_r$. A constant term giving the energy of the BEC ground state is omitted in Eq.~\eqref{BeyondFrohlich}, $M(\bq)=\sqrt{n_0\epsilon_{\mathbf qb}/E_{\mathbf q}}$, and all momenta $\bk$ and $\bq$ for the boson and Bogoliubov operators are different from zero. In the last term we kept the expression in terms of the untransformed boson operators for notational brevity. 

The second term in Eq.~\eqref{BeyondFrohlich} is the kinetic energy of the impurity shifted by a mean-field interaction term with the condensate, and the third term describes how it emits or absorbs Bogoliubov modes as it moves through the BEC. The first three terms correspond to the Fr\"ohlich model describing how an electron emits or absorbs phonons in a crystal, see Fig.~\ref{Fig:BosepolaronIllustration}, and it was the basis of early investigations of the Bose polaron \cite{Tempere2009,Cucchietti2006}. As first  noted in \cite{RathSchmidt2013} and later rigorously shown in a perturbative calculation \cite{Christensen2015}, the Fr\"ohlich Hamiltonian alone is however insufficient to describe impurities that interact strongly with atomic BECs such as realized by Feshbach resonances. In the rest of this review we will therefore not discuss in detail the many papers analyzing the Bose polaron within the Fr\"ohlich model and refer instead the reader to earlier reviews~\cite{devreese_Polaron_Review_1996,alexandrov_PolaronBook_1996,grusdt2016Rev}.

Note that even when the last term in Eq.~\eqref{BeyondFrohlich} is included, the Bogoliubov approximation omits terms involving three and four boson operators. These terms describe the interaction between Bogoliubov modes and can typically be neglected when the gas parameter $na_b^3$ is small ~\cite{pethick2002,pitaevskii2016bose}. They may however be important in the presence of an impurity, which can increase the surrounding density so much that the gas parameter becomes large thereby invalidating the Bogoliubov approximation as we shall discuss later~\cite{Christianen2023}.

\begin{figure}
\centering
\includegraphics[width=\columnwidth]{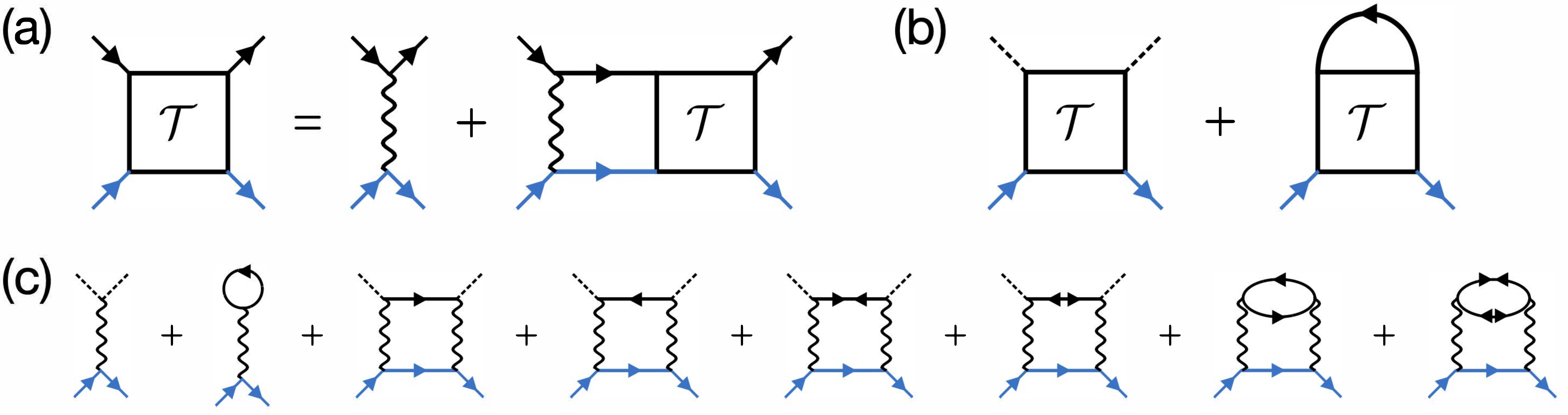}
\caption{\label{fig:FeynmanFig}
{\bf Feynman diagrams for Bose polarons.}
(a) Ladder approximation of the $\mathcal T$-matrix.
(b) Self-energy in the ladder approximation.
(c) Second order self-energy. 
Solid blue and dashed/solid black lines indicate, respectively, impurities and condensate/non-condensate bath bosons.}
\end{figure}
 
In Ref.~\cite{RathSchmidt2013} a quantum field theoretical resummation approach was applied to analyze the spectrum of Eq.~\eqref{BeyondFrohlich}. In analogy with the ladder approximation for the Fermi polaron one includes the Feynman diagrams shown in Fig.~\ref{fig:FeynmanFig}, which gives for the impurity self-energy 
\begin{align}\label{ladderBose}
\Sigma(\bk,\omega)=&n_0{\mathcal T}(\bk,\omega)+\sum_{\bq}u_{\bq}^2n_B(E_{\bq})\mathcal T(\bk+\bq,\omega+E_\bq)\nonumber\\
&+\sum_{\bq} v_{\bq}^2[1+n_B(E_{\bq})] \mathcal T(\bk+\bq,\omega-E_\bq)
\end{align}
where $n_B(E)=1/[\exp(E/T)-1]$ is the Bose distribution function. The in-medium scattering matrix is again given by Eq.~\eqref{Tmatrix} but now with  
\begin{equation} 
\Pi(\bk,\omega)=\sum_{\bp}\big\{ \frac{u_{\bp}^2[1+n_B(E_{\bp})]}{\omega-\epsilon_{\bk+\bp }-E_{{\mathbf p }}} +\frac{v_{\bp}^2n_B(E_{\bp})}{\omega-\epsilon_{\bk+\bp }+E_{{\mathbf p }}}\big\}
\label{PairpropagatorBEC}
\end{equation}
the boson-impurity pair propagator in the presence of the BEC.
The first term in Eq.~\eqref{ladderBose} describes the scattering of the impurity on bosons in the condensate with density $n_0$ as illustrated by the first diagram in Fig.~\ref{fig:FeynmanFig}(b), whereas the second and third terms describe the scattering on bosons excited out of the condensate either due to temperature or boson-boson interactions as illustrated by the second diagram in Fig.~\ref{fig:FeynmanFig}(b). The ladder approximation includes two-body correlations between the impurity and the bosons exactly at the vacuum level but ignores $n\ge3$-body correlations. Note that the last term in Eq.~\eqref{BeyondFrohlich} describing the scattering of the impurity on bosons outside the condensate is crucial for obtaining this, which technically is why the Fr\"ohlich Hamiltonian is insufficient to describe strong interactions~\footnote{A renormalization group analysis of the relevance of coupling constants was performed in \cite{von2023superconductivity}.}.

The first term for the self-energy in Eq.~\eqref{ladderBose} gives the following self-consistent equation or the energy $\varepsilon $ of a zero momentum Bose polaron at zero temperature,
\beq
\frac{\varepsilon}{\epsilon_n}=\frac{2m_b}{3\pi m_r}\frac{1}{(k_na)^{-1}-\Xi\left(\varepsilon/\epsilon_n,k_n\xi,m/m_b\right)}.
\label{eq:BosePolaronEnergyLadder}
\eeq
We define $k_n=(6\pi^2 n_0)^{1/3}$ and $\epsilon_n=k_n^2/2m_b$ as a characteristic momentum and energy, and the BEC healing length $\xi=1/\sqrt{2 m_b \mu}$. 
For $m=m_b$, the quantity $\Xi$ takes a remarkably simple analytic form,
\beq
\Xi(e,x,1)=\frac{1-e\, x^2{\rm arctanh}\left(\sqrt{1+e x^2}\right)}
{\sqrt{2}\,\pi\, x\, \sqrt{1+e x^2}}.
\eeq
The ladder approximation therefore predicts that the energy of the Bose polaron measured in units of $\epsilon_n$ depends only on the impurity-boson scattering length $a$, on the bath-bath scattering length $a_b$ through the BEC healing length $\xi$,
and on the mass ratio $m/m_b$. 

\begin{figure}[t]
\centering
\includegraphics[width=\columnwidth]{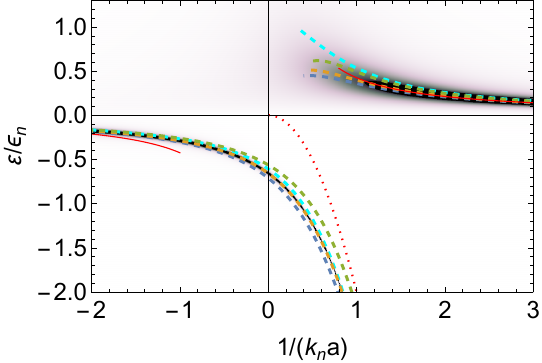}
\caption{\label{fig:BosePolaronSpectrumWithOneBogoliubovMode}
\textbf{\bf Impurity spectral function in a BEC} 
obtained from the ladder approximation for equal masses $m=m_b$, $k_n\xi=2$, and zero momentum. The dashed lines show the energies computed from Eq.~\eqref{eq:BosePolaronEnergyLadder} for $k_n\xi=1$ (green), 2 (yellow) and $\infty$ (blue). The last value corresponds to  an ideal BEC. For comparison, the cyan dashed lines are the attractive and repulsive Fermi polaron. The red solid and dotted lines are the mean-field energy and the energy of the vacuum dimer.}
\end{figure}

The spectral function $A(\bk,\omega)=-2\text{Im}G(\bk,\omega)$ of an impurity in a BEC obtained from the ladder approximation is plotted in Fig.~\ref{fig:BosePolaronSpectrumWithOneBogoliubovMode} for zero momentum ($\bk=0$), mass ratio $m/m_b=1$, and different boson-boson scattering lengths. We see that it is very similar to that of the Fermi polaron, also plotted in Fig.~\ref{fig:BosePolaronSpectrumWithOneBogoliubovMode}. From this we identify the sharp quasiparticle peak at negative energies with an attractive Bose polaron whose energy approaches the bound state energy $-1/2m_ra^2$ for $a\rightarrow 0_+$, and the broader peak at positive energy with a damped repulsive polaron. Contrary to the Fermi polaron, however, there is no crossing of the attractive polaron and molecule energies since they have the same quantum statistics and therefore can hybridize. The broadening of the upper branch comes from decay of the polaron into a continuum. Its correct description, however, requires the inclusion of $n>2$ body correlations ignored in the ladder approximation. This remains an open and challenging problem, as in the case of the Fermi polaron.

Figure \ref{fig:BosePolaronSpectrumWithOneBogoliubovMode} shows that the ladder approximation predicts the Bose polaron energy to depend rather weakly on the boson-boson scattering length $a_b$ via the BEC healing length. Taking an ideal BEC with $a_b=0$ gives the energy $\varepsilon=-(2\pi^2 n_0^2)^{1/3}/m_r = -0.7115\bar\epsilon_n$ at unitarity $1/a=0$ with $\bar\epsilon_n=k_n^2/(4m_r)$.\footnote{We introduce here $\bar\epsilon_n=k_n^2/(4m_r)$, defined in terms of the reduced two-body mass $m_r$. For $m=m_b$, $\bar\epsilon_n$ coincides with the previously-introduced $\epsilon_n=k_n^2/(2m_b)$.} For equal masses, this energy is slightly lower than the energy $\varepsilon=-0.61\epsilon_F$ of the Fermi polaron at unitarity as obtained from the ladder approximation. 
This difference comes about because scattering with the Fermi sea is suppressed due to Fermi blocking, while each boson in the BEC can freely scatter. Indeed, as we discussed in Sec.~\ref{TdependenceFermiPolaron}, the energy shift of the attractive Fermi polaron initially increases (i.e.,  the polaron energy becomes even more negative) as the temperature is raised from zero due to the reduction in Fermi blocking. The ladder approximation also gives 
\beq
Z=\frac2{3-\varepsilon/g n_0}\quad\text{and}\quad C=\frac{3\pi^2 k_n}{2}Z\left(\frac{\varepsilon}{\bar\epsilon_n}\right)^2
\eeq
for the residue and Tan's contact of the attractive polaron~\cite{Yan2020}. The effective mass is $m^*=m+z\, m_b$, where $z$ is a function that grows monotonically from $0$ in the BCS limit $1/k_na\rightarrow -\infty$ to $1$ in the BEC limit $1/k_na\rightarrow +\infty$. At unitarity one finds $Z=2/3$, $z=1/(3+2m_b/m)$ and $C=4.997 k_n$. These results for an ideal BEC are however an artifact of the ladder approximation. In fact, the polaron energy in an ideal BEC approaches the mean-field value (i.e., it diverges to $-\infty$ approaching unitarity) while its residue vanishes due to a macroscopic number of bosons in its dressing cloud as will be discussed in Sec.~\ref{sec:BOC}. This is well beyond the reach of the ladder approximation, which  describes correlations between the impurity and at most one boson at a time. The ladder approximation was also used to identify attractive and repulsive Bose polarons in 2D~\cite{CardenasCastillo23}.

All in all, the ladder approximation suggests that the behavior of Bose polarons should be rather similar to that of Fermi polarons where two well-defined quasiparticles exist, and where, as the only major difference, the polaron-to-molecule transition in the ground state is replaced by a smooth crossover. However, while these  features are qualitatively correct, it turns out that experimental observations pointed quickly to a much more involved problem which still presents many open questions, as we will be describe in the following. 

\subsection{Experiments}   \label{sec:BosePolaronExperiments}

\begin{figure}[t]
\centering
\includegraphics[width=\columnwidth]{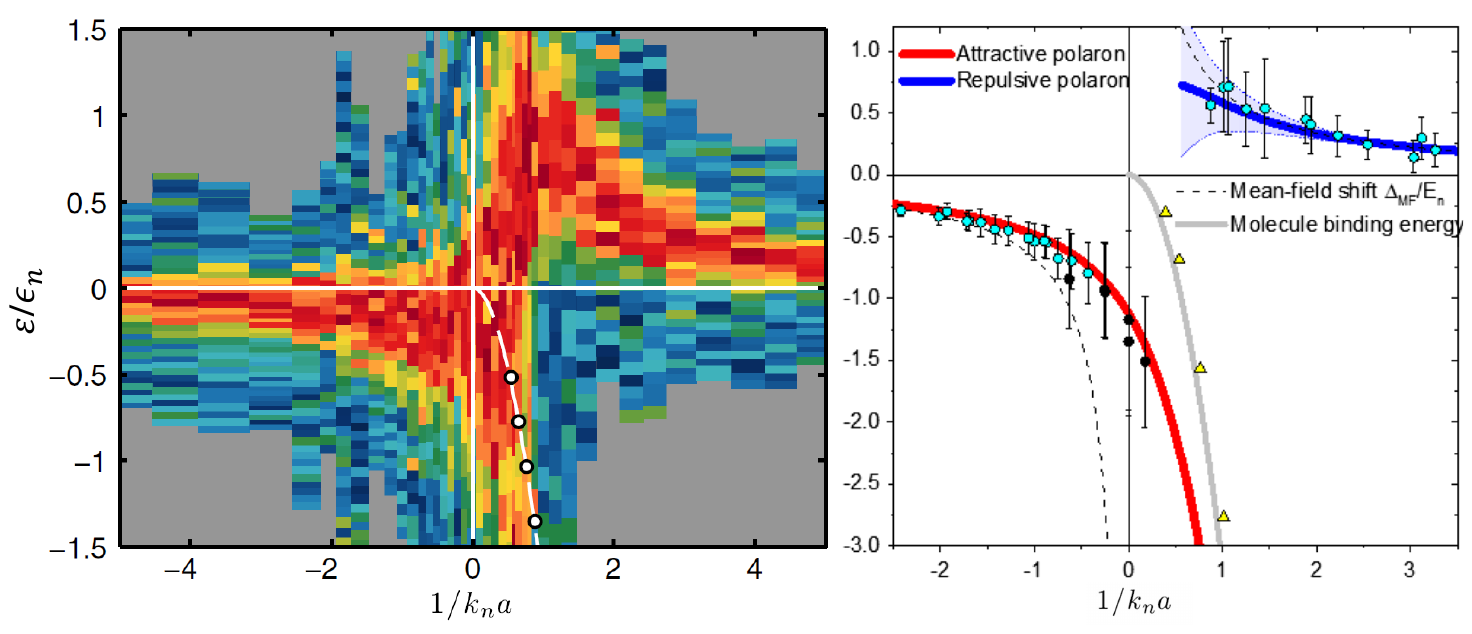}
\caption{\label{Fig:BoseSpectra}
\textbf{First observations of Bose polarons.} 
RF injection spectrum as a function of interaction strength $1/k_n a$, showing two branches corresponding to the repulsive and attractive Bose polaron. Left panel: $^{39}$K (spin) impurities in a $^{39}$K BEC. From~\citet{jorgensen2016}. Right panel: $^{40}$K impurities in a $^{87}$Rb BEC. Predictions from the ladder approximation Eq.~\eqref{BogExpansion} are shown as blue and red lines. From~\citet{hu2016}. In both panels, the molecular state was obtained from independent measurements (white dots and yellow triangles) and the corresponding lines show the two-body predictions for the energy of Feshbach molecules.}
\end{figure}

First signatures of large shifts in the RF spectrum of impurities immersed in a Bose gas were found in Ref.~\cite{Wu2012PRL}. While the large loss rate in BECs close to the unitary point $1/k_n a=0$ seemed to hinder further studies, a groundbreaking experiment showed that the momentum distribution was accessible even in this regime~\cite{makotyn_universal_2014}. The question arose whether there is a region of interaction strengths where a significant energy shift of the polaron state can be observed while losses remain moderate. This question was evaluated in detail for a single neutral impurity in a BEC by~\citet{Hohmann2015}.

The first experiments observing the Bose polaron as well-defined peaks in  the impurity spectral function measured using RF injection spectroscopy are shown in Fig.~\ref{Fig:BoseSpectra}~\cite{jorgensen2016,hu2016}. These works, and subsequent more refined ones~\cite{Ardila2019,Skou2021,Yan2020,Skou2022,morgen2023quantum,Etrych2024}, reported the presence of both an attractive and a repulsive polaron for weak to moderate interaction strengths, in agreement with the predictions of the ladder approximation and closely mimicking the picture known from the fermionic case. In Ref.~\cite{hu2016} fermionic $^{40}$K atoms were immersed in a BEC of $^{87}$Rb atoms, which has the advantage that impurity atoms are transferred between independent states, and that losses are reduced due to their fermionic character. In contrast, ~\citet{jorgensen2016} derived the impurity atoms directly from the BEC, whereas the detection relied on loss measurements of the BEC, so depletion had to be considered. In both cases the number of transferred atoms as a function of RF frequency was evaluated for interaction strengths $-3\lesssim 1/k_n a \lesssim 3$ and energy was extracted from fits to the data.

\begin{figure}[t]
\centering
\includegraphics[width=0.8\columnwidth]{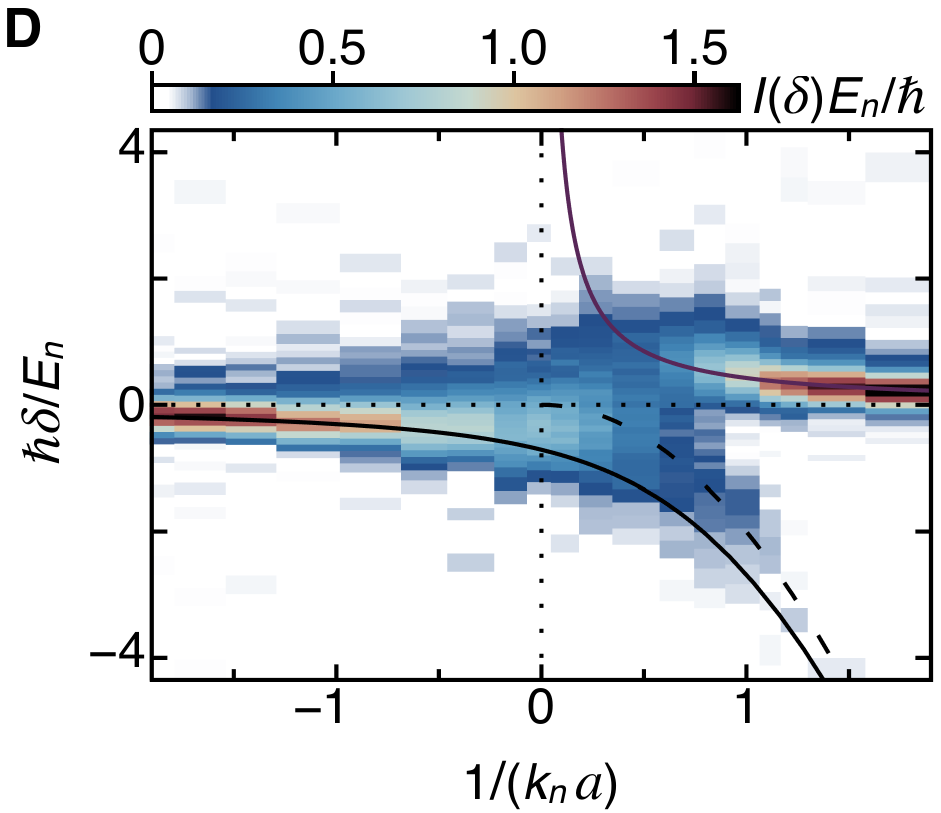}
\caption{\label{fig:BosePolaronSpectrumMeasurementsCambridge2024}
\textbf{Bose polaron spectrum in a homogeneous BEC}
for equal masses ($m=m_b$) with $^{39}$K atoms at $k_n a_b=0.008$ (i.e., $k_n\xi\approx 20$).
The solid black line is the energy of the attractive polaron given by the 1-mode Bogoliubov ansatz for an ideal BEC, the dashed black line is the energy of the dimer in vacuum, and the purple line is the mean-field  energy of the repulsive polaron. From \citet{Etrych2024}.}
\end{figure}

Figure~\ref{Fig:BoseSpectra} shows that the observed spectra are significantly broader in the strongly interacting region as compared to that of the Fermi polaron shown in Fig.~\ref{fig:FermiPolaronSpectrumExp}. Besides losses, a common source for spectral broadening is the presence of a harmonic trap causing an inhomogeneous bath density. This was addressed in a recent experiment using a box potential where the BEC is spatially homogeneous except at the trap edges~\cite{Etrych2024}. In Fig.~\ref{fig:BosePolaronSpectrumMeasurementsCambridge2024}, the corresponding impurity spectral function is shown as measured with injection RF spectroscopy. One again observes sharp spectral peaks at negative and positive energies for weak to moderate interaction strength corresponding to well-defined attractive and repulsive Bose polarons. Note, however, that the spectrum remains quite broad for strong interactions $k_n|a|>1$, with both spectral peaks acquiring a width larger than their energy in contrast to the Fermi polaron case. In fact, the spectrum has a similar width to that in the presence of a trap (compare with left panel of Fig.~\ref{Fig:BoseSpectra}), indicating that the broadening is mainly due to impurity-boson correlations and not simply the trap. The fact that all Bose polaron experiments so far have observed such broad spectra even in absence of trapping shows the challenging nature of the problem, and it even raises the basic question regarding whether well-defined Bose polarons exist for strong interactions. 

The ladder approximation for the  polaron is compared to experimental results in Fig.~\ref{Fig:BoseSpectra} (right panel) and Fig.~\ref{fig:BosePolaronSpectrumMeasurementsCambridge2024}. There is a fairly good agreement regarding the spectral function peaks even for 
strong impurity-boson interactions across different experiments especially for the attractive polaron, indicating that two-body correlations included by the ladder approximation are dominant in determining the observable impurity spectrum peaks. This holds even for an extremely weakly interacting and therefore highly compressible Bose gas with $a_b=9a_0$~\cite{Ardila2019}. Here, the discrepancies between experiment and the ladder approximation are however significantly larger than for the Fermi polaron, which points to an increased role of $n>2$-body correlations as expected due to the absence of the Pauli principle. 

To summarize, the much broader spectrum and decreased accuracy of the ladder approximation strongly  hint at physics  requiring theoretical  approaches beyond what have typically been applied for Fermi polarons. Some fundamental open questions include which length scales are important for determining the properties of the Bose polaron and which aspects are universal allowing for a unified description, whether the Bose polaron even exists as a well-defined quasiparticle for strong interactions, if there are observable states involving large numbers of bosons correlated with the impurity with lower energy, and what is the role of temperature and phase transition of the surrounding Bose gas. Progress towards answering some of these questions has been made in the past decade as we will review in the following.

\subsection{Static impurities and the  orthogonality catastrophe}   \label{sec:BOC}

We now consider the case of an infinitely massive impurity, which can be solved exactly in the case of an ideal BEC. The solution corresponds to a state involving a macroscopic number of bosons around the impurity which has zero overlap with the case of no impurity leading to an \emph{orthogonality catastrophe} (OC) in the thermodynamic limit.\footnote{The other exactly solvable case of a mobile impurity in a gas of infinitely heavy bosons was shown to connect Bose polarons and Anderson localization in the context of disorder physics~\cite{rose2022disorder}, see also \cite{breu2024impuritiestrapped1dbose}.}  Contrary to the case of fermions, in a Bose gas the OC arises also for mobile impurities. 

Treating the infinitely massive impurity as a static scattering potential, the ground state energy of an ideal BEC is simply determined by that of the lowest single-particle scattering state which yields the usual mean-field result~\cite{Huang1992,Collin2007}
\begin{equation}
\varepsilon=\frac{2\pi a}{m_r}n_0,
\label{meanFieldBosePolaronEnergy}
\end{equation}
 for $a\le 0$, where one should set $m_r=m_b$ for an infinitely massive impurity and $n_0$ is the BEC density in the absence of the impurity. The same calculation also yields  
\begin{equation}
Z=e^{-\kappa N^{1/3}(k_na)^2},
\end{equation}
for the residue for large $N$ with $\kappa=(1/3+1/4\pi^2)/3^{2/3}$. At unitarity, one obtains $Z=(8/3\pi)^{2N}$. Hence, the residue vanishes as a stretched-exponential with the number $N$ of bosons in the BEC~\cite{Guenther2021}. This gives rise to a bosonic OC~\cite{Sun2004orthogonality} in the sense that the non-interacting and interacting ground states have zero overlap in the thermodynamic limit. A similar OC was predicted by P.~W.~Anderson for a static impurity in an ideal Fermi gas, for which however the residue decays as a much slower power law~\cite{anderson_infrared_1967,combescot_1971}. The faster decay in the bosonic case arises because all bosons cluster around the impurity giving rise to a macroscopic number of particles in its dressing cloud, while only particles close to the Fermi surface participate in the fermionic case.

Unlike the Fermi polaron, the Bose polaron exhibits the OC when immersed in an ideal BEC also for a finite impurity mass \cite{shchadilova2016,Guenther2021}. This is  apparent already at the mean-field level. Indeed, the energy of an impurity in an interacting BEC with chemical potential $\mu_b=4\pi a_b n/m_b$ is for weak bath-impurity interactions given by Eq.~\eqref{meanFieldBosePolaronEnergy}. Taken together with Eq.~\eqref{DeltaN} this yields~\cite{Massignan2005}
\beq\label{DeltaN_GPE}
\Delta N=-\frac{m_b}{2m_r}\frac{a}{a_b}
\eeq 
for the number of bosons in the dressing cloud around the impurity. It follows that $\Delta N\rightarrow\infty$ when $a_b\rightarrow 0$, so that a macroscopic number of bosons are in the dressing cloud due to the infinite compressibility of an ideal BEC. Indications of this OC also emerge in perturbation theory, see Sec.~\ref{sec:weakinteractions}, and from a variational approach based on an expansion in Bogoliubov modes, see Sec.~\ref{sec:Bose_variational_Chevy}, but these schemes cannot fully describe this effect involving a diverging number of bosons in the dressing cloud. The variational approach based on the Gross-Pitaevskii equation described in Sec.~\ref{sec:VarGPE} instead recovers this mobile impurity OC, with $Z\rightarrow 0$, $\Delta N\rightarrow \infty$, and the energy approaching Eq.~\eqref{meanFieldBosePolaronEnergy} for $a_b\rightarrow 0$. 
 
\begin{figure}[t]
\centering
\includegraphics[width=\columnwidth]{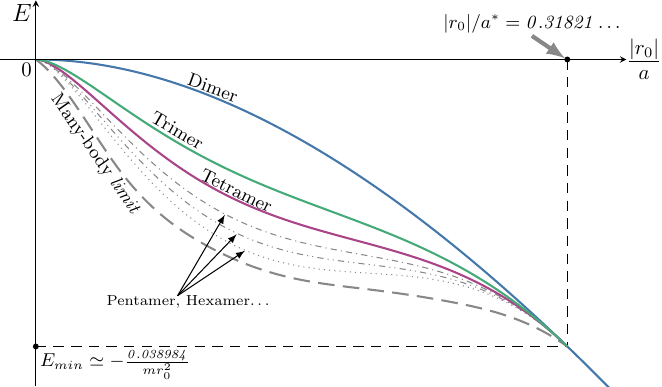}
\caption{\label{fig:MultibodyMJ}
{\bf Multi-body bound states.}
An ensemble of states consisting of $N$ non-interacting bosons is bound to an infinitely heavy impurity via a multichannel boson-impurity interaction with  effective range $r_0$ and scattering length $a$. Solid lines are microscopic calculations, whereas dashed and dotted lines are conjectures. 
From \citet{Shi2018}.}
\end{figure}

When $a>0$, there appears a two-body impurity-bath bound state with energy $\epsilon_B=-1/2m_ra^2$. For an infinite mass impurity in an ideal BEC, the ground state is then formed by all bosons in this state giving the energy $-N/2m_ba^2$. This result is a based on single channel model treating the impurity as a static potential. In atomic gases, a resonant interaction between the impurity and the bosons is however mediated by a Feshbach molecule in a different channel~\cite{chin_feshbach_2010}. The fact that an impurity can only form a Feshbach molecule with one boson at a time was argued to lead to correlations between the bosons equivalent to a repulsive 3-body force~\cite{Shi2018,Yoshida2018few}. Using a multichannel model reminiscent of the Anderson impurity model~\cite{anderson_localized_1961} for infinite impurity mass, it was shown that this repulsion significantly increases the binding energy of states involving two (trimer) and three (tetramer) bosons compared to a single channel model. As illustrated in Fig.~\ref{fig:MultibodyMJ}, close to unitarity there are states with an arbitrary number $N$ of bosons bound to the static impurity with an energy given by $\epsilon_{N+1}/\epsilon_B\simeq-N+N(N-1)\pi/\ln a$ for $a\rightarrow +\infty$ where the first term is the result of a contact single channel interaction. Within this framework, in the thermodynamic limit all bath bosons collapse within the same multi-body bound state with a finite energy independent of $N$.

\citet{Chen2018} obtained similar results for a static impurity in a gas of \emph{interacting} bosons. Using the Gross-Pitaevskii equation (GPE) as well as Path Integral Monte Carlo calculations, they found that under quite general conditions when there is a weakly bound two-body state the impurity can bind any number of bosons even when the boson-boson repulsion is non-zero. 

The time evolution of the condensate wave function for $T=0$ and the density matrix for $T>0$ ensuing the sudden insertion of a static impurity in an ideal BEC was calculated exactly by~\citet{Drescher2021}. The spectral function obtained by Fourier transforming the Ramsey signal $S(t)$ defined in Eq.~\eqref{GreaterGreens} proved to be very broad for strong attractive interactions $-1\lesssim1/k_na<0$ with the main peak given by Eq.~\eqref{meanFieldBosePolaronEnergy}. For repulsive interactions $k_na>0$, several peaks were found with an energy separation given by the $2$-body binding energy. 
 
In conclusion, the OC for an ideal BEC as well as the existence of multi-body bound states illustrate the challenging nature of the Bose polaron problem, which involves correlations between the impurity and a large number of bosons in a many-body environment. The multi-body states have zero density far away from the impurity, whereas the bath density in the Bose polaron problem remains finite everywhere with $n(r) \rightarrow n_0$ away from the impurity.  Since the size of the multi-body states rapidly increases with the number of bound bosons, one expects them to survive only provided that their size is smaller than the typical inter-particle distance $n_0^{-1/3}$.

\subsection{Weak interactions}   \label{sec:weakinteractions}

We now turn to the case when the impurity-boson interactions are weak. In this case one can apply diagrammatic perturbation theory to derive reliable results. As we shall see, signs of the importance of three-body correlations already show up at this level. 

To apply perturbation theory, we will assume that the bosonic bath is a weakly interacting BEC which can be described by the Bogoliubov approximation leading to Eq.~\eqref{BeyondFrohlich}, and we will calculate the impurity properties as an expansion in the impurity-boson interaction strength. The first- and second-order diagrams for the impurity self-energy are shown in Fig.~\ref{fig:FeynmanFig}. Note that the ladder approximation described in Sec.~\ref{sec:ladder} includes only the first second-order diagram in Fig.~\ref{fig:FeynmanFig}, which however is the lowest order in the BEC gas parameter and therefore dominates in a weakly interacting Bose gas. Including all diagrams up third order for the impurity self-energy $\Sigma(\bk,\omega)$ yields~\cite{Christensen2015}
\begin{align}\label{EnergyPert}
\varepsilon=&\frac{2\pi n_0a}{m_r}\left\{1+A(\alpha)\frac{a}{\xi}+\right.\\\nonumber
&\left[B(\alpha)\frac{a^2}{\xi^2}+\left.\tilde B(\alpha)\frac{aa_b}{\xi^2}\right]\ln(a^*/\xi)+\mathcal{O}(n_0{a^*}^3)\right\}
\end{align}
for the energy of a zero-momentum Bose polaron. Here 
$a^*=\max(a,a_b)$ and $\alpha=m/m_b$ is the mass ratio. The function $A(\alpha)$ is given in Refs.~\cite{Novikov2009,Casteels2014,Christensen2015}, whereas $B(\alpha)$ is given in Ref.~\cite{Christensen2015} and $\tilde B(1)$ in Ref.~\cite{Levinsen2017}. For later reference, we give the values $A(1)=8\sqrt{2}/(3\pi)$ and $A(\infty)=\sqrt{2}$. The energy is non-analytic in both the boson-boson scattering length $a_b$ and the impurity-boson scattering length $a$, and 
Eq.~\eqref{EnergyPert} recovers
the two leading terms for the chemical potential of a weakly interacting Bose gas including the Lee-Huang-Yang term~\cite{fetter_1971} when 
 $a=a_b$ and $m=m_b$. 

The logarithmic, third-order term in Eq.~\eqref{EnergyPert} arises from three-body correlations, and can be understood as a perturbative precursor of Efimov physics. Interestingly, it has the same form as that derived by Wu for the chemical potential~\cite{Wu1959}. Note that the Fr\"ohlich Hamiltonian recovers only the first and second order terms for the energy whereas the full Hamiltonian Eq.~\eqref{BeyondFrohlich} is needed to calculate third order terms and beyond. Likewise, perturbation theory yields  
 \begin{align}
 Z&=1-C(\alpha)\frac{a^2}{a_b\xi}+{\mathcal O}(n_0{a^*}^3)\label{ResiduePert}\\
\frac{m^*}m&=1+F(\alpha)\frac{a^2}{a_b\xi}+{\mathcal O}(n_0{a^*}^3)\label{MassPert}
\end{align}
for the polaron residue and effective mass. The functions $C(\alpha)$ and $F(\alpha)$ are given in Ref.~\cite{Christensen2015}. In particular, for a heavy impurity one finds $C(\infty)=1/(\sqrt{2} \pi)$ and $F(\alpha\rightarrow\infty)=1/(3\sqrt{2}\alpha)$.

While the expansion for the energy given by Eq.~\eqref{EnergyPert} indicates that perturbation theory is accurate for $|a|/\xi\ll 1$, Eq.~\eqref{ResiduePert} gives the additional condition $a^2/(a_b\xi)\ll 1$ in order for the residue to be close to unity. Since $\xi\propto 1/\sqrt{n_0 a_b}$, this may be written as $\sqrt{n_0 a^4/a_b}\ll1$, showing that perturbation theory becomes unreliable when $a_b\rightarrow 0$. Physically, this reflects that a weakly-interacting Bose gas is strongly affected around the impurity. Equation~\eqref{ResiduePert} is indeed a perturbative hint at the OC with  $Z\rightarrow 0$ arising for an ideal BEC irrespectively of the impurity mass, as discussed in Sec.~\ref{sec:BOC}. Although this prediction is, of course, well beyond the range of validity of perturbation theory, we shall see in Sec.~\ref{sec:VarGPE} that a variational theory taking into account large deformations of the BEC predicts that Eqs.~\eqref{EnergyPert}-\eqref{ResiduePert} hold in a surprisingly large range of interaction strengths.

Perturbation theory for the Bose polaron has also been performed in 2D~\cite{Pastukhov2018a,Pena2020}. Care has to be taken since the 2D scattering matrix in Eq.~\eqref{Tmatrices} depends logarithmically on the scattering energy. This results in a logarithmic dependence on the scattering length of the polaron energy, which at zero momentum and for $m=m_b$ reads 
\begin{equation}\label{2DBosePolaronEnergy} 
\varepsilon = \frac{4 \epsilon_n}{\ln(4\pi) - 2\ln(k_n a)},
\end{equation}
where $k_n = \sqrt{4 \pi n_0}$ in 2D. A similar logarithmic behavior emerges also for the effective mass and residue. As will see further in Sec.~\ref{TMDsection}, logarithmic dependencies are typical in 2D systems.

\subsection{Expansion in bath excitations}   \label{sec:Bose_variational_Chevy}

As we have seen, there is no reason to expect $n\ge 3$ correlations to be negligible for the Bose polaron. A systematic way to analyse such correlations is to use a variational polaron wave function based on expanding in the number of Bogoliubov excitations that the impurity creates in the BEC, in close analogy with the Chevy ansatz Eq.~\eqref{eq:ChevyAnsatz} for the Fermi polaron. Starting again from Eq.~\eqref{BeyondFrohlich}, this expansion reads
\begin{align} 
 |\Psi_{\mathbf p}\rangle=\bigg(&\sqrt{Z_{\mathbf p}}\hat c_{\mathbf p}^\dagger + \sum_{\bk\neq 0}\alpha_{\bp,\bk}\hat c_{\mathbf p-\mathbf k}^\dagger\hat\gamma_{\mathbf k}^\dagger+\nonumber\\
& \sum_{\bk,\bk'\neq 0}\alpha_{\bp,\bk,\bk'}\hat c_{\bp-\bk-\bk'}^\dagger\hat\gamma_{\bk}^\dagger\hat\gamma_{\bk'}^\dagger +\ldots\bigg)|\text{BEC}\rangle
\label{BogExpansion}
\end{align}
where $|{\rm BEC}\rangle$ is the ground state of a weakly interacting BEC in absence of the impurity defined by $\hat\gamma_\bk|{\rm BEC}\rangle=0$ and ${\mathbf p}$ is the total momentum. Correlations between the impurity and $n$ bosons can now be described by including terms with up to $n$ Bogoliubov modes in Eq.~\eqref{BogExpansion}. The variational parameters $\sqrt{Z_{\bp}},\alpha_{\bp,\bk}, \ldots$ are then determined by minimizing the energy $\langle\Psi_{\mathbf p}|H|\Psi_{\mathbf p}\rangle$. Equation~\eqref{BogExpansion} was first used in Ref.~\cite{Li2014}, truncating it after the first two terms (including a single Bogoliubov mode), which is equivalent to the ladder approximation discussed above.

\begin{figure}[t]
\centering
\includegraphics[width=\columnwidth]{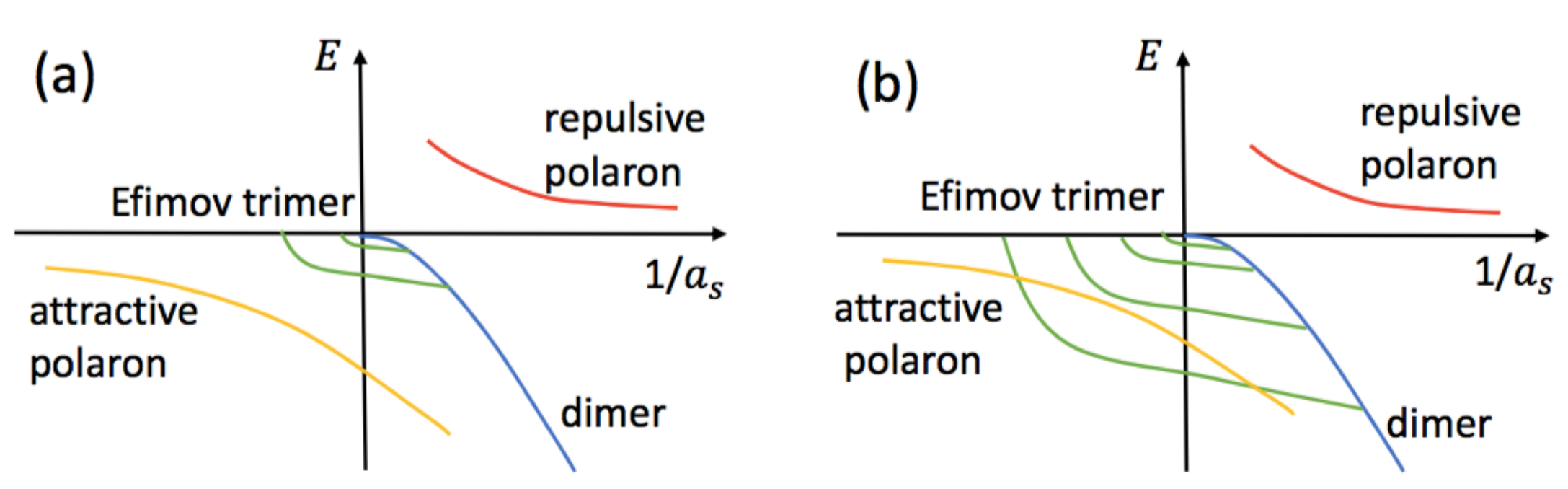}
\includegraphics[width=\columnwidth]{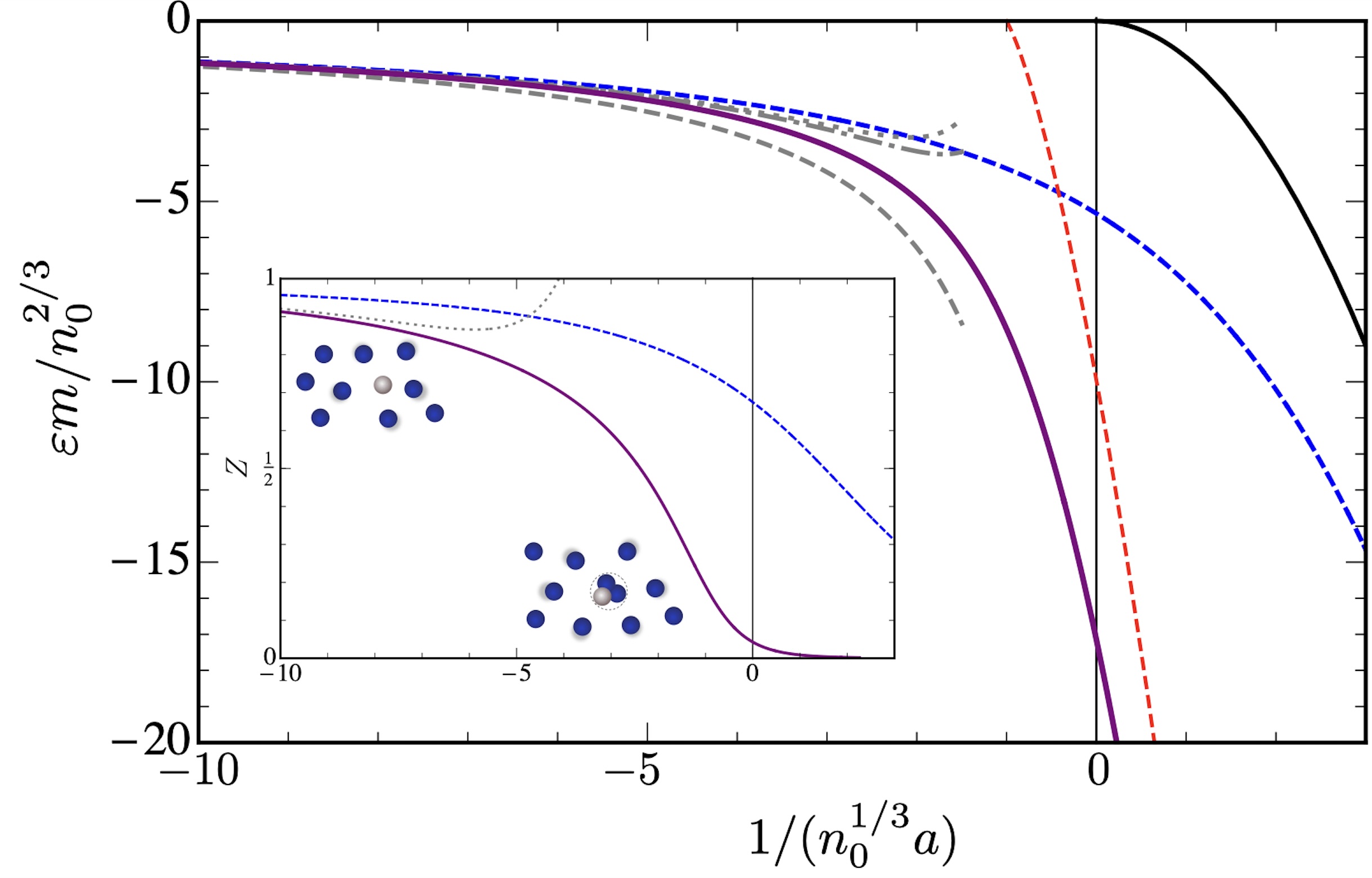}
\caption{\label{fig:Bose_polaron_hybridization}
\textbf{Efimov trimers and the Bose polaron.}
Top: When $n|a_-|^3\gg 1$ (left panel), all Efimov trimers lie  close to the free-particle continuum and have negligible effects on the attractive polaron, whereas for $na_-^3\simeq - 1$ (right panel) the lowest Efimov trimer hybridizes with the attractive polaron, lowering notably its energy. From~\citet{Sun2017}. Bottom: The energy of the attractive Bose polaron with $m=m_b$ for $na_-^3=-1$ and $a_-=-50a_b$ from Eq.~\eqref{BogExpansion} including one (blue dashed) and two (solid purple) Bogoliubov modes. The gray dashed, dot-dashed, and dotted lines are the perturbative results Eq.~\eqref{EnergyPert} to first, second, and third order. The red-dashed and black-solid lines are the energies of the ground trimer and dimer in vacuum. The inset shows the quasiparticle residue $Z$. From~\citet{Levinsen2015}.}
\end{figure}

To explore the effects of three-body correlations and Efimov trimers on the Bose polaron, the variational ansatz Eq.~\eqref{BogExpansion} was employed including up to two Bogoliubov modes~\cite{Levinsen2015,Sun2017,Sun2017b}. The ground Efimov trimer emerges from the continuum at the scattering length $a_-<0$, which introduces a three-body length scale depending on short range physics. Since the Efimov state has a size $\sim |a_-|$ and a binding energy $\sim 1/m_r a_-^2$, one would intuitively expect that it is destroyed by many-body effects when $n|a_-|^3\gg 1$ so that it has little influence on the Bose polaron, whereas its presence becomes relevant when $na_-^3\simeq -1$~\cite{Sun2017}, see top panel of Fig.~\ref{fig:Bose_polaron_hybridization}. Likewise, deeply-bound Efimov states for $n|a_-|^3\ll 1$ would likely have little effect on the visible spectrum. The bottom panel of Fig.~\ref{fig:Bose_polaron_hybridization} shows the attractive polaron energy calculated from the ansatz Eq.~\eqref{BogExpansion} including one and two Bogoliubov modes as well as the perturbative results up to third order given by Eq.~\eqref{EnergyPert} for $na_-^3=-1$~\cite{Levinsen2015}. One clearly sees an avoided crossing between the polaron and the Efimov trimer due to hybridization, which lowers its energy. Correspondingly, the inset shows that the residue becomes very small when the polaron hybridizes with the trimer.

\begin{figure}[t]
\centering
\includegraphics[width=\columnwidth]{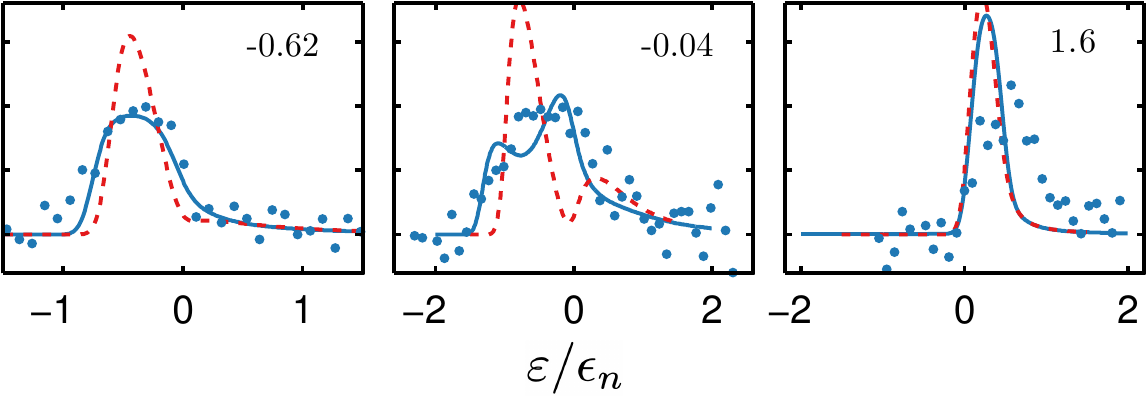}
\caption{\label{fig:Bose_polaron_twoBog}
\textbf{Expansion in Bogoliubov modes.} 
The impurity spectral function in the equal masses case as measured in the Aarhus experiment for three values of the interaction strength $1/k_na$. Lines show variational results obtained from the ansatz Eq.~\eqref{BogExpansion} including up to one (red dashed) and two (solid blue) Bogoliubov modes. The theory curves are trap-averaged and take into account the Fourier broadening of the RF pulse. From~\citet{jorgensen2016}.}
\end{figure}

So far, clear signatures of Efimov states on the Bose polaron spectrum remain unobserved, which may be changed by using light impurities~\cite{Sun2017,Sun2017b}. The inclusion of two Bogoliubov modes however turns out to improve the agreement with experimental data, especially in the spectral region between the two branches for strong interactions, as visible in Fig.~\ref{fig:Bose_polaron_twoBog}. The expansion of Eq.~\eqref{BogExpansion} was extended further to include up to three Bogoliubov modes assuming a resonant impurity-boson interaction with $1/a=0$~\cite{Yoshida2018}. Taking the limit $a_b\rightarrow 0$ of an ideal BEC, it was found that the polaron energy is strongly affected by the Efimov trimers even when $n|a_-|^3\gg 1$, at odds with the intuition above. We will discuss this point in more detail in Sec.~\ref{sec:BosePolaronUnitarity}. In Ref.~\cite{Nakano2024}, the variational wave function Eq.~\eqref{BogExpansion} was used to explore the Bose polaron in 2D and significant differences in the spectral function were found between including one and two Bogoliubov modes in the wave function indicating the importance of $n\ge 3$ body correlations.

In general, the variational ansatz in Eq.~\eqref{BogExpansion} yields valuable insights into the properties of the Bose polaron and the role of $n\geq 3$ body correlations. A disadvantage of this approach is that the truncation at a small number of Bogoliubov modes excludes the description of states involving a large number of bosons correlated with the impurity and in particular the OC. A related disadvantage is that the expansion assumes a uniform condensate and thus cannot include the back-action of the impurity on the condensate wave function. Also, the Bogoliubov approximation treats the boson-boson repulsion at the quadratic level, thereby neglecting phonon-phonon repulsion. This can lead to unphysical behavior for states where the gas density is large in the vicinity of the impurity. However, such states typically have a very small spectral weight and are therefore difficult to observe. Indeed, the variational ansatz  describes quite well most observable spectral features.

\subsection{Gross-Pitaevskii approach}   \label{sec:VarGPE}

We now discuss a different variational wave function closely related to the Gross-Pitaevskii (GP) mean-field theory~\cite{Gross1962,pitaevskii1961vortex}. Compared to the expansion in BEC excitations, Eq.~\eqref{BogExpansion}, this wave function has the advantage that it naturally describes the back-action of the impurity on the BEC. It also takes into account the boson-boson repulsion energy to quartic order at the mean-field level. It should thus be well suited to describe the large deformations of the BEC around the impurity involving many bosons in weakly interacting BECs 
leading to the OC for $a_b\rightarrow 0$, see Sec.~\ref{sec:BOC}. On the other hand, this wave function assumes that the BEC instantly adjusts to the motion of the impurity 
akin to the Born-Oppenheimer approximation and is therefore expected to be most accurate for heavy impurities. As we shall see, this variational approach indeed agrees very well with diffusion Monte Carlo (DMC) calculations for an infinite mass impurity. Finally, this mean-field approach neglects higher order correlations such as Efimov states.

As a warm-up, consider first a very heavy impurity with $m_r\approx m_b$ moving with constant velocity ${\bf v}$ through a weakly interacting BEC. The mean-field GP energy functional is given by~\cite{Astrakharchik2004}
\begin{equation}
E=\int\!\left[\frac{|\nabla\psi|^2}{2m}+
V({\bf r}-{\bf v}t)|\psi|^2+\frac{g_b}{2}|\psi|^4\right] d^3x.
\label{GPE:energy functional}
\end{equation}
Treating the impurity as a  perturbation, the condensate wave function can be split into an unperturbed part $\phi_0=\sqrt{n_0}$ plus
a small correction (dressing cloud): $\psi({\bf r},t)=\phi_0+\delta\psi({\bf r},t)$. For a static impurity (${\bf v}=0$), we have $\delta\psi_{{\bf k}}=-V({\bf k})\phi_0/[\frac{\hbar^2k^2}{2m}+2\mu]$ in momentum space. It follows that for a contact potential $V({\bf r})$ 
relevant for neutral atoms, the dressing cloud has the Yukawa form 
\begin{equation}\label{eq:Yukawa}
\delta\psi(r) 
= -a\frac{e^{-\sqrt{2}r/\xi}}{r}\phi_0
\end{equation}
with a size set by the healing length $\xi$ of the BEC. Likewise, the energy can be calculated from Eq.~(\ref{GPE:energy functional}) in a perturbative manner giving 
\begin{equation}
E=E_0+\frac{2\pi n_0 a}{m_b}\left(1 +\sqrt{2}\frac{a}{\xi}\right) + \frac12m_{\rm ind}v^2,
\label{EnergyGPstaticImpurityContactInteraction}
\end{equation}
where $E_0$ is the energy of the BEC in the absence of the impurity and $m_{\rm ind}=m_ba^2/(3\sqrt{2}a_b\xi)$ is the ``induced" mass of fluid moving with the impurity. The second term in Eq.~\eqref{EnergyGPstaticImpurityContactInteraction} agrees with Eq.~\eqref{EnergyPert} for the energy of an infinitely heavy impurity up to second order in the impurity-boson interaction. Likewise, by writing $m^*=m+m_{\rm ind}$ one recovers Eq.~\eqref{MassPert} 
taking the mass ratio to infinity. This illustrates how these perturbative results can be obtained using two different approaches. Interestingly, the induced mass is directly related to the suppression of the superfluid density $n_s$ of the BEC caused by the presence of impurities with concentration $n_i$ as $n_s=n-n_i\, m_\text{ind}/m_b$~\cite{Huang1992,Astrakharchik2002}. 

The GP treatment presented until now applies to a heavy impurity in the perturbative regime. We now consider the general situation of a mobile impurity and arbitrary interaction strengths. To do so, we switch to the reference frame where the impurity is at rest by applying the Lee-Low-Pines (LLP) transformation $\hat H_\text{LLP}=\hat U_\text{LLP}^\dagger \hat H_B\hat U_\text{LLP}$  with $\hat U_\text{LLP}=e^{-i\hat W}$ and $\hat W=\hat {\mathbf R}\cdot \sum_j\hat{\mathbf p}_j$, where $\hat{\mathbf R}$ is the position of the impurity~\cite{Lee1953,Girardeau1961,shchadilova2016}. Applying this to the  Hamiltonian, Eq.~\eqref{eq:HB}, one finds
\begin{align}\label{HLLP}
\hat H_\text{LLP}=&\sum_j\left[\frac{\hat {\mathbf p}_j^2}{2m_b}+V({\hat {\mathbf r}}_j)\right]+\sum_{i<j} V_b(\hat{\mathbf r}_i-\hat{\mathbf r}_j)
\nonumber\\
&+\frac{(\hat {\mathbf p}_0-\sum_j\hat {\mathbf p}_j)^2}{2m}
\end{align}
where $\hat{\mathbf p}_0$ is the total momentum operator.
In this way, the impurity motional degrees of freedom 
have been eliminated, giving rise to an entanglement of the boson momenta in the last term of Eq.~\eqref{HLLP}. Assuming a weak and short-ranged boson-boson repulsion parametrized by $g_b = 4\pi a_b/m_b$, one can now apply GP  theory. For zero total momentum $\bp_0=0$ and a spherically symmetric ground state the last term in Eq.~\eqref{HLLP} reduces to $\sum_j \hat\bp^2_j/2m$, and minimizing the energy functional yields \cite{Guenther2021,Schmidt2022}
\beq\label{eq:GPe}
\left[-\frac{\hbar^2\nabla^2}{2m_r}+V(r)+g_b|\phi(r)|^2\right]\phi(r) = \mu \phi(r).
\eeq 
This is similar to the standard GPE with the presence of the impurity entering through the static scattering potential $V(r)$ and the impurity-boson reduced mass $m_r=1/(m_b^{-1}+m^{-1})$. Equation~\eqref{eq:GPe} can also be derived directly from Eq.~\eqref{eq:HB} using the variational ansatz
\begin{equation}\label{GPEWavefn}
|\Psi\rangle=\int d^3r\,\hat{c}^\dagger_\br\,e^{\int \! d \bs\,[\phi(\br-\bs)\hat b^\dagger_\bs-\text{H.c.}]}|0\rangle,
\end{equation}
where $\hat{c}^\dagger_\br$ $(\hat{b}^\dagger_\br)$ creates an impurity (a boson) at position $\br$. Equation~\eqref{GPEWavefn} describes a BEC in a coherent state $\hat b^\dagger_\bs|\phi(\br)\rangle=\phi(\br-\bs)|\phi(\br)\rangle$ following instantaneously the impurity located at $\br$ in the spirit of the Born-Oppenheimer approximation. This shows explicitly that this approach should be most accurate for heavy impurities whose motion is much slower than that of the bosons. 

Due to the presence of the reduced mass in the kinetic term of Eq.~\eqref{eq:GPe}, balancing the kinetic and mean-field terms yields the ``modified healing length" $\bar \xi=1/\sqrt{2 m_r \mu}$ as a characteristic length in the problem. In the limit of infinite impurity mass, $\bar \xi$ reduces to the usual healing length $\xi$ of a weakly-interacting BEC [which was defined right after Eq.~\eqref{eq:BosePolaronEnergyLadder}]. Introducing ${\mathbf x}={\mathbf r}/\bar \xi$ and $\tilde\phi({\mathbf x})=\phi(x\bar \xi)/\sqrt{n_0}$, Eq.~\eqref{eq:GPe} can be written in the dimensionless form $[-\nabla^2_x+2m_r\bar\xi^2 V(x)+|\tilde \phi(x)|^2-1]\tilde\phi(x)=0$. For the polaron energy $\varepsilon$, i.e.\ the ground state energy measured with respect to the homogeneous solution $|\tilde\phi({\mathbf x})|^2=1$ in the absence of the impurity, we obtain~\cite{Massignan2005,Guenther2021,Schmidt2022} 
\begin{equation}
\varepsilon=-\frac12E_{\bar \xi}\int\!d^3r\,[|\tilde\phi(\br)|^4-1]
\label{EnergyGP}
\end{equation}
with the characteristic energy scale $E_{\bar \xi}=n_0\bar \xi/(2m_r)$. Likewise, the residue and number of particles in the dressing cloud around the BEC can be expressed as 
\begin{equation}
Z=e^{-N_{\bar \xi}\int\!d^3r|\tilde\phi(\br)-1|^2}\hspace{0.2cm}\text{,}\hspace{0.2cm}
\frac{\Delta N}{N_{\bar\xi}}=\int\!d^3r[|\tilde\phi(\br)|^2-1]\label{DeltaNGP}
\end{equation}
where $N_{\bar \xi}=n_0\bar \xi^3$. Equations \eqref{EnergyGP}-\eqref{DeltaNGP} show that within this variational approach the effects of the boson-boson interaction enter through the (modified) healing length in agreement with the ladder approximation.

The mean-field GP equation is reliable only when the gas parameter $n(r)a_b^3$ remains small everywhere, even in the vicinity of the impurity where the bath density $n(r)$ may grow rapidly. A detailed perturbative analysis showed that this condition is satisfied when $(n_0a_b^3)^{1/4}a_b\ll R$. Here $n_0$ is the bulk bath density and $R=4\pi/\int\!d^3r|\psi_0|^4$ is a typical range of the potential, with $\psi_0$ the zero energy solution of the single particle Schr\"odinger equation with the impurity-boson interaction potential $V(\mathbf r)$ normalized as $\psi_0\rightarrow 1/r$ for large $r$~\cite{Massignan2021a,Yegovtsev2022}. When $|a|^3\ll \xi^2R$ and for infinite impurity mass, Eq.~\eqref{EnergyGP} recovers the perturbative expression Eqs.~\eqref{EnergyPert} for the  polaron energy to second order, and $\Delta N$ in Eq.~\eqref{DeltaNGP} recovers Eq.~\eqref{DeltaN_GPE}. Likewise, the  residue and Tan's contact become~\cite{Guenther2021}
\beq\label{BosePolaronResidueLinearizedGPE}
Z=e^{-\sqrt{2}\pi n_0\bar \xi a^2}\quad\text{and}\quad C=16\pi^2 n_0 a^2.
\eeq
When expanded, the residue has the same functional dependence as Eq.~\eqref{ResiduePert} although with a slightly different prefactor. Remarkably, the condition $|a|^3\ll \xi^2R$, which may be rewritten as $n_0|a|^3\ll R/a_b$, suggests that when $a_b\rightarrow 0$ these expression hold even for large $|a|$, which is well beyond the expected range of perturbation theory. In particular, this variational ansatz recovers the OC where the residue vanishes and the number of particles in the dressing cloud diverges as $a_b\rightarrow 0$ also for a finite mass impurity mass as discussed in Sec.~\ref{sec:BOC}.

\begin{figure}
\centering
\includegraphics[width=\columnwidth]{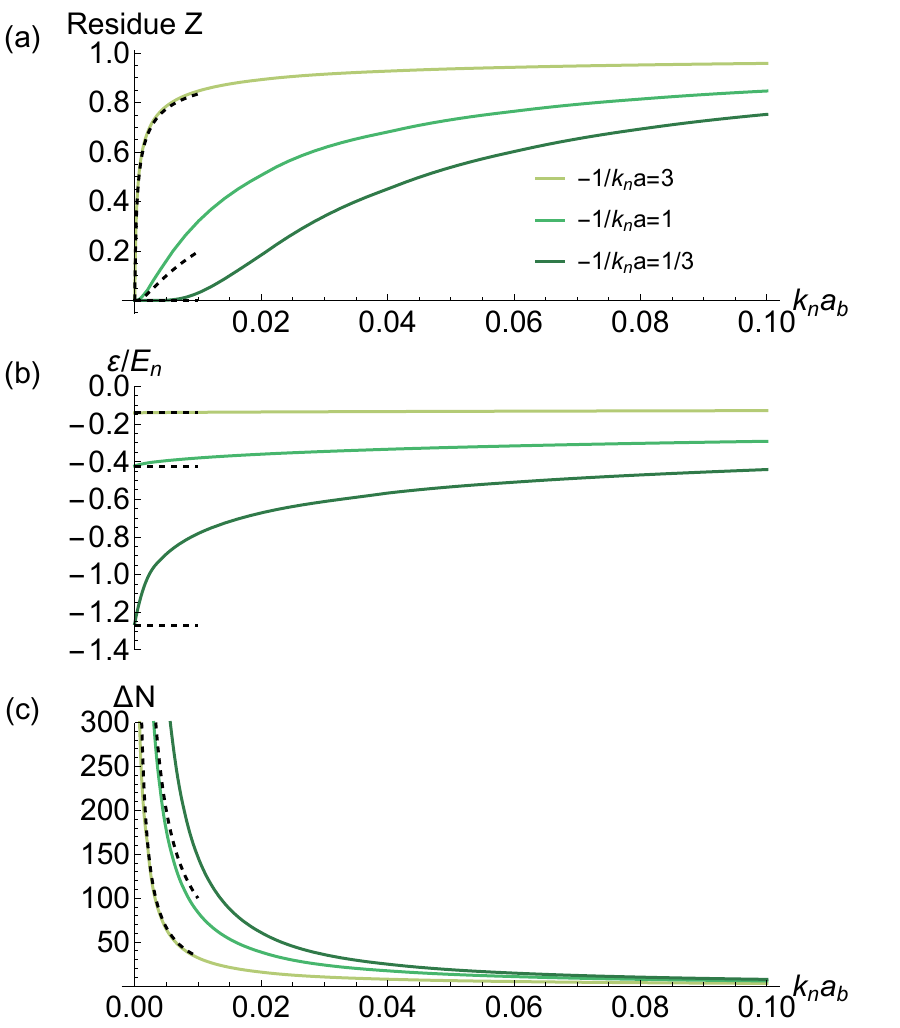}
\caption{\label{fig:quasiparticle_properties_within_GPE}
\textbf{Properties of the Bose polaron.} 
The residue (a), energy (b), and number of particles in the dressing cloud (c) of the Bose polaron as a function of $k_n a_b$ for $-1/k_n a = (3, 1, 1/3)$ (darker lines correspond to stronger boson-impurity attraction). In all panels, we consider mobile impurities with $m=m_b$ and $k_n r_e=0.05$, where $r_e$ is the effective range of $V(r)$. The dashed lines are the perturbative expressions Eqs.~\eqref{meanFieldBosePolaronEnergy}, 
\eqref{DeltaN_GPE}, and \eqref{BosePolaronResidueLinearizedGPE} valid in the region $|a|^3\ll \xi^2R$ where the OC takes place. From~\citet{Guenther2021}.}
\end{figure}

Figure~\ref{fig:quasiparticle_properties_within_GPE} plots the polaron residue, energy, and number of particles in the dressing cloud obtained from Eq.~\eqref{eq:GPe} as a function of $k_na_b$ for different boson-impurity scattering lengths. It clearly shows how the polaron becomes increasingly dressed with a smaller residue and lower energy as the BEC gets softer. In particular, when $a_b\rightarrow 0$ one finds $Z\rightarrow 0$, $\varepsilon \rightarrow 2\pi a n_0/m_r$ and $\Delta N\rightarrow \infty$, as expected for the bosonic OC. We have assumed a zero-range boson-boson interaction $V_b(\br)=g_b\delta(\br)$ when deriving Eq.~\eqref{eq:GPe}. This leads to a divergence in the mean-field energy if the impurity-boson interaction is also taken to be zero range, and therefore a non-zero ranged $V(r)$ impurity-boson interaction potential must be used. Nonetheless, it was found that a wide range of experimentally relevant interaction potentials $V(r)$ with the same values of $a\leq 0$ and effective range $r_e$ yielded the same results when $k_nr_e\ll 1$~\cite{Guenther2021,Schmidt2022}. Hence, within the experimentally-relevant potentials examined, there emerged an effective two-parameter universality, in the sense that it was enough to describe the interaction with $a$ and $r_e$. 

\begin{figure}
\centering
\includegraphics[width=\columnwidth]{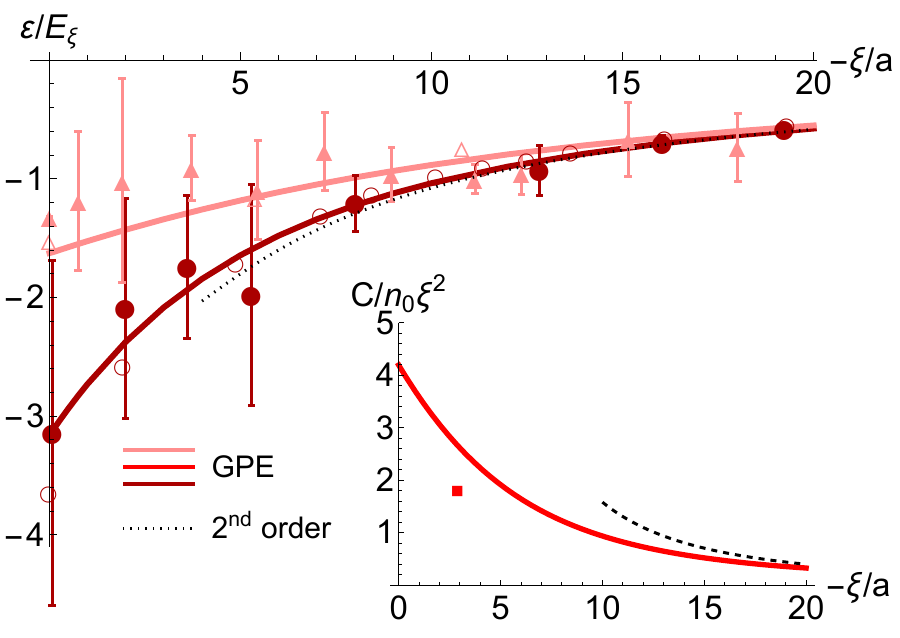}
\caption{\label{fig:energyVsaOverXi_comparison_with_many_experiments}
\textbf{Polaron energy and contact.} 
The main panel shows the polaron energy: filled triangles/circles experimental data from Aarhus~\cite{jorgensen2016} / JILA~\cite{hu2016}, whereas empty symbols are QMC data~\cite{Pena2016,Ardila2019}. The solid lines are obtained from  Eq.~\eqref{eq:GPe} for effective ranges $r_e/\xi=0.002$ (pink) and $0.02$ (dark red), corresponding to the conditions at Aarhus and JILA. The  dotted line is the perturbative result Eq.~\eqref{EnergyPert} up to second order. The inset shows Tan's contact $C$ with $r_e/\xi=0.01$. The red square is experimental data from MIT~\cite{Yan2020}, and the dashed line is the weak-coupling result Eq.~\eqref{BosePolaronResidueLinearizedGPE}. From \citet{Guenther2021}.}
\end{figure}

The polaron energy obtained from Eq.~\eqref{eq:GPe} is plotted in Fig.~\ref{fig:energyVsaOverXi_comparison_with_many_experiments} as a function of $1/k_na$ using mass ratios and effective ranges corresponding to the Aarhus and JILA experiments (where $m/m_b$ was equal to 1 and 40/87, respectively). There is excellent agreement between theory and these two experiments as well as with quantum Monte Carlo (QMC) calculations, even though the impurity is not heavy. The inset shows the contact (with $m/m_b=40/23$), again obtaining good agreement with the Massachusetts Institute
of Technology (MIT) experiment. Unlike the ladder and variational approaches, which also recover the experimental data as previously discussed, the GP ansatz however predicts large dressing clouds involving many bosons and small residues. This illustrates the general situation where different theories reproduce  experimental data, in particular for the attractive polaron, and where possible discrepancies are difficult to quantify since the observed spectra are  broad for strong interactions. Figure~\ref{fig:energyVsaOverXi_comparison_with_many_experiments} also shows that the GP approach predicts  the range of the impurity-boson interaction potential to be  a new relevant length scale for the polaron energy, as  will be discussed further in Sec.~\ref{sec:BosePolaronUnitarity}.

A complementary approach to the one discussed above is to consider a zero-range impurity-boson interaction $V$ and a non-zero range boson-boson interaction $V_b$. This leads to an integro-differential (i.e., non-local) GP equation~\cite{Drescher2020}. A later analysis which kept non-zero ranges for both interaction potentials~\cite{Yegovtsev2023} showed that the polaron energy depends most strongly on the range of the impurity-boson interaction; see Sec.~\ref{sec:BosePolaronUnitarity}. A coherent state Ansatz (with a Fr\"ohlich Hamiltonian) was  used to explore the momentum relaxation of impurities, showing that impurities injected in a bosonic bath with momentum larger than $p_c=mc$ (with $c$ the speed of sound) emit phonon shock waves akin to Cherenkov radiation, and slow down until they reach $p_c$~\cite{Seetharam2021}.

In general we assume that the condensate depletion is small so that $n_0\simeq n$ since $(na_b^3)^{1/2}\ll 1$ in  experiments so far probing the Bose polaron. However, the 
question regarding how the impurity affects this depletion is interesting. It was examined for a static impurity in a harmonic trap by extending the 
GP approach described in this section to include the depletion via the Bogoliubov-de Gennes equations~\cite{Takahashi2019}.

\subsection{Gaussian state approach}   \label{sec:CohGauss}

As already mentioned, the major challenge in describing the Bose polarons arises from the interplay between their formation as well-defined quasiparticles, the tendency towards decoherence and loss due to the scattering on an infinite number of low-energy excitations of the BEC, the  existence of few-body bound states, the formation of macroscopic dressing clouds that are ultimately stabilized  by Bose repulsion, with the latter beyond the reach of schemes based on the Bogoliubov approximation. In three consequent works~\cite{Christianen2021b,Christianen2022GaussianTheory,Christianen2023}, a theory capturing all these aspects including the self-stabilization of the dressing cloud was developed using a combination of the Lee-Low-Pines transformation with a Gaussian state variational ansatz. 

The main idea of the Gaussian state approach is derived from the Efimov effect, where three particles collectively suppress kinetic energy and bind even when two of them cannot bind~\cite{Naidon2017}. In Ref.~\cite{Christianen2021b}, it was investigated how this cooperative mechanism translates to the many-body regime. To this end, the authors included the possibility of infinitely many boson excitations, as well as the Efimov effect, by combining several canonical transformations in one variational ansatz
\begin{equation} \label{eq:BPol_varapproach}
    |\psi\rangle = \hat{U}_{n_0} \hat{U}_{LLP} \hat{A}[\mathbf{x}] |0\rangle.
\end{equation}
Here, $\hat{U}_{n_0}$ is a coherent state shift describing the presence of a background BEC, whereas $\hat{U}_{LLP}$ is the LLP transformation given in Sec.~\ref{sec:VarGPE}.  Finally, 
\begin{equation}\label{Eq_BPolGaussian}
\hat{A}(\mathcal{N},\phi,\xi)=\mathcal{N} e^{\int_\bk \ [\phi(\bm{k}) \hat{b}^{\dagger}_{\bm{k}} - \text{h.c.}]} 
e^{\frac12\int_{\bm{k}}\int_{\bm{k'}} \hat{b}^{\dagger}_{\bm{k}} \xi(\bm{k},\bm{k'}) \hat{b}^{\dagger}_{\bm{k'}}}
\end{equation}
with $\mathcal{N}$ a normalization constant, starts from a variational coherent state (that recovers the GPE solution discussed before) and extends it by a Gaussian state transformation. The ground state is obtained minimizing the Hamiltonian Eq.\ \eqref{eq:HB} over the variational parameters $\phi(\bm{k})$ and $\xi(\bm{k},\bm{k'})$. In particular, the correlation matrix $\xi(\bm{k},\bm{k'}) $ allows to fully include  three-body impurity-boson-boson correlations. Indeed, expanding the Gaussian state in the vacuum limit of $n_0\to 0$, the exact solution of the three-body problem, and thus the Efimov effect, is recovered. The ansatz describes exactly the case of an infinitely heavy impurity in a non-interacting BEC. Nonetheless, as discussed below, it works well also for very light impurities for typical values of Bose repulsion.

\begin{figure}[t!]
\centering
\includegraphics[width=\columnwidth]{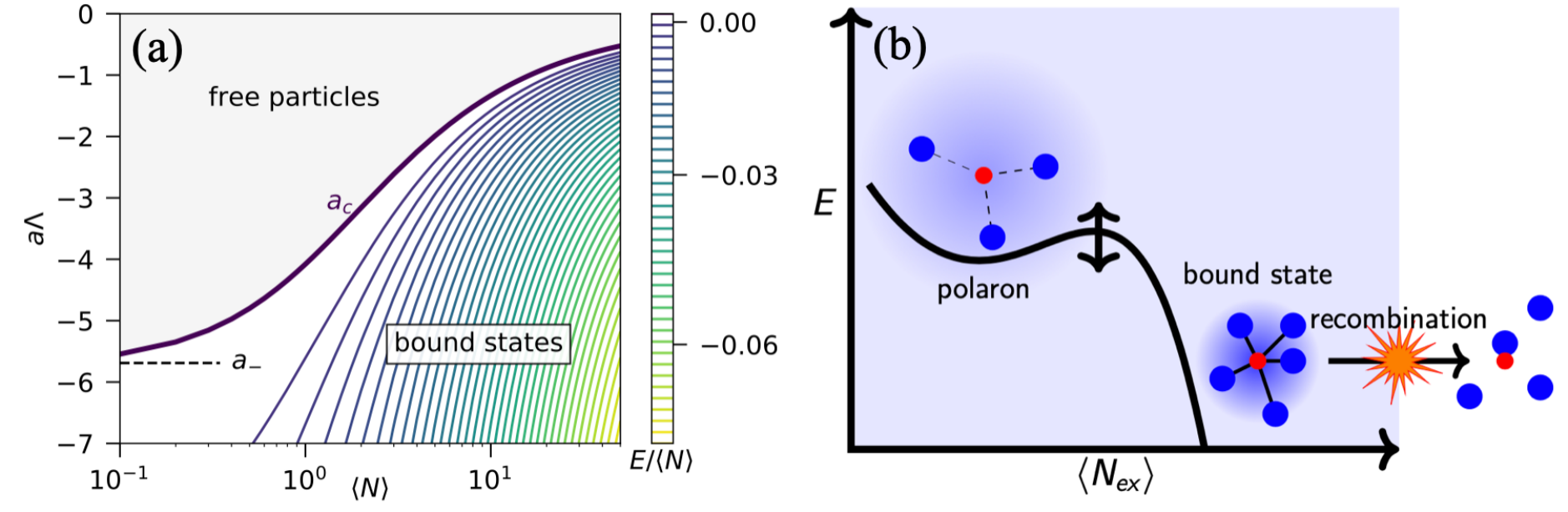}%
\caption{
\label{fig:BPolGaussEnergy}
\textbf{Gaussian state approach.} 
(a) A single light impurity having strong attractive interactions with an environment of average $\langle N\rangle$ bosons (in absence of a background condensate density) can form $n\ge 3$-body bound states (here the mass ratio is $m/m_b =6/133$, and $\Lambda$ is a three-body parameter). (b) Variational energy landscape given by the Gaussian state approach as a function of the average excitations $\langle \hat N_{\rm ex} \rangle$ over a background BEC with non-zero density. A local minimum supports the existence of a metastable Bose polaron, and an energy barrier separates it from decay to deeply bound Efimov-like clusters, which are rapidly destroyed by three-body recombination. From \citet{Christianen2021b}.}
\end{figure}

Applying the Bogoliubov approximation (i.e., keeping only terms up to quadratic order in the boson $\hat b_\veck$ operators in the Hamiltonian) and considering a cloud of finite extension containing on average $\langle N\rangle$ bosons (i.e., in absence of a background condensate), \citet{Christianen2021b} found that every boson will bind to the impurity, even if the impurity is mobile. As shown in Fig.~\ref{fig:BPolGaussEnergy}(a), when plotting the total energy per particle obtained using the wave function Eq.~\eqref{eq:BPol_varapproach} as a function of $\langle N\rangle$ and $a$, one finds that the binding energy per particle $|E|/\langle N \rangle$ monotonically increases with particle number. This indicates a cooperative mechanism where binding becomes stronger when more  particles participate in the formation of a deeply bound many-body cluster state. In other words, within the Bogoliubov approximation one finds that a single mobile impurity can trigger the  collapse of the Bose gas driven by the build-up of three-body correlations. It was also found that the thus enhanced effect of attractive impurity-bath interactions leads to a shift of the scattering length $a_-$ at which Efimov states appear in the three-body limit. As a result  three-body recombination is modified by many-body effects. 

Having shown that within this approximation the ground state technically corresponds to a deeply bound many-body cluster, one may ask what consequences this entails for the existence of Bose polarons. ~\citet{Christianen2021b,Christianen2022GaussianTheory} showed that, despite the existence of these deeply bound clusters, the Bose polaron indeed remains an excited, long-lived state protected by an energy barrier in the variational landscape; see Fig.~\ref{fig:BPolGaussEnergy}(b). This barrier gradually reduces as attraction increases, and it disappears at a critical scattering length $a^*$. Remarkably, up to that point, the Bose polaron is well described by the result of a coherent variational state. As the density of the background BEC $n_0$ is reduced, the scattering length $a^*$ converges to the value $a_-$ of the three-body problem, highlighting the intimate link of this Bose-polaron instability to the Efimov effect. Hence, the value of $a^*$ is effectively determined by the three-body recombination of two Bogoliubov modes and a Bose polaron~\cite{Christianen2021b}. Performing a similar analysis (using Gaussian variational states) for the repulsive polaron, \citet{mostaan2023unifiedtheorystrongcoupling} found multiple many-body states with energies in between those of the attractive and repulsive polaron.

\begin{figure}
\centering
\includegraphics[width=\columnwidth]{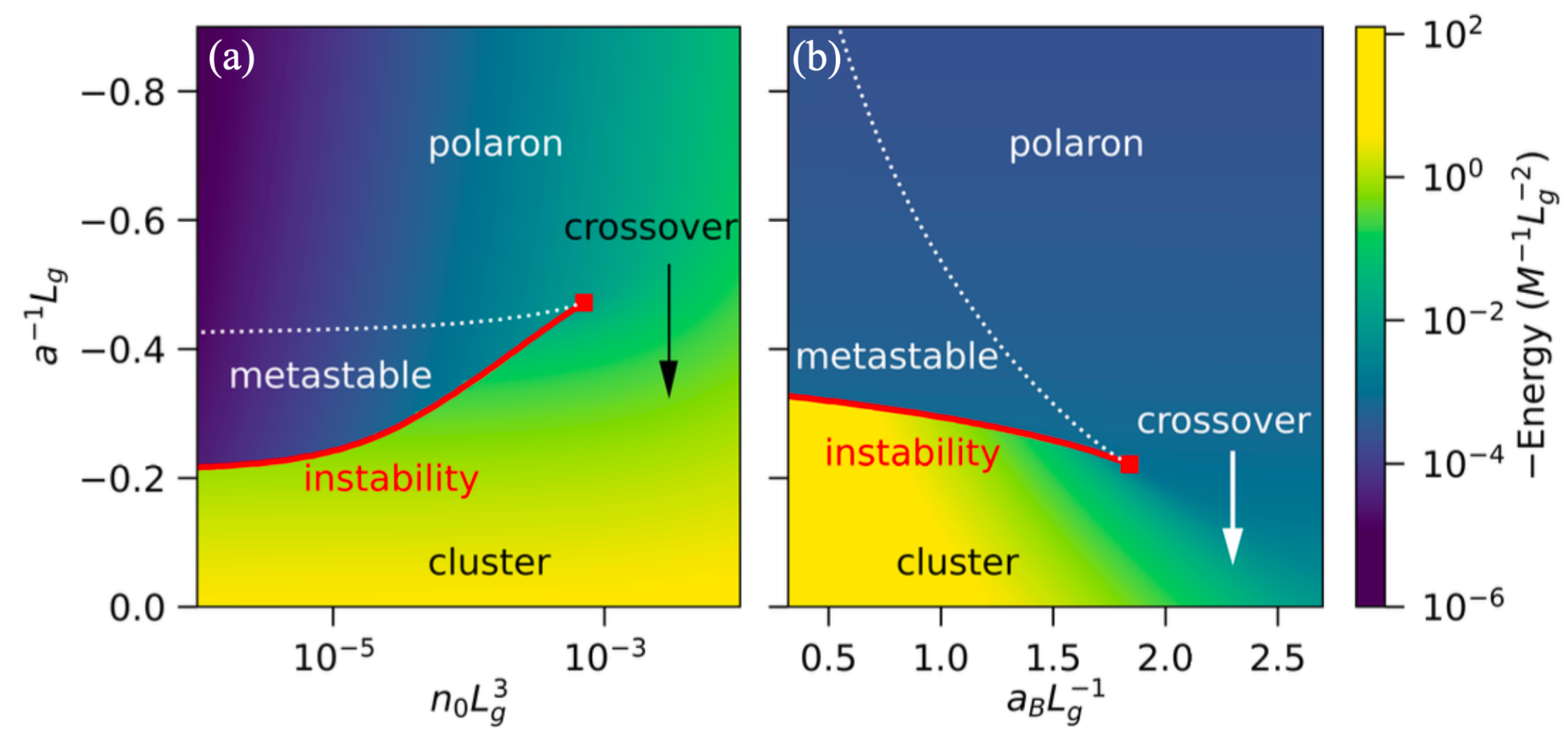}%
\caption{
\label{fig:BPolPhaseDiagram}
\textbf{Phase diagram for Bose polarons.} 
The Gaussian state ansatz predicts an effective phase diagram of Bose polarons reminiscent of a liquid-gas transition. Here $L_g$ is the range of the impurity-boson interaction. Depending on (a) boson density (for $a_B=1.2 L_g$) or (b) boson-boson scattering length $a_B$ (at $n_b L_g^3 \approx 3 \cdot 10^{-5}$, appropriate for typical experimental settings), a first-order phase transition occurs from a phase where the polaron is a stable quasiparticle to a phase where the ground state is a deep Efimov-like cluster. These phases are separated by a region where the polaron is a metastable state. The first-order transition terminates in a critical point where the transition turns second order, and beyond it a crossover occurs. The mass ratio for both plots is $m/m_b=6/133$. From \citet{Christianen2023}.}
\end{figure}

The above results are based on the Bogoliubov approximation, which neglects quartic terms in the boson-boson repulsion. As discusse above, however, accounting for those shall stabilize the dressing cloud of the cluster states. To investigate this, ~\citet{Christianen2023} applied the variational state~\eqref{Eq_BPolGaussian} to the full Hamiltonian~\eqref{eq:HB}. Finite range attractive impurity-boson and repulsive boson-boson interaction potentials were employed with potential ranges $L_g$ and $L_U$ fitted to the corresponding van der Waals lengths $l_\text{vdw}$ of the atomic species under consideration. Importantly, the ansatz~\eqref{Eq_BPolGaussian} treats the boson-boson repulsion beyond the Born approximation employed in the GPE discussed in Sec.~\ref{sec:VarGPE}. This analysis gives the ``phase diagram'' for strong coupling Bose polarons shown in Fig.~\ref{fig:BPolPhaseDiagram}. The Bose polaron is a stable quasiparticle up to a critical scattering length $a$, which depends on the density and the boson-boson scattering length $a_B$. At the critical value, the system undergoes a first-order ``phase transition'' to a Efimov-like cluster ground state, with a region of metastability (as usual in first-order transitions) of the Bose polaron in between. At a critical endpoint, this first-order transition turns second order, beyond which a continuous crossover from the Bose polaron into the Efimov-like cluster appears. A Landau energy functional was derived, which precisely recovers this phase diagram without the need of the full computationally-expensive numerical solution. 

\begin{figure}
\centering
\includegraphics[width=\columnwidth]{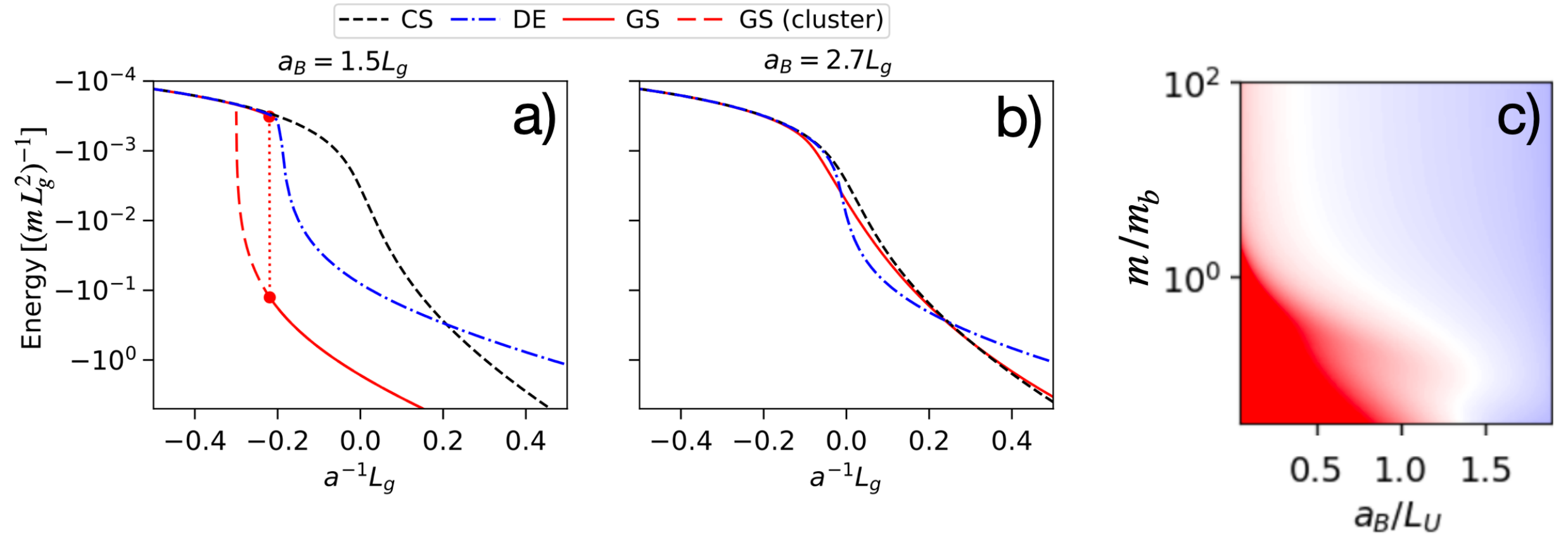}
\caption{\label{fig:BPolComparison}
\textbf{Comparison of variational functions.} 
(a,b) Ground state energies obtained from a coherent state (CS) which corresponds to the GPE solution, a double-excitation ansatz (DE), a Gaussian state (GS), and the energy of the Efimov-like cluster states (dashed red). 
In (a,b) the  mass ratio is $m/m_b=6/133$ and the ratio of boson-boson to impurity-boson interaction ranges is $L_U/L_g =2.3$. (a) For small Bose repulsion, the GS outperforms the DE ansatz and a region of metastability is found. (b) For larger Bose repulsion, the DE ansatz yields the lower  energy close to unitarity. (c) Ratio of the variational energies $E_{GS}/E_{DE}$ at unitarity ($a\to\infty$) as function of mass ratio and boson-boson scattering length (here $L_g = L_U$).  Red (blue) color indicates a lower energy of the GS (DE) ansatz.  From \citet{Christianen2023}.}
\end{figure}

The Gaussian state ansatz was also directly compared to a double-excitation (DE) ansatz as given by Eq.~\eqref{BogExpansion} keeping terms up to second order in Bogoliubov excitations, which is then applied to the full Hamiltonian including the quartic terms in the boson-boson repulsion. As can be seen in Fig.~\ref{fig:BPolComparison}, the Gaussian state ansatz yields an energy lower than the double-excitation ansatz for light impurities and weak Bose repulsion, but the trend is reversed for moderate mass ratios and stronger Bose repulsion, in particular close to unitarity. This behavior can be attributed to the ability of the DE  ansatz to describe the detailed structure of correlations close to the impurity, while the Gaussian state ansatz is better suited to describe large dressing clouds at the expense of requiring many bosons to follow the same correlation structure. The comparison of both approaches highlights that there is presently no unique best variational wave function for all cases. Instead, depending on the system parameters and questions addressed (e.g. static versus dynamic properties) different approaches have to be employed. This is much similar to the case one encounters in the theoretical description of electronic structure and dynamics in solid-state physics.

\subsection{Quantum Monte Carlo calculations}   \label{sec:QMC}

Given the complexity of the Bose polaron problem, it is very useful to have access to exact numerical results. The polaron energy $\varepsilon$ can be calculated as $\varepsilon= E_{\rm tot}(N;1)-E_{\rm tot}(N;0)$ where $E_{\rm tot}(N;M)$ is the energy of $N$ majority particles and $M$ impurities. These energies can be obtained using exact diagonalisation as has been done for $N\lesssim{\mathcal O}(10)$ fermions in 2D~\cite{Amelio2023,Amelio2024}. While this provides useful information both on the ground and excited states of mesoscopic systems, the computational cost of exact diagonalization typically increases exponentially with $N$ making the thermodynamic limit out of reach. 

Quantum Monte Carlo methods offer a reliable solution to this problem by evaluating multidimensional integrals with stochastic techniques. They are generally more effective for the Bose polaron than for the Fermi polaron, since the latter suffers from the infamous ``sign problem''. The Fermi polaron can be analysed using a fixed-node Diffusion Monte Carlo method based on a suitable guess for the nodal surface of the many-body wave function~\cite{Bombin2019,Bombin2021,Pilati2021,Pessoa2021}. It has also been explored with bold diagrammatic Monte Carlo methods based on evaluating Feynman diagrams by stochastic sampling~\cite{Mishchenko2014,prok2008,Prokofiev2008b,Vlietinck2013,Vlietinck2014,Goulko2016,VanHoucke2020}. Monte Carlo methods have also been applied to the Bose polaron problem both at zero and at finite temperatures, which we now discuss for the 3D and 2D cases. The case of polarons with long range interactions is discussed in Sec.~\ref{sec:long-ranged_Bose_polarons}.

\begin{figure}[t]
\centering
\includegraphics[width=0.48\columnwidth]{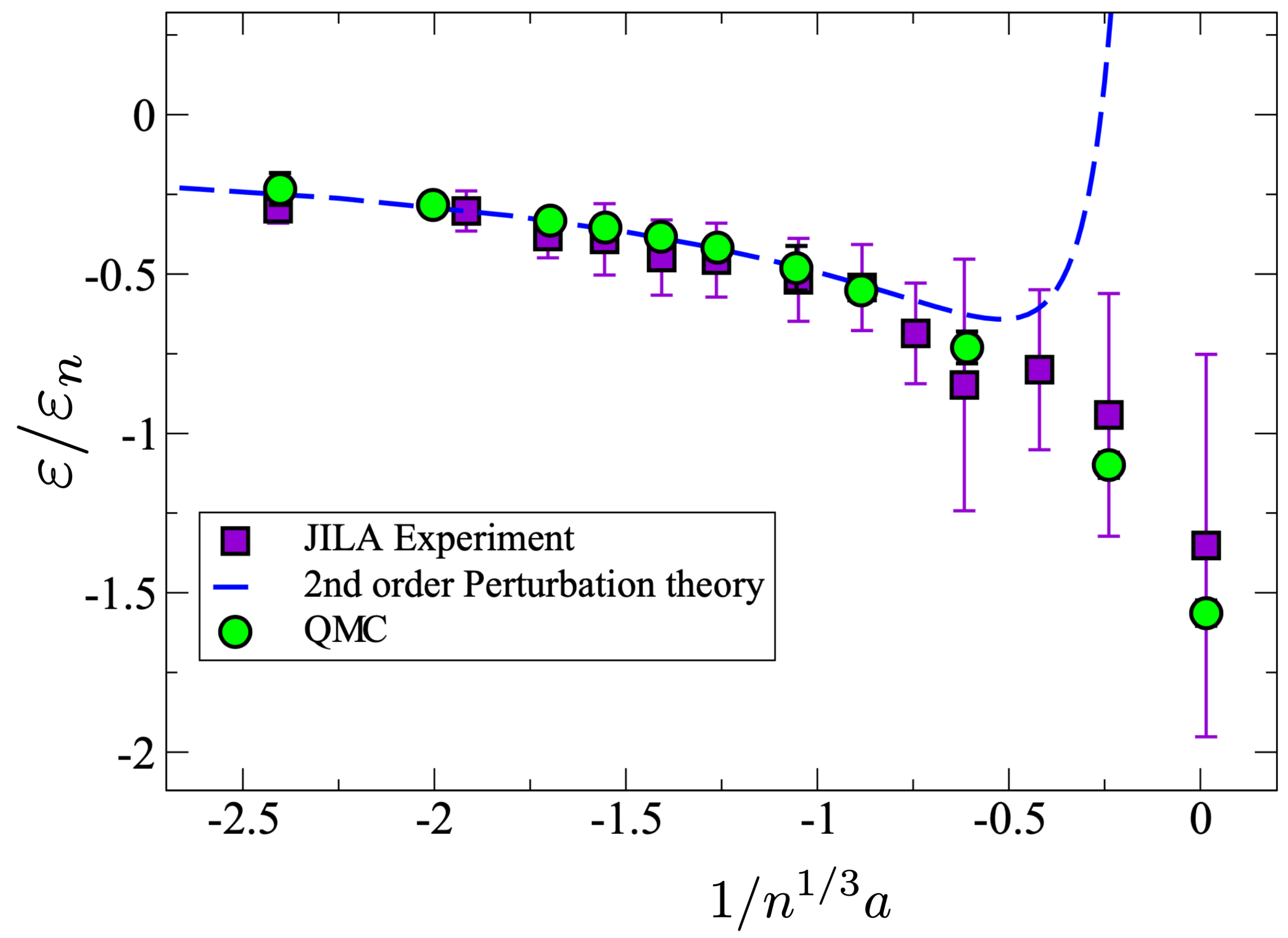}
\includegraphics[width=0.48\columnwidth]{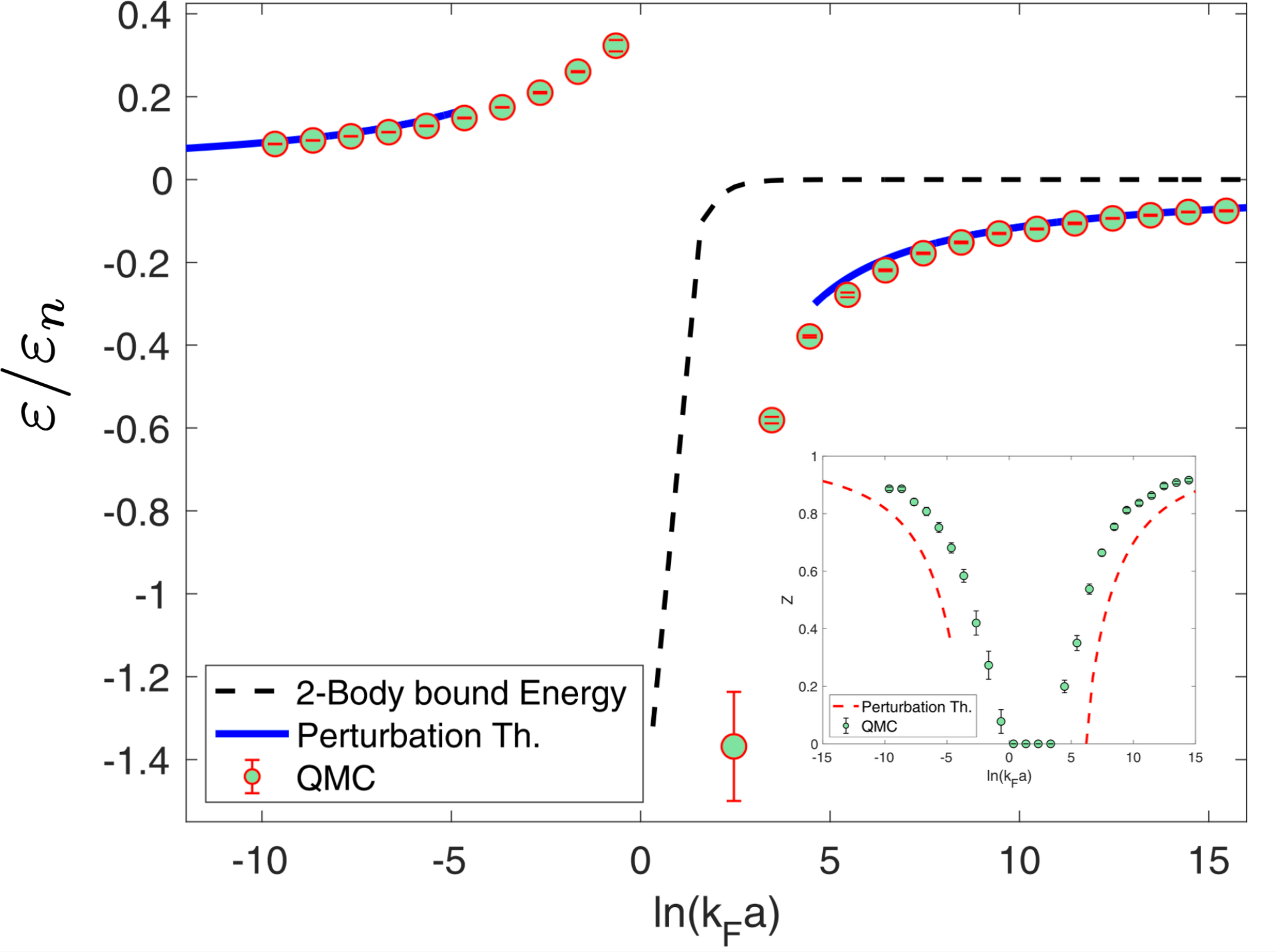}
\caption{\label{fig:energy2D_from_DMC}
{\bf Bose polarons in 3D and 2D.} 
Left panel: DMC calculation (green points) of the 3D attractive Bose polaron energy as a function of the impurity-boson scattering length for mass ratio $m/m_b=1/2$ and gas parameter $na_b^3=2.66\times10^{-5}$ close to the parameters in the JILA experiment (pink squares)~\cite{hu2016}. The dashed line is second order perturbation theory given by Eq.~\eqref{EnergyPert}. From \citet{Pena2016}. Right panel: DMC calculation of the 2D attractive and repulsive Bose polaron energy as a function of the interaction strength $\ln(k_{F}a)$  with $na_b^2 = 10^{-40}$. The dashed line shows the dimer binding energy and the solid line the perturbative result  Eq.~\eqref{2DBosePolaronEnergy}. The inset shows the residue. From \citet{Pena2020}.}
\end{figure}

In three dimensions, the zero-temperature properties of the attractive and repulsive Bose polaron for $m=m_b$ have been analyzed using the Diffusion Monte Carlo method, which solves the many-body Schrödinger equation in imaginary time. The repulsive boson-boson interaction was modeled by hard spheres of diameter equal to the scattering length $a_b$, and the impurity-boson interaction was modeled as either a square-well potential or as a hard sphere for $a>0$~\cite{Pena2015}. For weak impurity-boson interactions, the numerical results recovered the second-order perturbation theory results for the energy and effective mass given by Eq.~\eqref{EnergyPert} and Eq.~\eqref{MassPert}. This agreement remained for the repulsive polaron up to surprisingly strong interactions, whereas significant deviations from perturbation theory were found for the attractive polaron close to  unitary $1/a=0$. The effective mass was found never to exceed twice the impurity mass ruling out self-localization, which would be signaled by a diverging effective mass. In subsequent studies, the impurity-boson interaction was modeled by a zero-range pseudopotential and good agreement was found with the JILA experiment~\cite{hu2016} as shown in Fig.~\ref{fig:energy2D_from_DMC}~\cite{Pena2016}. Good agreement with the Aarhus experiment was also obtained~\cite{Ardila2019}. The dependence of the polaron energy on the mass ratio $m/m_b$ and the boson-boson scattering length $a_b$ at unitarity $1/a=0$ was explored, with the latter fitted to a polynomial function. This fit was later replaced by a logarithmic ansatz~\cite{Shi2018} as discussed in Sec.~\ref{sec:BosePolaronUnitarity}. 

The Bose polaron at non-zero temperature was explored using both the Path Integral Monte Carlo (PIMC) and Path Integral Ground State (PIGS) methods~\cite{Pascual2021}.  It was found that in the bulk, the energy of the attractive/repulsive branch increases/decreases with temperature and that the polaron ceases to exist above the critical temperature $T_c$ of the BEC. This technique was furthermore used for impurities in a harmonic trap in the case where the impurity-boson interaction $V$ is more repulsive than the intra-boson interaction $V_b$~\cite{pascual2024}. It  was further applied to impurities in a harmonic trap, exploring different scenarios where the impurity-boson interaction, $V$, could be either weaker or stronger than the intra-boson interaction, $V_b$~\cite{pascual2024}. At low temperatures, strong impurity-boson interactions caused the impurities to be expelled to the surface of the gas. In contrast, at higher temperatures, though still below the critical temperature $T_c$, the impurities remained mixed with the gas.

In two dimensions, the repulsive Bose polaron was examined with DMC method using repulsive hard disk boson-boson and boson-impurity interactions~\cite{Akaturk2019}. Good agreement was found with the perturbative result Eq.~\eqref{2DBosePolaronEnergy}~\cite{Pastukhov2018a} for weak interactions and low values of the gas parameter, $na_b^2=10^{-5}$. The effective mass was found to increase more than in the 3D case, reaching values as large as $m^*/m \simeq 2.5$ for the largest considered values of $a$. The 2D Bose polaron was also studied using the variational Monte Carlo method, where a variational wave function is optimized to obtain an upper bound of the ground-state energy, as well as DMC technique with soft-disk interactions between bosons and square-well boson-impurity interactions~\cite{Pena2020}. Good agreement with perturbation theory Eq.~\eqref{2DBosePolaronEnergy} was obtained for weak interactions and values of the gas parameter as small as $na_b^2 = 10^{-40}$, see right panel of Fig.~\ref{fig:energy2D_from_DMC}. Interactions were found to have a larger effect on the effective mass and residue of the polaron leading to significant disagreement with perturbation theory as shown in the inset in the right panel of Fig.~\ref{fig:energy2D_from_DMC}. Indeed, large values of the effective mass and a vanishing quasiparticle residue $Z$ were predicted for strong interactions, signaling a transition to a cluster state with no broken translational symmetry, i.e., no localization.

\subsection{Other methods}   \label{sec:FRG}

A complementary view on polaron physics is provided by the Functional Renormalization Group (FRG) approach, a non-perturbative method which allows to account systematically for many-body correlations~\cite{Dupuis20211}. Applications to the case of Fermi polarons were among the first to demonstrate the use of advanced approximation schemes that included full frequency- and momentum resolved correlation functions~\cite{schmidt_excitation_2011, vonmilczewski2023momentumdependent,Duda2023}, complementing approaches that relied on simpler approximations \cite{Kamikado2017,pawlowski2017physics,Milczewski2022}. FRG methods were applied to Bose polarons in later works. ~\citet{von2023superconductivity} provided a general RG analysis that compared the Fr\"ohlich model and the full Hamiltonian given by Eq.~\eqref{BeyondFrohlich} assuming an ideal BEC. In particular, ~\citet{Isaule2021} studied the problem of Bose polarons at zero temperature across all interaction strengths, in both 2D and 3D, while ~\citet{Isaule2022} studied also the case of balanced Bose-Bose mixtures. These works reported polaron energies that compare favorably with the perturbative expansion in Eq.~\eqref{EnergyPert}, with the variational approaches presented in Sec.~\ref{sec:Bose_variational_Chevy}, and with Monte Carlo simulations, especially when three-body correlations were explicitly included. The calculations were extended to consider finite temperature Bose polarons in both 2D and 3D~\cite{isaule2024functionalT}, finding the energy of the attractive polaron at unitarity to decrease with temperature for $T<T_c$, in agreement with the theoretical findings of~\citet{Guenther2018} and the MIT experiment~\cite{Yan2020}. 

The Jensen-Feynman variational path integral method was introduced in an early paper as a new powerful approach for analysing the Bose polaron within the Fr\"ohlich model~\cite{Tempere2009}. This was later applied to discuss Bragg spectroscopy~\cite{casteels_response_2011,casteels_many-polaron_2011} and reduced dimensional systems~\cite{casteels_polaronic_2012}. Importantly, the variational path integral method was extended beyond the Fr\"ohlich model to the full Hamiltonian~\cite{Ichmoukhamedov2019}. More recently, further progress was made by including higher-order corrections~\cite{ichmoukhamedov_general_2022}, which led to an improved agreement with Bold Diagrammatic Monte Carlo calculations. 

Starting from a quantum Brownian motion model and solving the emerging quantum Langevin equation, which describes impurities in a BEC  including memory effects, it was shown that the polaron can exhibit superdiffusive motion and position-squeezing~\cite{Lampo2017bosepolarons,Lampo2018}. When the bath is a coherently-coupled two-component BEC, this model even predicted a subdiffusive regime~\cite{Charalambous2020controlofanomalous}.

\subsection{Bose polarons at unitarity}   \label{sec:BosePolaronUnitarity}

In the previous sections we have seen that different theories for the Bose polaron generally agree for weak interactions but tend to give diverging predictions for strong interactions. In particular, there is no consensus yet regarding the number of parameters that are important in the strongly interacting region. To illustrate this, we now focus on the properties of the Bose polaron at unitarity where the impurity-boson scattering length $a$ diverges and therefore disappears from the problem. Unlike for the Fermi polarons described in Sec.~\ref{sec:FermiPolarons}, where the only remaining relevant parameters for a broad resonance are the interparticle spacing $\sim k_F^{-1}$ and the mass ratio $m/m_b$, here the importance of further length scales such as the boson-boson scattering length $a_b$, the ranges of $V$ and $V_b$ and other scales related to $n>2$-body correlations remains largely an open question. 
 
As discussed in Sec.~\ref{sec:ladder}, see Eq.~\eqref{eq:BosePolaronEnergyLadder} and Fig.~\ref{fig:BosePolaronSpectrumWithOneBogoliubovMode}, the ladder approximation predicts that the energy of the attractive Bose polaron at unitarity depends only on the mean distance between the bosons $\sim k_n^{-1}$, the BEC healing length $\xi$, and the mass ratio $m/m_b$. Since the ladder approximation only includes $2$-body correlations, it cannot capture effects arising from processes involving one impurity and $n\ge 2$  bosons such as three-body physics. As a consequence, it predicts rather well the maximum of the spectral density but it cannot describe the large dressing clouds of the ground state which arise for strong interactions and in a soft BEC. 

The variational wave function discussed in Sec.~\ref{sec:Bose_variational_Chevy} provides a systematic way to explore the effects of few-body correlations on the Bose polaron. In Ref.~\cite{Yoshida2018}, the Bose polaron at unitarity was analyzed using  Eq.~\eqref{BogExpansion} including up to three Bogoliubov modes. Focusing on an ideal BEC with $a_b\rightarrow 0$, the energies of the three- and four-body bound states were calculated using both a multi-channel model with non-zero effective range and a single channel model with a cut-off in the $3$-body sector. Tuning the two models such that they yielded the same 3-body parameter $a_-$, excellent agreement was found between the two models regarding the  Bose polaron energy. In particular, the energy $\varepsilon$ of the polaron obtained from the two models were found to coincide when $nr_0^3\ll 1\ll n|a_-|^3$, where $r_0$ is a short range scale of the interaction. This agreement lead the authors to conjecture that $\varepsilon$ is a universal function of $na_-^3$. Nonetheless, the energy $\varepsilon$ and the residue $Z$ were found to steadily decrease with the number of Bogoliubov modes included, which is consistent with the OC and the disappearance of the polaron for $a_b=0$. This instability might be cured by the multichannel nature of the interaction leading to an effective $3$-body repulsive force as discussed in Sec.~\ref{sec:BOC}, although close to a broad resonance this will likely happen only at a very low energy.

A different prediction was obtained from the modified GP Eq.~\eqref{eq:GPe}, which can be solved analytically for heavy impurities (so that $\bar\xi\approx \xi$) and short range impurity-boson interactions satisfying $(n_0a_b^3)^{1/4}a_b\ll R\ll \xi$. The first of these inequalities ensures that the local gas parameter $n(r)a_b^3$ remains small also in the vicinity of the impurity making a mean-field GP description reliable. The range $R$ defined in Sec.~\ref{sec:VarGPE} naturally emerges within this treatment and is (like the effective range $r_e$) typically of the order of the physical range $r_c$ of $V$.\footnote{For a unitary square well interaction $V=-\pi^2\Theta(r_c-r)/[8 m_r r_c^2]$, one finds $R=0.56r_c$ and $r_e=r_c$, while for a unitary P\"oschl-Teller  $V=-1/[m_r r_c^2\cosh^2(r/r_c)]$ interaction one finds $R=1.05r_c$ and $r_e=2r_c$. However, there exist ``shape-resonant'' potentials having $|r_e|\ll r_c$ even though $R\sim r_c$ and $|a|\gg r_c$. For these peculiar fine-tuned cases, the effective two-parameter universality of the GPE discussed in Sec.~\ref{sec:VarGPE} does not hold~\cite{Massignan2021a,Yegovtsev2022,Yegovtsev2023}.} At the unitary point where $|a|\rightarrow\infty$ one obtains~\cite{Massignan2021a,Yegovtsev2022}
\begin{align}
\varepsilon&=-2\pi E_\xi\left(3\delta^{1/3}-2^{3/2}\delta^{2/3}+4\delta\ln\delta+\ldots\right)\label{Eunitarity}\\
\Delta N&=4\pi N_\xi\left[\delta^{1/3}-5\delta^{2/3}/(3\sqrt2)+2\delta\ln\delta+\ldots\right]\label{DeltaNunitarity}\\
\ln Z&=-\sqrt 2\pi n_0\xi^3\delta^{2/3} \label{Zunitarity},
\end{align}
for the energy, number of particles in the dressing cloud, and residue of the attractive polaron. This predicts that the properties of the attractive Bose polaron at unitarity depend on the range of the impurity-boson interaction and the boson-boson interaction via the ratio $\delta=R/\xi$ with no more interaction parameters needed. Equations~\eqref{Eunitarity}-\eqref{Zunitarity} were later generalized to the neighborhood of the unitary point, finding that the first corrections to both $\varepsilon/E_\xi$ and $\Delta N/N_\xi$ scale as $\delta^{2/3}\xi/a$~\cite{Yegovtsev2022}. The same formalism was also used to compute the induced mass $m_{\rm ind}$ of a heavy polaron, and the interaction energy between two distant ones~\cite{Yegovtsev2023b}. It is at present unclear how to relate the predictions of Ref.~\cite{Yoshida2018} obtained from a variational wave function including Efimov correlations but unable to describe large dressing clouds (and in particular the OC), with Eqs.~\eqref{Eunitarity}-\eqref{Zunitarity} based on a wave function capable of describing large dressing clouds including the OC but excluding $n\ge 3$ correlations and finite impurity mass corrections. The variational wave functions discussed in Section \ref{sec:CohGauss} that include Efimov correlations \cite{Christianen2022GaussianTheory, Christianen2021b,Christianen2023} apply the theory on a model that features direct boson repulsion. In contrast, Ref.~\cite{Yoshida2018} studies a two-channel model that introduces an effective repulsion between bosons. Due to the differences in the model studied, a direct comparison of the predictions of these works is not straightforward, despite their predictions being similar. Note that there are no Efimov states for an infinitely heavy impurity. 

A different prediction for the energy of an infinitely heavy impurity resonantly interacting ($1/a=0$) with a Bose gas was obtained comparing DMC calculations with the variational ansatz Eq.~\eqref{BogExpansion}~\cite{Levinsen2021}. A fit of the DMC results, obtained using contact impurity-boson and hard-sphere boson-boson interactions, indicated that in the dilute limit the polaron energy depends logarithmically on the gas parameter: $\varepsilon\propto-(n^{2/3}/m)\ln(n_0a_b^3)$, consistent with the ``Anderson'' model discussed in Sec.~\ref{sec:BOC}, where boson-boson correlations however arise from the multichannel nature of the interaction rather than from a direct repulsion with $a_b>0$. From this it was argued that the $a_b$-dependence is due to a quantum blockade effect beyond the reach of mean-field GP theory, since only one boson at a time can interact with the impurity when $a_b\gtrsim r_0$, with $r_0$ a typical range of the impurity-boson interaction. 

\begin{figure}[ht]
\centering
\includegraphics[width=\columnwidth]{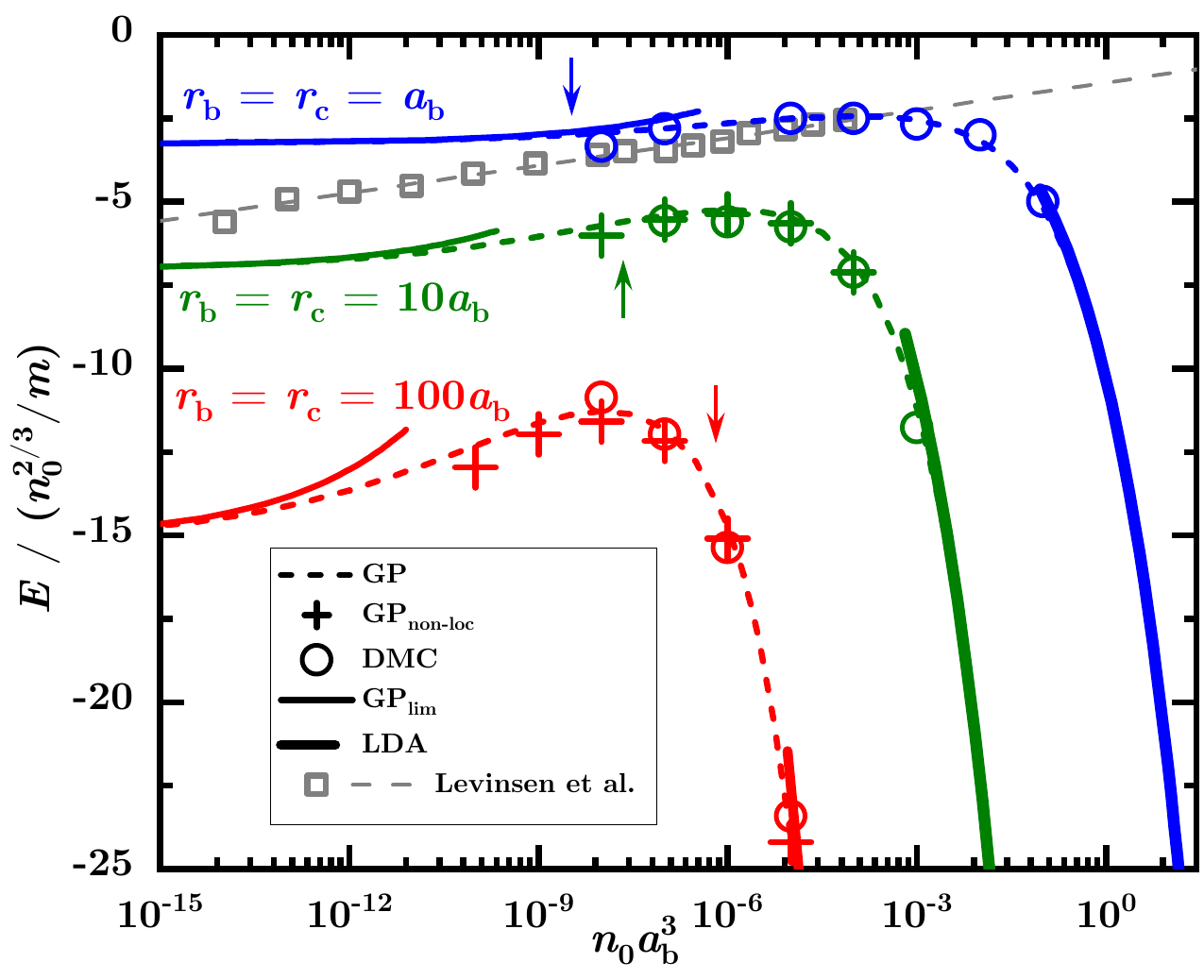}
\caption{\label{fig:energy_vs_gas_parameter_from_GPE}
{\bf Heavy attractive polaron at unitarity.} 
Energy of a static impurity interacting resonantly ($1/a=0$) with a BEC as a function of the gas parameter for three values of the ratio between the boson-boson scattering length $a_b$ and the boson-impurity interaction range $r_c$. The circles show DMC results (statistical error bars smaller than  symbol size) with arrows indicating when the polaron size becomes comparable to the box size and finite size effects set in. The dashed lines represent the energy obtained from a numerical solution of  Eq.~\eqref{eq:GPe} for $r_b=0$, whereas $+$'s is the energy obtained from a non-local GP equation with $r_b>0$. Thin solid lines at low density show the analytic result Eq.~\eqref{Eunitarity}, while the solid thick lines on the right side indicate the LDA prediction, see Sec.~\ref{sec:long-ranged_Bose_polarons}. The grey squares and grey dashed line show DMC data and their logarithmic fit from Ref.~\cite{Levinsen2021} where $a_b=r_b$ and $r_c=0$. From \citet{Yegovtsev2023}.}
\end{figure}

The energy of the attractive Bose polaron at unitarity for an infinitely heavy impurity was later studied further with DMC~\cite{Yegovtsev2023} method. Figure~\ref{fig:energy_vs_gas_parameter_from_GPE} shows the obtained energy as a function of the gas factor $n_0a_b^3$ using P\"oschl-Teller impurity-boson and Gaussian boson-boson interactions with ranges $r_c$ and $r_b$ respectively. Since the size of the polaron cloud is determined by the BEC healing length $\xi\propto 1/\sqrt{n_0 a_b}$ and contains $\Delta N\propto 1/a_b$ bosons, it increases with decreasing gas factor, rendering the DMC calculations more challenging with larger system sizes. The vertical arrows in Fig.~\ref{fig:energy_vs_gas_parameter_from_GPE} indicate the gas parameter below which finite size effects become important for the DMC calculations involving $N\sim 100$ particles, which is indeed the region where the DMC results of \cite{Levinsen2021} predicted the logarithmic dependence of $\varepsilon$ discussed above (gray squares and gray dashed line). In experiments one typically has $r_b\sim r_c \sim a_b$ and $n_0 a_b^3 \gtrsim 10^{-6}$, for which the two DMC calculations reassuringly agree. The energy obtained from the GP variational wave function Eq.~\eqref{GPEWavefn} (dashed lines) agrees very well with the DMC results in the region where finite-size effects are small, explicitly demonstrating the accuracy of this approach for heavy impurities. For low BEC densities where $(n_0a_b^3)^{1/4}a_b\ll R\ll \xi$, the GP energy converges to the analytical result Eq.~\eqref{Eunitarity} (thick solid lines). For high densities ($R\gtrsim \xi$), both the DMC and GP energies are well captured by the local density approximation (LDA) discussed in Sec.~\ref{sec:long-ranged_Bose_polarons}. Finally, the $+$'s give the energy from a non-local GP equation including a non-zero boson-boson interaction range $r_b$, which deviates negligibly from the energy obtained from Eq.~\eqref{eq:GPe}. This shows that the boson-boson interaction range has small effects on the polaron energy 
under experimentally relevant conditions where $r_b\sim r_c$.

This Section illustrated the challenges connected to understanding the properties of the Bose polaron at strong interactions. Open questions include the role of $n>2$ correlations (Efimov trimers, tetramers, $\ldots$) and their associated length scales, which are likely most important for light impurities, and the differences between a single channel and a multichannel interaction. On the repulsive side $a>0$ where there is a bound impurity-boson dimer state, the role of short range physics and $n>2$ correlations and bound states is likely even greater as we saw in Sec.~\ref{sec:BOC} for a static impurity, and it is presently unclear whether universal results for the Bose polaron exists in this region. Since the experimental spectra at strong interactions are all broad, they have unfortunately not been able to resolve these questions so far, and it is in fact not even clear when/if the Bose polaron is a well-defined quasiparticle for strong interactions.

\subsection{Temperature dependence}   \label{sec:Tempdependence}

The Bose polaron has a qualitatively new feature compared to the Fermi polaron in the sense that its environment undergoes a phase transition at the critical temperature $T_c$ of the BEC. Since the low energy spectrum of the Bose gas changes from linear for $T<T_c$ to quadratic for $T\ge T_c$, 
this phase transition should affect the Bose polaron significantly. Exploring the Bose polaron for non-zero temperature is  an even more challenging problem than at zero temperature and there are several different theoretical predictions, as we will now discuss. 

For high temperatures $T\gg T_c$, one can perform reliable calculations for all interaction strengths using a virial expansion~\cite{Sun2017,Sun2017b,MULKERIN201929}. When truncated to second order in the fugacity, this is equivalent to the ladder approximation, and it yields a polaron damping rate $\Gamma\propto a^2\sqrt T$ for weak interactions and $\Gamma\propto 1/\sqrt T$ at unitarity. This can easily be understood from the classical expression $\Gamma=n\sigma v$ with a thermal relative velocity $v\propto \sqrt T$, as discussed for the Fermi polaron in Sec.~\ref{TdependenceFermiPolaron}. For low temperatures $T\ll T_c$,  a calculation to second order in $a$ gives for a zero momentum polaron~\cite{Levinsen2017} 
\begin{align}
\varepsilon(T)\simeq \varepsilon(0)+\frac{\pi^2}{60}\frac{a^2}{a_b^2}\frac{T^4}{nc^3}
\label{TdependencePert}
\end{align}
when $m=m_b$. Interestingly, the $T^4$ increase in the energy is identical to the change in the chemical potential of a weakly interacting Bose gas due to a thermal population of the phonon branch when $a=a_b$~\cite{khalatnikov1989introduction}. Perturbation theory furthermore predicts a non-monotonic behavior of the energy and a large   damping close to $T_c$. 
The Bose polaron at $T>0$ has also been analysed using mean-field theory~\cite{Boudjemaa2014}.

A diagrammatic calculation based on a ladder approximation extended to take into account a thermally populated Bogoliubov mode predicts that in the strong coupling regime, the energy of the attractive polaron decreases with increasing temperature reaching a minimum at $T_c$~\cite{Guenther2018}. The same calculation shows that the polaron damping increases with temperature making it ill-defined above $T_c$, and that a second quasiparticle branch appears for strong interactions for $0<T<T_c$ in analogy with what has been found for quasiparticles in quark-gluon plasmas~\cite{Weldon1989}. The decrease in the polaron energy with temperature and the appearance of another quasiparticle branch  was also found using a functional RG approach~\cite{isaule2024functionalT}. The properties of the Bose polaron for non-zero temperature were later further explored using an operator form of the ansatz Eq.~\eqref{BogExpansion} generalized to take into account a thermal BEC~\cite{Field2020}. Assuming an ideal BEC, the attractive polaron was found to split into two and three branches for $T>0$ when two and three Bogoliubov modes were included respectively. From this it was argued that the predicted splitting of the attractive polaron is an artifact of the expansion in Bogoliubov modes, and that it should instead remain a single peak with a width $\propto T^{3/4}$. Using a dynamical variational approach based on the coherent state applicable to a thermal BEC combined with a Lee-Low-Pines transformation, \citet{Dzsotjan2020} obtained improved agreement with the data of the Aarhus experiment~\cite{jorgensen2016,Ardila2019} when a non-zero temperature was taken into account. No splitting of the attractive polaron branch was found. 

Path-integral Monte Carlo calculations~\cite{Pascual2021} found that the energy of the attractive/repulsive polaron increases/decreases with temperature in contrast to the theoretical work discussed above. A direct comparison is however not straightforward since in QMC calculations the impurity is in thermal equilibrium with the bath and therefore it has a non-zero kinetic energy. By developing a functional determinant approach to calculate the spectral properties of a static impurity (infinite mass) in an ideal Bose gas, the spectral width of the ground state was predicted to decrease as a function of temperature near unitarity, which somewhat surprisingly would correspond to an increasing life-time~\cite{drescher2024bosonic}. 

\begin{figure}[ht]
\centering
\includegraphics[width=\columnwidth]{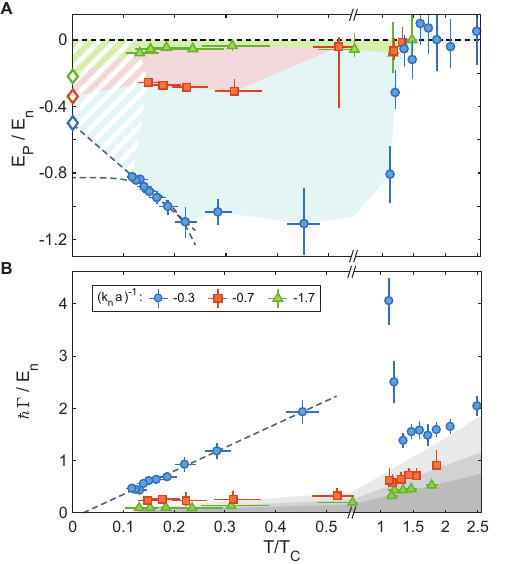}
\caption{\label{fig:experimentsOnBosePolarons_MIT}
{\bf Bose polarons at non-zero temperatures.}
The attractive polaron energy (a) and decay rate (b) (half-width at half-maximum of the spectral function) as a function of temperature measured by ejection RF spectroscopy using $^{40}$K impurity atoms in a $^{23}$Na BEC for various values interaction strengths. The dashed line in (b) is a linear fit to the data below $T_\mathrm{C}$. From~\citet{Yan2020}. }
\end{figure}

Experimentally, the properties of the attractive Bose polaron as a function of temperature were explored using RF ejection spectroscopy~\cite{Yan2020}. As shown in Fig.~\ref{fig:experimentsOnBosePolarons_MIT}, the energy was observed to decrease with temperature for strong interaction and the damping to increase in agreement with the theoretical predictions above~\cite{Guenther2018,Field2020,Dzsotjan2020}. The energy converged to that given by the ladder approximation as $T\rightarrow 0$ for different interaction strengths (open diamonds in Fig.~\ref{fig:experimentsOnBosePolarons_MIT}), whereas it jumped to zero at $T/T_c$, where the polaron becomes ill-defined due to large damping. The damping was found to increase linearly with $T$ at unitarity, as subsequently also found theoretically~\cite{Dzsotjan2020}, with a ``Planckian'' scale
$\sim k_BT/\hbar$, a signature of quantum critical behavior~\cite{sachdev_quantum_2011,Ludwig2011}. 

This section illustrates the challenges of understanding the Bose polaron at non-zero temperature. Moreover, an accurate theoretical description of the Bose polaron in the critical region $|T-T_c|/T_c\lesssim n_0^{1/3}a_b$ of the BEC is  lacking and  challenging, since it requires a formalism that includes fluctuations at all length scales such as the renormalisation group~\cite{Andersen2004}.

\subsection{Non-equilibrium dynamics}   \label{sec:NonequilBosepolaron}

As for the Fermi polaron discussed in Sec.~\ref{NonequilFermipolaron}, the relatively low density of atomic BECs and correspondingly long time-scales make them well suited to explore non-equilibrium many-body physics using interferometry with short pulses. This furthermore offers a useful alternative for measuring equilibrium properties, as the short lifetime of the Bose polaron sets an inherent limitation for the pulse duration and hence spectral resolution that can be achieved with RF spectroscopy~\cite{Spethmann2012,Spethmann2012a}.

The observed signal in Ramsey type experiments is proportional to  $S(t)$ given by Eq.~\eqref{GreaterGreens} precisely as for the Fermi polaron described in Sec.~\ref{NonequilFermipolaron}. Theoretically, $S(t)$ can be obtained by a Fourier transform of the exact expression for the impurity spectral function at high energies~\cite{Braaten2010} as discussed in Sec.~\ref{GeneralProb}. This yields~\cite{Skou2021}
\begin{equation}\label{S(t)_short}
  S(t)\simeq\begin{cases}
  1 - (1-i)\frac{16}{9\pi^{3/2}}\left(\frac{t}{t_n}\right)^{\frac32}&t\ll t_a \\
  1 +\frac{2}{3\pi}(k_n|a|)^3 - \sqrt{\frac{2it}{t_{\rm w}}}- iE_\text{mf}t& t\gg t_a
  \end{cases}
\end{equation}
where $t_a=ma^2$, $t_\text{w}=m/32\pi n^2a^4$, $t_n=2m/k_n^2$, and $E_\text{mf}=4\pi a n/m$ is the mean-field polaron energy. This result was later extended to second-order for $t\ll t_a$ by taking the presence of the attractive polaron branch into account~\cite{Skou2022}.

\begin{figure}[ht]
\centering
\includegraphics[height=0.34\columnwidth]{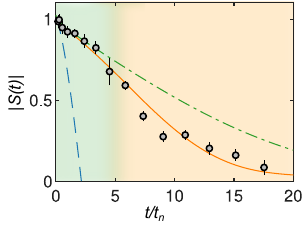}
\includegraphics[height=0.336\columnwidth]{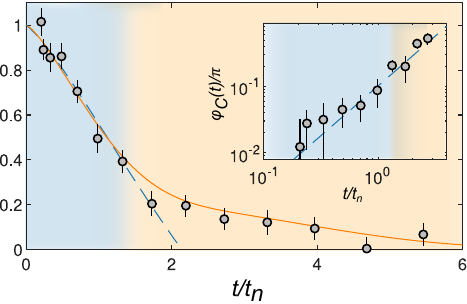}
\includegraphics[width=\columnwidth]{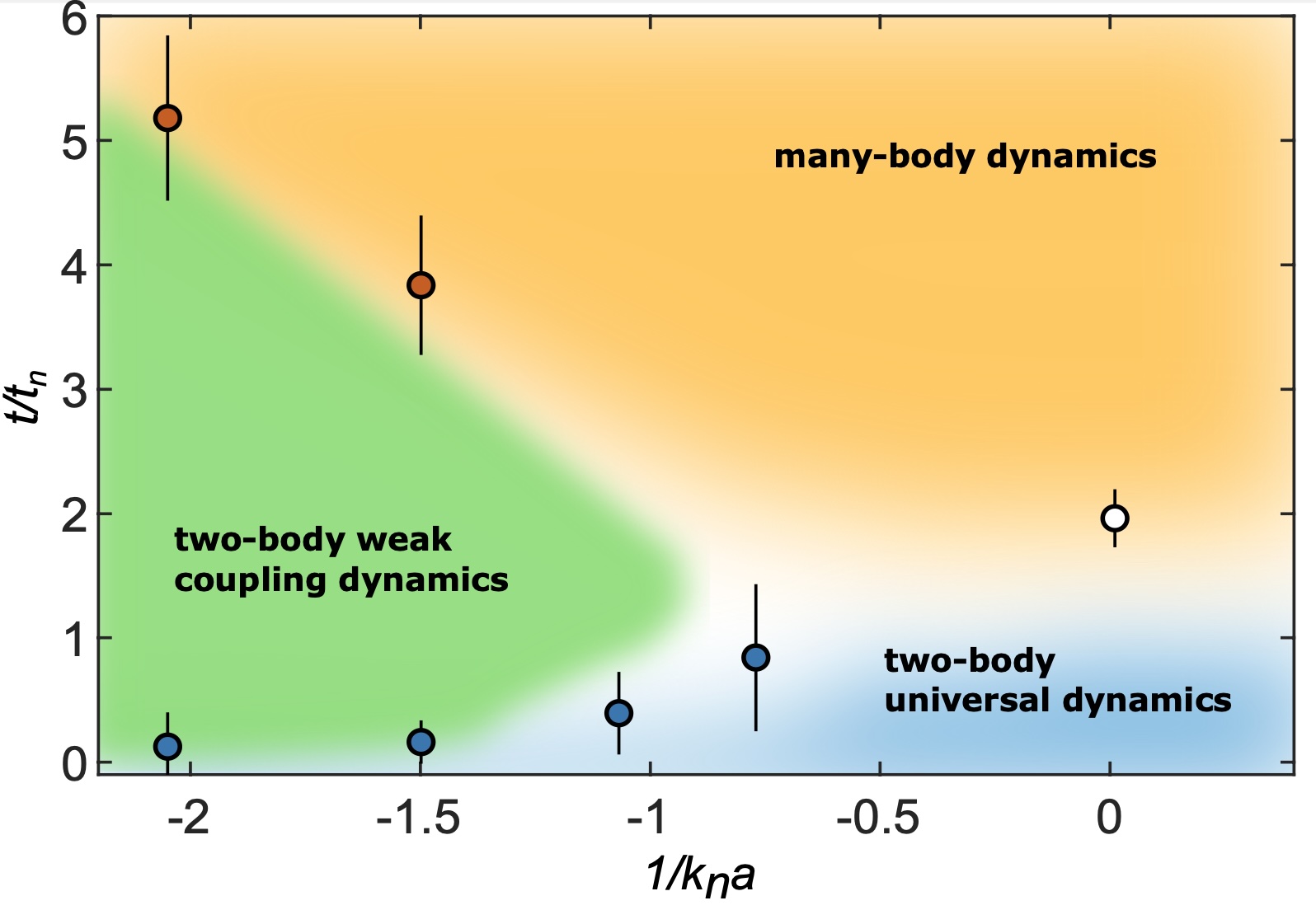}
\caption{\label{fig:JanArltBoseDynExp}
{\bf Formation of Bose polarons.} 
Top: Amplitude of the Ramsey signal measured for $1/k_na=-2$ (left) and $1/k_na=0$  with the inset showing its phase (right). The blue dashed and green dashed-dotted lines show the short- and long-time behaviors of Eq.~\eqref{S(t)_short}, and the solid orange line is the result of the ladder approximation. Bottom: Regimes of impurity dynamics as described by Eq.~\eqref{S(t)_short} and obtained from interferometric measurements. At short times the evolution is determined by universal high-energy two-body scattering (blue). For weak interactions low energy collisions dominate and the dynamics is governed by the mean field phase evolution (green). At longer times many-body correlations set in (orange).  From \citet{Skou2021}.}
\end{figure}

The dynamical regimes in Eq.~\eqref{S(t)_short}, shown as colored regions in Fig.~\ref{fig:JanArltBoseDynExp}, can be understood based on simple two-body scattering. The  rate of decoherence is determined by the scattering rate as $\dot S(t)=-n\sigma v$ with the collisional cross-section $\sigma$ and the typical relative velocity $v$. The  energy of the scattering processes contributing to the decoherence at time $t$ after inserting the impurity is $E\sim 1/t$, which gives 
$v\sim\sqrt{2E/m_r}$. Since the cross-section for high energies is given by its vacuum expression $\sigma(k)=4\pi a^2/[1+(ka)^2]$ with $k=m_rv$, one obtains $\sigma(k)\sim1/k^2$ for $t\ll t_a$ giving $\dot S(t)\sim-n\sqrt t/m_r^{3/2}$, and $\sigma(k)\sim4\pi a^2$ for $t\gg t_a$ giving $\dot S(t)\sim-na^2/\sqrt{m_r t}$. Integrating these expressions then yields Eq.~\eqref{S(t)_short}. Note that the short time $t^{3/2}$ dynamics  is universal in the sense that it is independent of the scattering length. 

Since the short-time dynamics is determined by high energy two-body scattering, it is independent of the quantum statistics of the bath. The initial $t^{3/2}$ dynamics in Eq.~\eqref{S(t)_short} has indeed also been derived in the context of the Fermi polaron discussed in Sec.~\ref{NonequilFermipolaron}, where it was shown that a non-zero range of the impurity-bath interaction gives rise to a $1-t^2$ dependence~\cite{parish2016}. Using the Master equation, the time evolution of $S(t)$ has been analyzed rigorously for weak interactions demonstrating a critical slow down of the formation of the Bose polaron when its velocity approaches the critical velocity of the BEC~\cite{Nielsen_2019}. The cooling dynamics via the emission of Cherenkov radiation was studied with the Boltzmann equation~\cite{Lausch_2018}. These works complemented earlier studies of $S(t)$ employing a LLP transformation combined with time-dependent coherent states~\cite{shchadilova2016}. Here it was found that the Bose polaron peak undergoes significant broadening as the Feshbach resonance is approached so that the Bose polaron indeed loses its meaning as a well-defined quasiparticle. A similar approach was at the basis of the analysis of the  spatially resolved formation dynamics of the Bose polaron~\cite{Drescher2019,Drescher2021}, and the formation of magnetic polarons was examined in Ref.~\cite{Ashida2018}.

The experimental points in Fig.~\ref{fig:JanArltBoseDynExp} were measured in a $^{39}$K BEC using a Ramsey scheme similar to that  described in Sec.~\ref{NonequilFermipolaron}: an  initial pulse created a small admixture of the BEC in the impurity state, and after  time $t$ a second pulse with variable phase then compared the evolved state with the initial BEC~\cite{Skou2021}. The observed signal was shown to be proportional to $S(t)$. The two upper panels of Fig.~\ref{fig:JanArltBoseDynExp} show the experimentally measured evolution of the amplitude $|S(t)|$ for two values of the impurity-boson interaction, together with the predictions of Eq.~\eqref{S(t)_short}. Based on these measurements, the transition times between the two dynamical regimes of Eq.~\eqref{S(t)_short} can be obtained, which is shown as blue points in the lower panel. At unitarity, $t_a$ diverges and the system transitions directly from two-body universal $t^{3/2}$ dynamics to the many-body regime. The solid lines in Fig.~\ref{fig:JanArltBoseDynExp} are obtained by Fourier transforming the impurity spectral function obtained from the ladder approximation, which includes the short time two-body dynamics exactly. This result agrees remarkably well with the experimental data even at unitarity, when an independently measured exponential decay due to three-body decay processes is included. The smooth crossover between the regimes of Eq.~\eqref{S(t)_short} was further analysed in Ref.~\cite{skou_initial_2021}. The impurity dynamics measured in Ref.~\cite{Skou2021} was later analysed using a time-dependent variational coherent ansatz obtaining good agreement~\cite{Ardila2021}. The 
dynamics of an impurity in an ideal Bose gas was also explored using the ladder approximation~\cite{Volosniev2015}. 

Based on these interferometric experiments, a detailed analysis of the time scales of polaron formation and loss was performed~\cite{Skou2022}. This indicated that significant decay at strong interaction indeed limits the RF spectroscopy resolution, and showed that the phase evolution at long times offers a useful alternative to measure the polaron energy in agreement with $S(t)\rightarrow Z\exp(-i\varepsilon t)$. Recently, interferometric investigations of the Bose polaron were extended to repulsive interactions $1/k_n a>0$~\cite{Skou2022,Etrych2024,morgen2023quantum}. In this case, one observed oscillations in $S(t)$ due to quantum beats between the attractive and repulsive Bose polaron like for the Fermi polaron discussed in Sec.~\ref{NonequilFermipolaron}, which was used to extract their energy difference. 

In another experiment, 
a weakly-interacting BEC of metastable Helium-4 ($^4$He*) atoms containing impurities expanded rapidly upon releasing it from a trap~\cite{Cayla2023}. The measured momentum distribution of the impurities at long times exhibited a remarkably clean $1/k^4$ tail, which disappeared in the absence of bath atoms, or when the bath was thermal. While these features resemble those expected from two-body interactions at equilibrium and Tan's contact, their origin must be clearly different, because the equilibrium $1/k^4$ tail is known to vanish over a very short time during an expansion in the presence of interactions~\cite{Qu2016}. Furthermore, the $1/k^4$ tails observed in this experiment have amplitudes which are orders of magnitude larger than the ones predicted at equilibrium. 

\subsection{Summary of theoretical methods for Bose polarons}
As discussed throughout this Section, the large compressibility of a bosonic bath makes the Bose polaron challenging to analyze both theoretically and experimentally. 
Contrary to the Fermi polaron case, there is no Fermi exclusion principle to suppress correlations between the impurity and $n\ge 2$ bosons, and the Bose polaron properties can generally depend on more parameters such as the boson-boson scattering length, the range of the impurity-boson interaction or a three-body parameter, and even $(n>3)$-body parameters. Since the bosonic bath is strongly affected by temperature, understanding the temperature dependence of the Bose polaron is moreover very complicated, in particular, in the critical region around $T_c$. 
In addition, experimental progress is complicated by three-body losses, which make strongly-interacting mixtures relatively short-lived. For these reasons, progress in understanding impurities in a Bose gas has been slower than in the case of a fermionic gas. As we have detailed in this Section, many different and sometimes conflicting results appear in the literature, and our understanding of even the cleanest experiments remains as of today for some aspects only partial.
We now provide a summary of the discussion in Secs.~\ref{sec:ladder}-\ref{sec:NonequilBosepolaron} regarding the most used theoretical models, highlighting their strengths and weaknesses.

Perturbation theory is rigorously valid for weak impurity-bath interactions and relatively stiff (i.e., not too compressible) BECs, so that the impurity induces only a small deformation of the bath. It therefore provides important benchmarks for other theories, numerical methods and experiments. However, it breaks down for strong interactions and cannot describe the formation of bound states such as Feshbach dimers.

The ladder approximation incorporates exactly the correlations between the impurity and one boson at a time and thus Feshbach physics. It  therefore permits to explore the unitary regime and beyond, agreeing well with most experiments studying equilibrium and non-equilibrium properties of the Bose polaron. This is likely because states dressed by a single bosonic mode are the ones with largest spectral weight (although they are possibly not the absolute ground state). The ladder approximation on the other hand ignores correlations with two or more bosons, which are important for small boson-boson repulsion and strong impurity-boson interaction.

 More general field theoretical methods like diagrammatic resummations and the functional renormalization group provide powerful ways to address non-perturbative effects beyond the ladder approximation. They naturally describe the effects of a non-zero temperature, a finite density of impurities and mediated interactions (Sec.~\ref{QPinteractions}), and the cross-over to mixtures (Sec.~\ref{sec:LimitManyBody}). These methods do not, however, give easy access to the wave function, and the approximation schemes typically become uncontrolled for strong interactions.
 
 Variational wavefunctions based on an expansion in Bogoliubov modes provide a systematic way to explore the correlations of the impurity with $n\ge 2$ bosons and effects such as the formation of Efimov trimers. Moreover, they naturally provide the wavefunction. Including such correlations generally improves the agreement with equilibrium and non-equilibrium experiments. However, these wave functions quickly become unwieldy when $n$ is increased, they ignore the back-action of the impurity on the condensate wave function, and they treat the boson-boson repulsion only within mean-field. They cannot therefore describe correctly cases where the number of bosons in the dressing cloud becomes large.  
 
 Coherent variational methods including those based on the GP equation account self-consistently for the deformation of the condensate wave function induced by the impurity, and therefore they describe correctly large dressing clouds and the Anderson orthogonality catastrophe in the limit of an ideal BEC. They give access to the wavefunction and agree well with most experiments, but they typically depend on the range of the impurity-boson interaction. The GPE is mostly reliable for heavy impurities, but it cannot describe correlations with $n\geq 2$ bosons. Three-body correlations have been included with more refined coherent Gaussian ansätze, but these may lead to an overestimation of $n>3$ correlations since the exponential ansatz contains terms up to infinite order. 
 
 Functional determinants have the advantage that they allow to solve the infinite-mass case exactly. However, they are by construction limited to Gaussian Hamiltonians and require the many-body system to be described by the grand canonical ensemble.  
 Since atomic BECs are not grand-canonical, functional determinants can only be applied to the thermal fraction of the gas. 
 Steps to address this challenge have been taken by combining functional determinants with coherent state wavefunctions. 
 
 Quantum Monte Carlo methods provide a numerically exact, non-perturbative description applicable to both model and realistic interaction potentials, including finite-range effects, and remain reliable in the strong-coupling regime. They naturally include correlations involving arbitrarily many particles, allowing the description of bound states and extended dressing clouds. Path-integral methods (PIMC, PIGS) provide a controlled framework for equilibrium calculations at finite temperatures.
 As QMC methods sample imaginary-time correlation functions, the extraction of spectral functions requires analytic continuation, which is mathematically an ill-posed problem. 
 In the ultradilute regime, where the healing length greatly exceeds the mean interparticle spacing, the dressing cloud contains many bosons, requiring large system sizes and making Monte Carlo calculations computationally demanding.
 
 The Master equation is designed to describe time-dependent properties, such as the formation dynamics of Bose polarons. In this framework it is however difficult to include strong impurity-boson interactions.

\section{Polarons with long-range interactions}   \label{sec:long-ranged_Bose_polarons}

The formation of polarons when the impurity-bath interaction is not short range involves an interesting and highly non-trivial interplay between few- and many-body physics, cold chemistry, and cluster physics. Experimentally, such systems can be created for example by immersing ions or Rydberg atoms in neutral atomic gases or by trapping atoms with magnetic/electric dipole moments.  

\subsection{Analytical results for long-ranged Bose polarons}
Before turning to these concrete cases, we first analyze an impurity interacting via a long-ranged potential with a dilute Bose gas, studying the GPE in the presence of a static potential with a range larger than the BEC healing length. We note that by using the Born-Oppenheimer approximation described in Sec.~\ref{sec:VarGPE}, see Eq.~\eqref{GPEWavefn}, the following analysis also applies to heavy but mobile impurities. When the range of the interaction potential is large, one may resort to the LDA to obtain an analytical expressions. This gives~\cite{Massignan2005}
\beq\label{eq:GPE_LDA_wf}
\psi(r)\approx \sqrt{n_0}\left[1-\frac{V(r)}{2gn_0}\left(1+\frac{6\xi^2}{r^2}\right)+C\frac{e^{-\sqrt{2}r/\xi}}{r}\right],
\eeq
for the wave function far away from the impurity, with $C$ a suitable constant.
The LDA term proportional to $V(r)$ dominates the long range behavior when the typical range of $V(r)$ is larger than $\xi$, while the Yukawa term proportional to $e^{-\sqrt{2}r/\xi}$ gives the leading contribution when the potential has a smaller range, but also whenever the BEC is sufficiently dilute. For smooth interaction potentials with a range $r_c$ much larger than the healing length $\xi$, one finds from Eq.~\eqref{eq:GPE_LDA_wf} the polaron energy 
\beq\label{eq:energy_LDA}
\varepsilon = \frac{n_0r_c}{2m_r}\int\! d{\bf y} \big[\mathcal{V}(y) - \frac{\mathcal{V}(y)^2}{2(r_c/\xi)^2} + \frac{(\nabla \mathcal{V}(y))^2}{4(r_c/\xi)^4}+\ldots\big],
\eeq
with $\mathcal{V}=V/(2m_r r_c^2)$, ${\bf y}=\br/r_c$, and a number of particles in the polaron cloud $\Delta N = -\frac{r_c}{8\pi a_b}\int d{\bf y} \,\mathcal{V}(y)+\ldots$ . Remarkably, these expressions hold also for strongly attractive potentials, and for shape-resonant ones, which are fine-tuned to have a vanishing effective range $r_e$. They accurately match DMC calculations in the regime $r_c\gg\xi$, see the Fig.~\ref{fig:energy_vs_gas_parameter_from_GPE}.
An explicit evaluation of these integrals for various unitary model potentials (Pöschl-Teller, Gaussian, exponential, and the simplest shape-resonant one) was given in  Ref.~\cite{Yegovtsev2023}. These expressions were shown to match accurately numerically-exact Diffusion Monte Carlo calculations in the regime $r_c\gg\xi$ (see the right hand-side of Fig.~\ref{fig:energy_vs_gas_parameter_from_GPE}).

\subsection{Ions in a BEC or Fermi gas}   \label{sec:Ions}

We now turn to the specific case of ions in neutral atomic gases. Experiments have explored atom-ion collisions and buffer gas cooling~\cite{EwaldGerritsma19,Feldker2020vl}, three-body recombination and molecule formation~\cite{Dieterle_IonBEC_2020,Mohammadi2021}, charge transport~\cite{Dieterle2021}, ions in a BEC~\cite{zipkes_trapped_2010,Kleinbach_IonBEC_2018}, Feshbach resonances~\cite{Weckesser2021}, and high resolution microscopy~\cite{Veit2021}. They have however not yet reached the quantum degenerate regime of polarons, which is the focus of this review. The following discussion will therefore present theoretical results regarding charged polarons; for excellent reviews giving a broader discussion of experimental and theoretical results regarding hybrid ion-atom systems see \citet{TomzaRMP2019} and \citet{Lous2022}.

At large distances, the interaction is attractive and arises from the electric field of the ion polarizing the atoms so that  $V(r) \to - C_4/r^4$, where $C_4$ is proportional to the polarizability of the atoms. The $1/r^4$ tail is longer range than the $1/r^6$ tail of the van der Waals interaction between neutral atoms. It sets the characteristic length $r_{ion}= \sqrt{2m_rC_4}$ and energy $\varepsilon_{ion} = 1/(2m_rr_{ion}^2)$. For a $^{87}$Rb$^{+}$ ion in a $^{87}$Rb BEC this gives $\epsilon_{ion}\approx80$nK and $r_{ion}\approx 260$ nm, which is the same order as a typical mean interparticle distance in atomic gases. This means that there is no separation of length scales so that the atom-ion interaction cannot in general be described by a contact pseudopotential. Its strength furthermore implies that the perturbation caused by an ion is much stronger than the one generated by a neutral impurity. 

At short distances, the electron clouds of the ion and the atom start to overlap giving rise to a strong repulsion. A model potential commonly used in the literature~\cite{KrychZbigniewPhysRevA.91.023430}
\begin{align}
\label{eq:ion:Vaireg}
V(r) = -C_4\frac{r^2 - c^2}{r^2 + c^2} \frac{1}{(b^2 + r^2)^2}.
\end{align}
features a $1/r^4$ attractive tail at large distances and strong repulsion at shorter ones, with a single minimum as a function of $r$. The parameter $c$ determines the onset of repulsion and typically $c\ll r_{ion}$. Different values of $b$ and $c$ give different short-range physics, but one typically assumes that the many-body physics is essentially the same as long as they give the same scattering length and energy of the highest bound state. 
The large strength of the atom-ion interaction may lead to a ``snow-ball'' state with many atoms bound to the ion as the dimer state energy decreases. As we shall see shortly,  calculations indeed predict this to occur for a bosonic bath in analogy with what has been experimentally observed for ions in liquid Helium~\cite{Atkins59,Chikina2007}. 

\begin{figure}[ht]
\centering
\includegraphics[width=\columnwidth]{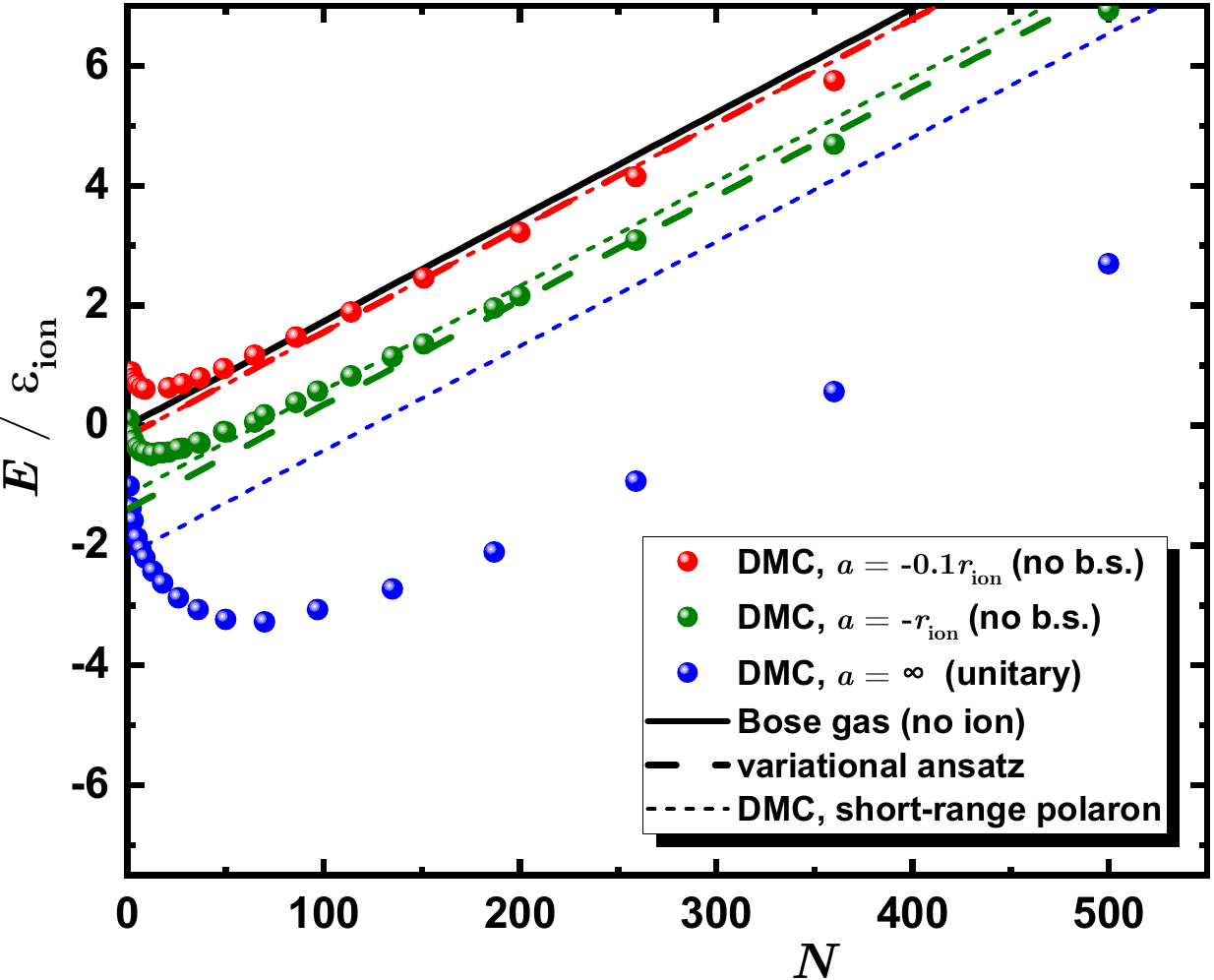}
\caption{\label{fig:ion:E}
{\bf Ionic Bose polarons.} 
The ground state energy of a single ion and $N$ bosons with equal masses in the regime where no two-body bound state is present. Symbols show DMC results, the long-dashed lines is Eq.~\eqref{eq:mu:variational}, and short-dashed lines show the DMC results for a short-ranged impurity-boson interaction~\cite{Pena2015}. The gas parameter is $n_0a_b^3=10^{-6}$, which for the interaction potential used here corresponds to $nr_{ion}^3=0.1288$. From \citet{Astrakharchik2021}.}
\end{figure}

Figure~\ref{fig:ion:E} shows the total ground state energy, $E$, of an ion in a bath of $N$ bosons of equal mass ($m=m_b$)  with the interaction potential given by Eq.~(\ref{eq:ion:Vaireg}). The results are obtained with DMC calculations, using a number of atoms $N$ in the range of a few hundreds~\cite{Astrakharchik2021}. For small negative scattering lengths, such as $a=-0.1r_{ion}$ (red) and $a=-r_{ion}$ (green), the system energy is close to the sum of the GP bulk energy and the polaron energy $\varepsilon$, i.e. $E(N) = \varepsilon + N g_bn/2$ with $g_b=4\pi a_b/m_b$ where $a_b$ is the boson-boson scattering length. Here, the polaron energy agrees well with the variational approximation~\cite{shchadilova2016}, 
\begin{equation}
\varepsilon=4\pi na_b^3\left(\frac{r_{ion}}{a_b}\right)^2\left(\frac{a_b}{a}-\frac{a_b}{a_0}\right)^{-1}\varepsilon_{ion},
\label{eq:mu:variational}
\end{equation}
where $a_{0}=\frac{32}{3\sqrt{\pi}}\sqrt{na_b^3}$ is the shift of the atom-ion scattering resonance position due to bath. In this weakly-interacting regime, the energy closely matches that of a short-range interaction with the same scattering length, as indicated by the dashed lines from the DMC calculations~\cite{Pena2015}. In contrast, at unitarity ($1/a=0$), Eq.~\eqref{eq:mu:variational} is no longer applicable and the energy  becomes of the same order as $\epsilon_{ion}$ significantly different from that of a neutral impurity.

\begin{figure}[ht]
\centering
\includegraphics[width=\columnwidth]{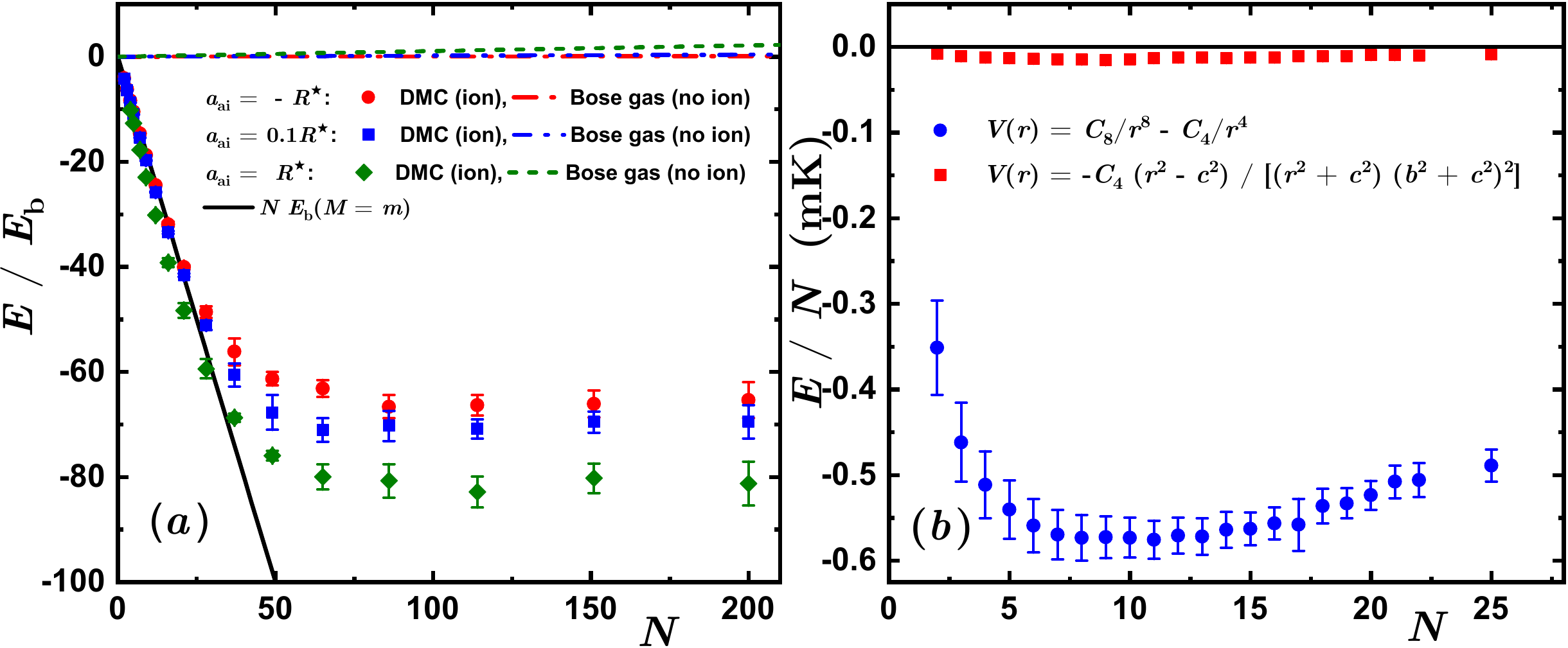}
\caption{\label{fig:ionPositivea:E}
{\bf Snowball states of ion-bosonic atom systems.}
(a) Ground state energy of a single ion and $N$ bosons for atom-ion attraction strong enough to support a two-body bound state of energy $E_b$. Symbols show DMC results and the solid line is $E(N) = N E_b$ where $E_b$ is the binding energy of an atom-ion dimer (specifically, $E_b/\varepsilon_{\rm ion} = -1.6,\, -9.0,\, -35$ for $a/r_{\rm ion} = 1,\,0.1,\,-1$). From \citet{Astrakharchik2021}. Right panel: The ground state energy of a $^{174}$Yb$^+$ ion in the presence of bosonic $^7$Li atoms as a function of the number of atoms obtained in DMC calculations for two characteristic models for atom-ion interaction potential. Adapted from \citet{Chowdhury2024}.}
\end{figure}

The effects of the strong atom-ion interaction become even more dramatic in the regime of positive scattering lengths $a>0$, where the atom-ion scattering problem supports a bound state. The left panel of Fig.~\ref{fig:ionPositivea:E} shows the energy as a function of the number of bosons in the bath for the same system as in Fig.~\ref{fig:ion:E}, but now in the region of stronger attraction, where the atom-ion potential supports a two-body bound-state. One clearly sees that the energy initially decreases linearly as $E(N) \approx N E_b$ with $E_b$ the energy of the bound state, indicating that the ion binds the bosons in the bath, creating a many-body bound state similar to the so-called snowballs formed by ions in liquid Helium~\cite{Atkins59,Chikina2007}. This reflects that the characteristic ion energy $\epsilon_{ion}$ is much larger than the chemical potential of the bath $g_b n$. Eventually, as the number of atoms is increased further, the energy flattens out reaching a minimum at a certain number of atoms $N_c\approx 140$, and then increases for larger $N$ as the ion cannot bind more atoms. 
Figure~\ref{fig:ionPositivea:E}(b) shows DMC calculations for the mass imbalanced case of a mobile $^{174}$Yb$^+$ ion in a gas of bosonic $^7$Li atoms for different atom-ion potentials either given by Eq.~\eqref{eq:ion:Vaireg} or by $V(r)=C_8/r^8-C_4/r^4$~\cite{Chowdhury2024}. These calculations predicted that the ion can bind around $8$ bosons. Also, the short range details of the potential were found to be important for determining the energy. This leaves an uncertainty regarding how many parameters are needed for a precise description of a mobile ion in a Bose gas.

\begin{figure}[ht]
\centering
\includegraphics[width=\columnwidth]{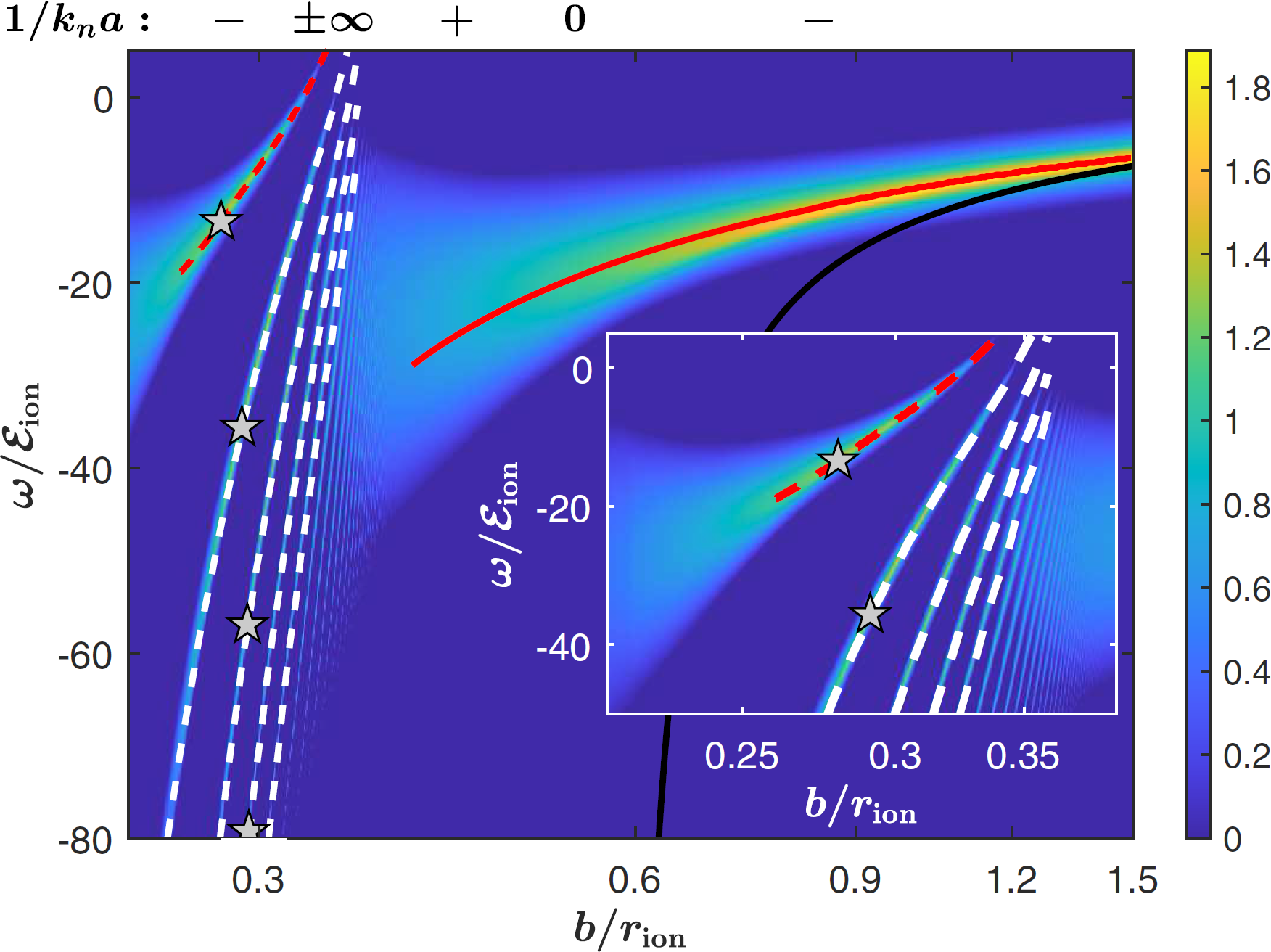}
\caption{\label{fig:ion:spectral_function}
{\bf Spectrum of a $p=0$ ion in a BEC}
as a function of $b$ (the corresponding $a$ is shown above the graph) with $n_0 r_{\rm ion}^3=1$, $c=0.0023r_{ion}$ and $m=m_b$. The red solid and dashed lines are polaron energies from the ladder approximation, the black line is the mean-field energy, and the white lines are ionic molecules containing an increasing number of bosons. The gray stars indicate where the BEC density at the ion reaches $n(0)=0.01a_b^{-3}$. From \citet{ChristensenCamachoGuardianBruun21}.}
\end{figure}

The spectral function of an ion in a BEC obtained from a coherent state ansatz (able to account for large dressing clouds) is shown in Fig.~\ref{fig:ion:spectral_function}: it features well-defined polarons, which are also captured by the ladder approximation (red lines). The white dashed lines are states containing an increasing number of bosons bound to the ion in agreement with the snowball states found in the DMC calculations described above. The gray stars indicate when the coherent state ansatz becomes unreliable due to a large local gas parameter close to the ion. Recently, a mobile ion in a BEC was explored using the LLP transformation combined with mean-field GP theory leading to Eq.~\eqref{eq:GPe} with the atom-ion interaction potential given by Eq.~\eqref{eq:ion:Vaireg}~\cite{Olivas2024}. The  polaron energy obtained from this approach agrees with that from the ladder approximation shown in Fig.~\ref{fig:ion:spectral_function} for weak to moderate interaction strengths, whereas deviations were found close to resonance. An alternative approach to describing a moving ion was proposed in Ref.~\cite{Oghittu2024}, where a master equation was derived to capture the system's dynamics.

\begin{figure}[ht]
\centering
\includegraphics[width=\columnwidth]{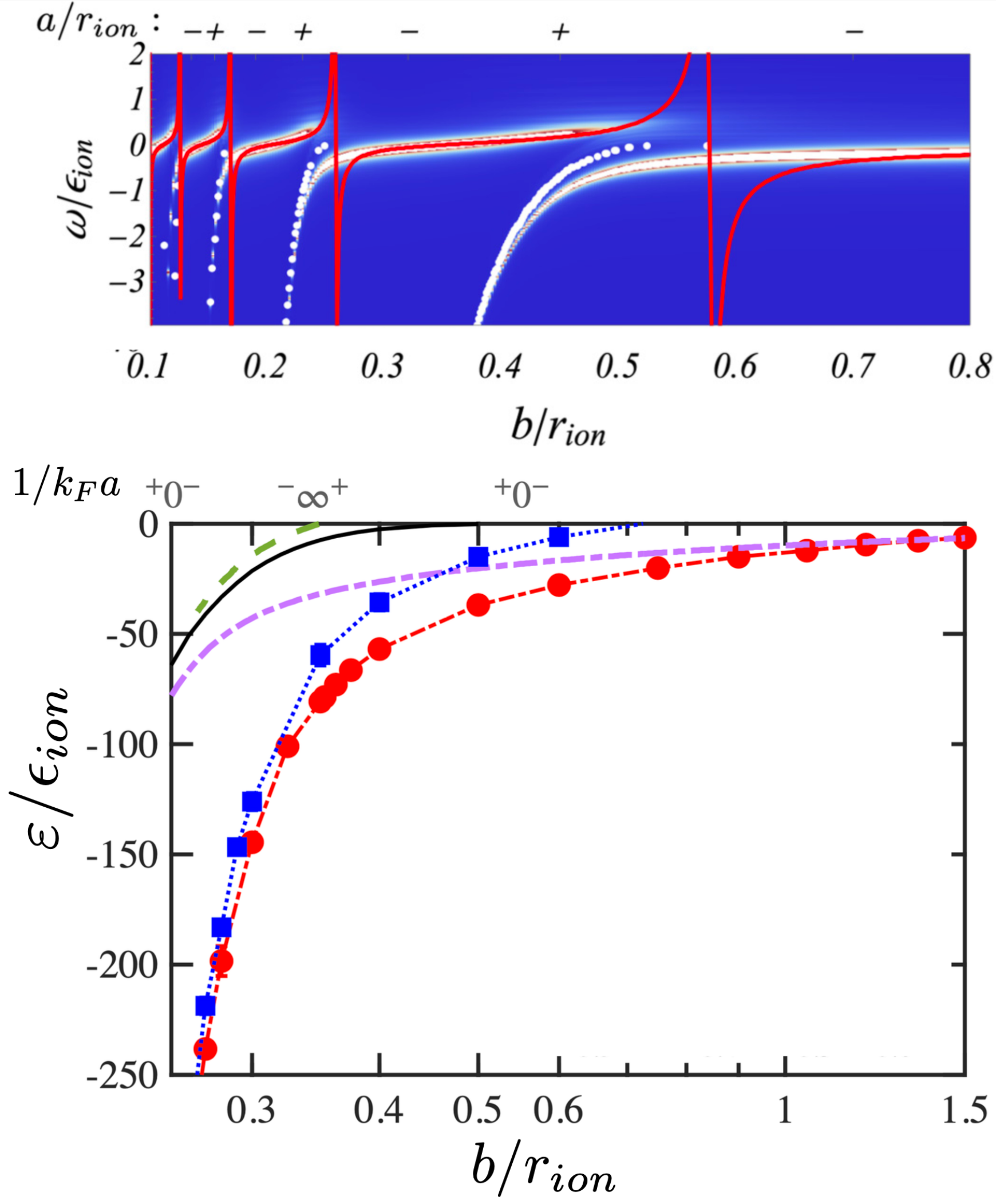}
\caption{\label{fig:ion:spectral_functionArdila}
{\bf Spectrum of a $p=0$ ion in a Fermi gas.}
Top: Spectrum in a dilute bath ($n_0r_{\rm ion}^3=0.01$) obtained from the ladder approximation, with $c=0.0023r_{\rm ion}$ and $m=m_b$~\cite{ChristensenCamachoGuardianBruun22}. The red line is the mean-field result. From \citet{ChristensenCamachoGuardianBruun22}.
Bottom: Spectrum in a dense bath ($nr_{ion}^3=1$). Red circles and blue squares are the polaron and dimer energies obtained by DMC method, purple dash-dotted and green dashed lines are the corresponding results from the ladder approximation, and the black line is the dimer energy in vacuum. From \citet{pessoa2024fermipolaronatomionhybrid}. The scattering length is shown above each graph.}
\end{figure}

We now turn to the properties of an ion in an ideal Fermi gas, which should be more robust towards the formation of snowball states due to the Pauli principle. The top panel of Fig.~\ref{fig:ion:spectral_functionArdila} shows the spectral function of an ion in a Fermi gas obtained from the ladder approximation. Another polaron branch emerges every time the interaction supports a new bound state so that there are $N+1$ branches when the interaction supports $N$ bound states in agreement with the results of Ref.~\cite{Massignan2005}.
For this low density, the polaron is well described by a short range interaction with the same scattering length. The ionic Fermi polaron was also explored using fixed node MC calculations with the interaction Eq.~\eqref{eq:ion:Vaireg}, and the results are shown in the bottom panel of Fig.~\ref{fig:ion:spectral_functionArdila}.
Large deviations between the predictions of the QMC and the ladder approximation were found for strong interactions. Using a Landau-Pekar energy functional and the Thomas-Fermi approximation, an ion interacting with a neutral ideal Fermi gas via a $-C_4/r^4$ potential was predicted to develop a diverging effective mass and therefore localize (self-trap)~\cite{Mysliwy2024}. 

The picture emerging from these works is that the number of bosonic atoms bound to an ion can be much larger than for neutral impurities leading to the formation of snowball states in agreement with what is found using different methods~\cite{Lukin2002,Massignan2005,Schurer2017}. Also, contrary to the case of a neutral Fermi polaron, the accuracy of the ladder approximation in the ionic case remains uncertain, since the suppression of $n\ge 2$ correlations is less efficient for the long-range atom-ion interaction.

\subsection{Dipolar polarons}   \label{sec:DipolarPolarons}

\begin{figure}[ht]
\centering
\includegraphics[width=\columnwidth]{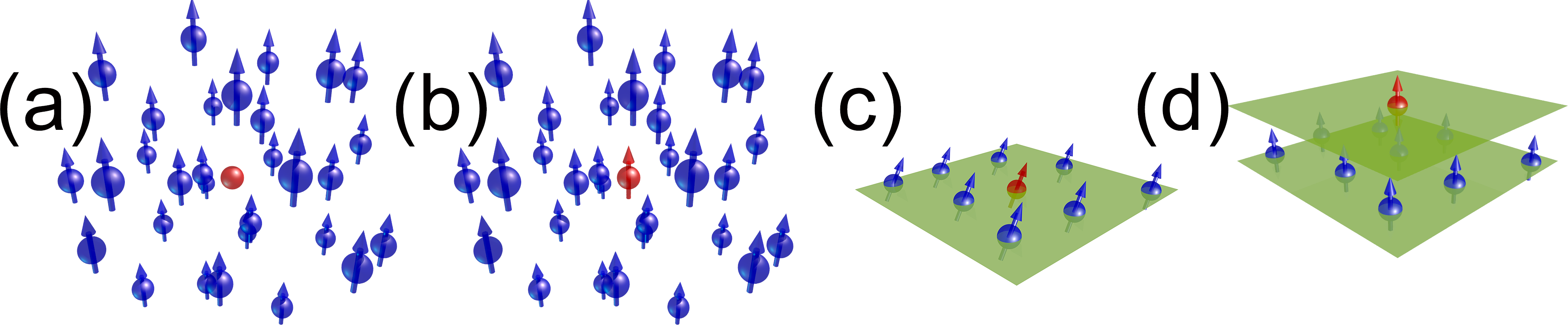}
\caption{\label{fig:dipoles:sketch}
{\bf Dipolar polarons.} 
A variety of configurations may be obtained with fully-aligned dipoles:
(a) non-dipolar impurity (red) in a dipolar gas (blue);
(b) dipolar impurity in a dipolar gas;
(c) dipolar impurity in a dipolar gas, all trapped in the same 2D layer.
(d) dipolar impurity in one layer interacting with a dipolar gas in another parallel layer.}
\end{figure}

Significant progress has been made on the trapping and cooling of atoms with a permanent magnetic or electric dipole moment~\cite{Chomaz_2023,Valtolina2020,Schindewolf2022,Bigagli2024}. These advances provide promising experimental platforms to explore the rich physics of dipolar interactions including novel quantum many-body physics~\cite{Trefzger_2011,Baranov2012}. So far there are however no experimental results regarding polarons in dipolar systems, so in the rest of this Section we will focus on the theoretical predictions for the cases illustrated in Fig.~\ref{fig:dipoles:sketch}. 

The interaction between two parallel dipoles is
\begin{equation}
\label{Eq:dipolar_interaction}
V({\bf r}) = \frac{D^2}{4\pi r^3} \left(1-3\cos^2\theta\right)\;,
\end{equation}
where $\theta$ is the angle between their dipole moments ${\bf d}$ and their relative separation ${\bf r}$. For magnetic dipoles one has $D^2=d^2\mu_0$ whereas $D^2=d^2/\epsilon_0$ for electric dipoles. Crucially, the dipolar interaction has attractive and repulsive regions: when two dipole moments are oriented head to tail they attract each other, whereas they repel when they are oriented side by side. This anisotropy leads to polarons with unique properties.

The cases of a non-dipolar and dipolar impurity immersed in a dipolar Bose gas, see Fig.~\ref{fig:dipoles:sketch}(a)-(b), were considered using a coherent-state variational ansatz. 
~\citet{Volosniev2023} computed the anisotropic dressing and relaxation dynamics of the impurity after a sudden switching on of the dipolar interaction, and they found an effective mass that depends on the direction of motion relative to the polarization axis. The case of a static impurity in a dipolar gas was explored in Ref.~\cite{shukla2024anisotropicpotentialimmerseddipolar}. The spectral function of a non-dipolar impurity in a dipolar Bose gas, Fig.~\ref{fig:dipoles:sketch}(a), was calculated using second order perturbation theory revealing an anisotropic dispersion~\cite{KainLing14}. It was furthermore analysed how the anisotropy of the phonon spectrum affects the Cherenkov radiation when the impurity moves faster than the speed of sound~\cite{Cherenkov37}. Using a time-dependent GPE, the density profile and breathing dynamics of a dipolar condensate in a harmonic trap interacting with a non-dipolar impurity was calculated~\cite{Guebli19}. The breathing modes were also analysed using a variational approach~\cite{Hu14}. A Fermi polaron formed by a non-dipolar impurity in a dipolar Fermi gas was shown using the Chevy ansatz Eq.~\eqref{eq:ChevyAnsatz} to have anisotropic properties stemming from the Fermi surface being deformed by the dipolar interaction~\cite{Nishimura21}. 

The energy, effective mass, and quasiparticle residue of a Bose polaron formed by a dipolar impurity in a quasi-2D dipolar Bose gas [as shown in Fig.~\ref{fig:dipoles:sketch}(c)], were calculated using second order perturbation theory~\cite{Ardila_2018}. Analysing dipolar mixtures using a Lee-Huang-Yang energy functional, it was shown that quantum fluctuations of a dipolar Bose gas can strongly modify the miscibility of dipolar impurities~\cite{Bisset2021}. The same 2D problem was examined using DMC for an arbitrary angle of the dipoles with respect to the plane~\cite{SanchezBaena24} [see Fig.~\ref{fig:dipoles:sketch}(c)], finding that the polaron energy and the quasiparticle residue follow a universal behavior with respect to the angle when scaled in terms of the $s$-wave scattering length. The properties of a Fermi polaron formed by a dipolar impurity interacting with a gas of dipolar fermions were calculated using DMC method in 2D~\cite{Bombin2019}. By comparing with the case of short range interactions, it was found that polaron properties are universal depending only on the scattering length $a$ for low densities, while for larger densities the specific shape of the interaction becomes important. 

Dipoles in one or more layers give rise to new and interesting effects. Even the problem of one dipole in one layer interacting with a dipole in another parallel layer with both dipole moments perpendicular to the layers is quite delicate, since the  integral of their mutual interaction  is exactly zero. It follows that one cannot use the usual criterion for a bound state in 2D - namely that the integral of the interaction is negative~\cite{LandauLifshitz_iii}. Eventually, it was shown that a two-body bound state in fact always exists although its energy can be exponentially small~\cite{Baranov11}. Trimers and tetramers were later predicted for interlayer distances beyond a certain threshold~\cite{Guijarro20,Guijarro21}.

\begin{figure}
\centering
\includegraphics[width=\columnwidth]{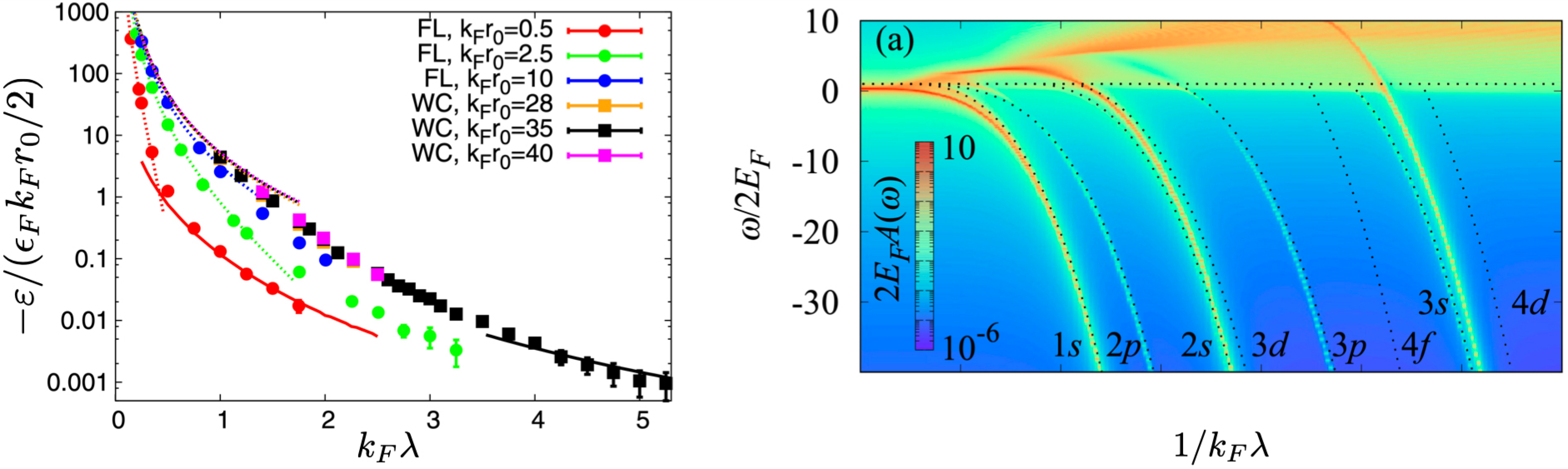}
\caption{\label{fig:dipoles:spectralFunction}
{\bf Dipolar Fermi polaron in a bi-layer.}
Left: Energy of a dipolar impurity confined in a layer interacting with a gas of dipolar fermions in a different layer as a function of the distance between layers $\lambda$ for different values of in-plane interaction strength $ r_0=mD^2/4\pi$. Dashed lines are two-body binding energies and solid lines correspond to perturbation theory. Circles/squares refer to the fermions forming a Fermi liquid/Wigner crystal. 
From \citet{MatveevaGiorgini13}. 
Right: Zero momentum spectral function of a dipolar impurity in the same setup  using the Chevy ansatz. The (black) dotted lines are $1s$ to $4d$ dimer energies with unbinding occurring at $\omega=2\epsilon_F$ (horizontal dotted line). From \citet{Tiene24}.}
\end{figure}

A Fermi polaron formed by a dipole in one layer interacting with a dipolar Fermi gas in another parallel layer with all dipole moments perpendicular to the planes was analysed using fixed-node DMC~\cite{MatveevaGiorgini13}. The left panel of Fig.~\ref{fig:dipoles:spectralFunction} shows the polaron energy 
as a function of the interlayer separation $\lambda$, for different values of the dipole strength $r_0$. In the regime where the dipolar Fermi gas forms a Wigner crystal, the properties of the polaron can be understood as an impurity coupled to lattice phonons much like the Fr\"ohlich model for electrons interacting with crystal phonons, see Sec.~\ref{sec:BosePolarons}. The effective mass of this polaron was found to become very large for small layer distances indicating  self-trapping. 
\citet{Tiene24} studied the same bilayer geometry using the Chevy variational wave function  Eq.~\eqref{eq:ChevyAnsatz} neglecting interactions between the fermions. The impurity spectral function is shown in the right panel of Fig.~\ref{fig:dipoles:spectralFunction}. Multiple polaron branches are found, which arise from two-body interlayer bound states with different angular momenta. The branch with the strongest spectral weight and lowest energy is the one associated with the $s$-wave scattering, and its energy is consistent with the DMC calculations of Ref.~\cite{MatveevaGiorgini13}.

In conclusion, dipolar interactions give rise to interesting  effects for Bose and Fermi polarons not present for short range interactions. This includes anisotropic properties, multiple bound states and polaron branches. Many open questions remain such as a complete understanding of the possible self-localisation for dipolar polarons~\cite{Mysliwy2024}, and the effects of a non-zero temperature.

\subsection{Rydberg polarons}


\begin{figure}[t]
\centering
\includegraphics[width=\columnwidth]{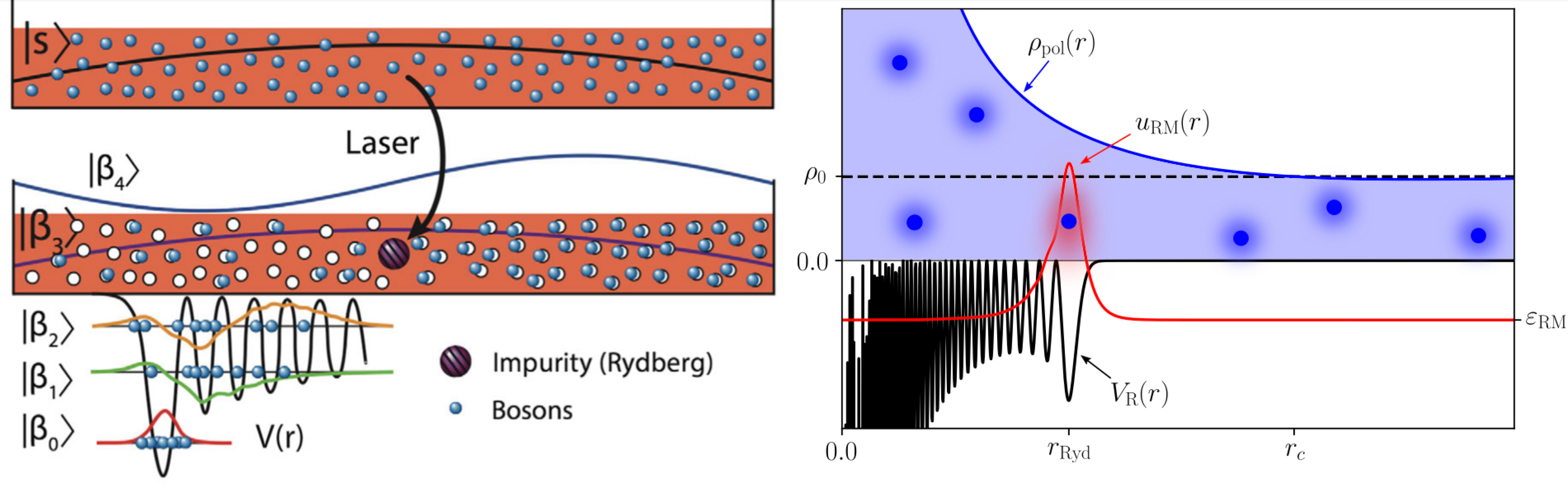}
\caption{\label{fig:RydPolCartoon}
\textbf{Rydberg polarons.} 
Left: Exciting an impurity to a Rydberg state inside a Bose gas leads to the formation of a very large quasiparticle. From \citet{Camargo2017}.
Right: Rydberg impurities can be used to observe the real-time formation of polaron clouds on suboptical lengths scales. From \citet{GieversSchmidt_Probing_2024}.}
\end{figure}

The excitation of atoms to Rydberg states results in the formation of atoms of greatly enhanced size and increased polarizability, generating strong and long-range interactions, which have been explored in a wide range of studies, from polariton physics to spin models. Shifts and broadening of the atomic Rydberg lines have been observed in BECs and dense atomic gases, arising from the scattering of ground-state atoms with the outer electron of the excited Rydberg atom. The underlying electron-atom interaction has the typical attractive $1/r^4$ form, which for many alkali species gives rise to a negative scattering length $a_e<0$. As a result, the neutral atoms experience an effective attractive potential generated by the Rydberg atom that is an image of the Rydberg wave function $\psi_{n}(\vecr)$ and given by
\begin{equation}
  V_\text{Ryd}(\vecr) = \frac{2\pi \hbar^2 a_e}{m_e}|\psi_{n}(\vecr)|^2.
\end{equation}
This interaction can be extremely long-ranged and supports a multitude of bound states (so-called Rydberg molecules) between the Rydberg atom and atoms from its environment, see left panel of Fig.~\ref{fig:RydPolCartoon}. The physical mechanism responsible for the shifts and the specific shape of the Rydberg lines in bosonic baths
was attributed by \citet{SchmidtDem2016} to the formation of a ``superpolaron'' state, in which the impurity atom is dressed by a large number of bound states of the Rydberg potential. This state was later observed, confirming the underlying bound-state physics as the driving mechanism behind the formation of Rydberg polarons~\cite{Camargo2017}. In order to describe the formation of Rydberg polarons in a BEC,  \citet{SchmidtDem2016,SchmidtR2018} employed a hybrid approach using a time-dependent coherent state and functional determinants, 
and \citet{drescher2024bosonic} later adapted this method to describe the excitation of heavy impurities in a Bose gas.


Describing Rydberg polarons in a Fermi gas requires accounting for the competition between bound state formation   with high-angular momentum and Pauli blocking~\cite{Sous2019}. \citet{GieversSchmidt_Probing_2024} proposed to use Rydberg excitations as an in-situ, spatially resolved, real-time probe of the formation of the dressing cloud 
 on suboptical length scales reaching down to the 50nm regime, see right panel of Fig.~\ref{fig:RydPolCartoon}.

\subsection{Rotating molecules in a quantum bath: angulons}

In most parts of this review we discuss how the translational motion of an impurity is changed by its interaction with a quantum many-body environment. One can, however, also raise the question how \textit{rotation} may be modified by similar polaronic effects. This question is indeed of central importance for the spectroscopy of molecules in solvents, in particular superfluid He$^\mathrm{4}$ nanodroplets, and it was observed in quantum chemistry experiments that the rotational constant of molecules is  changed by interaction with a many-body environment. In Ref.~\cite{Schmidt2015ang}, it was proposed that this effect can be understood in terms of the formation of a dressed quantum rotor similar to a Bose polaron, where  a rotating polaron cloud  inhibits the bare molecular rotation~\cite{SchmidtLem2016}.  It was later shown how the dynamics of such an "angulon" formation can be used for rotational cooling~\cite{Will2019}. Using variational ans\"atze~\cite{Zeng2023VarAngulon}, the concept of angulons was  adapted to molecular ions \cite{Midya2016} as well as inter-angulon interactions \cite{LiSchmidt_Mediated2020}. Moreover, it was shown how bound state formation requires accounting for the full Bose polaron model extended to rotation~\cite{Dome_LinRotThreshold_2024}. Angulons were also studied in real space using the GPE~\cite{suchorowski2024quantumrotortwodimensionalmesoscopic}. For an in-depth review of the progress on angulons we refer to the review \cite{lemeshko2017}.

\section{Polarons in 2D materials}   \label{TMDsection}

With advances in fabrication of atomically thin TMDs such as MoSe$_2$, MoS$_2$, MoTe$_2$, WSe$_2$, and WS$_2$ and their van der Waals heterostructures, a whole new class of  2D quantum materials can be designed and fabricated with applications in fundamental science and technology~\cite{Schaibley:2016aa, TMD-exciton-review}. As discussed in this section, this includes  recent realisations of both the  Fermi polaron formed by excitons interacting with electrons, and Bose the polaron formed by excitons in two spin states. At first glance, TMDs appear very different from ultra-cold atoms due to their  non-equilibrium nature involving a short-lifetime quantum impurity (exciton) coupled to interacting fermions (electrons), as well as their much higher densities and smaller particle masses. In a way, the success of the polaron model in describing the optical excitation spectrum of these 2D materials is a manifestation of the power and universal applicability of this framework. 

Before proceeding with a detailed discussion of polarons in TMDs, we highlight some of the unique features of these materials: 1) they are truly 2D since  the layers are atomically thin with even the size of the quantum impurity (exciton)  larger than the layer thickness $d$; 2) they are very cold, reaching $T/T_F\lesssim0.01$ in dilution refrigerators
; 3) the hybridisation of excitons with photons in an optical cavity gives rise to polaritons acting as impurities with an extremely light mass; 4) the strongly interacting nature of the electrons and holes imply that a plethora of many-body phenomena affect the polaron spectra, and therefore the latter may be used as optical probes of interesting strongly-correlated electronic states (such as Wigner crystals, kinetic magnetism, or integer and fractional Chern insulators); 5) at high exciton densities, it is possible to reach a degenerate Bose-Fermi regime, and the ability to tune the exciton decay rate using hybrid excitons ensures that one can study both driven-dissipative and equilibrium regimes.

\subsection{Excitons in TMDs}

\begin{figure}[ht]
\centering
\includegraphics[width=0.8\columnwidth]{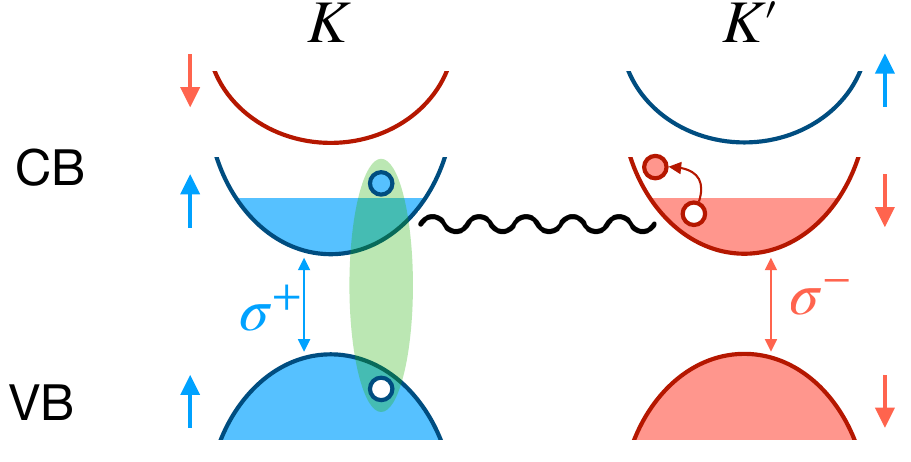}
\caption{\label{fig:TMDcartoon}
{\bf Exciton-polarons.}
TMD layers have a bipartite honeycomb lattice structure and the minima (maxima) of the conduction (valence) band are located at the K and K' points of the Brillouin zone. Absence of inversion symmetry ensures that the electronic spectrum is gapped at these points. Electrons in the K (K') valley can be optically excited using right (left) hand circularly polarized light. Large spin-orbit splitting in turn ensures that, by choosing appropriate light polarization and energy, it is possible to generate electrons with a well defined spin-valley quantum number. 
An exciton (green ellipse) excited in the $K$ valley by light with $\sigma^+$ circular polarization interacts strongly with electrons of opposite spin in the $K'$ valley, leading to the formation of an {\it exciton-polaron}.}
\end{figure}

TMDs are direct band gap semiconductors with extrema of their dispersion located at the so-called $K$ and $K'$ points in the hexagonal Brilllouin zone, which are connected by time-reversal symmetry. There is a large spin-valley splitting  as well as valley selective light-matter coupling using circularly polarized light~\cite{Mak2012,Zeng2012}, see Fig.~\ref{fig:TMDcartoon}. The optical excitation spectrum is dominated by a strongly bound exciton state of a conduction band (CB) electron and a valence band (VB) hole  due to the relatively heavy carrier mass and reduced dielectric screening in 2D~\cite{Chernikov2014}. As a consequence, the exciton has a small radius, which we refer to  as the Bohr radius even though the attractive potential yielding the bound state deviates from Coulomb's law at short distances and is better described as Rytova-Keldysh potential~\cite{Keldysh1979}. It follows that, to a good approximation, the excitons in the present context can be regarded as bosonic impurity particles, which is a crucial feature for realising polarons in these systems. 

The relevant length and energy scales include~\cite{Goryca2019,TMD-exciton-review}: 
\begin{enumerate}
  \item The lattice constant and layer thickness are $a_L \simeq 0.3$ nm and $d \simeq 0.7$ nm. These are the smallest length-scales in the problem, which justifies a continuum 2D treatment.
  \item The exciton radius $a_x$ ranges from $1.2$~nm in MoSe$_2$ to $1.7$~nm in WSe$_2$. Even though $a_x$ is only approximately four times larger than the lattice constant $a_L$, the excitons can be described as Wannier excitons. 
  \item The  measured binding energy is $|\epsilon_x|\simeq 200$~meV for the ground state $1s$ exciton when the monolayers are encapsulated by hexagonal boron nitride (hBN). For a suspended monolayer, the binding energy is predicted to be $|\epsilon_x|\simeq500$meV.
  \item The binding energy of an exciton-electron or exciton-hole state, i.e.\ a trion, is $|\epsilon_T|\sim 20-30\text{meV}$. 
  \item The rms size of the trion is $a_T \simeq 2.0 - 2.5$ nm. 
  \item The  Fermi momentum is in the range $1/k_F \simeq 1.5 - 20$ nm (corresponding to a Fermi energy $\epsilon_F\simeq 0.1 - 20$~meV), and it is readily tunable using applied gate voltages. The lower limit for $k_F$ arises either from disorder or many-body correlations that compete with the formation of a Fermi liquid. The upper limit is determined by the breakdown of the description of an exciton as a point particle.
\end{enumerate}

We note that the fundamental parameters of TMD excitons, such as $\epsilon_x$ and $a_x$ cannot be measured directly. Instead, the diamagnetic shift of the 1s exciton $\Delta \epsilon=e^2 a_x^2 B_z^2/8m_r$ at high magnetic fields ($B_z$) can be used to determine $a_x$; this method relies on an estimate of the reduced mass $m_r$ from a combination of angle-resolved photoemission spectroscopy (ARPES) and transport measurements or from density-functional theory (DFT) calculations. In turn, $\epsilon_x$ can be determined through measurement of  Rydberg exciton resonances (principal quantum numbers $2 \le n \le 5$) together with calculations based on the Rytova-Keldysh potential. In contrast, the trion binding energy $\epsilon_T$ is directly accessible in optical spectroscopy~\cite{Xu2014}; an estimate of $a_T$ from $\epsilon_T=-1/2m_ra_T^2$ is in good agreement with calculations based on exact diagonalization~~\cite{Fey2020}.

\subsection{Exciton-polarons}   \label{Sec:excitonpolarons}

The chemical potential $\mu$ and hence the itinerant electron or hole density in TMD layers can be tuned by gate voltages. When $\mu$  is set in between the valence band maximum and conduction band minimum, the material is charge neutral. In this case, the optical excitation spectrum is dominated by the 1s-exciton, with a linewidth determined by the radiative decay rate in high quality samples. Upon increasing (decreasing) $\mu$ above (below) the conduction (valence) band minimum (maximum), one injects itinerant electrons (holes) into the monolayer. Experimentally, this leads to the observation of a red-detuned resonance that appears in the spectrum upon injection of charged carriers, which initially was attributed to formation of trions~\cite{Mak2013,Ross2013,Zhang2014,Chernikov2015,Zhu2015,Cadiz2016,Courtade2017}. The large oscillator strength and  narrow linewidth of this red-shifted resonance is however inconsistent with a trion-based description, since the trion has only a small oscillator strength. This in close analogy with the small spectral weight of the dimer state in the atomic impurity spectral function as discussed in Sec.~\ref{sec:FermiPolarons}. Pioneering theoretical and experimental work then demonstrated that for $\epsilon_F\lesssim|\epsilon_T|\ll|\epsilon_x|$, the Fermi-polaron framework instead provides a more appropriate description of the elementary optical excitations~\cite{Sidler2016,Efimkin2017,Rana2020,Fey2020}. A similar framework was previously developed to describe the interacting exciton-electron problem~\cite{suris_03}. 

Even though the validity of Fermi-polaron description was initially demonstrated experimentally for a MoSe$_2$ monolayer embedded inside an optical cavity~\cite{Sidler2016}, the majority of the experiments are carried out  where excitons couple to free-space radiation field modes. In this case, the coupling between the photons and the excitons can be treated as a weak perturbative probe and one is left with the problem of analysing an exciton interacting with itinerant electrons, which will be discussed in this section. The case of a strong photon-exciton coupling achieved by immersing the semiconductor in an optical cavity is discussed in Sec.~\ref{Sec:MicroPolaritons}. 

Assuming that the effects of interactions between the electrons can be included in their band energy $\epsilon_{b\mathbf k}$ 
and treating the excitons as point bosons, the Hamiltonian describing an exciton interacting with itinerant electrons is given once more by Eq.~\eqref{eq:HF}, where the operators $\hat{c}^\dagger_\bk$ and $\hat{f}^\dagger_\bk$ now create an exciton and an electron with momentum $\bk$. The exciton-electron interaction has a range $\sim a_x$ following the classical charge-dipole interaction $-\alpha_0 e^2/(4 \pi \epsilon r^4)$ for distances $r\gg a_x$, where $\alpha_0$ is the polarizability. Thus, one should expect it for most present purposes to be well described by a contact interaction when $a_x$ and  $r_0=\sqrt{\alpha_0 m_e e^2/(2 \pi \epsilon)}$ are much smaller than the interparticle distance. Microscopic calculations indeed show that for most purposes a phenomenological contact potential gives accurate results~\cite{Efimkin2021,Fey2020,Kumar2024}. 

Within these approximations, the problem becomes identical (apart from being 2D) to that discussed in Sec.~\eqref{sec:FermiPolarons}: A mobile impurity (exciton) interacting via a short range potential with a Fermi sea (electrons). This means that these quasiparticles can be studied by means of the approaches developed for Fermi polarons. Like in 3D, a momentum independent interaction $g_{ex}$ gives rise to an ultraviolet divergence, which can be cured by relating the energy $\epsilon_T$ of a bound exciton-electron or exciton-hole state, i.e.\ a trion, to $g_{ex}$. Indeed, a bound state corresponds to a pole in the scattering matrix, $1-g_{ex}\Pi_v(0,\epsilon_T)=0$, which gives~\cite{Randeria1990,Wouters2007,Carusotto_2010,Zoellner2011,Levinsen2013,Schmidt2012b}
\begin{align}\label{LSE}
  \frac 1{g_{ex}}=\sum_{|\B k|<\Lambda} \frac{1}{\epsilon_T- \epsilon_{x\B k} - \epsilon_{e\B k} },
\end{align}
where $\Lambda$ is an UV cut-off related to $1/a_x$. This relation can be used to eliminate the coupling constant $g_{ex}$ and the cut-off in favor of the trion energy. Since $a_x$ is smaller than all other relevant length-scales, one may take the limit $\Lambda \to \infty$ at the end of the calculation. Using this approach, the 2D scattering matrix can be written as 
\begin{equation}\label{Tmatrix2D}
{\mathcal T}(\bk,\omega)=\frac{1}{\Pi_v(0,\epsilon_T)-\Pi(\bk,\omega)}
=\frac{g_{2}}{1-g_{2}\Delta \Pi(\bk,\omega)}
\end{equation}
where the pair propagators are given by Eq.~\eqref{Pairpropagator} for 2D, 
$\Delta \Pi(\bk,\omega)=\Pi(\bk,\omega)-\text{Re}\Pi_v(0,\epsilon_F)$, and 
\begin{equation}
g_2=-\frac{\pi}{m_r}\frac1{\ln(k_Fa_Tm_r/m)}.\label{Coupling2D}
\end{equation}
The second equality in Eq.~\eqref{Tmatrix2D} is obtained by adding and subtracting the vacuum pair propagator evaluated at a typical many-body energy $\text{Re}\Pi_v(0,\epsilon_F)$ in the denominator.  In this way, we can extract the typical interaction strength given by Eq.~\eqref{Coupling2D}, which should be compared with the corresponding 3D parameter $k_Fa$.  This shows that the strongly interacting regime is for $k_Fa_T\sim 1$~\cite{Bloom1975,Engelbrecht1992}, and that there is no unitarity regime $1/a_T=0$ since a bound state always exists in 2D. 

\begin{figure}[ht]
\centering
\includegraphics[width=\columnwidth]{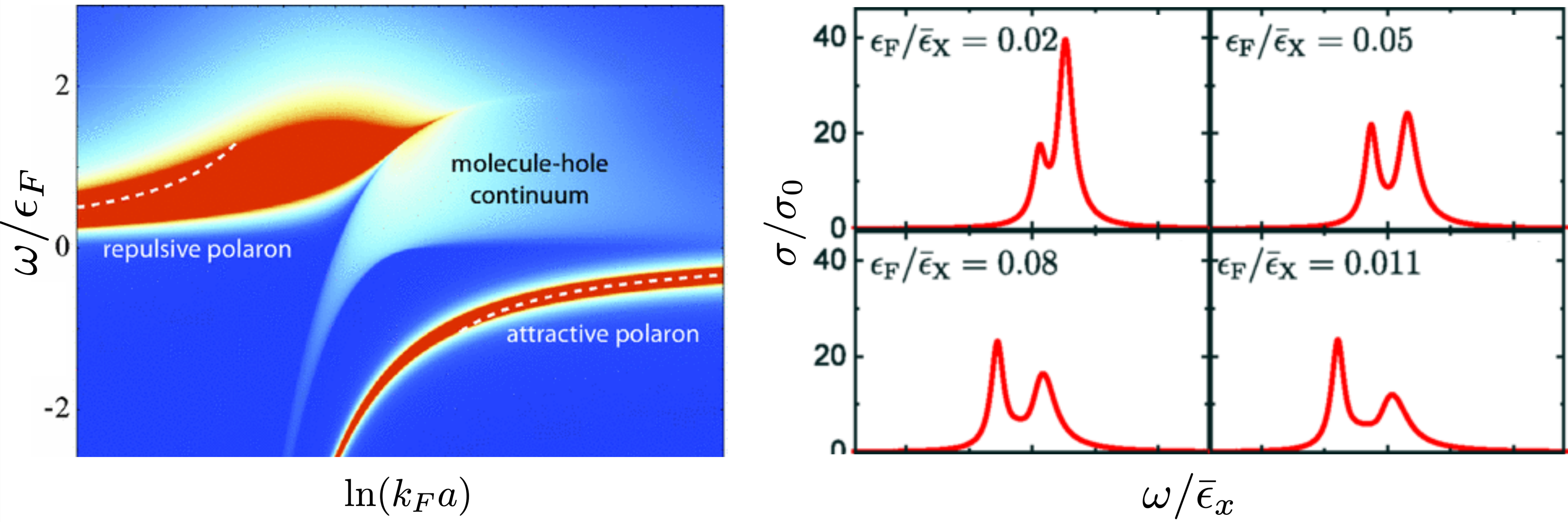}
\caption{\label{fig:Schmidt2D}
\textbf{Exciton spectra.} 
Left: Zero momentum exciton spectral function for $m_x=m_e$ at fixed electron density. From~\citet{Schmidt2012b}. Right: Optical conductivity of a 2D semiconductor (proportional to the $p=0$ exciton spectral function) for different electron densities. Here $\bar \epsilon_X$ is the exciton binding energy in the absence of the Fermi sea, and the trion binding energy is fixed at $|\epsilon_T|=0.07\bar \epsilon_X$. Recall that $\epsilon_T\propto 1/a_T^2$. From \citet{Efimkin2017}.}
\end{figure}
 
The left panel of Fig.~\ref{fig:Schmidt2D} shows the exciton spectral function obtained from the ladder approximation  Eqs.~\eqref{eq:ChevyAnsatz}-\eqref{eq:SelfenergyLadder}~\cite{Schmidt2012b}. We clearly see two branches corresponding to  an attractive and repulsive polaron. The energy unit is $\epsilon_F$ since this is typically constant in atomic gases while $a$ is tuned. The right panel shows the optical conductivity of 2D semi-conductors, which is proportional to the zero momentum exciton spectral function calculated using a a similar approach~\cite{Efimkin2017}. This shows a characteristic prediction of the polaron theory: the spectral weight of the attractive polaron  and its energy splitting to the repulsive polaron increases with the electron density, where the energy unit is $|\epsilon_x|$ since in semiconductors one typically tunes the Fermi density at fixed trion binding energy $\epsilon_T$ 
(recall that $\epsilon_T\propto 1/a_T^2$). The ladder approximation should be accurate when $\epsilon_F \lesssim |\epsilon_T|\ll |\epsilon_x|$, whereas three-body correlations not included in the theory may become important for $\epsilon_F\ll|\epsilon_T|$. It has however been shown that the polaron picture yields oscillator strengths almost indistinguishable from a description based on independent trion excitations in the regime of a Fermi energy smaller than the line broadening~\cite{Glazov2020}. We also remark that Fermi polarons within the ladder approximation can be mapped to the formation of bright polaritons within the Tavis-Cummings model, which allows new insight into the nature of Fermi polarons~\cite{imamoglu_TCModel_2020}. 

We emphasize that the Hamiltonian Eq.~\eqref{eq:HF} assumes that the excitons are robust mobile bosonic impurities in the $1s$-state of the quantized relative electron-hole motion. An experimental validation of this assumption is discussed in Sec.~\ref{Sec:ExcitonPointParticle}. Equation~\eqref{eq:HF} also suppresses any valley and spin degrees of freedom since we assume that excitons generated in the K (K') valley interact predominantly with electrons in the lowest energy spin state of the K' (K) valley as illustrated in Fig.~\ref{fig:TMDcartoon}. This is a consequence of (i) the large spin-orbit interaction ensuring that only the lowest energy spin state of either valley is occupied for $\epsilon_F \lesssim 20$~meV, and (ii) Pauli exclusion ensuring that a bound trion state exists only if the exciton and the electron occupy opposite valleys so that they have different spins, see Fig.~\ref{fig:TMDcartoon}. The interaction between an exciton and an electron (or hole) in the same valley does not generally support a bound state, since the Pauli exclusion principle prohibits the electron bound in the exciton from coming close to those in the Fermi sea~\cite{Tiene2022}. Since attractive and repulsive polaron physics mainly arise from interactions between excitons and electron (or holes) occupying opposite valleys, we will mostly ignore intra-valley interaction in the following. These assumptions hold for all hole-doped TMD monolayers, as well as for electron-doped MoSe$_2$ and MoTe$_2$ layers. The opposite sign of spin-orbit interaction in W-based TMD monolayers leads to a richer polaron spectrum \cite{TMD-exciton-review}. 

The intrinsic radiative decay of the excitons is also neglected in a description based on the Hamiltonian in Eq.~\eqref{eq:HF}, which is justified when the  decay rates are small compared to the relevant energies of the problem. In practice, the radiative decay rate can be reduced to $\Gamma_{\rm rad} \simeq 0.5$~meV by properly choosing the thickness of the hBN layers, which is comparable to a thermal energy $T\approx 4$K. Finally, we have ignored Coulomb interactions between electrons so far, which completes the mapping to the Fermi polaron problem in atomic gases. At a first glance this looks problematic since this interaction is not a-priori small. In most cases however, the dominant effect of electron-electron interactions can be captured through an
interaction modified band structure $\varepsilon_{\mathbf k}$ in Eq.~\eqref{eq:HF}. In the limit of strongly correlated electrons, the appropriate starting point for the mobile impurity problem would be the dressing of an excitons via collective (bosonic) excitations out of the correlated electronic ground state
as will be discussed in Sec.~\ref{Excitonsasprobes}. Interestingly, the charge of the electrons in the dressing cloud also makes it possible to manipulate the polarons via a Coulomb drag effect~\cite{Efimkin2018,Cotlet2019}, which has been experimentally realised using the polaron-polaritons~\cite{Chervy2020}. In a similar context it has recently been shown that the dressing of excitons by electrons can lead to a striking change in the diffusion of excitons~\cite{upadhyay2024giant}, see also~\cite{zerba2024tuning}.

\begin{figure}[ht]
\centering
\includegraphics[width=0.50\columnwidth]
{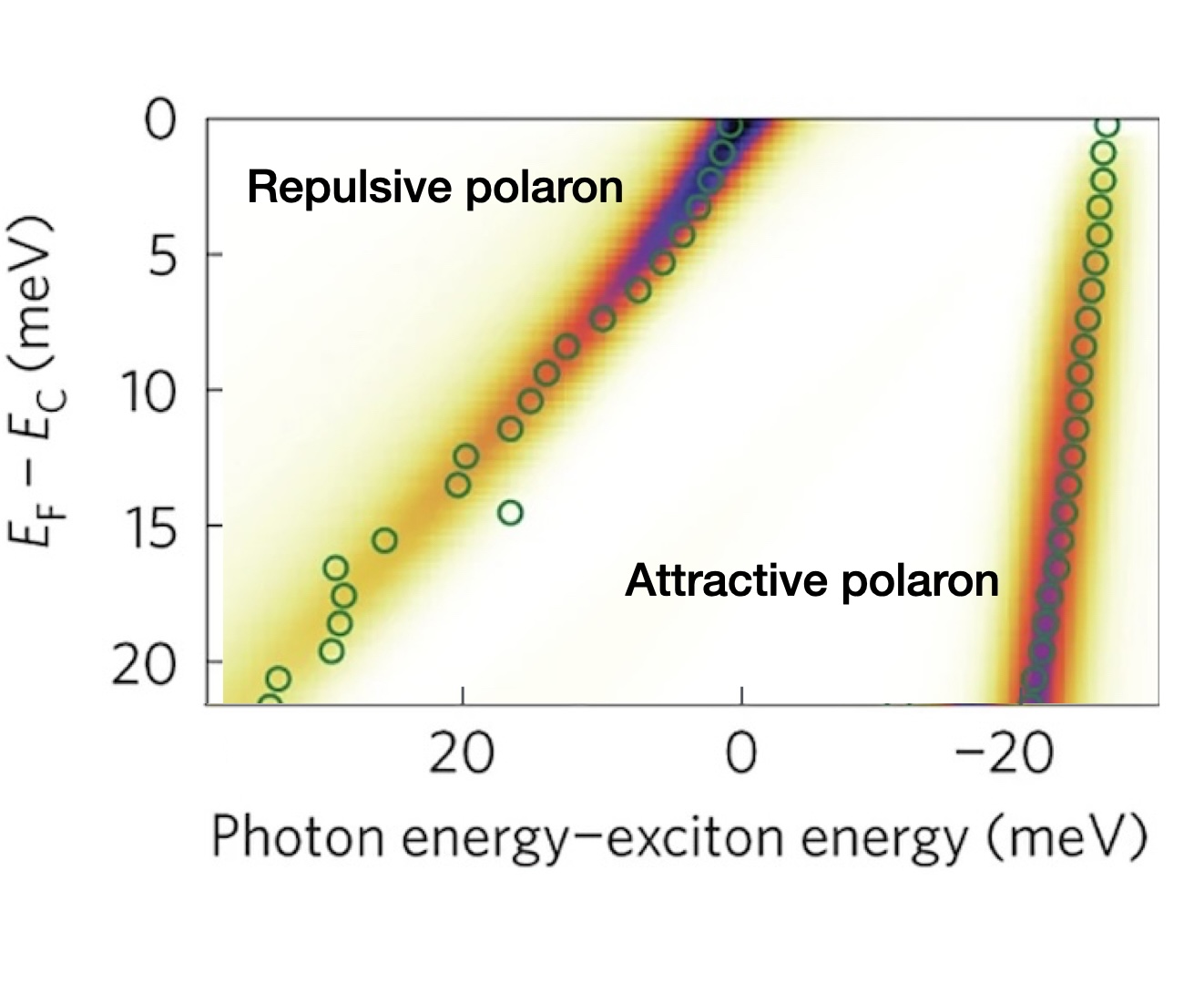}
\includegraphics[width=0.46\columnwidth]{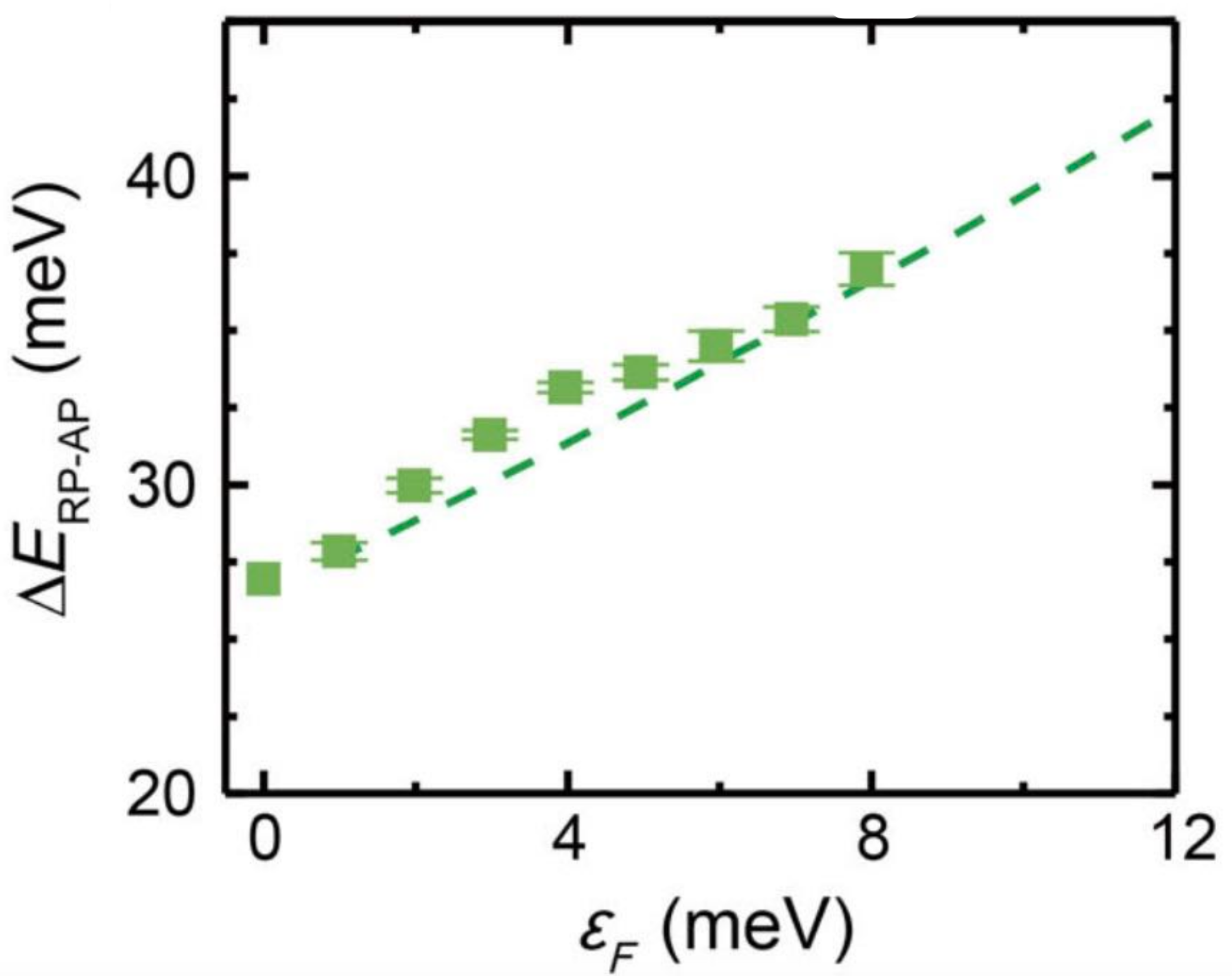}
\caption{\label{fig:TMDPolaron}
\textbf{Observation of Fermi polarons in TMDs.} 
Left: Exciton spectral function as a function of the Fermi energy of the electrons from the Chevy ansatz. Greens dots are experimental data for the attractive and repulsive polarons obtained from differential reflection spectra for a MoSe$_2$ monolayer. From \citet{Sidler2016}. Right: energy splitting between the attractive and repulsive polaron measured in a MoSe$_2$ monolayer. The dashed line is the prediction of the Chevy ansatz. From \citet{Huang2023}.}
\end{figure}

The left panel of Fig.~\ref{fig:TMDPolaron} shows the measured optical spectrum of monolayer MoSe$_2$ as a function of electron Fermi energy controlled by gating~\cite{Sidler2016}, compared to a theoretical calculation based on the Chevy ansatz given by the first two terms of Eq.~\eqref{eq:ChevyAnsatz}. The excellent agreement between theory and experiment is obtained with only one fitting parameter: A density dependent blue shift $\beta\epsilon_F$ has been added to the calculated spectrum with $\beta$ a free parameter. The origin of this blue shift that is not captured by Eq.~\eqref{eq:HF} is a combination of (i) the repulsive interactions between excitons and electrons occupying the same valley (intra-valley interaction) as the gating injects electrons both in the K and K' valleys, (ii) phase-space-filling effects that render low momentum states unavailable for exciton formation in the same valley as degenerate electrons, and (iii) band-gap renormalization due to finite electron density. 

Figure \ref{fig:TMDPolaron} illustrates how the polaron model recovers the energies of the two peaks as well as several other key features, which confirms its validity. First, the high energy peak is continuously blue-shifted from the bare exciton peak with increasing electron concentration while it gradually looses spectral weight. Second, a new well-defined low energy peak emerges with an oscillator strength that increases linearly with electron density $n_e$ for $n_e \le 1 \times 10^{12}$~cm$^{-2}$. These features are captured by the Chevy ansatz  and allow the identification of the low (high) energy resonance as the attractive (repulsive) polaron. The right panel of Fig.~\ref{fig:TMDPolaron} shows the energy splitting between the upper and lower polaron as a function of the Fermi energy $\epsilon_F$  as measured in a doped MoSe$_2$ monolayer~\cite{Huang2023}. The dashed line is a  calculation based on the Chevy ansatz predicting a linear  increase in the splitting as a function of $\epsilon_F$. 
 
Finally, \citet{Tiene2023} analyzed the temperature dependence of the 2D Fermi polaron in TMDs, finding that the attractive polaron is replaced by a trion-hole continuum at high temperatures, and the radiation induced transition between the attractive and repulsive polarons was explored in Ref.~\cite{shentsev2025terahertzradiationinducedattractiverepulsive}.

\subsection{Polaron-polaritons} 
\label{Sec:MicroPolaritons}
\begin{figure}[ht]
\centering
\includegraphics[width=\columnwidth]{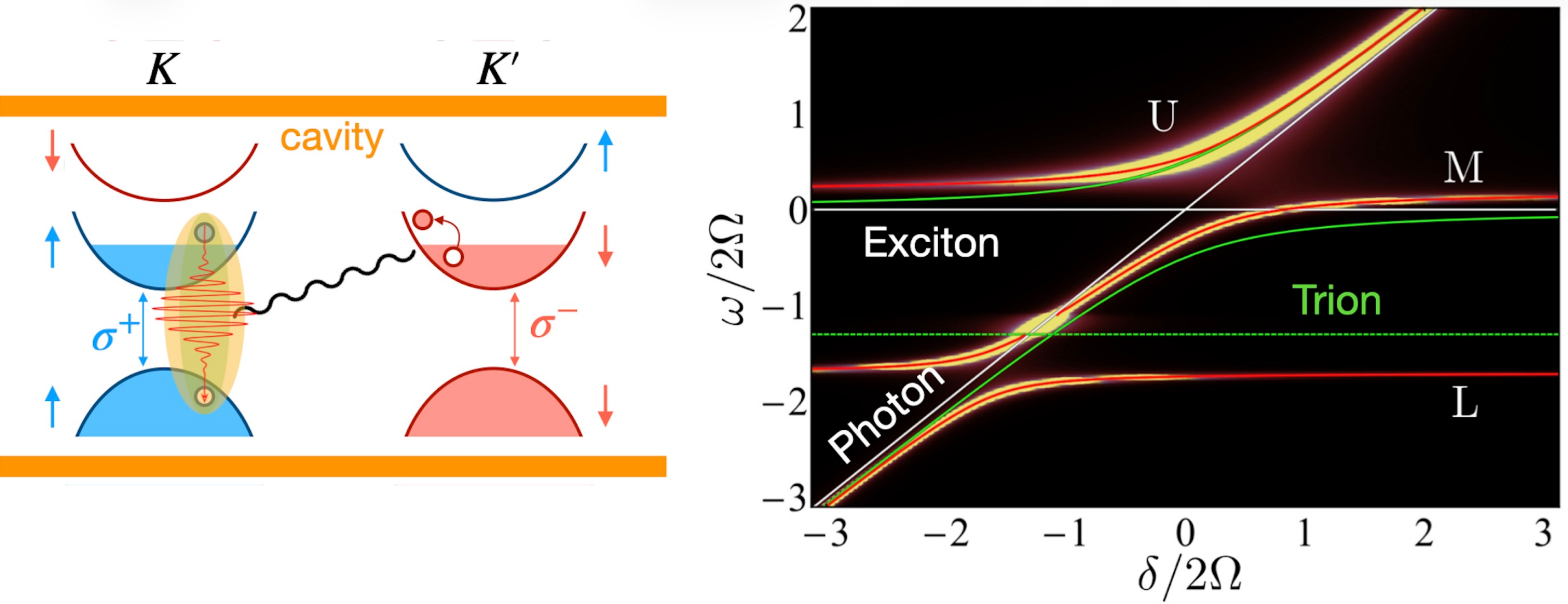}
\caption{\label{fig:TMDcartoon_polaritons}
{\bf Polaron-polaritons.} 
Left: Within a TMD material in an optical cavity, excitons hybridize with the cavity photons to
form polaritons (orange ellipse). Polaritons interact with electrons of opposite spin in the opposite valley to form {\it polaron-polaritons}. 
Right: Zero momentum cavity photon spectral function as a function of energy $\omega$  and detuning $\delta$. White lines are the uncoupled photon and exciton energies, green lines are the upper and lower polariton in the absence of electrons, the horizontal dashed green line is the trion energy, and the red lines are solutions to Eq.~\eqref{PolaronPolaritonEnergy}. Here $m_x=2m_e$, $m_c=10^{-5}m_e$, $\epsilon_F/2\Omega=0.23$, and $|\epsilon_T|/2\Omega=1.56$. From \citet{BastarracheaMagnani2021}.}
\end{figure}

A useful feature of 2D materials such as TMDs is that one can embed them inside an optical cavity. This realises a coupling $\Omega$ between the cavity photons and the excitons, which can be tuned to be very strong by changing the cavity length. When $\Omega$ is comparable to or larger than the photon-exciton detuning for a given cavity mode (i.e.\ the difference in their energies) as well as the cavity photon and exciton decay rates, the excitons hybridise with the photons to form polaritons~\cite{carusotto_quantum_2013}. In the presence of an electron gas, this gives rise to an interesting interplay between polaron and polariton physics,  as illustrated in Fig.~\ref{fig:TMDcartoon_polaritons}. In addition, since the polariton energy is tunable by changing the cavity length this opens up the possibility to realise Feshbach resonances to increase the interaction strength using a bi-exciton state as we shall discuss in Sec.~\ref{FeshbachTMD}. 

The minimal Hamiltonian describing the interacting exciton-electron system in a TMD monolayer coupled to a cavity mode can be written as
\begin{eqnarray}
  H_{xe} &=& \sum_{\B k}
  \begin{bmatrix}
\hat x^\dagger_{\mathbf k}& \hat a^\dagger_{\mathbf k}
\end{bmatrix}
  \begin{bmatrix}
\epsilon_{x\mathbf k} & \Omega \\ \Omega &\epsilon_{c\mathbf k}
\end{bmatrix}
\begin{bmatrix}
\hat x_{\mathbf k}\\ \hat a_{\mathbf k}
\end{bmatrix}+\sum_{\B k}\epsilon_{e\B k }e^\dagger_{\mathbf k}e_{\mathbf k}\nonumber\\
&&+g_{ex}\sum_{\B {k,k',q}}  \hat x^{\dagger}_{\B {k+q}} \hat x_{\B k} \hat e^{\dagger}_{\B {k'-q}} \hat e_{\B k'}.
\label{eqn:PolaronPolaritonhamiltonian}
\end{eqnarray}
Here, $x^\dagger_{\mathbf k}$, $a^\dagger_{\mathbf k}$, and $e^\dagger_{\mathbf k}$ are the creation operators of excitons of mass $m_x$, cavity photons of mass $m_c$, and electrons of mass $m_e$, all with in-plane momentum $\B k$. The corresponding dispersions are $\epsilon_{x\B k }=\B k^2/2m_x$, $\epsilon_{c\B k }=\B k^2/2m_c+\delta$, and $\epsilon_{e\B k }=\B k^2/2m_e$, where $\delta$ is the cavity detuning. Compared to Eq.~\eqref{eq:HF}, we have  added the coupling $\Omega$ (assumed real) to the cavity photons. Diagonalising the non-interacting part of Eq.~\eqref{eqn:PolaronPolaritonhamiltonian} using the transformation $\hat x_{\mathbf k}={\mathcal C}_{\B k}\hat L_{\B k}-{\mathcal S}_{\B k}\hat U_{\B k}$ and $\hat a_{\mathbf k}={\mathcal S}_{\B k}\hat L_{\B k}+{\mathcal C}_{\B k}\hat U_{\B k}$ yields the eigenoperators $\hat L_{\B k}^\dagger$ and $\hat U_{\B k}^\dagger$ creating lower and upper polaritons, which are hybrid light-matter particles~\cite{carusotto_quantum_2013,Al-Ani_2022}. Their energies are $\epsilon^{\pm}_{\B k}=(\epsilon_{x\B k}+\epsilon_{c\B k}\pm\sqrt{\delta_{\mathbf k}^2+4\Omega^2})/2$, where $\delta_{\mathbf k}=\epsilon_{c\B k}-\epsilon_{x\B k}$ is the photon-exciton detuning and ${\mathcal C}_{\B k}^2=1-{\mathcal S}_{\B k}^2=(1+\delta_{\B k}/\sqrt{\delta_{\B k}^2+4\Omega^2})/2$ are the corresponding Hopfield coefficients~\cite{Hopfield1958}. 

When  electrons are present, interactions  lead to the formation of polaron-polaritons, which like polarons can be analysed with many methods. Here, we shall use field theory providing a convenient way to include a non-zero exciton concentration and temperature. Defining a $2\times2$ retarded Green's function $G(\bp,t)=-i\theta(t)\langle[\hat\Psi_{\B k}(\tau)\hat\Psi_{\B k}^\dagger(0)]_-\rangle$ for  excitons coupled to  cavity photons with $\Psi_{\B k}=\begin{bmatrix}\hat x_{\B k}& \hat c_{\B k}\end{bmatrix}^T$, the Dyson equation is
\begin{equation}
{G}^{-1}(k)=\begin{bmatrix}\omega-\epsilon_{x\B k}& 0\\0&\omega-\epsilon_{c\B k}\end{bmatrix}-\begin{bmatrix}\Sigma_x(k)& \Omega\\\Omega&0\end{bmatrix}
\label{GreenPolaronPolariton}
\end{equation}
in momentum and frequency space $k=(\B k,\omega)$ with $\Sigma_x(k)$ the exciton self-energy coming from interactions with the electrons~\cite{Levinsen2019,Bastarrachea-Magnani2021Polaritons,wasak_QuantumZeno_2019,Tan2020}. Equation \eqref{GreenPolaronPolariton} is a matrix generalisation of the impurity retarded Green's function introduced in Sec.~\ref{GeneralProb}, and its poles give the energies $\varepsilon_{\B k}$ of the quasi-particles. From Eq.~\eqref{GreenPolaronPolariton} we obtain the self-consistent equations
\begin{equation}
\varepsilon^{\pm}_{\B k}=\frac12\left[\epsilon_{x\B k}+\Sigma_x(\B k,\varepsilon^{\pm}_{\B k})+\epsilon_{c\B k}\pm \sqrt{\tilde\delta_{\B k}^2+4\Omega^2}\right]
\label{PolaronPolaritonEnergy}
\end{equation}
where $\tilde\delta_{\B k}=\epsilon_{c\B k}-\epsilon_{x\B k}-\Sigma_x(\B k,\varepsilon^{\pm}_{\B k})$. Importantly, since the energy $\epsilon_{c\B k}$ of the cavity photons depend on the cavity length, it is possible to tune the quasiparticle energy.

Equations~\eqref{GreenPolaronPolariton}-\eqref{PolaronPolaritonEnergy} illustrate the interplay between polaron physics (entering via the self-energy $\Sigma_x$) and polariton physics (entering via the coupling $\Omega$ to cavity photons). Indeed, the dispersion of the quasiparticles given by Eq.~\eqref{PolaronPolaritonEnergy} is identical to that of polaritons except for the replacement $\epsilon_{x\B k}\rightarrow \epsilon_{x\B k}+\Sigma_{x\B k}(\B k,\varepsilon_{\B k})$. Likewise, the  Hopfield coefficients are given by the vacuum expressions above with the replacement $\delta_{\B k}\rightarrow \tilde \delta_{\B k}$. The small photon mass $m_c\sim10^{-5}m_x$ moreover means that the light coupling has only small effects on the self-energy $\Sigma_x(\B k,\omega)$ even for a small detuning $\delta$ and large $\Omega$, since the  electrons scatter the excitons predominantly to states where the photon is off-resonant (large $\delta_{\bf k}$). This  can be seen explicitly by using the ladder approximation to calculate the self-energy $\Sigma_x$ entering Eq.~\eqref{PolaronPolaritonEnergy}, which again gives Eq.~\eqref{eq:SelfenergyLadder} with a scattering matrix identical to that in Eq.~\eqref{Tmatrix2D}. The coupling to the cavity photons enters only through the pair propagators, which are however essentially the same as in the absence of light, due to the small photon mass~\cite{Bastarrachea-Magnani2021Polaritons,BastarracheaMagnani2021,wasak_QuantumZeno_2019,Tan2020}. 

It follows that the quasiparticle emerging from the exciton being strongly coupled to light while simultaneously interacting with electrons can be understood as a \emph{polaron-polariton}, i.e.\ a coherent superposition of a cavity photon and a polaron. Since the Rabi coupling between the cavity photon and the polaron is  $\sqrt{Z_{\B k}}\Omega$ with $Z_{\B k}$  the polaron residue, the minimal splitting between the upper and lower polaron-polariton branch is reduced by a factor $\sqrt{Z_{\B k}}$ as can explicitly be shown from Eq.~\eqref{PolaronPolaritonEnergy}. As we shall see in Sec.~\ref{Sec:ExcitonPointParticle}, this intuitive picture of polaron-polaritons is corroborated by experimental findings. 

The right panel of Fig.~\ref{fig:TMDcartoon_polaritons} shows the zero momentum cavity photon spectral function as a function of detuning $\delta$ obtained from Eq.~\eqref{GreenPolaronPolariton} using the ladder approximation. The avoided crossings of the attractive and repulsive polarons with the cavity photon lead to the formation of three polaron-polariton branches. The Feshbach resonance realized when the lower polariton (L) is tuned into resonance with the trion (horizontal green line) was analyzed in detail by \citet{Kumar2023}, taking into account the composite electron-hole nature of the exciton.

\begin{figure}[ht]
\centering
\includegraphics[width=\columnwidth]{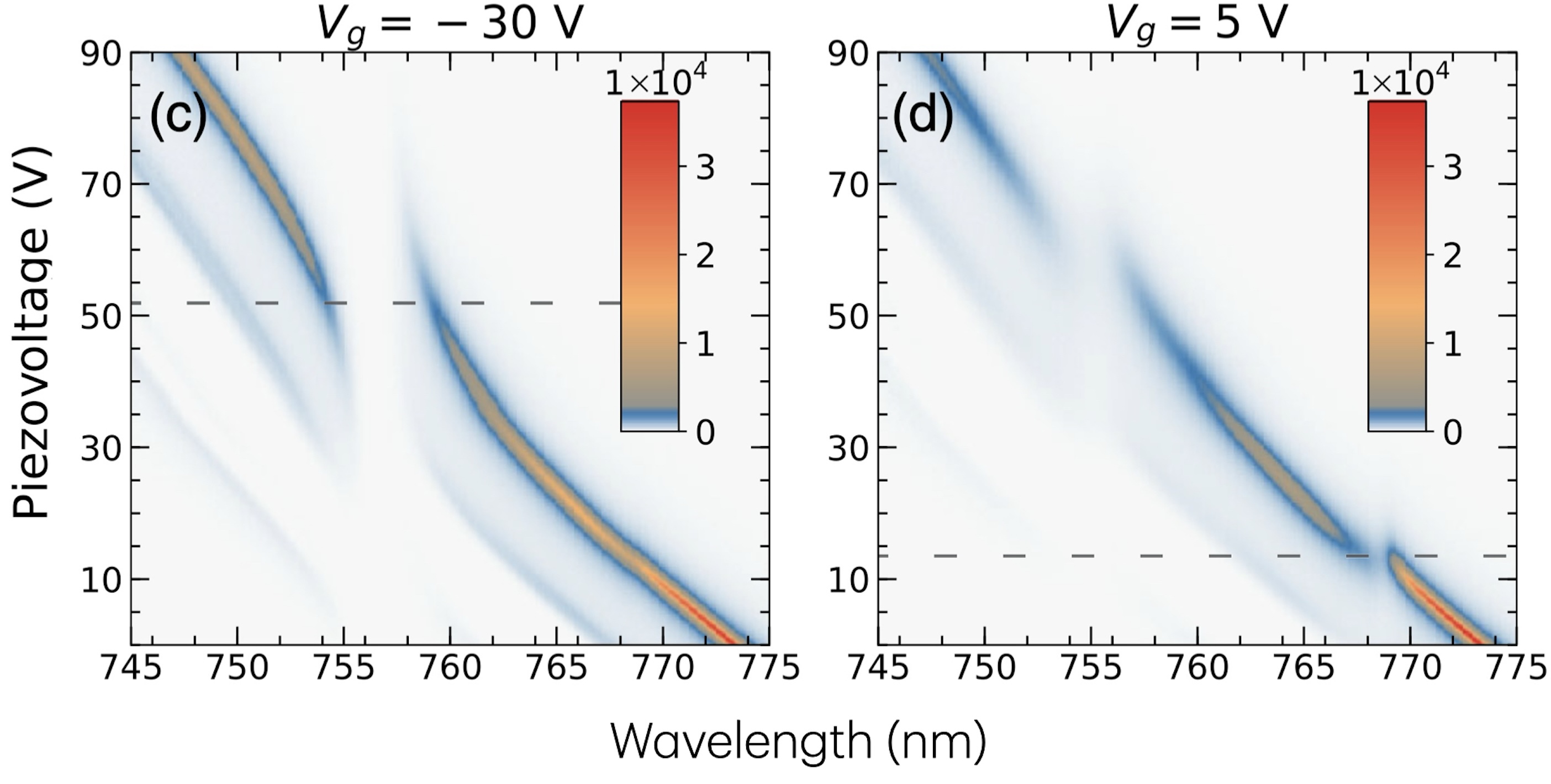}
\caption{\label{fig:TMDPPolaronPolariton}
{\bf Observation of polaron-polaritons.}
Light transmission spectrum of a MoSe$_2$ monolayer in an optical cavity as a function of the cavity mode energy. 
In the left panel, measurements with no itinerant electrons yield a characteristic polariton spectrum with an avoided crossing between the cavity and exciton modes. 
In the right panel, measurements with itinerant electrons show the presence of two avoided crossings of the photon with the repulsive and attractive polarons and the formation of three polaron-polaritons  branches. From \citet{Tan2020}.}
\end{figure}

Polaron-polaritons were first observed in the pioneering experiment by~\citet{Sidler2016}.  In Fig.~\ref{fig:TMDPPolaronPolariton} we show results from a later experiment measuring the light transmission spectrum of a MoSe$_2$ monolayer in a zero-dimensional optical cavity as a function of cavity length~\cite{Tan2020}. The energy of the cavity photon and thereby the detuning $\delta$ depends on the cavity length. In the left panel, the gate voltage is such that there are no itinerant electrons and one clearly observes the typical polariton spectrum with an avoided crossing between the  cavity photon and the exciton as  described by Eq.~\eqref{PolaronPolaritonEnergy} with $\Sigma_x=0$, giving $\epsilon_{\bk}^{\pm}$. In the right panel, the gate voltage is increased so that the conduction band is partially filled with electrons that interact with the excitons. This gives rise to two avoided crossings and three polaron-polariton branches in agreement with  Fig.~\ref{fig:TMDcartoon_polaritons}. 
The first branch is continuously blue shifted with increasing electron density from the crossing  in the left panel, and it arises from the crossing of the repulsive polaron with the cavity mode. The second crossing has no analogue for zero electron density, and it emerges at low energy with increasing electron density from the crossing of the attractive polaron with the cavity mode. Consistent with this picture, the mode splittings at the avoided crossings are reduced compared to the bare splitting when there are no electrons. These experimental results thus clearly demonstrate the formation of polaron-polaritons. 

Interactions with electrons cause a range of non-linear effects for coherent states of  polaron-polaritons~\cite{Julku2021}, and one can  use light to probe polaron-polariton physics in a non-demolition manner~\cite{Camacho-Guardian2023}. Polaron-polaritons also emerge for light propagation in atomic gases under the condition of electromagnetically induced transparency  leading to a number of interesting effects such as self-trapping~\cite{Grusdt2016}, a cross-over from a bare polariton to a polaron-polariton~\cite{Camacho-Guardian2020}, and superfluid flow above Landau's critical velocity~\cite{Grusdt2016,Camacho-Guardian2020,Nielsen2020}.

\subsection{Bose polarons in TMDs}   \label{BosepolaronTMD}
\begin{figure}[ht]
\centering
\includegraphics[width=\columnwidth]{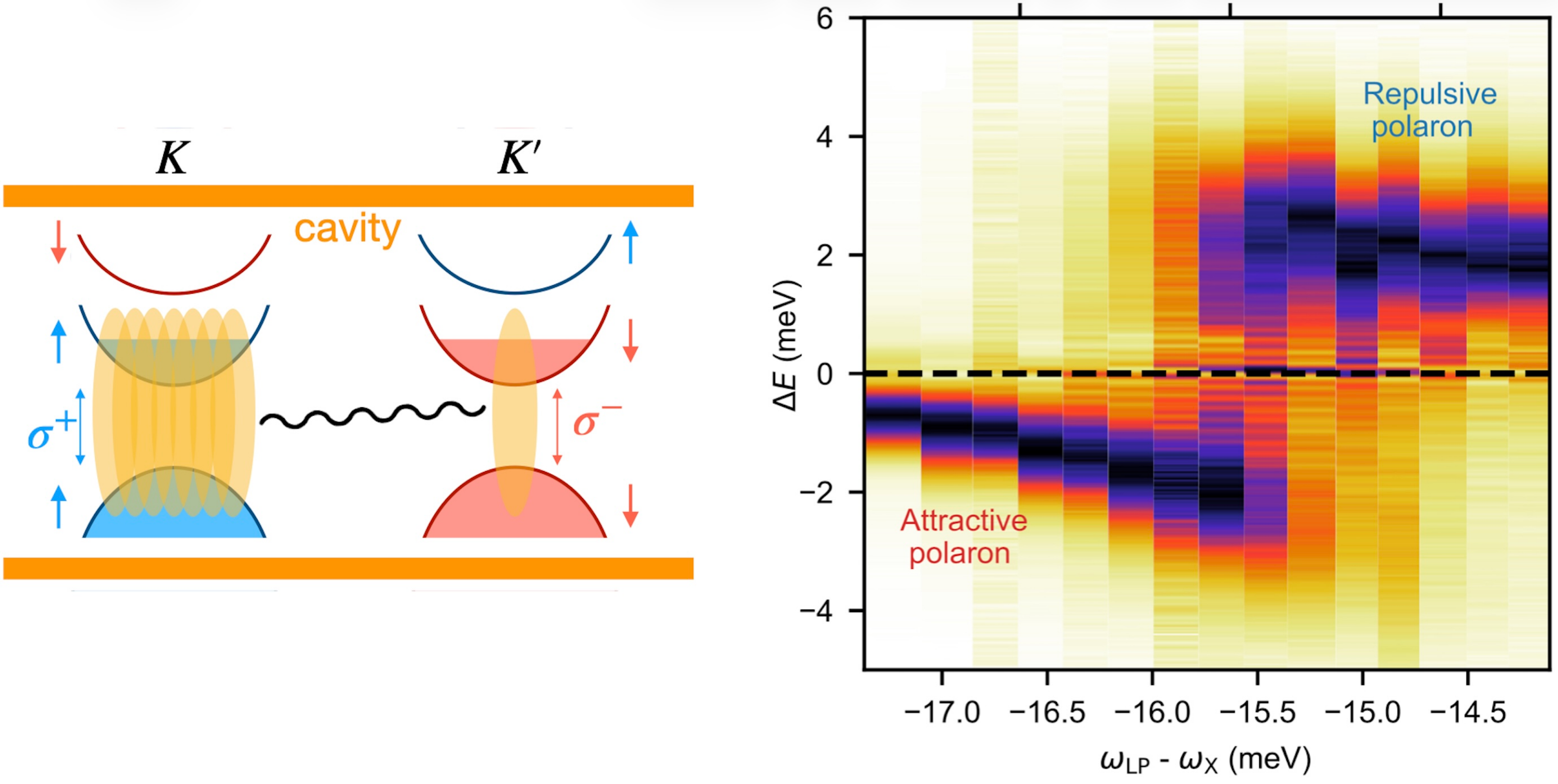}
\caption{\label{fig:TMDcartoon_BosePolarons}
{\bf Bose polarons in TMDs.} 
Left: A pump beam creates a bath of polaritons in the $K$ valley and a probe beam creates a  few ``impurity'' polaritons in the $K'$ valley, leading to the formation of Bose polarons.
Right: Transmission spectrum of the probe beam 
in such a system, realized in a monolayer MoSe$_2$.
The two branches are identified as attractive and repulsive Bose polarons. The energy $\Delta E$ is relative to that of a bare polariton $\omega_{LP}$, and $\omega_X$ is the energy of an exciton. From \citet{tan2022bose}.}
\end{figure}

While the Fermi polaron naturally emerges from the interactions between excitons and electrons as discussed above, 
a solid state realisation of the Bose polaron requires a different mechanism. Early experiments used a pump beam to create a bath of spin $\uparrow$ polaritons in a GaAs microcavity, and a probe beam to create spin $\downarrow$ polaritons as sketched in the left panel of  Fig.~\ref{fig:TMDcartoon_BosePolarons}~\cite{Takemura:2014vr,Takemura2017,NavadehToupchi2019}. The interaction potential between the two kinds of excitons supports a bound state, i.e.\ a biexciton, which can be brought into resonance by tuning the energy of the polaritons via the cavity length. A Feshbach resonance between the polaritons was in this way realised giving rise to significant shifts in the transmission spectrum of the probe pulse~\cite{Takemura:2014vr,Takemura2017,NavadehToupchi2019}. Subsequent theoretical works based on a Chevy type variational function~\cite{Levinsen2019} and a diagrammatic ladder approach~\cite{BastarracheaMagnani2019} argued that the experiment could be interpreted in terms of the spin $\downarrow$ polaritons forming a Bose polaron by interacting with the bath of $\uparrow$ polaritons. Fitting the theory to the experimental data however indicated a large damping rate of the bi-exciton, which strongly suppresses the Feshbach resonance.

Recently, clear signatures of the Bose polaron were reported in an experiment where polaritons consisting of cavity photons and excitons on the K'-valley of a monolayer MoSe$_2$ were created by a probe beam, while polaritons in the K valley were created by a stronger pump beam~\cite{tan2022bose}. The polaritons in the K' valley served as impurities whereas the polaritons in the K valley formed the bosonic bath, and the interaction between the two kinds of polaritons supported a bound state (a biexciton). By changing the cavity length, the energy of a pair of polaritons in the two valleys was tuned to that of the bi-exciton thereby realising a Feshbach resonance, leading to two quasiparticle branches in the transmission spectrum of the probe beam, see right panel of Fig.~\ref{fig:TMDcartoon_BosePolarons}. Good agreement was obtained comparing to a theory for the Bose polaron based on the Chevy ansatz Eq.~\eqref{BogExpansion} generalised to include light coupling and with the bosonic bath of polaritons described as a Fock state.
This allowed to identify the two branches in the transition spectrum as the attractive and repulsive Bose polaron. The observed spectra depended strongly on the delay time between the pump and probe pulses reflecting the inherently non-equilibrium nature of the experiment due to the rapid decay of the polaritons and biexcitons.



Attractive and repulsive Bose polarons formed by intralayer excitons  in a degenerate bath of interlayer excitons were recently observed in the photoluminescence spectrum of a 2D semiconductor heterostructure, with an energy splitting increasing with the density of the interlayer excitons in agreement with theory~\cite{szwed2024excitonicbosepolaronselectronholebilayers}. The Bose polaron formed by an impurity interacting with a BEC of polaritons in a microcavity was investigated in Ref.~\cite{Vashisht2022}. Several dynamical regimes were identified by calculating the effective mass and the drag force acting on the impurity.

\subsection{Feshbach resonances}   \label{FeshbachTMD}

Given their tremendous utility in cold atomic gases, it is highly desirable to have Feshbach resonances available for tuning the interaction also in TMDs. In Sec.~\ref{Sec:MicroPolaritons}, we saw how this can be achieved for the electron-polariton interaction  by tuning a polariton into resonance with a trion~\cite{BastarracheaMagnani2021,Kumar2023}. The very steep polariton dispersion however means that the resonance condition is only valid for a small momentum range $\sim 0.2$ times the photon momentum. In Sec.~\ref{BosepolaronTMD} we discussed a Feshbach resonance between two excitons using bi-excitons, whose typically  short lifetime  however broadens and suppresses the  resonance significantly. One way to avoid this is to use a bi-layer setup where the direct (intralayer) excitons are hybridized with long-lived interlayer excitons, which can then form bound states~\cite{Camacho-Guardian2022}.

\begin{figure}[ht]
\centering
\includegraphics[width=\columnwidth]{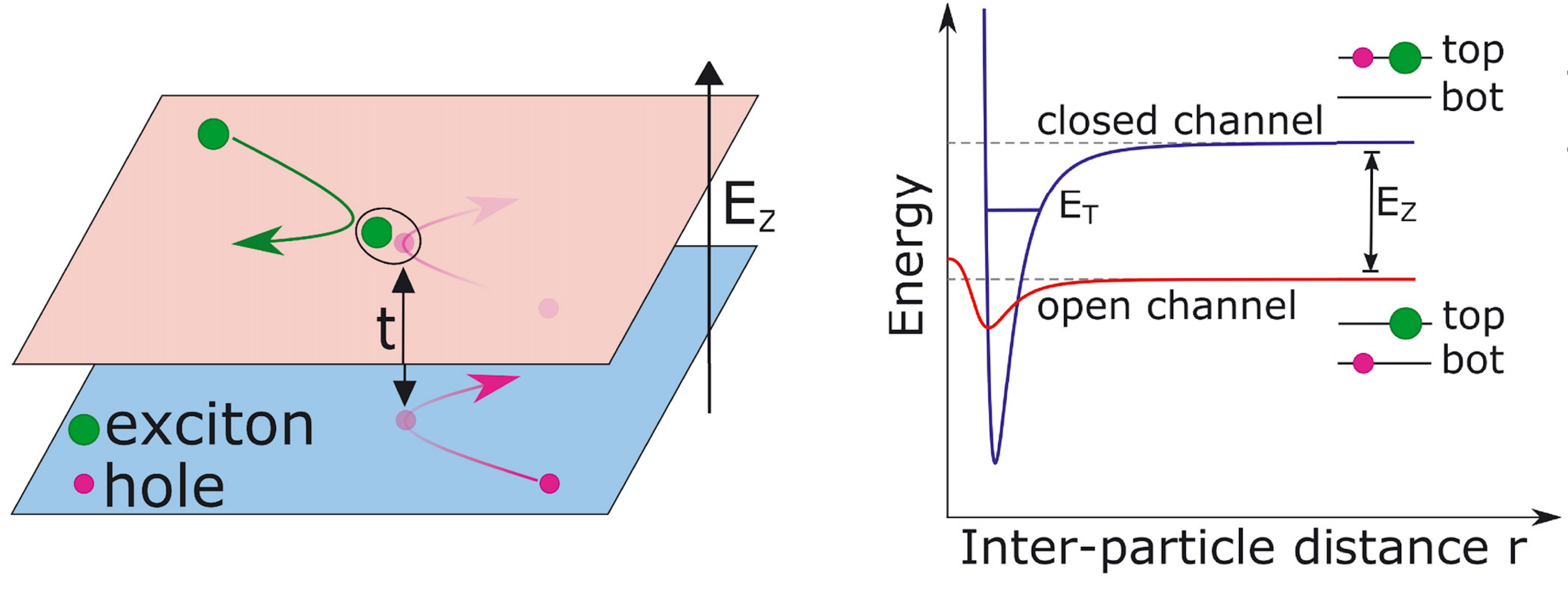}
\caption{\label{Fig:FeshbachFigTMD}
{\bf Feshbach resonances in TMDs.}
A Feshbach resonance between a hole in the bottom layer and an exciton in the top layer is realised when the hole can tunnel to the top layer and bind with the exciton forming a trion, whose energy $E_T$ matches that of the free exciton and hole in the top and bottom layers respectively. From \citet{schwartz2021}.}
\end{figure}
 
In order to realize a Feshbach resonance between excitons and electrons/holes, a different approach based on a bi-layer setup has been implemented. Investigating an exciton in one MoSe$_2$ layer (top) interacting with holes in the other MoSe$_2$ layer (bottom), avoided crossings were observed for the  polaron branches as a function of an electric field~\cite{schwartz2021}. This experiment provides a direct observation of a Feshbach resonance arising from holes tunneling to the top layer where they can bind to the exciton forming a trion. As was later analysed in detail theoretically~\cite{kuhlenkamp2021}, when the trion energy equals that of the exciton in the top layer and a hole in the bottom layer, scattering between a bottom layer hole and a top layer exciton is resonant leading to strong interaction effects and avoided crossings, see Fig.~\ref{Fig:FeshbachFigTMD}. \citet{wagner2023feshbach} explored this further by solving the full three-body problem of two holes and one electron (or two electrons and one hole).

\subsection{Excitons as robust impurities}   \label{Sec:ExcitonPointParticle}

In most optically active semiconductors that form the backbone of the optoelectronics devices, the exciton binding diminishes in the presence of itinerant carriers, due to screening of the electron-hole attraction. However, the polaron framework requires  the quantum impurity to be robust in the presence of a degenerate Fermi or Bose gas, implying that the exciton wave function remains unchanged. Therefore, it is important to determine the range of electron densities where this holds.

To assess the modification of the electron-hole exciton wavefunction, one can use the fact that the optical oscillator strength of an exciton resonance is proportional to $1/a_x^2$. If we assume that the optical excitation spectrum can be described using the Fermi-polaron model, then the total oscillator strength of the attractive and repulsive polaron plus the trion-hole continuum should be the same as that of the bare exciton in the absence of free electrons. Two independent experiments have been used to check the limits of validity of this assumption.

In the first set of experiments, a monolayer MoSe$_2$ was embedded inside a zero-dimensional cavity, leading to the formation of polaritons when the monolayer is devoid of carriers, and of attractive or repulsive polaron-polaritons when free electrons are introduced~\cite{TanThesis2022}. The minimal polariton (normal mode) splitting is given by
\begin{equation}
\Omega \simeq \frac{ e v_D}{a_x} \frac{\hbar}{\sqrt{|\epsilon_x| \varepsilon_r L_\text{cav}}}
\label{SplittingFormula}
\end{equation}
where  $v_D$ is the Dirac velocity assuming that MoSe$_2$ can be described using a massive Dirac model, $\varepsilon_r$ is the dielectric constant and $L_\text{cav}$ is the cavity length. Consequently, measuring $\Omega$ allows  to determine the product $a_x \sqrt{L_\text{cav}}$. 

\begin{figure}[ht]
\centering
\includegraphics[width=0.53\columnwidth]{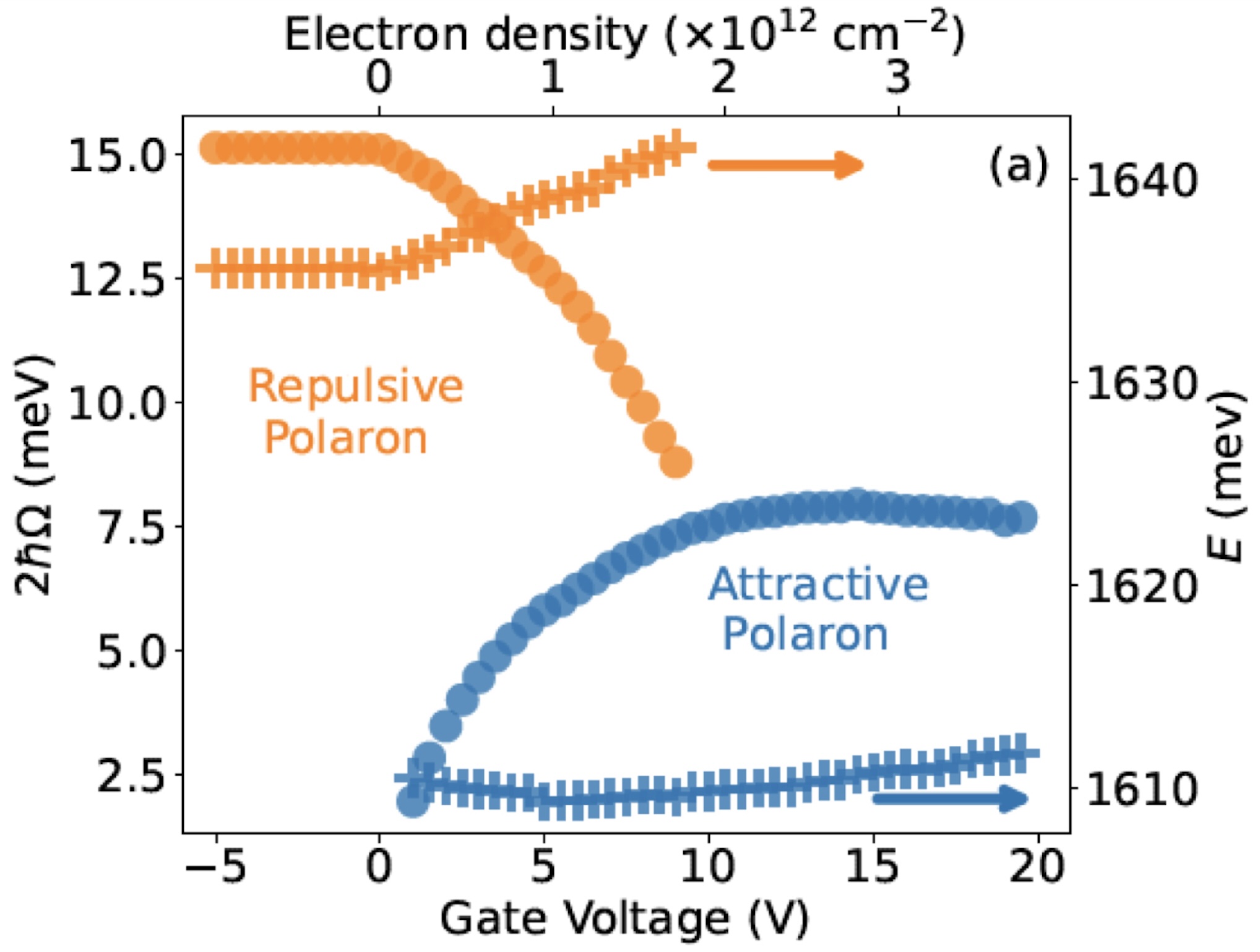}
\includegraphics[width=0.45\columnwidth]{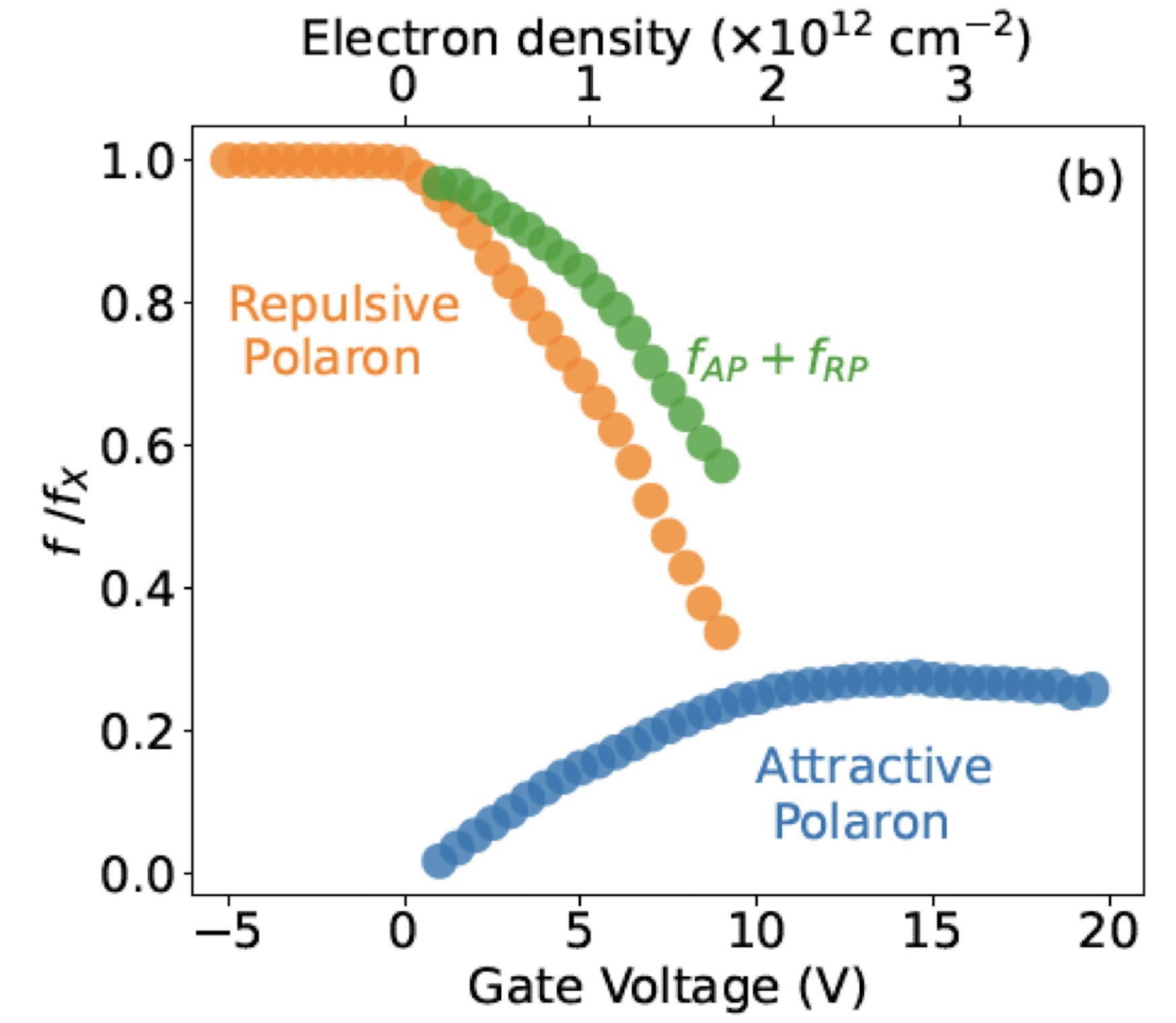}
\caption{\label{Fig:oscstr}
\textbf{Polaron energies and oscillator strengths.}
Left: Energy (circles, left axis) and normal-mode splitting (crosses, right axis) of the attractive (AP) and repulsive (RP) polaron as a function of electron density and extracted polariton splittings. 
Right: Normalized oscillator strength of the AP, RP and their sum as a function of doping density. The oscillator strength is extracted from the normal-mode splitting of the polarons. From \citet{TanThesis2022}.}
\end{figure}
 
In the presence of free electrons, both repulsive and attractive polarons were observed. As described in Sec.~\ref{Sec:MicroPolaritons}, if the exciton remains a well-defined impurity particle and the polaron model applies, then the respective normal-mode couplings are $\Omega_{RP} = \sqrt{1-Z} \Omega$ and $\Omega_{AP} =\sqrt Z \Omega$ with $\Omega_0$ the splitting in the absence of the electrons and $Z$ is the residue of the attractive polaron. It follows that $\Omega_{AP}^2+\Omega_{RP}^2=\Omega_0^2$ provided we can ignore the weight of the trion-hole continuum. By experimentally measuring the normal mode splittings, one can test this prediction. The left panel of Fig.~\ref{Fig:oscstr} shows the measured energy and normal splitting of the attractive and repulsive polaron as a function of electron density $n_e$. The splitting of the attractive/repulsive polaron increases/decreases with  $n_e$, which is consistent with an increasing/decreasing residue as predicted by the polaron model. The right panel shows the normalized oscillator strengths $f=\Omega^2/\Omega_0^2$ of the attractive and repulsive polaron, together with their sum $\Omega_{RP}^2+\Omega_{AP}^2$. The decrease in this sum corresponds to a reduction of $a_x$ via Eq.~\eqref{SplittingFormula} of the order of $20\%$ for $n_e = 1 \times 10^{12}$~cm$^{-2}$ (Fermi energy $\epsilon_F = 3$meV). This is an upper bound since the trion-hole continuum has been ignored, and the relatively small reduction confirms that the exciton wave function to a good approximation is unaffected by the electrons. 
In a second set of experiments, the oscillator strengths of the attractive and repulsive polarons were determined using a transfer-matrix fit to the observed reflection lineshapes. The results of these measurements are in full agreement with the experiments based on polaritons~\cite{TanThesis2022}.

\subsection{Exciton-polarons as quantum probes}   \label{Excitonsasprobes}

\begin{figure}[t]
\centering
\includegraphics[width=\columnwidth]{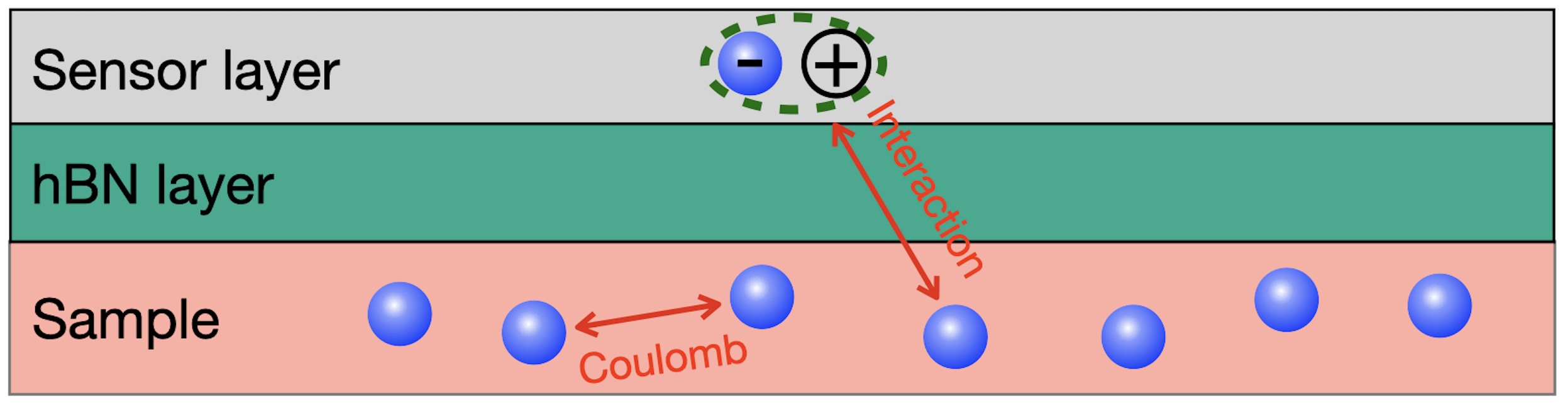}
\caption{\label{Fig:Sensor}
{\bf Quantum sensing with excitons.} 
An exciton in a probe TMD layer acts as a quantum sensor by interacting with electrons in an adjacent material of interest. The exciton can also be in the same layer as the material to be probed.}
\end{figure}

As  mentioned above, a major difference between the Fermi polaron in atomic gases and in TMDs is that the electrons surrounding  exciton interact. This 
in turn influences the dressing of the exciton and thereby  the exciton-polaron spectrum due to strong exciton-electron interactions. While at first sight these multiple interactions sound like a major complication, they can be turned around to be a feature since they provide an invaluable tool for optically probing strongly correlated electron states via the exciton spectrum, see Fig.~\ref{Fig:Sensor}. 

Such probing is important as mono- and multi-layer semiconductors can realise strongly correlated 2D phases, since the large electron and hole band masses together with reduced dielectric screening of Coulomb interactions lead to very large interaction-to-kinetic energy ratios. Of particular interest are semiconductor moir\'e materials composed of TMD bilayers~\cite{Bistritzer2011,Mak2022}. A moir\'e superlattice potential for the electrons emerges when the two TMD layers have a lattice mismatch, or when they are stacked with a non-zero twist angle. Typical superlattice constants are $\sim10$~nm and moir\'e potentials have strengths in the $50-100$~meV range, generically resulting in almost-flat electronic bands in the reduced Brillouin zone. Semiconductor moir\'e systems provide a very high degree of tunability of the lattice parameters relevant for electron correlations, such as the carrier density and the ratio of interaction energy to the hopping strength~\cite{Wu2018,Pan2020}. They therefore realise a powerful quantum simulation platform for many-body physics. However, even though strongly-correlated electrons have been traditionally explored using transport spectroscopy, the difficulty in making good electrical contacts to TMDs renders such measurements challenging. Likewise, $X$-rays and neutrons couple weakly to the layers, rendering spectroscopy difficult. All this leaves an urgent need for new sensors, which exciton-polarons address.

\begin{figure}[ht]
\centering
\includegraphics[width=0.49\columnwidth]{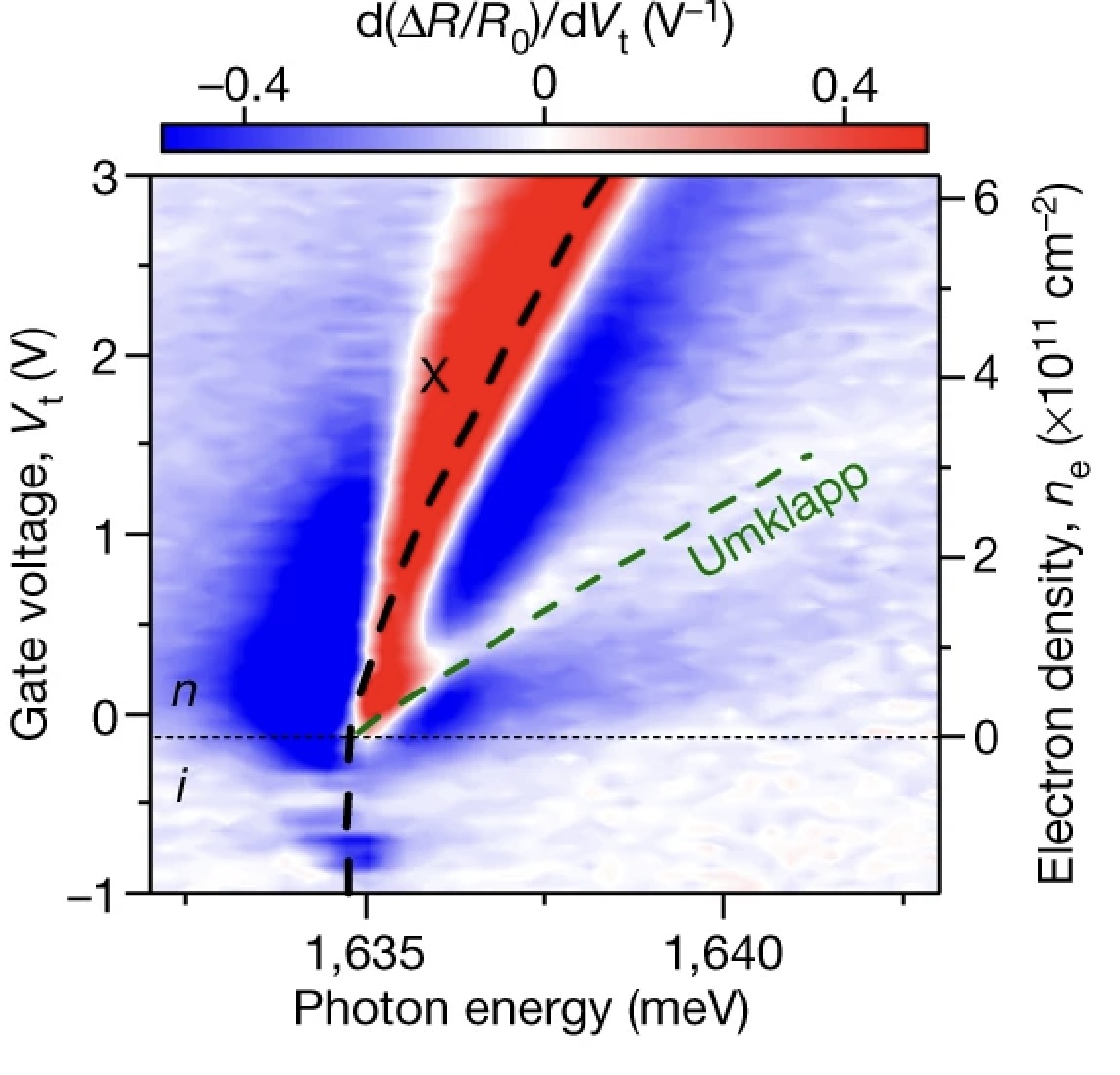}
\includegraphics[width=0.49\columnwidth]{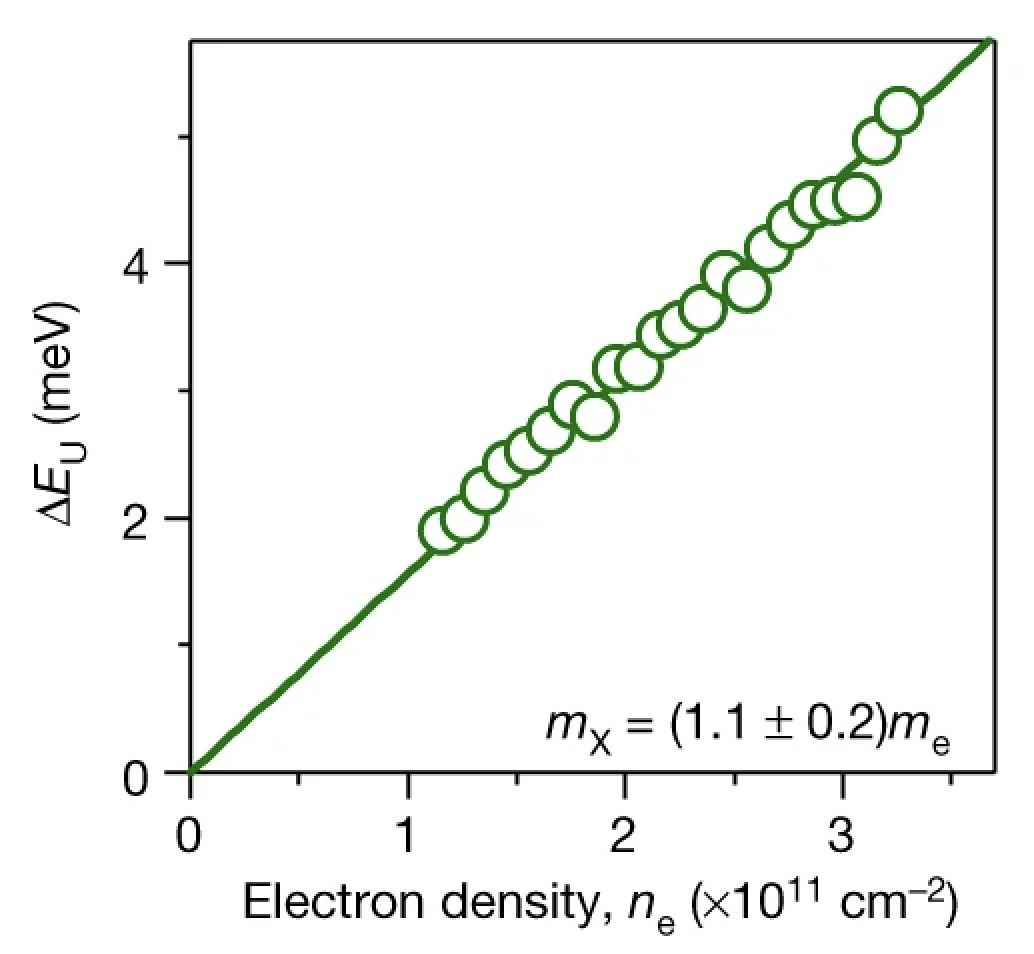}
\caption{\label{Fig:WignerObs}
\textbf{Observation of a Wigner crystal.}
Left: Derivative of the reflectance contrast spectrum of a  MoSe$_2$ monolayer as a function of gate voltage (or equivalently electron density) and photon energy. In addition to the main exciton branch  "X", which is blue shifted corresponding to a repulsive polaron with increasing electron density, an extra high energy umklapp branch appears. Right: The energy difference between the umklapp and  polaron branch increases linearly with the electron density. From~\citet{Smolenski2021}.}
\end{figure}

As a first example, exciton-polarons have been used to detect broken symmetry states of electrons in the charge sector. In a charge-tunable MoSe$_2$ monolayer, one can make the ratio of Coulomb interaction energy to kinetic energy very large so that it is favorable for the electrons to break translational symmetry and form a Wigner crystal~\cite{Smolenski2021}. The excitons in turn feel a periodic mean-field potential from this Wigner crystal, which leads to a folding of the exciton spectrum into the Brillouin zone of the Wigner lattice. The result is a new optically active Umklapp-Bragg resonance at the $\Gamma$ point as shown in the left panel of Fig.~\ref{Fig:WignerObs}. The energy of this new branch was observed to depend linearly on the electron density $n_e$, which can be understood simply from the fact that the kinetic energy of the exciton folded into the $\Gamma$ point ($\bk=0$)  is given by $k_w^2/2m_x$, where $k_w\propto \sqrt{n_e}$ is the Wigner crystal reciprocal vector, see right panel of Fig.~\ref{Fig:WignerObs}. The appearance of umklapp terms in the exciton-polaron spectrum has also been used to detect incompressible Mott-like correlated states in a MoSe$_2$/MoSe$_2$ bilayer~\cite{Shimazaki_2020,Shimazaki_2021}.
Recently, two experiments have observed so-called Wigner polarons formed when an exciton binds to an electron in the Wigner crystal~\cite{zhang2025wignerpolaronsrevealwigner,wang2025spectroscopywignercrystalpolarons}. 
Two theories have been presented for describing Wigner polarons: \citet{adlong2025theoryexcitonpolarons2d} discussed how they may arise from the coupling of the exciton to an electron vibrating in a harmonic oscillator potential formed by the other electrons in the Wigner crystal, while \citet{pichler2025purelyelectronicmodelexcitonpolaron} used a  purely electronic model. \citet{nyhegn2025excitoninteractingphononselectronic} furthermore showed that the coupling of an exciton to the phonons of the Wigner crystal can be mapped onto a multiband Fr\"ohlich model, leading to broadening and energy shifts.

\begin{figure}[ht]
\centering
\includegraphics[width=0.32\columnwidth]{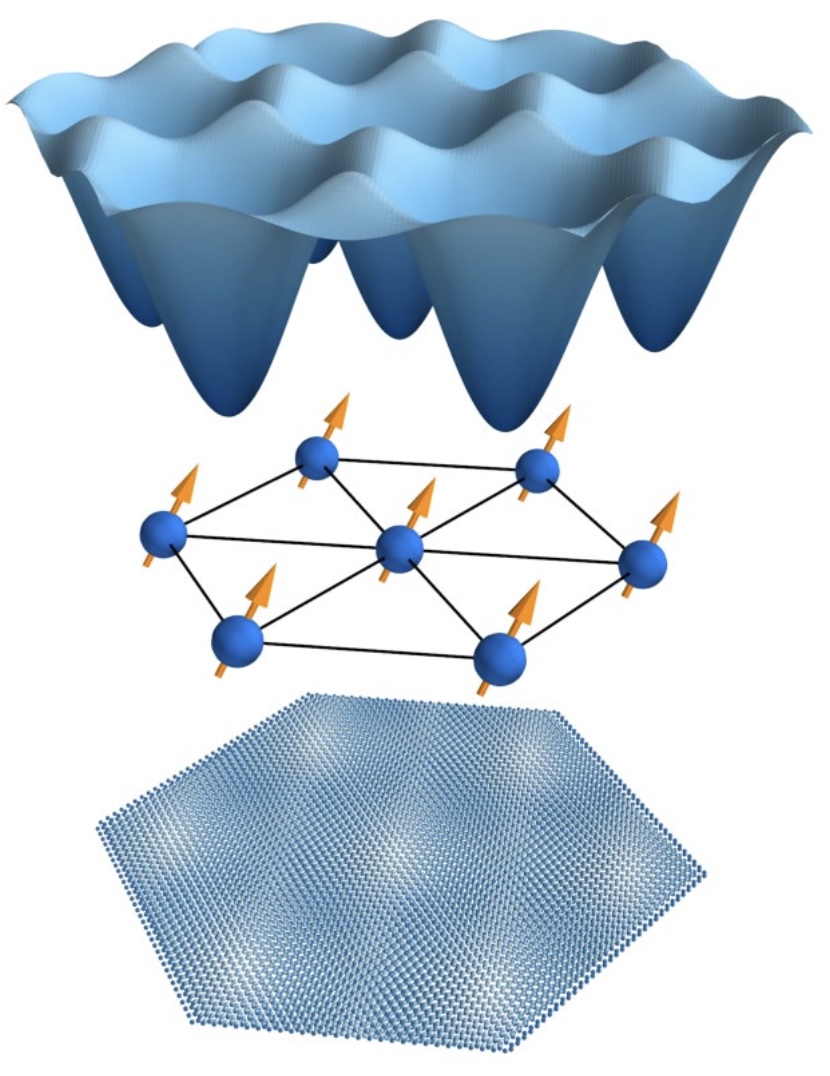}
\includegraphics[width=0.56\columnwidth]{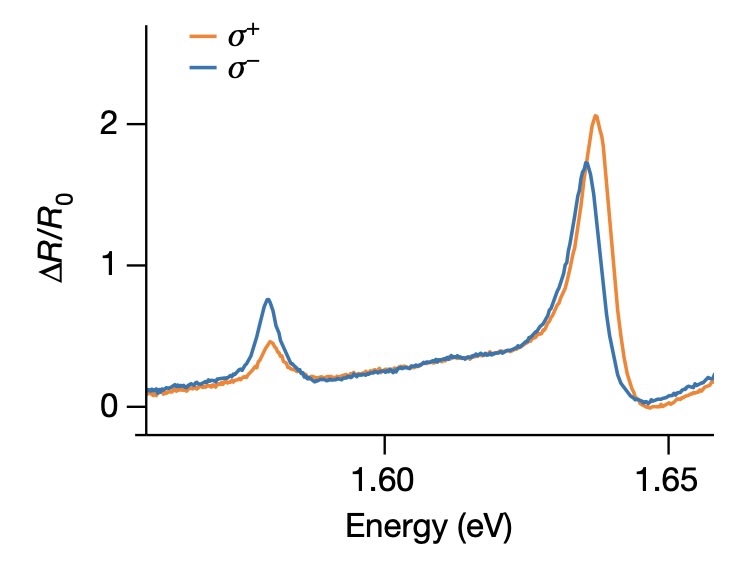}
\caption{\label{Fig:FerromagnetismObs}
\textbf{Observation of spin ordering.} 
Left: Electrons in a triangular moir\'e lattice formed by a MoSe$_2$/WS$_2$ bi-layer. Right: Polarization-resolved reflection spectrum at unit filling. The low energy peak of the blue (orange) line corresponds to the attractive polaron formed by a $\downarrow$ ($\uparrow$) exciton interacting with $\uparrow$ ($\downarrow$) electrons. From~\citet{Ciorciaro2023}.}
\end{figure}

Exciton-polarons have also been used to detect kinetic magnetism in a triangular moir\'e lattice formed by a MoSe$_2$/WS$_2$ bi-layer~\cite{Ciorciaro2023}, see left panel in Fig.~\ref{Fig:FerromagnetismObs}. This observation is based on the fact that excitons in the K/K' valley, which have spin $\uparrow$/$\downarrow$ due to spin-valley locking respectively, form trions only with electrons with the opposite spin in the K'/K valley as we discussed in Sec.~\ref{Sec:excitonpolarons}. It follows that the spectral weight (peak area) of the attractive exciton-polaron in the K/K' valley, which arises from these trion states, is roughly proportional to the density of $\downarrow$/$\uparrow$ electrons. The experimental results shown in the right panel of Fig.~\ref{Fig:FerromagnetismObs} demonstrate that the spectral weight of the attractive polaron formed by a spin $\downarrow$ exciton interacting with a spin $\uparrow$ electrons is larger than that of the attractive polaron formed by a spin $\uparrow$ exciton interacting with spin $\downarrow$ electrons. 
From the spectral weight of $\sigma^+$ and $\sigma^-$ polarized attractive polarons as a function of small applied magnetic field, the spin susceptibility was measured, revealing Nagaoka-type ferromagnetic correlations of moiré electrons \cite{Ciorciaro2023}. This experiment demonstrates the more general fact that the degree of circular polarization of the attractive polaron resonance provides a way to determine the spin-susceptibility and magnetic properties of strongly correlated electrons. 

\begin{figure}[t]
\centering
\includegraphics[width=0.9\columnwidth]{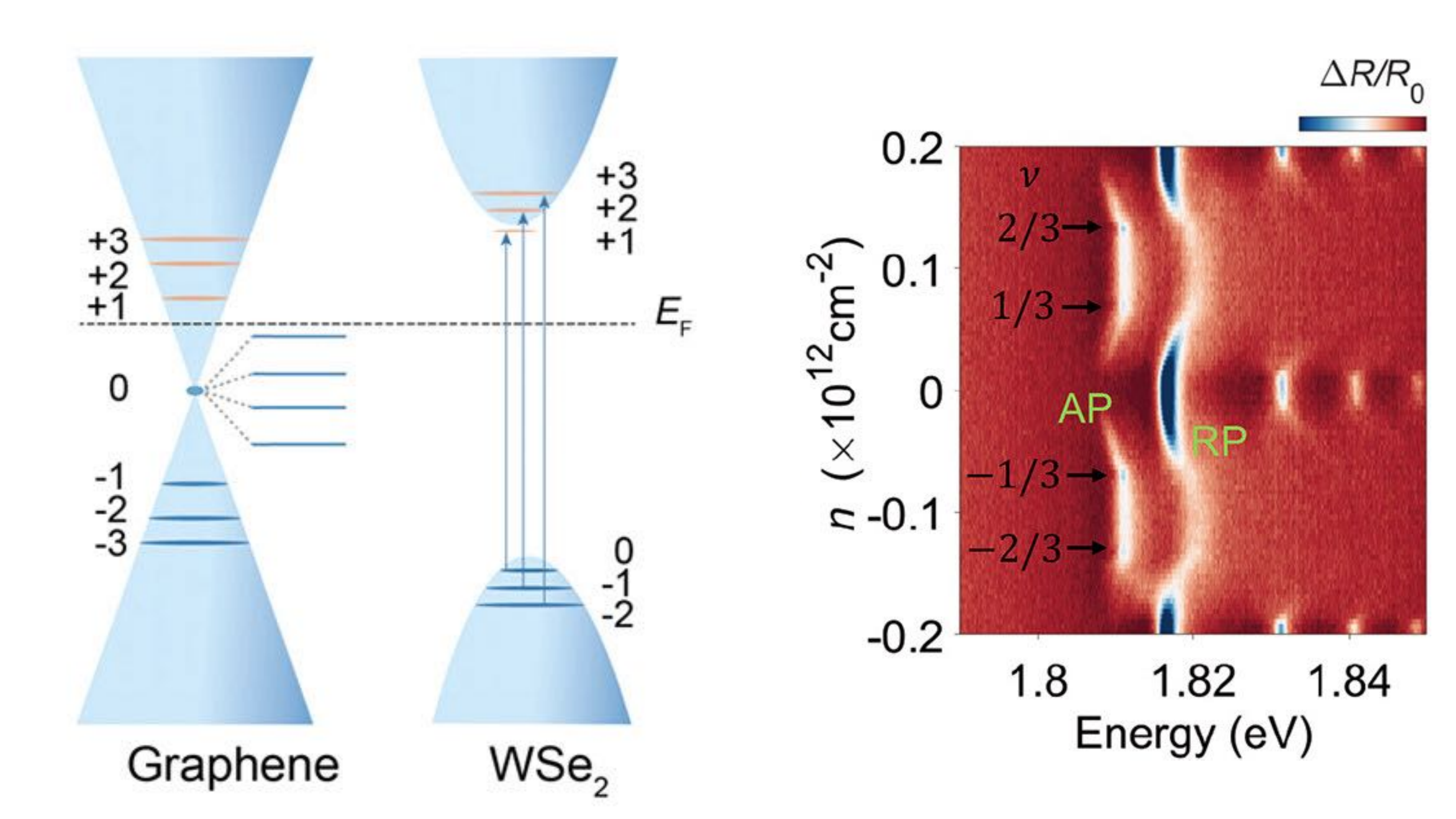}
\caption{\label{Fig:InterlayerPolaron}
\textbf{Observation of incompressible states.} 
Left: Excitons formed in a WSe$_2$ layer interact with electrons in Landau levels in an adjacent graphene layer. Right: Corresponding attractive and repulsive polaron branches observed in reflection contrast spectra as a function the electron density. The attractive branch is enhanced at fractional fillings. From~\citet{Cui2024}.}
\end{figure}

When electrons undergo a phase transition from a compressible to an incompressible state, their ability to dynamically screen excitons is partially suppressed. If the energy gap of the electronic state is small compared to the trion binding energy, both polaron energies are to first order unmodified. Even in this regime however, the phase transition  results in a cusp-like blue shift in the attractive polaron resonance, whereas the repulsive polaron resonance is narrowed due to lack of low energy electronic excitations. These features have been observed upon application of moderate magnetic fields in a charge tunable MoSe$_2$ whenever the electrons form an integer quantum Hall state~\cite{Smolenski2019}. Recently, the formation of interlayer attractive and repulsive polarons was observed in a bi-layer setup, where excitons in a WSe$_2$ layer are dressed by electrons in an adjacent graphene layer~\cite{Cui2024}. When the graphene layer was doped away from the incompressible states of  filled Landau levels, attractive and repulsive  polarons were observed, see Fig.~\ref{Fig:InterlayerPolaron}. 

The emergence of exciton-polaron peaks in optical spectroscopy has also been used to argue for the presence of a dipolar exciton insulator in a moir\'e lattice formed by a WSe$_2$/WS$_2$ bilayer~\cite{Gu2022}. Changes in the exciton spectrum were taken as signs of various correlated phases such a Wigner crystals in moir\'e lattices. However, these observations were not interpreted directly in terms of exciton-polaron physics~\cite{Miao2021,Chenhao2021,Xu2020,Zhou2021}.

In addition to these experiments, several theoretical works explored the use of excitons as sensors. For example, \cite{Salvador2022} studied the umklapp branches emerging from excitons interacting with out-of-plane ferromagnetic and antiferromagnetic order. Furthermore, \cite{julku2023exciton} showed that the dressing of an exciton in a probe layer by spin-waves in an adjacent moir\'e (anti-)ferromagnet leads to the formation of a new kind of "magnetic" polaron in analogy with the dressing of holes in anti-ferromagnet described in Sec.~\ref{Sec:MagneticPol}, which can be used to detect magnetic order in an arbitrary direction. \cite{HuangPRB2023} demonstrated that the coupling of an exciton to spin waves of an anti-ferromagnet in the same layer leads to polaron formation and observable spectral shifts. \cite{Amelio2023b} showed that the properties of an interlayer polaron formed by an exciton interacting with an adjacent excitonic insulator are affected by the hallmarks of the spectrum of the insulating layer. Considering an exciton interacting with electrons in a moir\'e lattice, \cite{Mazza2022} showed that the presence of Wigner metals and Mott insulators can be identified by means of the emergence of double peak structures in the exciton spectrum. \cite{Sorout_2020} explored the dynamics of immobile and mobile impurities interacting with the low energy Dirac fermions as well as the surface states of 2D and 3D solid-state topological insulators, and showed that in specific conditions the impurity spectral function exhibits power law features indicating the breakdown of the polaron picture.

\section{Polaron-polaron interactions}   \label{QPinteractions}

An inherent feature of quasiparticles is that they interact with each other, since the changes induced by one quasiparticle on its environment are felt by the other quasiparticles. This interaction mediated by the environment plays a key role for equilibrium as well as non-equilibrium properties of many-body systems, including collective modes~\cite{baym2008landau}, conventional and high temperature superconductivity~\cite{Schrieffer1983,SCALAPINO1995329,Lee2006}, and giant magnetoresistance~\cite{Baibich1988}. At a fundamental level, all interactions between elementary particles are mediated by gauge bosons~\cite{Weinberg1995}. In this section, we discuss interactions between polarons and how they can be explored using the great flexibility of atomic gases and TMDs. Since we focus on mediated  interactions, any direct interaction between the impurities is assumed to be weak. Further details can be found in recent perspective articles~\cite{Paredes2024,Levinsen2025roleimpuritystatisticsmedium}.

\subsection{Mobile impurities}

Taking the first derivative of Eq.~\eqref{LandauEnergy} gives the energy of a quasiparticle with momentum $\bf p$
\begin{equation}
\label{QPenergy}
\varepsilon_{\mathbf p}=\frac{\delta E}{\delta n_{\mathbf p}}=\varepsilon_{\mathbf p}^0 + \sum_{\mathbf p'}f_{\mathbf p,\mathbf p'}n_{\mathbf p'}\hspace{0.2cm} \text{,}\hspace{0.2cm} f_{\mathbf p,\mathbf p'}=\frac{\delta^2 E}{\delta n_{\mathbf p}\delta n_{\mathbf p'}}
\end{equation}
where $\varepsilon_{\mathbf p}^0$ is the energy of a \emph{single} quasiparticle and $\varepsilon_{\mathbf p}$ its energy for a non-zero quasiparticle concentration. Using Eq.~\eqref{QPenergy} together with second-order perturbation theory, one can rigorously show that the interaction between two Fermi and Bose polarons is~\cite{Yu2012}
\begin{equation}
\label{QPinteractionPert}
f_{\mathbf p,\mathbf p'}=\pm g^2\chi(\mathbf p-\mathbf p',\epsilon_{\mathbf pa}-\epsilon_{\mathbf p'a}),
\end{equation}
where the upper/lower sign is for bosonic/fermionic impurities, and the density-density response function is
\begin{equation} \label{Lindhard}
\chi(\mathbf p,\omega)=
\begin{cases}
\frac{2n_0\epsilon_{\mathbf pb}}{\omega^2-E_{\mathbf p}^2}&\text{Bose polarons}\\
\sum_{\mathbf k}\frac{n_{\mathbf kb}-n_{\mathbf k+\mathbf p b}}{\omega+\epsilon_{\mathbf kb}-\epsilon_{\mathbf k+\mathbf pb}}&\text{Fermi polarons}\\
\end{cases}
\end{equation}
for a weakly interacting BEC and an ideal Fermi gas, respectively (in the latter case, $\chi$ is called Lindhard function). 
In Eq.~\eqref{QPinteractionPert} $g=(\partial \mu_2/\partial n_1)_{n_2}$ is the Landau interaction between a dressed polaron and the surrounding medium, which is taken to be independent of momentum. For weak impurity-fermion interactions, we have $g=2\pi a/m_r$. The frequency dependence of $\chi(\mathbf p,\omega)$ reflects the fact that density fluctuations propagate with a finite speed through the medium, which leads retardation effects. In the limit of small momentum exchange and zero temperature, the quasiparticle interaction becomes 
\begin{equation} \label{RKKYzeromomentum}
\lim_{|\mathbf p|\rightarrow |\mathbf p'|}f_{\mathbf p,\mathbf p'}=\mp g^2 
\begin{cases}
1/g_b&\text{Bose polarons} \\
\mathcal N(\epsilon_F)&\text{Fermi polarons} \\ 
\end{cases}
\end{equation}
where $\mathcal N(\epsilon_F)=m_bk_F/2\pi^2$ is the density of states at the Fermi energy~\cite{Yu2012,Yu2010,Mora2010,Giraud2012}. The $1/g_b$ dependence for the Bose polaron shows that the interaction increases with the compressibility of the BEC. For Fermi polarons with arbitrarily strong impurity-bath interactions, Eq.~\eqref{RKKYzeromomentum} can be written as $\lim_{|\mathbf p|\rightarrow |\mathbf p'|}f_{\mathbf p,\mathbf p'}=\mp (\Delta N)^2/\mathcal N(\epsilon_F)$ where  $\Delta N$ is the number of fermions in the dressing cloud  given by Eq.~\eqref{DeltaN}. 

The $\pm$ sign in Eq.~\eqref{QPinteractionPert}, and the corresponding $\mp$ in Eq.~\eqref{RKKYzeromomentum}, explicitly shows the fundamental role of the quantum statistics of the quasiparticles: The quasiparticle interaction is generally repulsive/attractive for fermionic/bosonic quasiparticles. This sign difference arises because the interaction comes from an exchange term~\cite{Yu2012}, or equivalently because in a Fermi sea there are less available scattering states for the impurities due to Fermi blocking, which increases their energy~\cite{Mora2010}. We note that when taking the derivative in Eq.~\eqref{QPenergy}, the majority particle distribution function is assumed to be constant, which corresponds to keeping the majority density constant. Assuming instead a constant chemical potential for the majority particles yields an additional Hartree term for the interaction between the quasiparticles~\cite{Mora2010}. Also, the mediated interaction is zero to second order in the special case where the momenta of the two (bosonic) impurities are strictly identical, since the majority particles would have to change their density to mediate a zero momentum wave. In this case, which arises for example when  bosonic impurities form a BEC, the interaction mediated by a medium at constant density is given by a higher order process that can be repulsive~\cite{levinsen2024medium,Levinsen2025roleimpuritystatisticsmedium}. 

For strong impurity-medium interactions, one has to resort to approximations when calculating interactions between polarons. This is more challenging than the single polaron problem, since the theory now has to take into account the effects of a non-zero polaron concentration as seen from Eq.~\eqref{QPenergy}. One way to proceed is to compare Eq.~\eqref{QPenergy} with Eq.~\eqref{PolaronEnergyGreens} giving the energy of the polaron from the impurity self-energy. This yields 
\begin{equation} \label{landausigma}
f(\mathbf p,\mathbf p')=Z_{\mathbf p}\frac{\delta \text{Re}\Sigma_a(\mathbf p,\varepsilon_{\bf p})}{\delta n_{\mathbf p'}}. 
\end{equation}
Equation~\eqref{landausigma} completes the link between Landau's quasiparticle theory and microscopic  many-body theory, showing that the quasiparticle interaction can calculated from how the self-energy depends on the impurity concentration~\cite{Giuliani2005}.

\begin{figure}[ht]
\centering
\includegraphics[width=0.9\columnwidth]{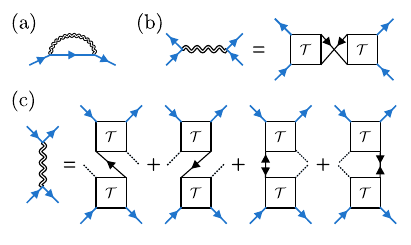}
\caption{\label{Fig:QPintFeynman}
\textbf{Interaction mediated by a Fermi gas and a BEC.} 
(a) The mediated interaction between polarons enters via an exchange (Fock) term for the impurity self-energy. (b) The interaction between Fermi polarons obtained from the ladder approximation generalised to non-zero impurity concentrations. (c) The interaction between Bose polarons obtained from a diagrammatic theory taking into account strong impurity-boson interactions via the ${\mathcal T}$-matrix. Solid black/blue lines are majority/impurity particle Green's functions and dashed lines are condensate bosons.}
\end{figure}

Given its accuracy for describing single Fermi polarons, a natural approach is to use the ladder approximation for the self-energy in Eq.~\eqref{landausigma} generalized to a non-zero impurity concentration. This gives rise to the quasiparticle interaction shown diagrammatically in the top panel of Fig.~\ref{Fig:QPintFeynman}~\cite{Baroni:2023aa}. This approximation recovers the perturbative result Eq.~\eqref{QPinteractionPert} for weak impurity-fermion interactions where $\mathcal T\rightarrow g$, and it was recently used to explain experimental results for the interaction between Fermi polarons as discussed in Sec.~\ref{Sec:DetectionQPinteraction}. The ladder approximation has also been employed to calculate the interaction between Fermi polaron-polaritons in TMDs using Eq.~\eqref{landausigma} with the replacement $Z_{\mathbf p}\rightarrow Z_{\mathbf p}{\mathcal C}^2_{\B p}$. The Hopfield coefficients ${\mathcal C}^2_{\B p}$ appear because it is only the exciton part of the polariton that interacts with the surrounding electrons~\cite{BastarracheaMagnani2021,Bastarrachea-Magnani2021Polaritons}. It was found that the quasiparticle interaction can be much stronger than the direct interaction between excitons, which is small due to their small radius. 

A diagrammatic expression for the interaction between two Bose-polarons in the regime of strong impurity-boson interaction including effects such as retardation and momentum dependence was developed from Eq.~\eqref{QPenergy}~\cite{Camacho2018b}. In order to recover the second order result given by Eq.~\eqref{QPinteractionPert} for weak interactions, it turns out that one has to go beyond the ladder approximation for the impurity self-energy. The result is illustrated in the lower panel of Fig.~\ref{Fig:QPintFeynman}, and it predicts significant energy shifts of the Bose polaron with its concentration via Eq.~\eqref{QPenergy},  which however have not been observed so far. A mixed dimensional setup with two Fermi gases separated by a BEC was shown to offer a promising alternative way to unambigously observe this interaction via the sizeable frequency shift it causes on their out-of-phase dipole mode~\cite{Suchet2017}. The interaction between Bose polarons was also considered using a perturbative Hugenholtz–Pines formalism as well as QMC methods~\cite{Ardilaatoms10010029}. \citet{SantiagoGarcia2023} studied the interaction between two mobile impurities mediated by collective spin excitations of bosons with a hard core repulsion in a lattice following a path integral methods. Finally, \citet{Charalambous2019} explored the entanglement between two distinguishable impurities due to an interaction mediated by a BEC using a quantum Brownian motion approach.

\subsection{Static impurities}   \label{MediatedStatic}
As for the case of single impurities, the limit of infinitely heavy impurities with $m/m_b\gg 1$ gives major simplifications, since there is no impurity recoil. A popular approach is to regard the impurities as static scattering potentials, although this makes them distinguishable so that the role of their quantum statistics is lost and their quasiparticle residues vanish as discussed in the previous sections. The interaction then arises because the two scattering potentials change the spectrum of the surrounding medium, just like in the case of the Casimir effect~\cite{Casimir-1948}. 
Using the Born approximation 
for two short range potentials with strength $g$ separated by a distance ${\bf r}$ yields 
\begin{equation}
V_m(\bf r)=
\begin{cases}
-g^2\frac{n_0m_b}{\pi}\frac{e^{-\sqrt{2}r/\xi}}{r}&\text{BEC}\\
g^2\frac{m_b}{16\pi^3}\frac{2k_Fr\cos2k_Fr-\sin2k_Fr}{r^4}&\text{Fermi gas}\\
\end{cases}
\label{MediatedRealSpace}
\end{equation}
for the Yukawa and the Ruderman–Kittel–Kasuya–Yosida (RKKY)~\cite{Ruderman1954,Kasuya1956,Yosida1957} interactions, mediated by density modulations in a BEC and a Fermi gas, respectively. 
Equation~\eqref{MediatedRealSpace} can be obtained by Fourier transforming Eq.~\eqref{QPinteractionPert} taking $\omega=0$, since the impurities are static. 

\begin{figure}
\centering
\includegraphics[width=0.9\columnwidth]{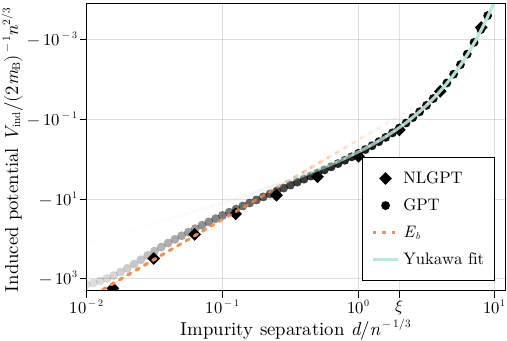}
\caption{\label{Fig:EnssMediated}
{\bf Interaction between two static impurities in  a BEC.}
The impurities interact resonantly $(1/a=0)$ with bosons, and the gas parameter is $na_b^3=10^{-6}$. Circles (diamonds) show results with zero (non-zero) range of the boson-boson interaction, and the dashed orange line is the energy of the Efimov trimer. From~\citet{drescher2023medium}. }
\end{figure}

The interaction between two static impurities mediated by an ideal Fermi gas was explored for a short range impurity-medium interaction of arbitrary strength by solving the scattering problem exactly~\cite{Nishida2009,enss2020scattering}. It was shown that the interaction can be quite different from the RKKY form in Eq.~\eqref{MediatedRealSpace} for strong interactions due to the presence of (Efimov) states where one fermion is bound between the two impurities. These bound states lead to resonances and sign changes in the scattering length~\cite{enss2020scattering}. Likewise, the interaction between two static impurities mediated by a BEC was obtained from the GP equation~\cite{drescher2023medium}. For distances short compared to the interparticle spacing, the interaction was shown to be dominated by a single boson bound between the two static impurities giving rise to an Efimov scaling $\propto 1/r^2$. For distances larger than the healing length of the BEC, the interaction is of the Yukawa form. This crossover between Efimov and Yukawa scalings for the mediated interaction, shown in Fig.~\eqref{Fig:EnssMediated},  was found for any impurity-boson interaction strength, contrary to results obtained from a variational approach~\cite{Naidon2018}. The same approach was used in Ref.~\cite{jager2022effect}. 
The interaction between two static impurities mediated by a superfluid with a linear low energy phonon dispersion was explored  microscopically first in the 
context of neutron stars~\cite{Kobyakov2016} and later in the context of cold atoms~\cite{fujii2022universal}.
Assuming weak and short range impurity-superfluid interactions, it was shown that the mediated interaction is dominated by the exchange of two phonons for 
$r\gg \xi$ giving rise to a $1/r^7$ interaction instead of the Yukawa interaction mediated by one phonon exchange for shorter distances. 

The quasiparticle interaction is affected by a non-zero range of the impurity-medium interaction. One example is the case of ionic impurities discussed in Sec.~\ref{sec:long-ranged_Bose_polarons} where the atom-ion interaction has the  charge-dipole $1/r^4$ form for large separations. Using perturbation theory as well as a diagrammatic $\mathcal T$-matrix approximation, the interaction between two static ions mediated by a BEC was shown to be proportional to $1/r^4$ for large distances and have a Yukawa form for short distances, whereas it switches from an RKKY to a $1/r^4$ form when mediated by a Fermi gas~\cite{Ding2022}. The same  $1/r^4$ behavior was found by solving the GPE with two static ion potentials~\cite{Olivas2024,drescher2023medium}. This interaction gives rise to measurable changes in the phonon spectrum of the two ions in a typical linear RF trap. Quantum Monte Carlo calculations exploring two static ions in a BEC obtained similar results with corrections for strong atom-ion interaction due to large distortions of the BEC around the ions and the presence of bound states in the atom-ion interaction potential~\cite{Astrakharchik2023}.

\subsection{Experimental detection}   \label{Sec:DetectionQPinteraction}

While pioneering experiments probed the interaction between bosons mediated by a Fermi gas in the perturbative regime~\cite{Edri2020,DeSalvo2017}, the interaction between two Fermi polarons was systematically measured for all coupling strengths only recently by \citet{Baroni:2023aa}. The interaction between Bose polarons has so far only been observed in TMDs, which is somewhat surprising since it should be stronger  due to the large compressibility of a  BEC. 

\begin{figure}[ht]
\centering
\includegraphics[width=\columnwidth]{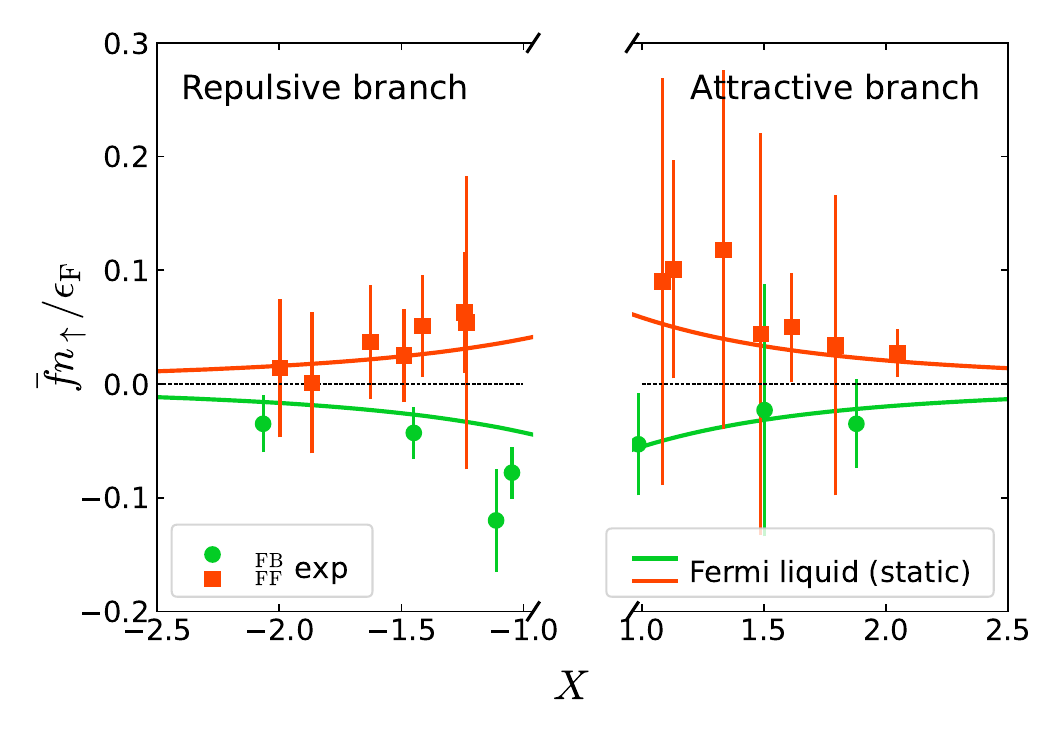}
\caption{\label{Fig:QPintGrimm}
{\bf Mediated interactions in a Fermi gas.} 
Measurement of polaron-polaron interaction between K impurities in a Li Fermi gas $(X=-1/k_Fa)$. Red squares are for fermionic $^{40}$K impurity atoms and green circles for bosonic $^{41}$K impurity atoms. The solid lines are the perturbative result Eq.~\eqref{RKKYzeromomentum}. From~\citet{Baroni:2023aa}.}
\end{figure}

\citet{Baroni:2023aa} measured the energy of Fermi polarons formed by $^{40}$K (fermion) or $^{41}$K (boson) atoms in a bath of $^6$Li atoms as a function of the impurity concentration using RF spectroscopy. The polaron interaction was then extracted by fitting to a momentum average of Eq.~\eqref{QPenergy}, $\varepsilon=\varepsilon^0+\bar fn_i$ (the experiment had no momentum resolution). Using a Li-K Feshbach resonance, $\bar f$ was measured as a function of impurity-medium scattering length $a$ and by comparing results for $^{40}$K or $^{41}$K atoms keeping everything else fixed the role of quantum statistics was probed directly. Figure~\ref{Fig:QPintGrimm} shows the interaction $\bar f$ as a function of $X=-1/k_Fa$. It shows that the quasiparticle interaction is repulsive/attractive for fermionic/bosonic quasiparticles. For weak to moderate interaction strengths, the results agree well with the second order expression given by Eq.~\eqref{RKKYzeromomentum} (solid lines), and the experiment therefore confirms two landmark predictions of Landau's Fermi liquid theory: The strength of the effective interaction and its sign dependence on the quantum statistics of the quasiparticles. For stronger interactions across the Li-K Feshbach resonance where $a$ diverges, Eq.~\eqref{RKKYzeromomentum} does not agree with the experimental results. Here, a non-perturbative diagrammatic theory for the quasiparticle interaction illustrated in Fig.~\ref{Fig:QPintFeynman}(b) explained the experimental results for strong and attractive interactions ($a<0$), whereas the results for strong and repulsive interactions require further analysis. 

Since the exciton radius in TMDs is small, the direct exciton-exciton interaction is weak, which limits their use for optical devices. This motivates the study of mediated interactions between exciton-polarons (or polariton-polarons) with the aim of increasing non-linear effects. Quasiparticle interactions were explored between the Bose polarons formed by exciton-polaritons in one valley immersed in a bath of exciton-polaritons in the other valley in monolayer MoSe$_2$ as discussed in Sec.~\ref{BosepolaronTMD}~\cite{tan2022bose}. Using  pump-probe spectroscopy, the energy of the polarons was measured as a function of their density and the interaction extracted from the slope using a momentum averaged Eq.~\eqref{QPenergy}. Attractive interactions were found between Bose polarons on the repulsive branch, whereas repulsive interactions were found between polarons on the attractive branch. Such repulsive interactions are not expected between bosonic quasiparticles in equilibrium, and they may be due to the inherent non-equilibrium nature of the experiment. 

Exploiting the spin-orbit splitting  in the K and K' valleys, the interactions between excitons mediated by a surrounding electron gas was explored~\cite{Muir:2022aa}. Evidence was found that these interactions mainly occur when they are dressed by electrons in the same valley. The interactions were found to be repulsive, which was attributed to the excitons competing for the same electrons during the  time span of the experiment, which was comparable to the time scale $1/\epsilon_F$ for the formation of Fermi polarons. Also, the probe transmission spectrum of the lower polaron-polariton branch of an electron doped MoSe$_2$ monolayer in an optical cavity was observed to exhibit a blueshift due to the presence of other polaron-polaritons created by a pump beam. This was interpreted as a repulsive  interaction airising from non-equilibrium phase-space filling effect~\cite{Tan2020}. An alternative explanation was given in terms of the interaction mediated by the electron gas shown in Fig.~\ref{Fig:QPintFeynman}(b)~\cite{BastarracheaMagnani2021,Bastarrachea-Magnani2021Polaritons}. An earlier experiment observed energy shifts of the transmitted light intensity of a doped MoSe$_2$ monolayer depending on the pump intensity creating the excitons. This was however interpreted in terms of the composite nature of trion-polariton wave functions and not in terms of polaron-polaron interactions~\cite{Emmanuele2020}.

\subsection{Bi-polarons}   \label{BipolaronSec}

A striking effect of the interaction between quasiparticles is that it can support bound states. Bound states of two polarons, called bi-polarons, are proposed as a  mechanism for superconductivity~\cite{Alexandrov_1994}, for charge transport in polymer chains~\cite{Bredas1985,Mahani2017}, and for magnetoresistance in organic materials~\cite{Wohlgenannt2007}. We now discuss various theoretical predictions for the existence of bi-polarons in atomic gases and TMDs. Very recently, experimental signs consistent with the presence of such bi-polarons were reported by \citet{Morgen2026}. 

A general theory of bound states of two quasiparticles in a many-body environment is very challenging. Their energy is given by the poles of the polaron-polaron scattering matrix, which obeys the Bethe-Salpeter equation~\cite{fetter_1971}. While this is very complicated to solve in general, one can use its close resemblance to the Lippmann-Schwinger equation in the quasiparticle approximation to derive an effective Schr\"odinger equation for the bound states of two polarons 
\begin{equation}
\varepsilon_\text{bp}\psi(\bk)=2\varepsilon_\bk\psi(\bk)+\sum_{\bk'}V_\text{m}(\bk,\bk')\psi(\bk'),
\label{EffectiveSchrodingerBipolarons}
\end{equation}
which is much simpler to solve~\cite{Camacho2018}. Here, $\psi(\bk)$ is the relative wave function of the bi-polaron in momentum space with energy $\varepsilon_{\rm bp}$, $\varepsilon_\bk$ is the energy of an isolated  polaron, and $V_\text{m}(\bk,\bk')$ is the interaction between two  polarons with momenta $\bk$ and $-\bk$ scattering into $\bk'$ and $-\bk'$. This interaction in general is non-local [$V_\text{m}(\bk,\bk')\neq V_\text{m}(\bk-\bk')$], which is typical for effective two-body Schr\"odinger equations in many-body systems such as the Skyrme force in nuclear matter~\cite{ring2004nuclear}. The relative wave function $\psi(\bk)$ has to be symmetric for two bosonic impurities, whereas it is anti-symmetric for fermionic impurities.

\begin{figure}[t]
\centering
\includegraphics[width=\columnwidth]{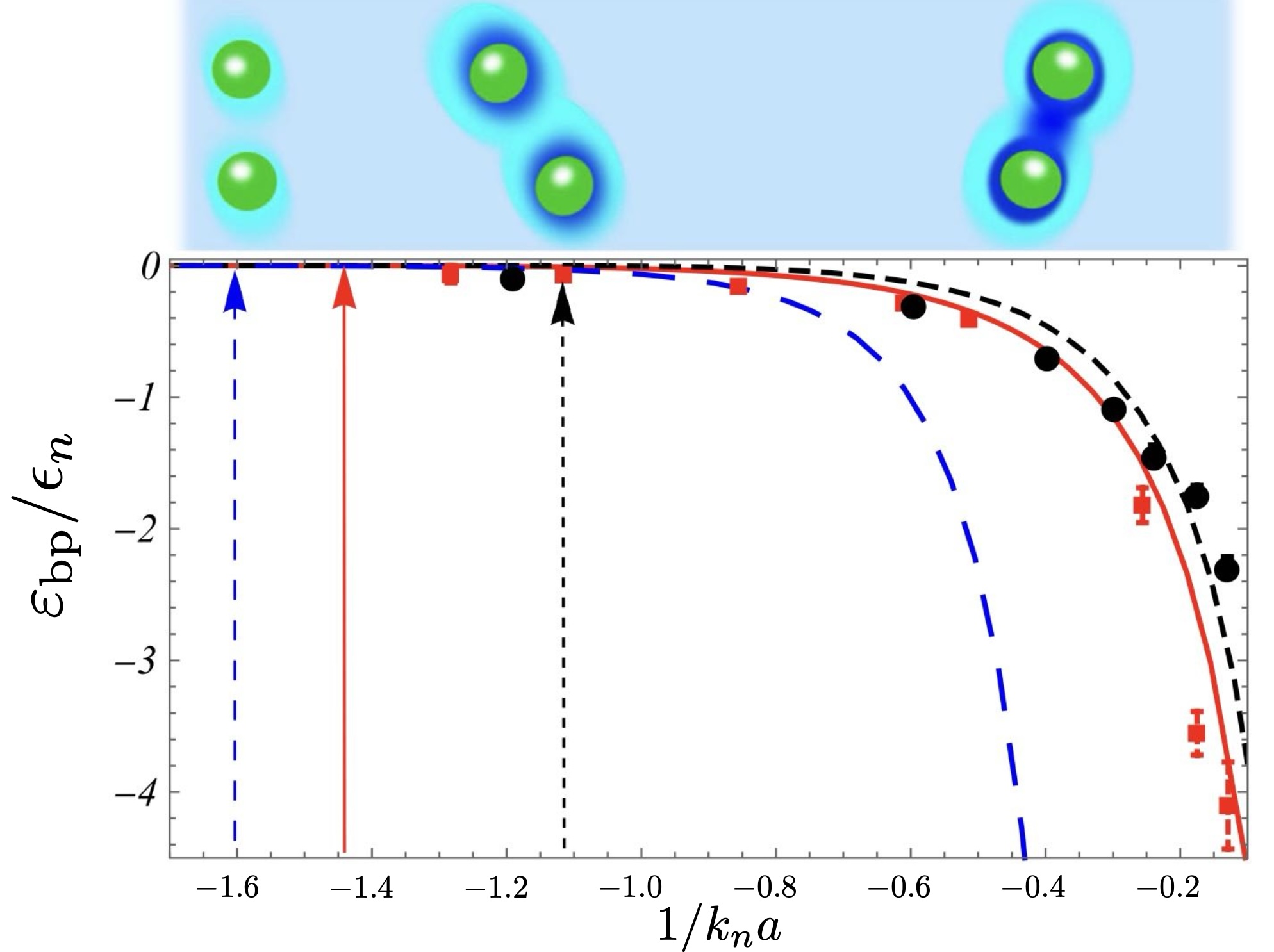}
\caption{\label{Fig:BipolaronCamacho}
{\bf Bi-polarons in a BEC.}
Top: binding of two polarons due to a mediated interaction. Bottom: binding energy $\varepsilon_\text{bp}$ of the bipolaron (with $m=m_b$). The solid red (black dashed) line show the result  obtained from Eq.~\eqref{EffectiveSchrodingerBipolarons} with the mediated interaction shown in Fig.~\ref{Fig:QPintFeynman} for a bath density of $n_ba_b^3=10^{-6}$ ($10^{-5}$), and the red squares (black dots) are the corresponding DMC results. Vertical arrows indicate the emergence of the bi-polaron. The blue dashed line gives the energy using the static Yukawa interaction  Eq.~\eqref{MediatedRealSpace}. From \citet{Camacho2018}.}
\end{figure}

Using an effective interaction $V_\text{m}(\bk,\bk')$ between two Bose-polarons derived by ignoring retardation effects in the diagram shown in Fig.~\ref{Fig:QPintFeynman}(c), bound states of Eq.~\eqref{EffectiveSchrodingerBipolarons} with an energy  below that of two isolated polarons $2\varepsilon_0$ were found in a weakly interacting BEC as shown in Fig.~\ref{Fig:BipolaronCamacho}. The bi-polarons emerge beyond a critical impurity-boson interaction strength with a binding energy that increases with decreasing boson-boson repulsion. This reflects the fact that the BEC becomes more compressible and thus mediates a stronger interaction. For weak impurity-boson interactions $k_n|a|\ll 1$, the interaction is given by Eq.~\eqref{QPinteractionPert} and the bi-polaron energy recovers analytical results for a Yukawa potential~\cite{Harris1962,Rogers1970,Edwards2017}. This method was also used to predict the existence of bi-polarons of different symmetries in a 2D lattice containing a BEC~\cite{Ding2023}. The effective Schr\"odinger equation Eq.~\eqref{EffectiveSchrodingerBipolarons} was also used to predict the presence of bi-polaron in a hard core boson gas in a lattice~\cite{SantiagoGarcia2023}. 

Bi-polarons were also found using Diffusion Monte Carlo calculations~\cite{Camacho2018}. As can be seen in Fig.~\ref{Fig:BipolaronCamacho}, these energies agree well with those obtained from Eq.~\eqref{EffectiveSchrodingerBipolarons} even for strong interactions $k_n|a|\gtrsim1$, which is remarkable since there is no small parameter in this regime and indicates the accuracy of the effective Schr\"odinger equation approach. 

Equation \eqref{EffectiveSchrodingerBipolarons} was generalised to the case of two polaritons in a TMD  interacting via the exchange of phonon modes in a condensate of polaritons in the other valley~\cite{CamachoGuardian2021}. It was found that this interaction supports dimer states, which due to the hybrid nature of polaritons corresponds to a bound state of photons. These bound states were predicted to give rise to new transmission lines of the TMD with photon-photon correlations determined by the dimer wave function. The formation of bi-polarons was also explored by solving a two-body Schr\"odinger equation using a mediated interaction extracted from the GPE treating the impurities as static potentials~\cite{jager2022effect}. 

Bound states of two Fermi polarons were considered in Ref.~\cite{Guo2024}. When the fermion mediated interaction illustrated in the top part of Fig.~\ref{Fig:QPintFeynman} was combined with a variational wave function, bi-polarons were found to be stable for a range of interactions strength for different atomic mixtures, and it was pointed out that such atomic experiments may shed light on the properties of $\alpha$ clusters in neutron matter.

\begin{figure}
\centering
\includegraphics[width=\columnwidth]{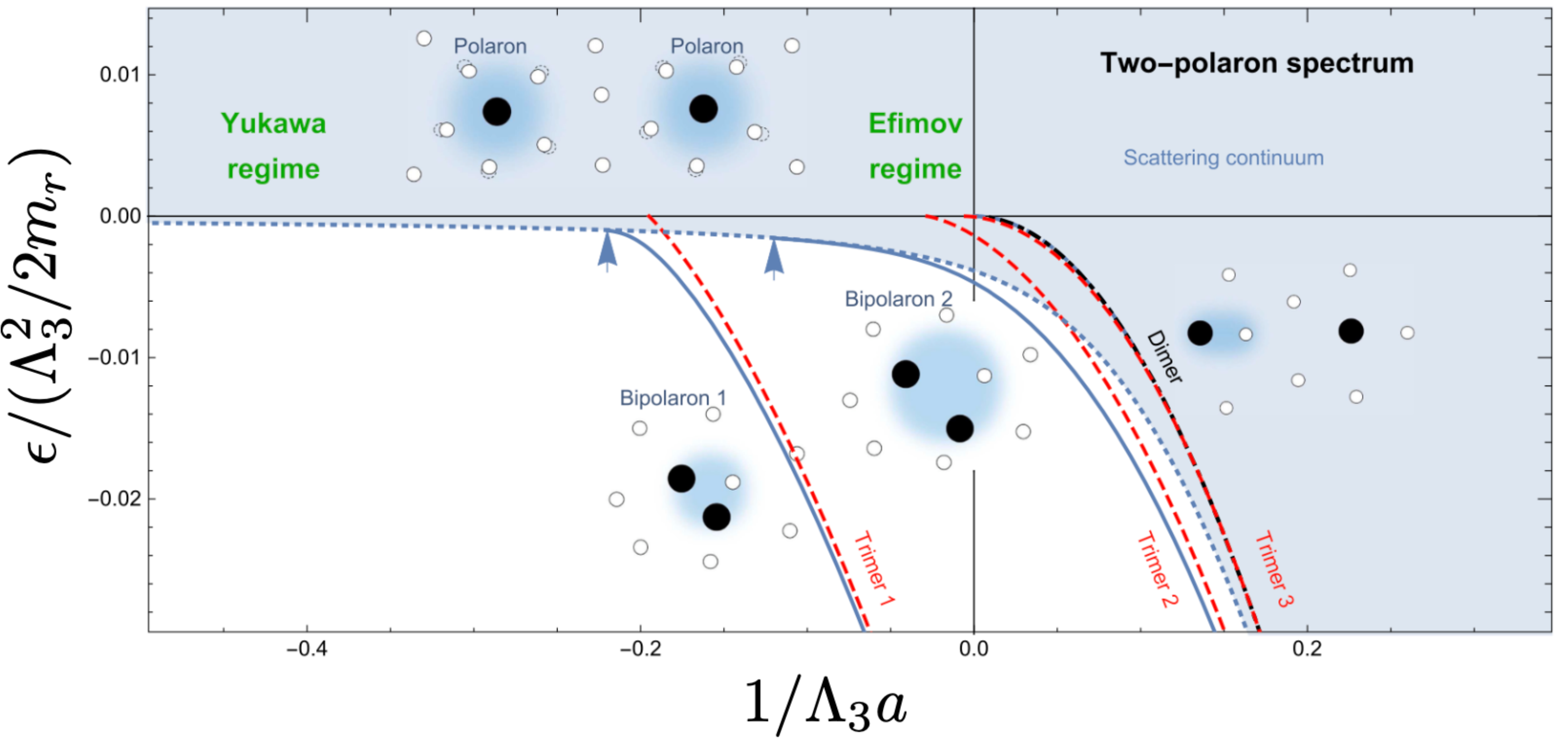}
\caption{\label{Fig:BipolaronNaidon}
{\bf Bi-polarons and Efimov states}. Spectrum of two impurities in a BEC as a function of the boson-impurity scattering length (solid blue lines). $\Lambda_3\simeq3.2k_n$ is a three-body cut-off parameter related to the range of the impurity-boson interaction. The mass ratio is $m/m_b=19$ and $k_n|a_{-}|\simeq 1.6$. The bi-polarons become stable at vertical arrows, red lines show the energies of Efimov trimers, and the black line is the boson-impurity dimer. The shaded area is the scattering continuum of two impurities. From \citet{Naidon2018}.}
\end{figure}

Bi-polarons are a many-body effect due to an attractive interaction mediated by a surrounding bath. As such, they are distinct from few-body states such as Efimov trimers, which exist also in a vacuum. As discussed in Sec.~\ref{MediatedStatic}, the presence of Efimov trimers can however affect the mediated interation at short range, and it was furthermore shown in Sec.~\ref{sec:Bose_variational_Chevy} that they can have large effects on the Bose polaron when $k_n|a_-|\sim 1$, see Fig.~\ref{fig:Bose_polaron_hybridization}. Efimov trimers may therefore also influence bi-polarons in this regime, which was explored using the variational ansatz given by Eq.~\eqref{BogExpansion} generalized to the case of two bosonic impurities in a  BEC~\cite{Naidon2018}. It was predicted that the bi-polaron, which for weak interactions is bound by a Yukawa potential, smoothly evolves into an Efimov trimer bound by a $1/r^2$ potential for strong interactions $1/k_n|a|\lesssim 1$ as illustrated in the left panel of Fig.~\ref{Fig:BipolaronNaidon}. The same problem was considered for two heavy $^{133}$Cs impurities in a $^6$Li Fermi sea, where the mass ratio ensures the existence of Efimov trimers~\cite{Sun2019}. Bound states with an energy well below that of two uncorrelated Fermi polarons were found for strong interactions. The bound states between two polarons in a dipolar Fermi gas were explored in Ref.~\cite{nakano2024twobody}. Using an RKKY form of the interaction as in Eq.~\eqref{QPinteractionPert} generalized to the dipolar gas in an effective Schr\"odinger equation, the regions of stability and binding energy of the bipolarons were analyzed. 

\section{Polarons as a limit of many-body phases}   \label{sec:LimitManyBody}

The Fermi and Bose polarons discussed in this review define the low-density limit  of two-component Fermi-Fermi, Fermi-Bose and Bose-Bose mixtures. Indeed, the Fermi polaron was originally introduced to constrain the equation of state of spin-imbalanced, strongly interacting Fermi mixtures~\cite{Chevy2006, Combescot2007, prok07signproblem, Prokofiev2008b}. The limit of Fermi polarons also constrains the nuclear equation of state, in particular for neutron matter~\cite{Forbes2014}. In this Section, we discuss how polaron physics is connected to and can provide important insights into the more general  setting of quantum mixtures. This connection has previously been reviewed in the context of the repulsive Fermi polaron and itinerant ferromagnetism~\cite{Massignan2014}, which we will therefore not discuss here. For a recent comprehensive review on atomic quantum mixtures, see Ref.~\cite{Baroni2024review}.

\subsection{Fermi mixtures}

In an equal population mixture of fermions in two attractively interacting spin states, the ground state is a superfluid of Cooper pairs~\cite{zwie05vort,Zwerger2011BECBCS,Zwierlein2014NovelSuperfluids}. With spin imbalance, some majority spins remain unpaired. The question of the fate of superfluidity in the presence of spin imbalance has a long history. In condensed matter, this relates to the stability of superconductors in strong magnetic field, and one generally has neutron-proton (isospin) asymmetry in nuclear physics. In the core of neutron stars, neutral superfluids of unequal densities of quarks are predicted to exist~\cite{alfo00color}. While imbalanced pairing is difficult to study in conventional superconductors since magnetic fields are typically expelled by the Meissner effect, the population in the two (hyperfine) spin states can be freely tuned in atomic gases.  This enabled the experimental investigation of the phase diagram of spin-imbalanced Fermi gases~\cite{zwie05imbalance,part06phase,shin06phase,shin07phasediagram,shin08eos,schirotzek_observation_2009,nascimbene_collective_2009,nasc10thermo}.

\subsubsection{Chandrasekhar-Clogston limit}
If magnetic fields do enter a superconductor, the superconducting state of electron pairs should be fragile as the field tends to align the spins, and there must be a critical field beyond which the normal state has lower free energy than the superfluid. 
Chandrasekhar~\cite{chan62} and  Clogston~\cite{clog62} derived an upper (CC) limit for the critical magnetic field of a superconductor. To evaluate this,  one compares the free energy $F(h)$ of the normal and the superfluid state in the presence of a ``magnetic field" $h = (\mu_\uparrow - \mu_\downarrow)/2$, where $\mu_\uparrow$ and $\mu_\downarrow$ are the chemical potentials of the majority and minority atoms, respectively. Consider first the BCS regime and ignore
the attractive interaction between opposite spins present also in the normal state. This will lead to an overestimate of the critical field, as it  neglects the formation of attractive Fermi polarons. The   free energy of a balanced fermionic superfluid is  $F_S = F_N(0) - \frac{1}{2}\mathcal N(\epsilon_F) \Delta^2$, which is lower than the free energy of the balanced normal gas at $h=0$ by the condensation energy $\frac{1}{2} \mathcal N(\epsilon_F) \Delta^2$. Here $\mathcal N(\epsilon_F)$ is the density of states at the Fermi energy, and $\Delta$ the superfluid gap. The free energy of the normal state as a function of $h$
is $F_N(h) = -\frac{4}{15} \mathcal N(\epsilon_F) (\mu_\uparrow^{5/2}+\mu_\downarrow^{5/2})\simeq F_N(0) - \mathcal N(\epsilon_F) h^2$, where $\mu_\uparrow = \epsilon_F+h$, $\mu_\downarrow = \epsilon_F-h$ and $F_N(0) = -\frac{8}{15} \mathcal N(\epsilon_F)\epsilon_F^2$. From this, one obtains $h_{\rm CC} = \Delta/\sqrt{2}$ for the critical magnetic field. In conventional superconductors, this corresponds to  $h_{\rm CC}\sim 18.5\, {\rm T}$ for $T_c\sim 10\,\rm K$, which is much larger than the typical critical field $H_{c2}$ where superconductivity breaks down.
Heavy fermion or layered superconductors may however attain this CC regime~\cite{Pfle2009heavy}.

The first-order superfluid-to-normal transition at the critical field was studied by Sarma~\cite{sarm63}. Fulde and Ferrell~\cite{FF64}, and  Larkin and Ovchinnikov~\cite{LO64} then found that not all the pairs necessarily break at once, but that there exists a novel superfluid state with 
 a certain amount of majority spins and the remaining Cooper pairs having a non-zero momentum. The order parameter  thus corresponds to a traveling (Fulde-Ferrell state) or standing (Larkin-Ovchinnikov state) wave, and majority spins can reside in its nodes without energy penalty. The number of nodes is given by the number difference between the spin states. The true ground state of spin-imbalanced superfluidity is however still not known. The problem arises in condensed matter for exotic superconductors  that are essentially Pauli limited~\cite{casa04fflo,Rado03,Bian03},  and in the study of superfluid pairing of quarks at unequal Fermi energies~\cite{alfo00color}. For strongly interacting atomic Fermi gases, where $\Delta$ approaches the Fermi energy, the critical field is a substantial fraction of $\epsilon_F$, and the window of superfluidity as a function of the field $h$ is  wide, which presents a new opportunity to study imbalanced superfluidity.

\begin{figure}
\begin{center}
\includegraphics[width=0.8\columnwidth]{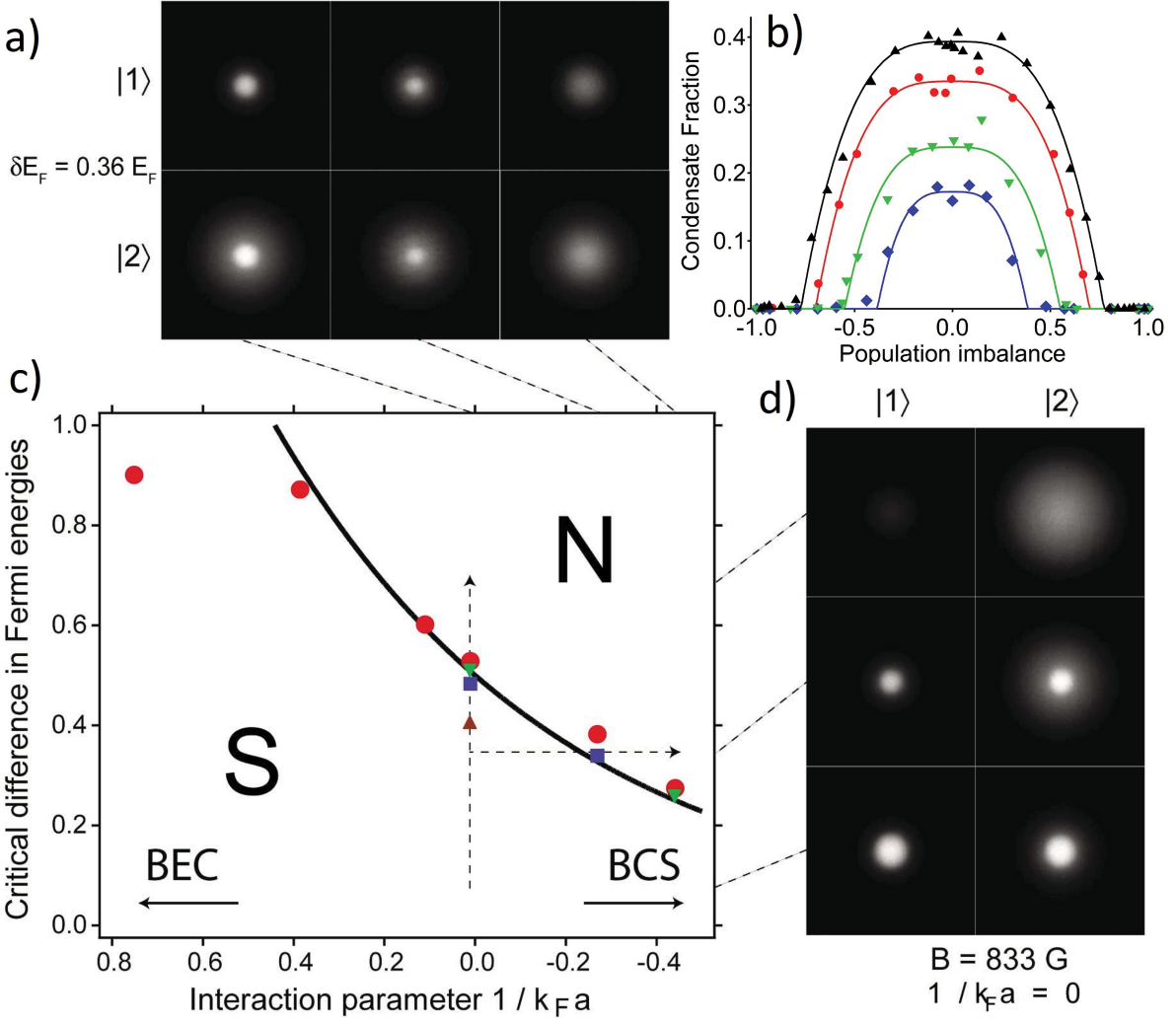}
\caption{\label{f:clogstonlimit}
{\bf Chandrasekhar-Clogston limit of superfluidity.} 
The critical Fermi energy mismatch $\delta \epsilon_F$ between the two spin states at the superfluid-to-normal transition shown in c) is observed in the condensate fraction for varying interaction strength at fixed $\delta \epsilon_F$ (a), and at fixed interaction strength and varying $\delta \epsilon_F$ (d). b) Window of superfluidity as obtained from the condensate fraction at $1/k_F a = 0.11$ (triangles pointing up), $1/k_F a = 0$ (resonance, circles), $1/k_F a = -0.27$ (BCS-side, triangles pointing down), $1/k_F a = -0.44$ (diamonds). The normal phase competing with the superfluid is a Fermi liquid of polarons. From~\citet{zwie05imbalance}.}
\end{center}
\end{figure}

The MIT group studied spin imbalanced Fermi mixtures in the presence of a stirring beam~\cite{zwie05imbalance}. The part of the mixture that was still superfluid despite the imbalance revealed a lattice of quantized vortices. The normal Fermi mixture above the critical imbalance for superfluidity is well-described as a Fermi liquid of polarons. The window of superfluidity was determined from the number of vortices as a function of imbalance, as well as from condensate fraction measurements~\cite{kett08varenna,Zwierlein2017}, see  Fig.~\ref{f:clogstonlimit}. At unitarity, superfluidity was robust up to a critical population imbalance $P = (N_\uparrow - N_\downarrow)/(N_\uparrow + N_\downarrow)$ of about $P_c = 75\%$, which agrees with the phase diagram later obtained by the ENS group~\cite{navo10eos}, and with a Monte Carlo study obtaining $P_c = 77\%$~\cite{lobo06}. BCS  theory overestimates this to $P_c = 92\%$ as it neglects  polaron formation  in the normal state, which is thus favored at large fields $h$ compared to the superfluid state. Indeed, at high imbalance, the Fermi mixture is normal down to the lowest temperatures realized thus far, and behaves as a Fermi liquid of polarons~\cite{schirotzek_observation_2009,nascimbene_collective_2009,nasc10thermo,nasc2011fermiliquid}.

\subsubsection{Phase separation}
While the BCS ground state is fully paired and therefore perfectly balanced, 
the normal state beyond the CC limit has imbalanced spin densities.  The first order transition from the balanced superfluid to the imbalanced normal state is therefore signaled by a jump in the density difference. First hints of a phase separation between the normal and superfluid phase were seen~\cite{zwie05imbalance,part06phase}, and using tomographic techniques a sharp separation between a superfluid core and a partially polarized normal phase was observed~\cite{shin06phase}. 
A jump in the density difference was observed thereby directly demonstrating the first order nature of the phase transition~\cite{shin07phasediagram}. At higher temperatures the signature of the first order transition disappears at a tricritical point, in good agreement with theoretical calculations~\cite{lobo06,gubb08}.


\subsubsection{Equation of state at unitarity}
At unitarity and  zero temperature, the energy of the gas can only depend on the two Fermi energies $\epsilon_{F\sigma}$. 
This allows to write for the energy density ${\cal E}$
\begin{equation}
	{\cal E}(n_\uparrow,n_\downarrow) = \frac{3}{5} n_\uparrow \epsilon_{F\uparrow} \, g\left(x\right)^{5/3}=\frac{3}{5} \left(n_\uparrow \mu_\uparrow + n_\downarrow \mu_\downarrow\right)
    \label{EqnofState}
\end{equation}
with  $g(x)$ a universal function of the density ratio $x = n_\downarrow/n_\uparrow$~\cite{bulg07asym}. The second equation  follows from  $\mu_\sigma = \partial {\cal E}/\partial n_\sigma$, and 
one has $g(x)^{5/3} = \frac{\mu_\uparrow}{\epsilon_{F\uparrow}}(1 + xy)$ with $y=\mu_\downarrow/\mu_\uparrow$. Within the local density approximation, the local chemical potentials are $\mu_\sigma(\vec{r}) = \mu_{0,\sigma} - U(\vec{r})$
 with the global chemical potentials $\mu_{0,\sigma}$ for each species and $U(\vec{r})$ the trapping potential. In the outer wings of the  mixture resides a non-interacting Fermi gas of only majority atoms. One can therefore directly obtain the majority global chemical potential from the radius of the majority cloud $R_\uparrow$ as $\mu_{0,\uparrow} = U(R_\uparrow)$. The minority global chemical potential $\mu_{0,\downarrow}$ can be obtained by noting that the last minority atom at the outermost wing of the minority cloud is a Fermi polaron, and thus $\mu_{0,\downarrow} = A \epsilon_{F\uparrow}(R_\downarrow) = A [U(R_\uparrow) - U(R_\downarrow)]$, where $\varepsilon=A\epsilon_{F\uparrow}$ is the energy of a single polaron in a uniform bath. This method was employed in Refs.~\cite{Chevy2006,bulg07asym} to estimate the polaron energy from the  cloud radii measured in~\cite{zwie06direct}. 
Alternatively, if the central part of the mixture is a balanced superfluid, we can write ${\cal E} = \frac{3}{5} \xi_B\epsilon_F(0)(n_\uparrow+n_\downarrow)$ with $\xi_B$ the Bertsch parameter. Since $n_\uparrow = n_\downarrow$ this implies from Eq.~\eqref{EqnofState} $\mu_{0,\uparrow} + \mu_{0,\downarrow} = 2 \xi_B \epsilon_F(0)$, where $\epsilon_F(0)$ is the Fermi energy in the center of the trap. This   provides a link between the polaron energy  $A=\varepsilon/ \epsilon_{F\uparrow}$, the Bertsch parameter $\xi_B$, and the experimental quantities $\epsilon_F(0)$, $R_\uparrow$ and $R_\downarrow$. One can also obtain  $\xi_B$  directly from the normalized compressibilities $\tilde{\kappa} = \kappa/\kappa_0 = {\rm d}\epsilon_{F\uparrow}/{\rm d}U$ ($\kappa_0$ is the ideal gas compressibility) of the majority species in the fully polarized normal wings where $\tilde{\kappa} = 1$ and in the superfluid region where $\tilde{\kappa} = 1/\xi_B$.

\begin{figure}
\begin{center}
\includegraphics[width=\columnwidth]{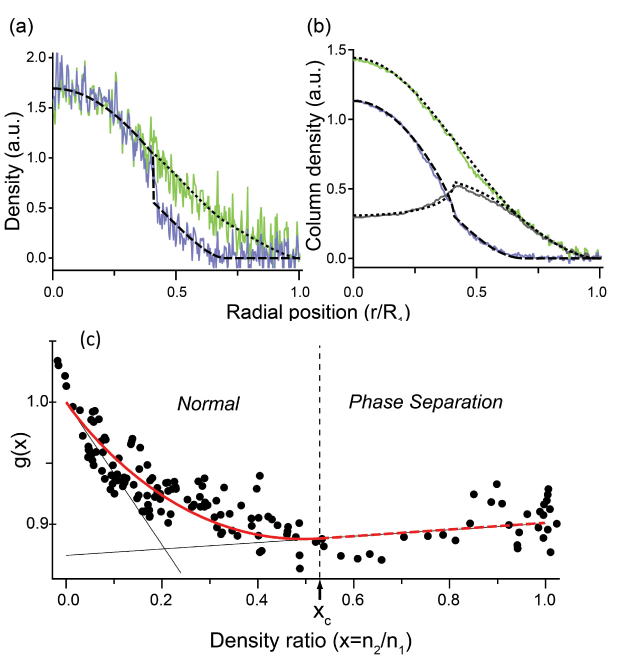}
\caption{\label{f:EOSYong}
{\bf Equation of state of a spin-imbalanced Fermi mixture at unitarity.} 
The 3D density profiles (a) obtained via an inverse Abel transform from the measured column density profiles (b) (at population imbalance of $44\%$) directly yield the equation of state (c) for spin-imbalanced Fermi gases. The normal-to-superfluid transition takes place at $n_\uparrow/n_\downarrow \equiv n_2/n_1 = 0.53(5)$. From~\citet{shin08eos}.}
\end{center}
\end{figure}

The  equation of state of spin-imbalanced Fermi gases   Eq.~\eqref{EqnofState} was measured from the density profiles of the trapped gas~\cite{shin08eos}, see Fig.~\ref{f:EOSYong}. Three distinct phases were found: A superfluid region at equal spin densities at small distances from the trap center, followed by a normal mixed region at unequal densities, and beyond the minority cloud radius $R_\downarrow$ a region of a fully polarized normal gas of majority atoms. In Fig.~\ref{f:EOSYong}a, the normal-to-superfluid transition is directly visible as the boundary between the spin-balanced region  and the imbalanced region. The jump in the density difference marks the first-order transition. The form of $g(x)$ is constrained by the limiting cases: In the superfluid region where $(\mu_\uparrow+\mu_\downarrow)/2 = \xi_B \epsilon_{F\uparrow}$, $\mu_\uparrow/\epsilon_{F\uparrow} = 2\xi_B/(1+y)$ and thus $g(1) = (2 \xi_B)^{3/5}$; in a fully polarized $(x=0)$ non-interacting Fermi gas one has $\mu_\uparrow = \epsilon_{F\uparrow}$ and $g(0)=1$. The critical chemical potential ratio $y_c$ above which the superfluid forms was found to be $y_c = 0.03(2)$, at a critical density ratio $x = 0.53(5)$. The polaron energy was estimated to be $y_c = A =-0.58(5)$ from a Thomas-Fermi fit to $g(x)$, and $y_c = A = -0.69(8)$ from the measured cloud radii, assuming $\xi_B = 0.42(1)$, which is however slightly larger than the present value $\xi_B=0.37(1)$~\cite{ku2012thermodynamics}. The values for $A$ are in good agreement with later studies via RF spectroscopy~\cite{schirotzek_observation_2009}. The experiment found good agreement with a Fermi liquid description of the normal mixed state.

\begin{figure}
\begin{center}
\includegraphics[width=\columnwidth]{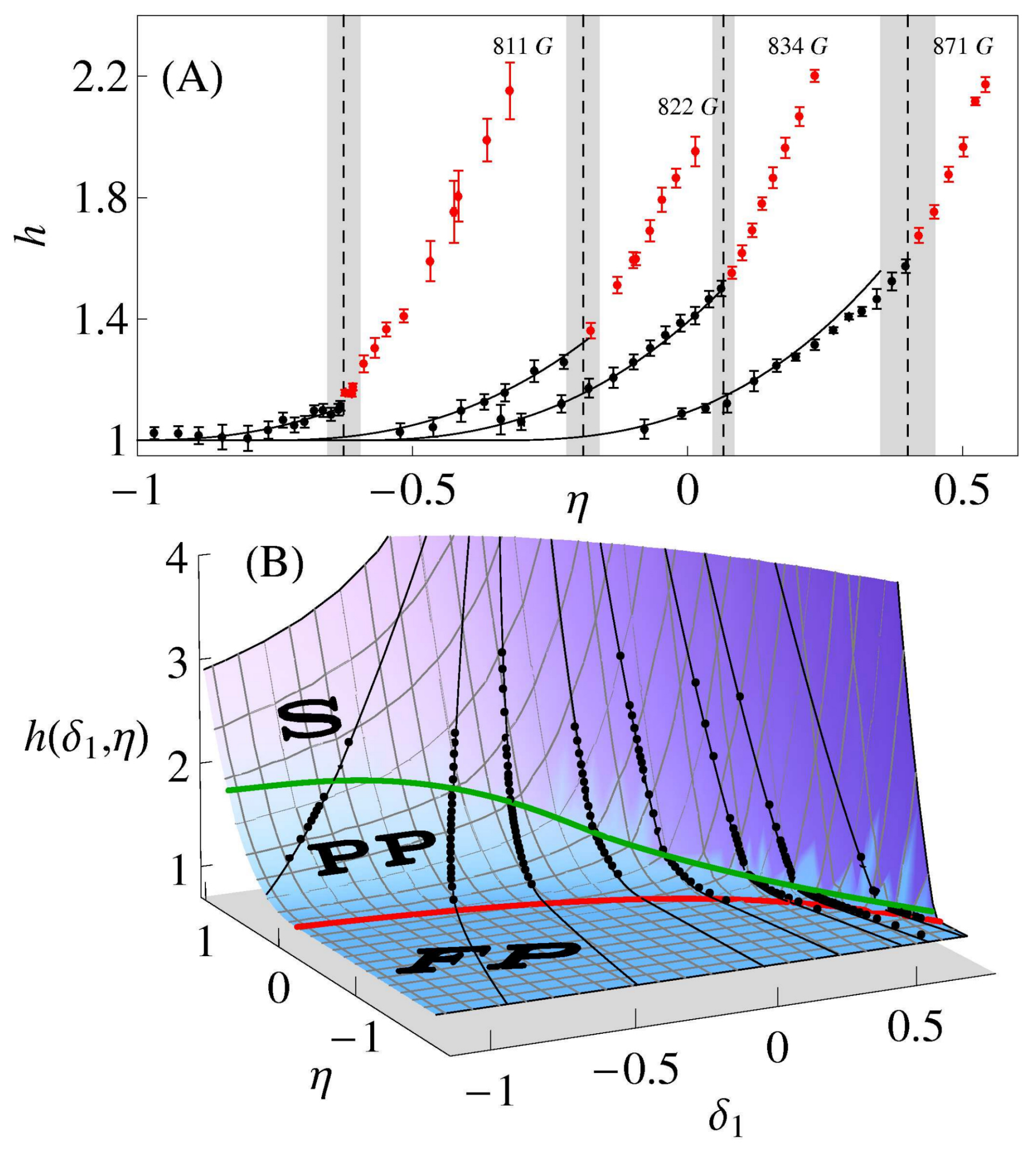}
\caption{\label{f:EOSENS}
{\bf Equation of state of a Fermi mixture as a function of interaction strength and spin imbalance.} (A) Normalized pressure $h = P(\mu_\uparrow,\mu_\downarrow)/P_0(\mu_\uparrow)$ versus the chemical potential ratio $\eta=\mu_\downarrow/\mu_\uparrow$ for several magnetic fields in the BEC-BCS crossover. Black (red) datapoints are in the normal (superfluid) regime, and solid lines the polaron Fermi liquid equation of state. (B) Normalized pressure $h$ versus $\eta$ and interaction parameter $\delta_1 = \hbar/\sqrt{2m\mu_\uparrow}a$ with $a$  the scattering length. The red line is the boundary between a fully polarized Fermi gas (FP) and the partially polarized Fermi liquid of polarons (PP), and the green one marks the superfluid-to-normal phase transition.
From~\citet{navo10eos}.}
\end{center}
\end{figure}

A later experiment yielded a low-noise equation of state for imbalanced gases making use of a direct relation between the pressure of the gas and the doubly integrated density~\cite{nasc10thermo}. In the superfluid region with $\mu_s =(\mu_\uparrow+\mu_\downarrow)/2$ the pressure is $	P(\mu_\uparrow,\mu_\downarrow) = \frac{1}{15\pi^2}\left(\frac{m}{\xi_B \hbar^2}\right)^{3/2}(\mu_\uparrow+\mu_\downarrow)^{5/2}$. In the normal region 
good agreement was found, as in~\cite{shin08eos}, assuming a non-interacting Fermi gas of majority atoms coexisting with a Fermi liquid of polarons with renormalized energy and mass. The corresponding pressure is $	P(\mu_\uparrow,\mu_\downarrow) = \frac{1}{15\pi^2}[\left(\frac{m}{\hbar^2}\right)^{3/2}\mu_\uparrow^{5/2} + \left(\frac{m^*}{\hbar^2}\right)^{3/2}\left(\mu_\downarrow - \varepsilon_\downarrow\right)^{5/2}]$ where  $m^*$ is the Fermi polaron effective mass. In terms of the energy density, this can be expressed in the canonical form as a Landau-Pomeranchuk functional~\cite{lobo06,pila08phase,Mora2010}
\begin{equation}\label{eq:LandauPomeranchukFunctional}
	{\cal E}(n_\uparrow,n_\downarrow) = \frac{3}{5} n_\uparrow \epsilon_{F\uparrow}\,\left(1 + \frac{5 A}{3} x + \frac{m}{m^*} x^{5/3} + F x^2\right)
\end{equation}
where the first term is the energy of the majority fermions, the second is the polaron energy shift, the third is the energy of a non-interacting gas of polarons, and the fourth is the  interaction between polarons discussed in Sec.~\ref{QPinteractions}. From Eq.~\eqref{RKKYzeromomentum} we have $F = \frac{5}{9} \left(\frac{{\rm d}\varepsilon_\downarrow}{{\rm d}\mu_\uparrow}\right)^2$ as confirmed by MC calculations~\cite{pila08phase}. 

The polaron energy given by Eq.~\eqref{eq:LandauPomeranchukFunctional}, $\epsilon = [{\cal E}(n_\uparrow,n_\downarrow) - \frac{3}{5} n_\uparrow \epsilon_{F\uparrow}]/n_\downarrow$ proved to be well reproduced by experiments~\cite{schirotzek_observation_2009} using the MC value $A = -0.615$ \cite{prok2008}, the analytic result $m^* = 1.2m$~\cite{comb08fullpolaron}, and a weak repulsion between polarons with $F=0.14$~\cite{pila08phase}. Assuming $\xi_B = 0.42(1)$, also the experiment in~\cite{nasc10thermo} agreed excellently with the theoretical value for the polaron energy $A = -0.615$ and mass $m^*/m = 1.20(2)$. The simple expression Eq.~\eqref{eq:LandauPomeranchukFunctional} works well even close to the superfluid-to-normal transition. The work was extended to  the BEC-BCS regime~\cite{navo10eos}, see Fig.~\ref{f:EOSENS}, and the Fermi liquid picture for the mixed region was confirmed in detail~\cite{nasc2011fermiliquid}.
Surprisingly, accurate estimates for key parameters such as the polaron's energy, its effective mass and the contact at unitarity could be extracted applying the local density approximation to QMC calculations which studied very few particles ($N\leq 10$) in a harmonic trap \cite{Rakshit2012,Yin2015Trapped,Levinsen2017Universality}.

\subsection{Bose-Fermi mixtures}
\begin{figure}
\begin{center}
\includegraphics[width=0.8\columnwidth]{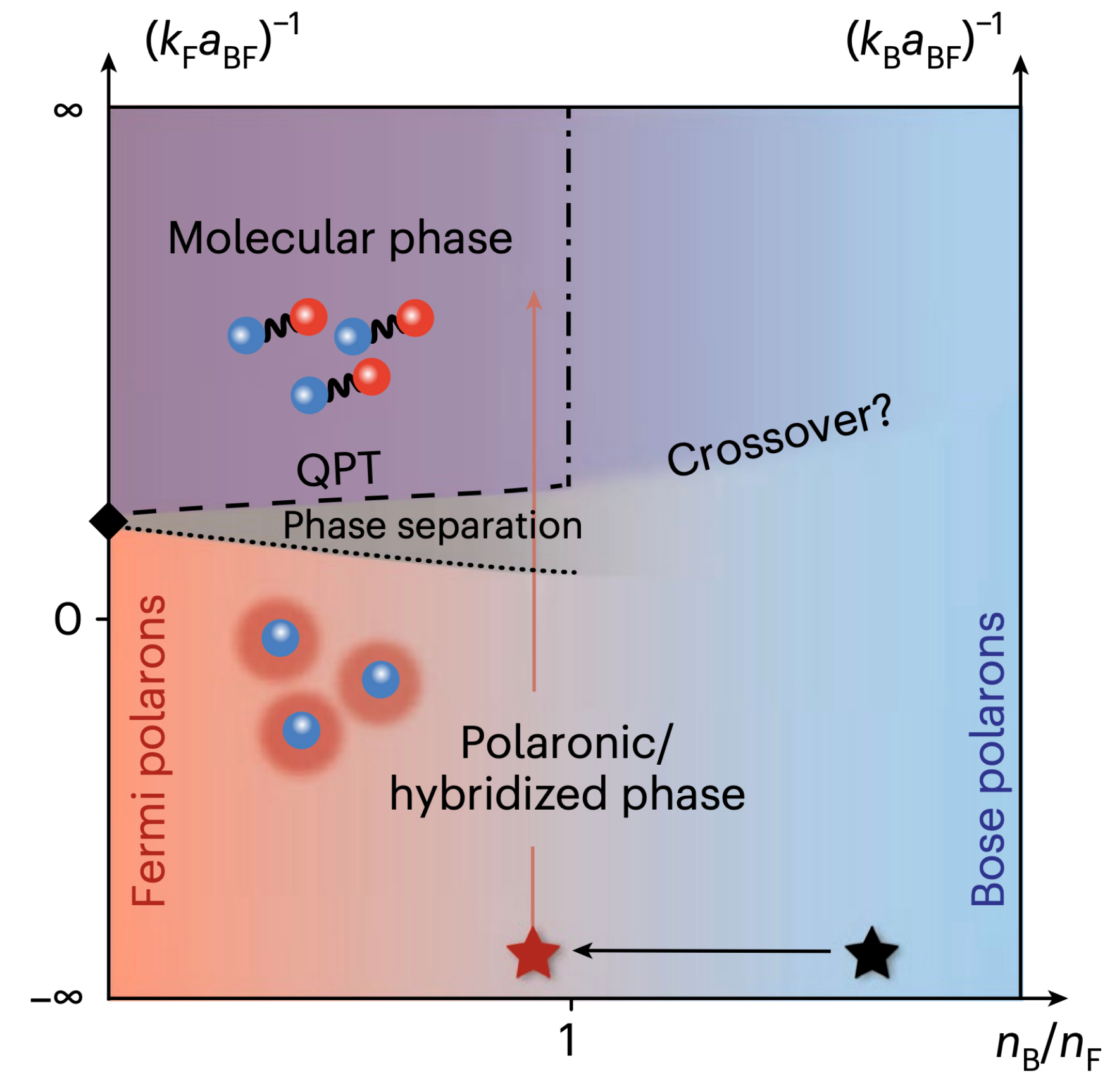}
\caption{\label{f:phasediagramBoseFermi}
{\bf Phase diagram of a  Bose-Fermi mixture.}
In the limit $n_{\rm B}/n_{\rm F} \rightarrow 0$ bosons are impurities in the Fermi sea and  form Fermi polarons, 
whereas the fermions form Bose polarons for  $n_{\rm B}/n_{\rm F} \rightarrow \infty$. For $n_{\rm B}< n_{\rm F}$, a quantum phase transition between a polaronic and a molecular phase is expected. The long-dashed line marks the complete depletion of the condensate and, in the case of phase separation, the dotted line marks its onset. The dash-dotted line marks a possible quantum phase transition. From \citet{Duda2023}.}
\end{center}
\end{figure}

The conditions for thermodynamic stability of a Bose-Fermi mixture were first considered in the context of dilute solutions of He$^3$ in He$^4$ \cite{LandauPomeranchuk1948,Pomeranchuk1949,Anderson66thermal,Bardeen67effective}, and later discussed  for atomic gases~\cite{Viverit2000}.
Bose-Fermi mixtures give rise to many phenomena connected to polarons. Indeed, in the regime of the boson density $n_{\rm B}$ 
much smaller than the fermion density $n_{\rm F}$, the bosons form Fermi polarons or, for strong enough attraction, bind to a fermion into a molecule. 
In the other extreme $n_{\rm F}\ll n_{\rm B}$, the fermions form Bose polarons, see Fig.~\ref{f:phasediagramBoseFermi}. Recently, this transition from Fermi to Bose polarons was observed: A small number of thermal $^{41}$K atoms in a $^6$Li Fermi sea formed Fermi polarons, whereas for larger concentrations the $^{41}$K atoms became a dense BEC in which the $^6$Li atoms formed Bose polarons~\cite{fritsche2021stability}. 

For  balanced densities, one has at weak interactions a mixture of a BEC and a Fermi gas. As interactions increase, bosons  bind with fermions into molecules, which themselves are fermionic, leading at a quantum critical point to the complete vanishing of the BEC and the emergence of a Fermi sea of molecules~\cite{powe05bosefermi,Ludwig2011}. This  was recently observed by sweeping across a Feshbach resonance~\cite{Duda2023}. The role of three-body correlations may be important, as such sweeps potentially also yield trimers or larger clusters instead of only dimers as discussed in Sec.~\ref{sec:BosePolarons}. Recently, the theory underlying the analysis of this experiment was extended to  describe  mixtures of excitons and  electrons in TMDs~\cite{von2023superconductivity}. It was found that the interplay of Bose and Fermi polaron formation combined with exciton exchange between the electrons can induce an emergent BEC-BCS crossover in such  mixtures with critical temperatures up to $T_c/T_F\simeq0.1$. In general,  an attractive interaction between electrons mediated by excitons can  give rise to Cooper pairing and  
superconductivity~\cite{bighin2022resonantlyenhancedsuperconductivitymediated,Zerba2024}, which may be of topological nature and with a critical temperature enhanced by strong coupling to light leading to the formation of exciton-polaritons~\cite{Julku2022,Zerba2024}. Also, Bose-Fermi mixtures  of long-lived dipolar inter-layer excitonic insulators interacting with degenerate itinerant electrons can be realised~\cite{Chenhao2023doi:10.1126/science.add5574,Mak-Shan2023degenerate,FengWang2023degenerate, mhenni2024}.

The highly imbalanced regime hosts a  quantum phase transition from zero to non-zero boson or fermion density~\cite{Nikolic2007,sachdev_quantum_2011}. Already the non-interacting Fermi and weakly interacting Bose gas can be discussed from this viewpoint. Neglecting complications from three-body correlations, a Bose-Fermi mixture is described by four parameters: the boson and fermion chemical potentials $\mu_{\rm B}$ and $\mu_{\rm F}$ and the boson-boson and boson-fermion interaction strengths $g_{\rm BB}$ and $g_{\rm BF}$, with the ratio of fermion to boson mass an additional parameter. This gives rise to several different phases~\cite{Ludwig2011}, and Ref.~\cite{Yan2020} explored the case of a quantum phase transition occurring in the presence of a BEC, separating the vacuum of fermions $n_{\rm F}=0$ from the Fermi liquid phase with $n_{\rm F}>0$. This phase transition is shifted from $\mu_{\rm F}=0$ to $\mu_{\rm F}^*=\varepsilon$ given by the energy of Bose polaron, which  is the energy needed to inject a single fermion into the BEC. The critical chemical potential $\mu_{\rm F}^*$ depends on $\mu_{\rm B}$, $g_{\rm BB}$ and $g_{\rm BF}$. This quantum critical line at $T=0$ determines the behavior of the polaron gas also at finite temperature. In particular, for unitarity limited Bose-Fermi interactions $a \rightarrow \infty$, the polaron lifetime $\Gamma$ will take on a "Planckian" limit, just given by temperature, $\Gamma \simeq k_B T/\hbar$.

\section{Complex environments and sensing}   \label{InteractingBaths}

While the majority of investigations of polarons so far have considered cases where the environment can be treated as an ideal Fermi gas or a weakly interacting BEC as described in Secs.~\ref{sec:FermiPolarons} and~\ref{sec:BosePolarons}, there is an increasing focus on situations where the environment has strong quantum/thermal correlations or non-trivial topology. In addition to being conceptually interesting, such studies are also motivated by the idea of using the impurities as probes in the spirit of quantum sensing~\cite{Degen2017}. For TMDs, this idea is discussed in Sec.~\ref{Excitonsasprobes} and this section therefore focuses on the atomic case. 
  
The investigation of the impurity regime was in fact recognized as an important tool for probing BECs even before focus was on polarons. Impurities were realized by a blue-detuned laser beam and thus mimicked the movement of a macroscopic object through a superfluid~\cite{Onofrio2000}. These results were corroborated in an experiment where atoms in a  BEC were accelerated using Raman transitions, which showed a strongly reduced collision rate below the speed of sound~\cite{chikkatur2000}. In Ref.~\cite{catani_quantum_2012}, K impurities probed a  Rb BEC by expanding within it. A superposition of motional states of Li atoms in a Na BEC allowed for an interferometric  observation of a phononic Lamb shift~\cite{scelle2013} in agreement with Fr\"ohlich model of the polaron~\cite{Rentrop2016a}. An in-situ interferometric technique was used to explore the effective mass and dispersion  of the Bose polaron~\cite{Marti2014} (at that time called a magnon) in Rb BECs. These results showed first indications of effects beyond a mean-field description. Finally, impurities have been used to measure the temperature~\cite{Olf2015,Hohmann2016,Bouton2020} and density~\cite{Adam2022} of a BEC and the interaction mediated by an ideal Fermi gas~\cite{Edri2020}.

\subsection{Polarons in optical lattices}   \label{PolaroninLattice}

By trapping atomic gases in optical lattices formed by standing laser waves,  the famous Hubbard model is realised in a pristine and highly tunable way~\cite{bloch_many-body_2008}. This has provided a wealth of insights into quantum magnetism, topological matter,  phase transitions and non-equilibrium physics~\cite{Gross2017}. 

\begin{figure}
\centering
\includegraphics[width=\columnwidth]{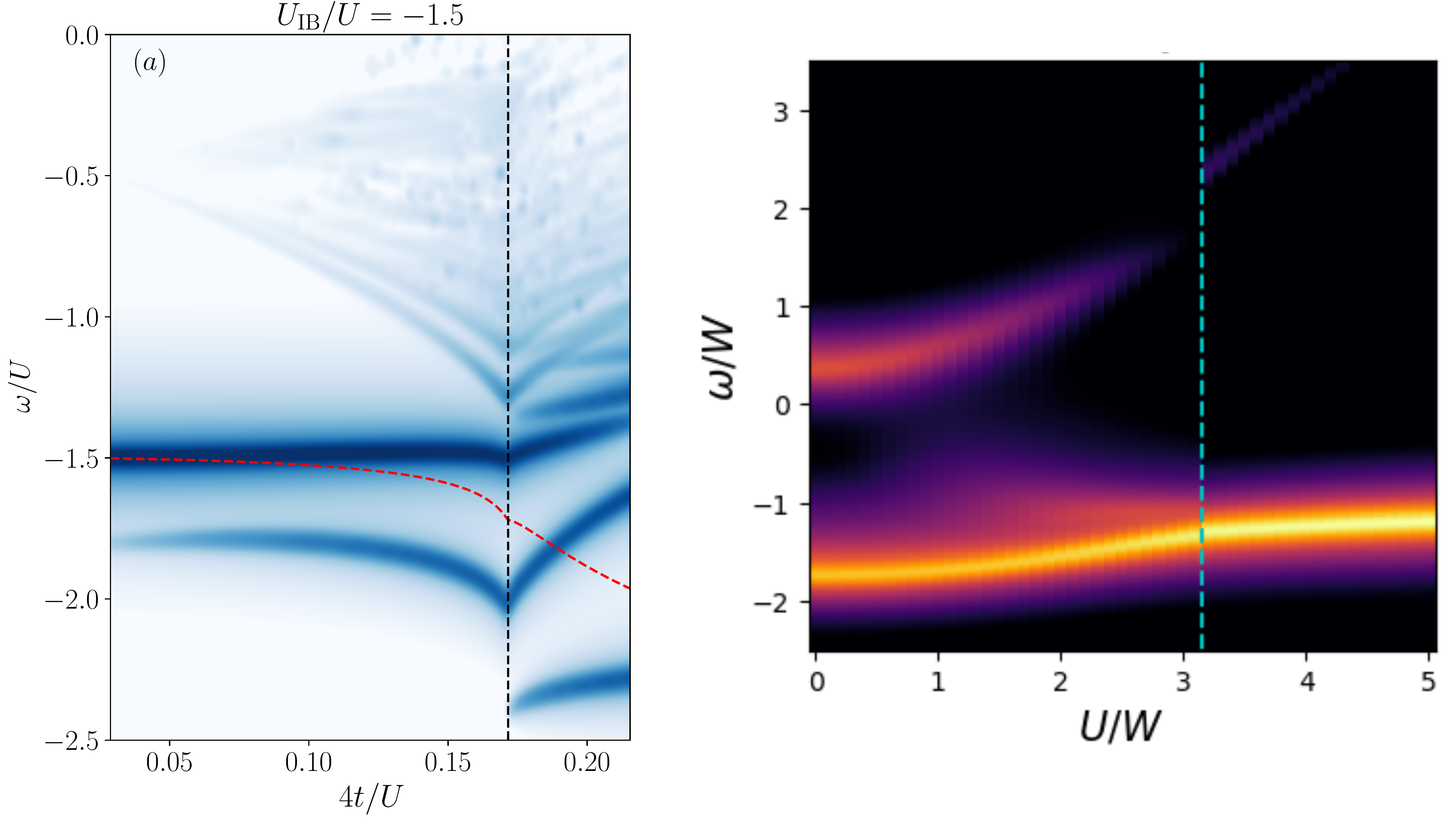}
\caption{\label{PolMottFig}
\textbf{Polaron and the Mott transition.} 
Left: The spectral function of an impurity in a bosonic bath as a function boson-boson interaction strength $U$ with $U_{IB}$ the impurity-boson interaction strength and $t$ the hopping. The bosons undergo a Mott insulator to superfluid transition at the vertical line. From \citet{alhyder2024latticebosepolaronsstrong}. Right: The spectral function of an impurity in a two-component fermionic bath with repulsive interaction $U$ and bandwidth $W$.  The fermions are in a metallic (insulating) phase to the left (right) of the vertical line. From \citet{amelio2024polaronformationinsulatorskey}.}
\end{figure}

In a breakthrough result, the superfluid to Mott insulator transition was observed in the Bose-Hubbard model~\cite{greiner_quantum_2002}, which  enabled the investigation of the dynamics of impurities and magnon bound states in 1D~\cite{fukuhara_quantum_2013,fukuhara_microscopic_2013}. The properties of a mobile impurity in a Bose gas in the quantum critical regime of the Mott insulator to superfluid transition at integer filling were explored using a quantum Gutzwiller approach combined with second-order perturbation theory~\cite{Colussi2023}. 
Extending this  to strong interactions with a generalised diagrammatic ladder approximation as well as QMC calculations, the polaron spectrum was shown to exhibit non-analytic features at the Mott transition such as a cusp and the emergence of a new branch, coming from  gapless Goldstone and Higgs modes, see Fig.~\ref{PolMottFig}. 
Effective field theory combined with the Boltzmann equation was used to calculate the spin diffusion of an impurity in a bath of bosons in a 2D lattice near the quantum critical point between the superfluid and insulating phases~\cite{Punk2013}.

A mobile impurity in the attractive 2D Fermi Hubbard model was explored using diagrammatic Monte Carlo calculations~\cite{Pascual2024OnPolarons}. It was shown that beyond majority filling fractions of about 20\% of the lattice sites, the Fermi polaron remains the ground state even in the limit of large attraction. This is in stark contrast to the continuum case, where eventually a polaron to molecule transition takes place for strong enough attraction. This has important implications for the phase diagram of the spin-doped Hubbard model, which should necessarily feature a Fermi liquid of polarons at large imbalance, instead of a mixture of bosonic molecules and excess fermions. 
Indeed, a recent experimental study of the spin-doped attractive Hubbard model in 2D with $^{40}$K atoms finds strong evidence for such a Fermi liquid of polarons~\cite{hartke2025}.

The infinite impurity mass case was explored using auxiliary free fermions mimicking properties extracted from dynamical mean-field theory~\cite{amelio2024polaronformationinsulatorskey}. The polaron energy was predicted to exhibit a cusp at the Mott insulator to metal transition, see Fig.~\ref{PolMottFig}. Polaron formation in Bloch bands and fermionic charge-density waves were also explored. Recently, it was demonstrated that the band geometry of the majority particles affects the exponents describing the Fermi edge singularity of an impurity in a flat lattice band~\cite{pimenov2024polaron}.

Considering mobile impurities in a lattice containing bosons in the hard core limit of strong repulsion away from half filling, it was shown that the resulting polarons are strongly affected by the properties of the surrounding bath~\cite{santiagogarcia2024lattice}. In the opposite limit of mobile impurities in a weakly interacting BEC, it was shown that in addition to the attractive and repulsive polaron, a third type of polaron emerges for repulsive impurity-boson interactions~\cite{Ding2023}. 
This is in analogy with the repulsively bound states  observed for bosons in an optical lattice~\cite{Winkler2006}.

Using a weak coupling approach, the transport properties of an impurity in an optical lattice interacting with a homogeneous BEC were shown to change from coherent to diffusive with increasing temperature~\cite{Bruderer2007}. Its non-equilibrium dynamics including Bloch oscillations and incoherent drift  were studied within the Fr\"ohlich model~\cite{GrusdtBloch2014}. This was later extended to the case when the impurity occupies two bands in a lattice~\cite{Yin2015}, and when also the bosons are in a lattice~\cite{Privitera2010}. 

\subsection{Polarons in spinor quantum gases}   \label{TwocomponentFermi}

We now turn to the question of what happens when the environment is composed by atoms that feature internal spin degrees of freedom. For a continuum system, this is reminiscent of the Kondo effect \cite{kondo1964}, and  it has been shown how Rydberg impurities in a BEC allow for the extension of the Kondo model to include atomic bound state formation~\cite{Ashida2019,Ashida2019b}. Experimentally, it was shown how spin-exchange interactions served to engineer impurities that worked as quantum probes of a surrounding spinor BEC~\cite{Bouton2020}. In Ref.~\cite{Ashida2018} it was found that spin-flip excitations can dominate the dressing of  impurities leading to the formation of magnetic polarons in continuum systems. The corresponding dynamics of impurities in two-component Fermi gases was shown to allow for the study of quantum spin transport at the single atom level \cite{You2019}. Ref.~\cite{Wang2019} considered impurities in a BEC with a synthetic spin-orbit-coupling between two of its hyperfine states, and discussed how the resulting polarons get dressed by roton excitations, and therefore acquire acquires a non-zero momentum and an anisotropic effective mass.

The experimental observation of a smooth cross-over between a weakly interacting BCS superfluid, a strongly correlated superfluid and a BEC of weakly interacting dimers (as the interaction between two hyperfine components of fermionic atoms is tuned across a Feshbach resonance) stands out as a major success of quantum simulation with cold atoms~\cite{Giorgini2008}. Adding impurities to such a two-component Fermi superfluid  as done experimentally with bosonic atoms~\cite{FerrierBarbut2014,Yao2016,Roy2017}, opens the possibility to explore polarons in the celebrated BCS-BEC crossover, and in particular how they change from Fermi to Bose polarons.

Theoretically, crossovers from polaron to a trimer states were predicted using variational wave functions~\cite{Nishida2015,Yi2015}. Using a generalised Chevy ansatz combined with BCS theory, \citet{Amelio2023} calculated the spectral function of an impurity across the full BCS-BEC crossover and predicted avoided crossings due to the coupling to a Higgs mode. Using the ladder approximation combined with BCS theory, \citet{Hu2022} predicted that the superfluid gap would stabilize the attractive polaron and would introduce additional damping for the repulsive polaron. Second order perturbation theory was used to identify UV divergencies related to $3$-body physics in analogy with the case of Bose polarons, see Sec.~\ref{sec:weakinteractions}, which can be regularized using effective field theory~\cite{Pierce2019}. Finite values for the polaron energy were obtained in the whole BEC-BCS crossover when the density-density correlation function of the Fermi gas (which enters the second order impurity self-energy) was calculated using the RPA approximation~\cite{Bigue2022,alhyder2023exploring}. The problem of a static impurity in a fermionic superfluid was analysed using a functional determinant approach together with BCS theory, and the gap was predicted to protect the polarons against Anderson's orthogonality catastrophe with in-gap Yu-Shiba-Rusinov bound states present for magnetic impurities~\cite{Hu2022b,Wang2022}. 
Experimentally,  the impurity-fermion interaction was however too weak to see polaron effects beyond the mean-field level \cite{FerrierBarbut2014,Yao2016,Roy2017}, and an observation of these predictions remains an interesting topic for future investigations. 

In 2D, a superfluid undergoes a Kosterlitz-Thouless (KT) phase transition to a normal phase at a critical temperature with a discontinuous jump in the superfluid density~\cite{Kosterlitz_1973}. For an impurity in a  superfluid Fermi gas, \citet{alhyder2022} used perturbation theory to predict a rapid increase in the polaron energy at  the transition temperature, reflecting that the normal phase is less compressible than the superfluid one. Using stochastic classical-field methods, a low energy polaron branch was predicted to emerge at the KT transition connected to the binding of the impurity to vortex cores~\cite{comaron2024quantumimpuritiesfinitetemperaturebose}. The same problem was studied at zero temperature using exact diagonalisation for  up to 10 particles on a square lattice at zero temperature, predicting attractive and repulsive polarons  with avoided crossings~\cite{Amelio2024}. These results may become relevant for experiments exploring mesoscopic atomic gases, where few-body precursors of polaron physics, pairing and Higgs modes have already been observed~\cite{wenz2013,Bayha2020}.

\subsection{Polarons in baths with non-trivial topology}

The realization that phases with non-trivial topological properties are frequent in nature has sparked an intense research effort~\cite{Hasan2010, Qi2011, Cooper2019}. While non-interacting topological phases are well understood by now, many questions remain regarding the interplay between interactions and topological states. Polarons in a bath in a topological phase provide an interesting ``bottom up" approach for exploring this scenario, and can be used as new probes for topological order. 

\citet{Grusdt:2016aa,Munoz2020,Grass2020,Baldelli2021} examined theoretically the regime of strong impurity-environment interactions where the impurity binds to the quasiparticle excitations in a surrounding fractional quantum Hall phase. They showed that the resulting molecules can acquire the properties of the quasiparticles in the topological phase such as fractional quantum statistics. This may provide new ways to address the challenging problem of probing non-local topological order by e.g.\ Ramsey spectroscopy or scattering experiments with the impurities.

In the opposite limit of weak impurity-environment interactions, it was shown using diagrammatic perturbation theory that the polaron inherits some of the non-trivial topological properties of the majority particles in its dressing cloud, leading to a discontinuous jump in the polaron Hall conductivity at the topological phase transition boundary~\cite{CamachoGuardian2019,Pimenov2021}. This jump is however not quantized according to the Thouless-Kohmoto-Nightingale-den Nijs relation~\cite{Thouless1982}, since the polaron partly consists of a topologically trivial impurity and partly of a topological dressing cloud. \citet{Qin2019} studied a mobile impurity in a 2D fermionic superfluid and proved that a discontinuity in the second derivative of its energy should appear when its $p_x+ip_y$ pairing undergoes a phase transition from a trivial to a topological symmetry. The polaron properties were shown to closely reflect also the phases of an environment described by a non-Abelian Aubry-Andr\'e-Harper model, which exhibits an interplay between localisation and topological order~\cite{Bai2018}. The dressing of a mobile impurity interacting with the chiral edge modes circulating around both non-interacting Chern insulators and strongly correlated fractional Chern insulators was explored using exact diagonalisation, the Chevy ansatz, as well as tensor network techniques~\cite{vashisht2024chiralpolaronformationedge}. The resulting chiral polarons were found to exhibit two characteristic features: An asymmetric spectrum and a splitting into two damped states for momenta larger than a critical momentum determined by the velocity of the chiral edge modes times the impurity mass. 

\section{Magnetic polarons}   \label{Sec:MagneticPol}

\begin{figure}
\centering
\includegraphics[width=\columnwidth]{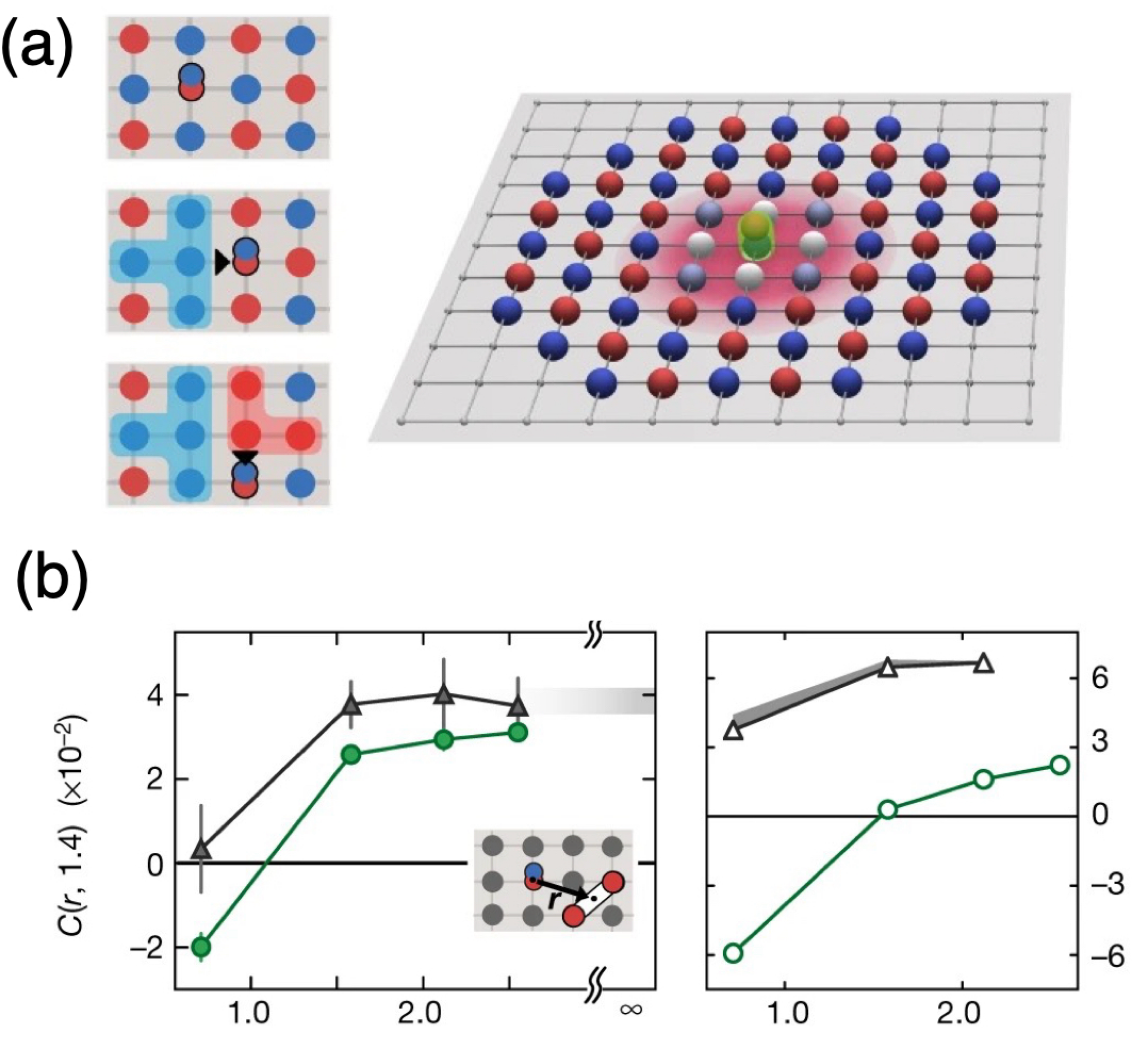}
\caption{\label{KoepsellFig}
{\bf Magnetic polarons.} 
(a) A doublon consisting of a spin $\uparrow$ and a spin $\downarrow$ fermions (blue and red balls) moves through an AFM lattice leaving behind itself a ``string" of ferromagnetic correlations (blue and red shading). This leads to the formation of a magnetic polaron, i.e., a doublon surrounded by a cloud of magnetic frustration. (b) Correlations between diagonal spins as a function of the distance $r$ to a mobile/static doublon (green/black). The left panel shows experimental results and the right panel theoretical calculations from a string model (mobile doublon) and exact diagonalisation (static doublon). From \citet{Koepsell2019}.}
\end{figure}

In this section, we discuss a different incarnation of the polaron, which is closely related to those discussed in the rest of this review: the so-called magnetic polarons, also known as spin polarons. Magnetic polarons arise when dopants such holes or extra particles move in a lattice with spin $1/2$ fermions close to half-filling (one fermion per lattice site), which due to a strong repulsive interaction form an antiferromagnet at zero temperature. The motion of the dopants destroy this antiferromagnetic (AFM) order leading to a competition between kinetic and magnetic energy, and to the formation of a polaron consisting of the dopant surrounded by a dressing cloud of magnetic frustration, see Fig.~\ref{KoepsellFig}(a)~\cite{Brinkman1970,SchmittRink1988,Shraiman1988,Kane1989,Martinez1991,Sachdev1989}.

Magnetic polarons have been intensely studied in the condensed matter community because  unconventional superconducting phases such as those in cuprates~\cite{highTc}, pnictides~\cite{Wen2011}, layered organic metals~\cite{Wosnitza2012}, and twisted bilayer graphene~\cite{Cao2018} emerge when doping an AFM away from half-filling. This suggests that the properties of magnetic polarons may cast light on these intriguing phases, and we refer the reader to excellent condensed matter oriented reviews on this vast topic~\cite{Manousakis1991, Dagotto1994}. Here, we will give a brief overview of the main features of magnetic polarons highlighting the close connections to the Bose polarons discussed in Sec.~\ref{sec:BosePolarons}. We will also discuss the new and remarkably detailed spatial information regarding the spatial properties of magnetic polarons obtained with quantum gas microscopes~\cite{Koepsell2019,Ji2021,Koepsell2021}. 

Consider spin $1/2$ fermions in a square lattice close to half-filling. For strong repulsion between the spin $\uparrow$ and $\downarrow$ fermions, they form an AFM ground state, which can be described by the $t-J$ model~\cite{Dagotto1994} \begin{equation}
\hat H=-t\sum_{\langle i,j\rangle\sigma}(\tilde f_{i\sigma}^\dagger\tilde f_{j\sigma}+\text{h.c.})
+J\sum_{\langle i,j\rangle}(\hat {\mathbf S}_i\cdot\hat {\mathbf S}_j-\hat n_i\hat n_j/4).
\label{tJmodel}
\end{equation}
Here $\tilde f_{i\sigma}^\dagger=\hat f_{i\sigma}^\dagger(1-\hat n_{i\bar \sigma})$, and the factor $1-\hat n_{i\bar \sigma}=1-\hat f_{i\bar \sigma}^\dagger \hat f_{i\bar \sigma}$ (with $\bar \sigma$ denoting the opposite spin of $\sigma$) ensures that there is maximally one fermion per lattice site. We furthermore have $\hat n_i=\sum_\sigma\hat n_{i\sigma}$ and $\langle i,j\rangle$ denotes nearest neighbors. Also, $ \hat {\mathbf S}_i=\frac 12\sum_{\sigma\sigma'}\hat f_{i\sigma}^\dagger\boldsymbol{\sigma}_{\sigma\sigma'}\hat f_{i\sigma'}$ with $\boldsymbol{\sigma}=(\sigma_x,\sigma_y,\sigma_z)$ a vector of Pauli matrices. When the $t-J$ model is derived from the Hubbard model, $J=4t^2/U$ is the superexchange interaction where $U\gg t$ is the onsite repulsive interaction between opposite spin fermions.

Using a Holstein-Primakoff transformation generalised to include the presence of holes, Eq.~\eqref{tJmodel} becomes
\begin{equation} \label{tJmodel2}
\hat H=\sum_{\bf k}\omega_{\bf k}\hat \gamma_{\bf k}^\dagger \hat \gamma_{\bf k}+
\sum_{\bf q,k}g({\bf q,k})[\hat h_{\bf k+q}^\dagger\hat h_{\bf q}\hat \gamma_{-\bf q}^\dagger+\text{h.c.}].
\end{equation}
Here, $\bf k,\bf q$ are crystal momenta inside the first Brillouin zone of the lattice, $\hat \gamma_{\bf k}^\dagger$ is a bosonic operator creating a spin wave with energy $\omega_{\bf k}=2J\sqrt{1-(\cos k_x+\cos k_y)^2/4}$ (unit lattice constant), and $\hat h_{\bf k}^\dagger$ is a fermionic operator creating a holon. We refer to Refs.~\cite{SchmittRink1988,Kane1989,nielsen2021spatial} for an expression of the vertex $g({\bf q,k})$. One has used linear spin wave theory to derive Eq.~\eqref{tJmodel2}, which is known to be very accurate for a square lattice~\cite{Manousakis1991}. Note  that while this model naturally describes the case of hole doping, it can also describe particle doping (doublons) using a particle-hole transformation~\cite{Jiang2021}. 

Equation~\eqref{tJmodel2} is a so-called slave-fermion representation of the $t-J$ model and describes a fermionic holon emitting or absorbing bosonic spin waves as it moves through the lattice. Comparing with Eq.~\eqref{BeyondFrohlich} explicitly demonstrates the close connection between magnetic and Bose polarons. Indeed, the two Hamiltonians have the same structure, apart from two differences: First, Eq.~\eqref{tJmodel2} has no scattering term, i.e. the last term in Eq.~\eqref{BeyondFrohlich} and therefore corresponds to the Fr\"ohlich Hamiltonian; second Eq.~\eqref{tJmodel2} has no bare kinetic energy of the impurity (holon). The lack of this term reflects that the hole cannot move in a square lattice without destroying magnetic order as illustrated in Fig.~\ref{KoepsellFig}(a). For other geometries such as the triangular lattice, a bare kinetic term is present~\cite{Trumper2004,Kraats2022}. We note that the strongly interacting regime $t\gg J$ physically corresponds to the hole moving rapidly and destroying magnetic order, which in terms of the underlying Hubbard model is equivalent to $U\gg t$. 

\begin{figure}
\centering
\includegraphics[width=0.8\columnwidth]{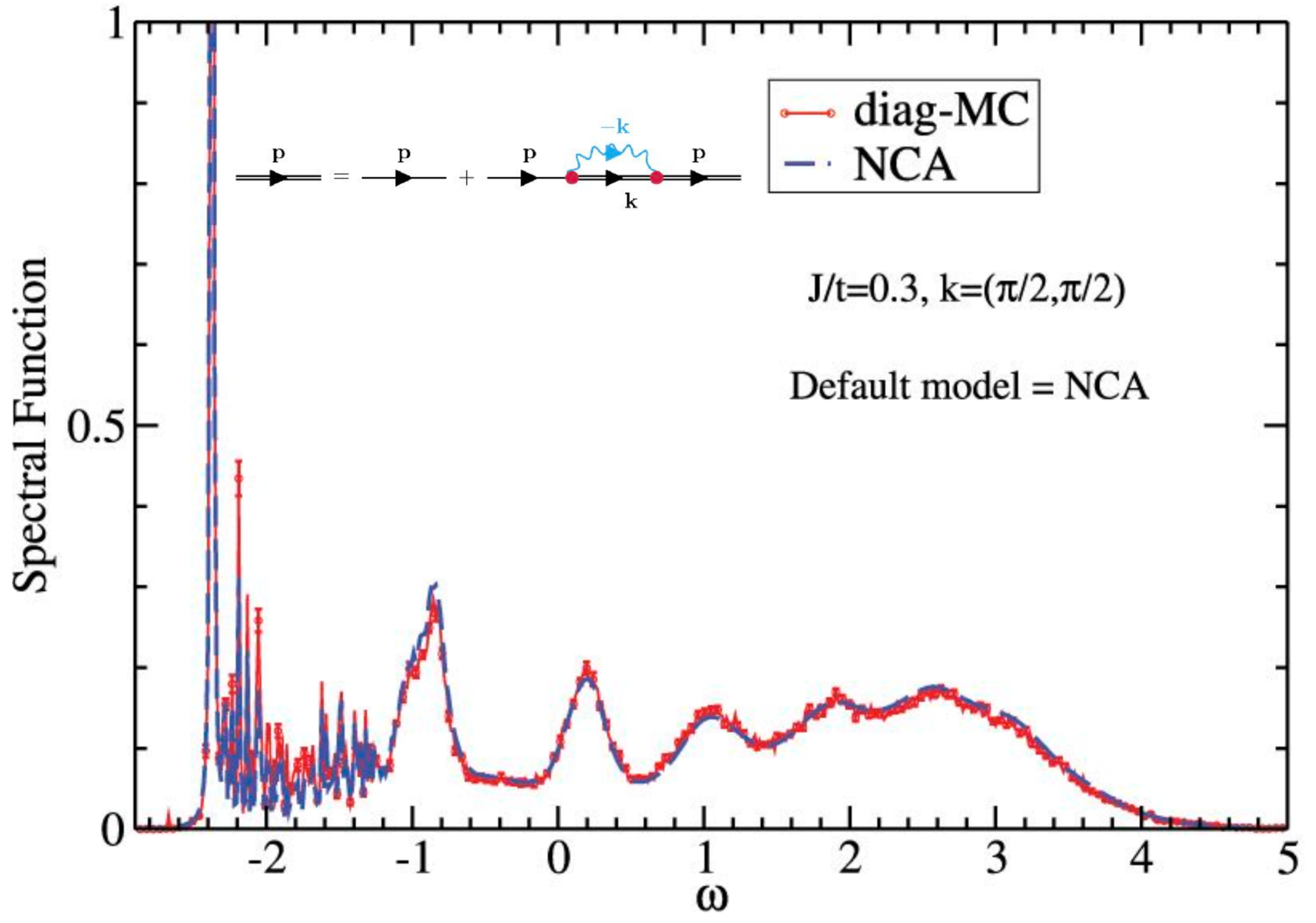}
\caption{\label{SpecFigMagPol}
{\bf Spectral function of a hole in an AFM} computed at momentum $\bk=(\pi/2,\pi/2)$, with $J/t=0.3$ and the frequency in units of $t$. The inset shows the diagrammatic structure of the SCBA Green's function. From \citet{Diamantis_2021}.}
\end{figure}

The spectral function of a single hole obtained from Eq.~\eqref{tJmodel2} is presented in Fig.~\ref{SpecFigMagPol}, comparing results from the diagrammatic Monte Carlo method and the self-consistent diagrammatic Born approximation (SCBA) [also known as non-crossing approximation (NCA)]. One clearly sees a well-defined quasiparticle peak at low energy, which is the magnetic polaron. The minimum energy of this polaron turns out to be at the crystal momenta $(\pm \pi/2,\pm \pi/2)$ for a square lattice, which is consistent with ARPES experiments in copper oxides~\cite{Kim1998,Damascelli2003}. There is a remarkable agreement between the diagrammatic Monte Carlo calculation and the SCBA, the approximation corresponding to summing the class of non-crossing diagrams shown as an inset in Fig.~\ref{SpecFigMagPol}~\cite{SchmittRink1988,Kane1989,Martinez1991,Liu1992,Chernyshev1999}. A similar accuracy of the SCBA was found in other QMC calculations~\cite{Brunner2000, Mishchenko2001}.

One can intuitively understand the basic physics of magnetic polarons using a so-called geometric string picture, where the hole moves at the fast time-scale $t$ leaving a string of magnetic frustration in its wake, see Fig.~\ref{KoepsellFig}(a), which creates a linear potential that is only repaired on the much slower time-scale $J$. Hence, the hole is effectively bound by a linear string potential whose ground state is the magnetic polaron and excited states are string excitations, which can be seen as broader peaks at higher energies in Fig.~\ref{SpecFigMagPol}. This string picture predicts a characteristic energy scaling $(J/t)^{2/3}$ of the polaron and its excited states and is widely used in the literature~\cite{Bulaevski1968,Barnes1989,Dagotto1990,Shraiman1988,Liu1992,Mishchenko2001,BERAN1996707}. It has recently been revisited and extended in the cold atom context~\cite{Grusdt2018b,Grusdt2019,Bohrdt2021,Bermes2024}. 

In direct analogy with the Bose polaron, the wave function of the magnetic polaron can be written as an expansion in the number of spin waves that the dopant creates in the AFM background. Formally, one can simply replace the impurity operator $\hat c_{\bk}$ in Eq.~\eqref{BogExpansion} with the holon operator $\hat h_{\bk}$ and the ground state $|\text{BEC}\rangle$ with $\ket{\text{AFM}}$. Closed expressions for the first three expansion coefficients have been derived using the SCBA~\cite{Reiter1994,Ramsak1998}, and later extended to infinite order to capture strong interactions $J/t\ll 1$~\cite{nielsen2021spatial,Nyhegn2023}. 

In a groundbreaking experiment, the spin and density of $^6$Li atoms in two internal spin states in a 2D square lattice close to half filling was measured with single site resolution~\cite{Koepsell2019}. Due to a strong repulsive interaction between the two spin components, the atoms exhibit strong AFM correlations at low temperature. A few sites were occupied with both spin components thereby creating doublons, and a significant reduction and even sign reversal of the AFM correlations was observed in their neighborhood, see Fig.~\ref{KoepsellFig}(b), in qualitative agreement with a string model for the magnetic polaron~\cite{Grusdt2018b}. The mobility of the doublon was shown to be key for this, as no sign reversal was measured when it was pinned. This experiment therefore provided a direct observation of the microscopic spatial structure of a magnetic polaron formed by the doublon and its surrounding dressing cloud of magnetic frustration. The transition between a gas of magnetic polarons and a Fermi liquid was studied in a subsequent experiment~\cite{Koepsell2021} and theoretically analysed using variational functions containing polaronic correlations~\cite{muller2024polaroniccorrelationsoptimizedancilla,shackleton2024emergentpolaroniccorrelationsdoped}.

\begin{figure}
\centering
\includegraphics[width=\columnwidth]{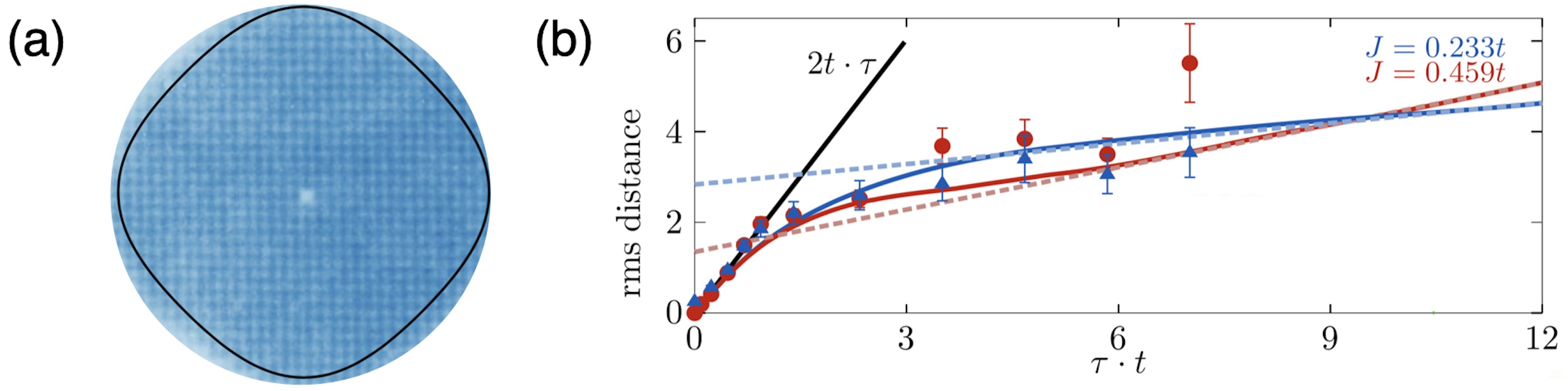}
\caption{\label{ExpansionFigMag}
{\bf Hole dynamics in an AFM.} 
Left: in-situ image of a hole created in an AFM formed by two repulsively-interacting spin states of $^6$Li atoms at half-filling. From \citet{Ji2021}. Right: root-mean-square distance of a hole from its initial position as a function of time $\tau$ measured by \citet{Ji2021} compared to a quantum walk of a free particle (black line), and to a non-equilibrium SCBA calculation (red and blue lines). From \citet{Nielsen2022}.}
\end{figure}

The non-equilibrium dynamics of a hole released from a given site in a square optical lattice was measured using quantum gas microscopy~\cite{Ji2021}. The hole was moving in a background of antiferromagnetically correlated spins formed by two repulsively interacting spin states of $^6$Li at half-filling, see Fig.~\ref{ExpansionFigMag}. For short times, the hole moved ballistically with a velocity $2t$ in agreement with DMRG simulations on a $18 \times 4$ lattice strip~\cite{Bohrdt_2020} and short-time analytics~\cite{Nielsen2022}. For longer times, the hole slowed down as it became increasingly dressed by magnetic frustration, as shown in Fig.~\ref{ExpansionFigMag}, which can be phenomenologically explained by mapping the dynamics onto a free quantum walk in a Bethe lattice~\cite{Ji2021}. Using a time-dependent wave function for the hole derived from the SCBA, these experimental observations were quantitatively explained at all time scales by \citet{Nielsen2022}, see Fig.~\ref{ExpansionFigMag}. This extends the use and accuracy of the SCBA also to non-equilibrium dynamics, and demonstrate that the slowdown of the hole at long times quantitatively agrees with a theory for polaron formation. The theory furthermore showed that oscillations at intermediate times can be interpreted as quantum beating between string states. 

\begin{figure}
\centering
\includegraphics[width=\columnwidth]{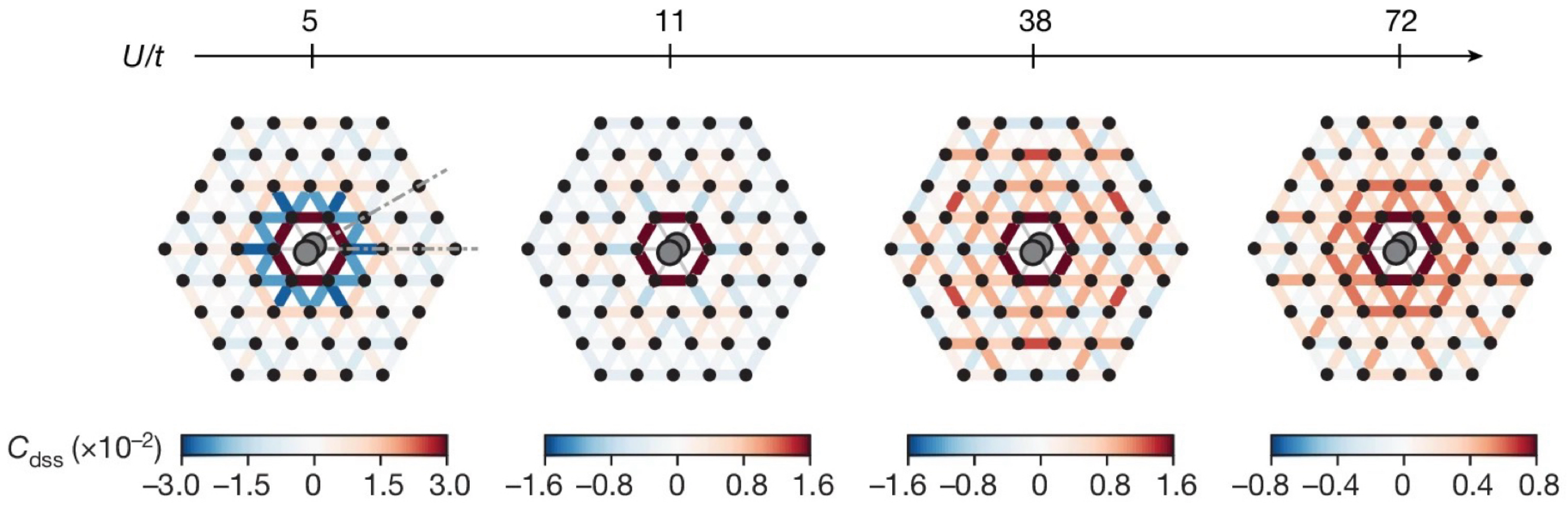}
\caption{\label{NagaokaFig}
{\bf Nagaoka polarons.} 
Ferromagnetic correlations (red) replace AFM correlations (blue) around a doublon with increasing interaction strength $U/t$ corresponding to increasing $t/J$. From \citet{lebrat2023observation}.}
\end{figure}

In a seminal paper, \citet{Nagaoka1966} showed that the motion of a single dopant induces a ferromagnetic ground state in a wide range of different lattices when $t/U\rightarrow 0$ in the Hubbard model ($J/t\rightarrow 0$). This effect emerges in the extreme limit $t/U<N$ with $N$ the number of lattice sites, as a result of the dopant minimizing its kinetic energy, which is only possible in the fully ferromagnetic state. For a very large but finite $U/t$, it has similarly been proposed that a ferromagnetic bubble forms around the dopant. This so-called Nagaoka polaron has remained elusive in condensed matter systems, but it was recently observed in two experiments exploring doublons with a two-component $^6$Li gas in a triangular optical lattice. Such a lattice was chosen to enhance the formation of the Nagaoka polaron by suppressing AFM order by geometric frustration~\cite{lebrat2023observation,Prichard2024}. Figure~\ref{NagaokaFig} shows the experimentally observed increasing size of the ferromagnetic bubble around a doublon with increasing interaction strength, which is consistent with a $(t/J)^{1/4}$ scaling obtained using variational arguments~\cite{White2001}. 

These results demonstrate that optical lattice experiments can shed light on the spatial properties of magnetic polarons, which complements the spectral information typically obtained in  condensed matter experiments. They however provide somewhat indirect evidence of the magnetic polaron and it would be very useful in future experimental work to measure the spectral function of dopants in an optical lattice to confirm the presence of a quasiparticle peak~\cite{Bohrdt2020, nielsen2021spatial}. This would provide a confirmation of the existence of magnetic polarons without complications from disorder, doping dependent screening, and sample sensitivity typical of condensed matter experiments. Also, the topic of mobile dopants in spin backgrounds is very rich and there are many interesting open questions not discussed here. This includes the effects of non-zero temperature~\cite{Shen2024}, polarons in layered systems~\cite{Nyhegn2022,Nyhegn2023}, spin liquids~\cite{Kadow2022, Kadow2024,nyhegn2024probingquantumspinliquid,Jin2024b}, bound states of polarons and pairing~\cite{Hartke2023,Grusdt2023b,Bohrdt2023b}, and mixed dimensional systems~\cite{Hirthe2023, nielsen2024pairingdisorderdopantsmagnetic}, which may improve our understanding of pairing in unconventional superconductors~\cite{Lange2024}.

\section{Perspectives}   \label{sec:Perspectives}

This review provided a comprehensive description of polarons as realized in cold atomic gases and 2D semiconductors. We highlighted the many common properties characterizing polarons in these two seemingly very different systems, showcasing the power and universal applicability of this concept. With this review we hope to bridge the gap between different communities and foster collaborations in this rapidly evolving topic. Indeed, while many properties of polarons are by now well understood, there are many exciting research directions open for future studies, as we will now briefly outline. 

As is clear from Sec.~\ref{sec:BosePolarons}, there are several  questions and different theoretical predictions regarding Bose polarons for strong interactions. These include the number of relevant parameters and the influence of $n>2$-body correlations and few-body states, which may evolve from low energy cluster states that are hard to observe, and the role of the bosonic OC. Indeed, a clear cut experimental confirmation of the existence of well-defined polarons in the unitarity region is still lacking. Also, the role of the multichannel nature of the impurity-boson interaction is not clear as is the  temperature dependence of the Bose polaron and its fate in the critical region of the BEC. Regarding magnetic polarons, the experimental evidences of their existence in optical lattices are rather indirect and related to real space observables, see Sec.~\ref{Sec:MagneticPol}, and a smoking gun observation of a  quasiparticle peak in frequency space is highly desirable. 
    
The non-equilibrium properties of both Bose and Fermi polarons is another interesting topic. This is of particular relevance for the lossy atomic Bose polaron where the competition between scales such as its formation time, decay time, and experimental times complicates its observation. Polarons in TMDs are moreover intrinsically of non-equilibrium nature due to the electron-hole recombination of excitons and photon leakage discussed in Sec.~\ref{TMDsection}, and a systematic theory describing this, for instance based on the Keldysh formalism,  would be very useful. The fate of atomic polarons under strong RF driving, see Sec.~\ref{NonequilFermipolaron}, or polarons in strongly-pumped TMDs is another intriguing open issue. 

There are many questions regarding the experimental exploration of mediated interactions between polarons as discussed in Sec.~\ref{QPinteractions}. For instance, beyond-mean-field medium-induced interactions between Bose polarons have not yet been observed, even though they are expected to be stronger than those between Fermi polarons (due to the higher compressibility of the Bose gas). The puzzling  observations of repulsive mediated interactions between polarons in TMDs and the influence of non-equilibrium effects (see Sec.~\ref{QPinteractions}) also call for further analysis. An unambiguous experimental observation of the predicted bound states between two polarons would be a major breakthrough, as bipolarons are precursors for Cooper pairs and superfluidity, which in the case of magnetic polarons may be closely connected to high $T_c$ superconductivity, as discussed in Sec.~\ref{Sec:MagneticPol}. This is furthermore important for the exploration of quantum mixtures and their polaronic limit (Sec.~\ref{sec:LimitManyBody}), where such mediated  interactions play a key role in their many-body phase-diagram. Regarding polaron-polaritons in TMDs discussed in Sec.~\ref{Sec:MicroPolaritons}, such interactions may give rise to entirely new hybrid light-matter quantum phases and strong photon-photon interactions with applications in opto-electronics. 

A systematic investigation of the role of the composite electron-hole nature of the exciton for polarons in 2D semiconductors, as well as effects of the Coulomb interactions in the electron bath is highly relevant. This would improve our microscopic understanding of polarons  in TMDs and likely  lead to a better agreement between theory and experiment, which as explained in Sec.~\ref{TMDsection} is generally less satisfactory than in atomic gases. 

Finally, the use of polarons as sensors for many-body correlations discussed in Secs.~\ref{Excitonsasprobes} and \ref{InteractingBaths}  is still in its infancy with many exciting perspectives for both fundamental science and technology. In particular, there is an urgent need for sensors to probe the properties of the rapidly growing class of TMDs with many possible applications.   One can for instance imagine using several pinned polarons in a moir\'e lattice to create a spatially resolved sensing of multi-point correlation functions. Also, using entangled polarons may lead to 
much higher sensitivity and entirely new  capabilities. 
This is likely to be a major research topic as many important quantum phases are characterized by subtle many-body correlations, which are hard to detect by conventional means, in particular in TMDs.

\begin{acknowledgments}
We wish to warmly thank I.~Amelio, C.~Baroni, M.~Caldara, A.~Camacho-Guardian, O.~Cotlet, X.~Cui, E.~Demler, R.~Grimm, F.~Grusdt, M.~Knap, M.~Kroner, C.~Kuhlenkamp, J.~Levinsen, A.M.~Morgen, A.~Negretti, M.~Parish, F.~Scazza, T.~Shi, L.B.~Tan, and M.~Zaccanti for insightful discussions.

\vspace{.4cm}
P.M.~acknowledges support of the {\it ICREA Academia} program, the Institut Henri Poincaré (UAR 839 CNRS-Sorbonne Université) and the LabEx CARMIN (ANR-10-LABX-59-01).
P.M.~and G.E.A.~further acknowledge support by the Spanish Ministerio de Ciencia, Innovación y Universidades (grant PID2023-147469NB-C21, financed by MICIU/AEI/10.13039/501100011033 and FEDER-EU).
R.S.~was supported by the Deutsche Forschungsgemeinschaft under Germany's Excellence Strategy EXC 2181/1 - 390900948 (the Heidelberg STRUCTURES Excellence Cluster) and Project-ID 273811115 - SFB 1225 ISOQUANT.
A.I.~acknowledges support by the Swiss National Science Foundation (SNSF) under Grant Number 200020\_207520.
M.Z.~acknowledges support by the NSF Center for Ultracold Atoms and NSF PHY-2012110, AFOSR (FA9550-23-1-0402), ARO (W911NF-23-1-0382), and the Vannevar Bush Faculty Fellowship (ONR N00014-19-1-2631).
J.A.~acknowledges support by the Novo Nordisk Foundation NERD grant (Grant no.~NNF22OC0075986).
G.B.~and J.A.~were supported by the Danish National Research Foundation through the Center of Excellence “CCQ” (DNRF152). 
\end{acknowledgments}

\bibliography{RMP_on_polarons}

\end{document}